\documentclass[11pt,a4paper,twoside]{book}
\usepackage[a4paper,top=2.5cm, bottom=2.5cm, left=3cm, right=3cm]{geometry}
\usepackage[svgnames]{xcolor}
\usepackage{graphicx}

\let\oldFootnote\footnote
\newcommand\nextToken\relax

\renewcommand\footnote[1]{%
    \oldFootnote{#1}\futurelet\nextToken\isFootnote}

\newcommand\isFootnote{%
    \ifx\footnote\nextToken\textsuperscript{,}\fi}

\usepackage{fancyhdr}
\usepackage[Sonny]{fncychap} 
\usepackage{hyperref}
\hypersetup{
    colorlinks,
    citecolor=blue,
    filecolor=black,
    linkcolor=blue,
    urlcolor=blue
}

\usepackage{pdfpages}
\usepackage[square,numbers,sort&compress]{natbib}
\usepackage{enumitem}

\usepackage{amsmath,amssymb,amsfonts}
\usepackage{physics}
\usepackage{dsfont}

\usepackage{afterpage}

\newenvironment{dedication}
  {
   \thispagestyle{empty}
   \vspace*{\stretch{2}}
   \itshape             
   \raggedleft          
  }
  {\par 
   \vspace{\stretch{3}} 
   \clearpage           
  }

\usepackage[toc,page]{appendix}

\begin{document}

\frontmatter 

\begin{titlepage}

\begin{minipage}[c]{0.45\textwidth}
\begin{flushleft}
\includegraphics[width=0.8\textwidth]{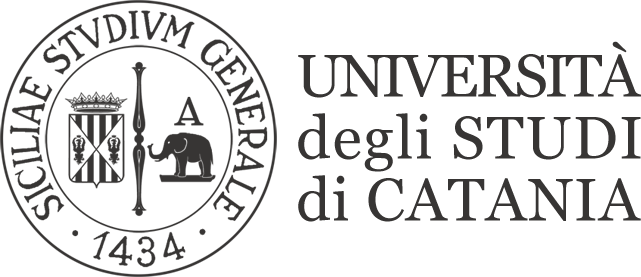}
\end{flushleft}
\end{minipage}
\hfill           
\begin{minipage}[c]{0.45\textwidth}
\begin{flushright}
\includegraphics[width=\textwidth]{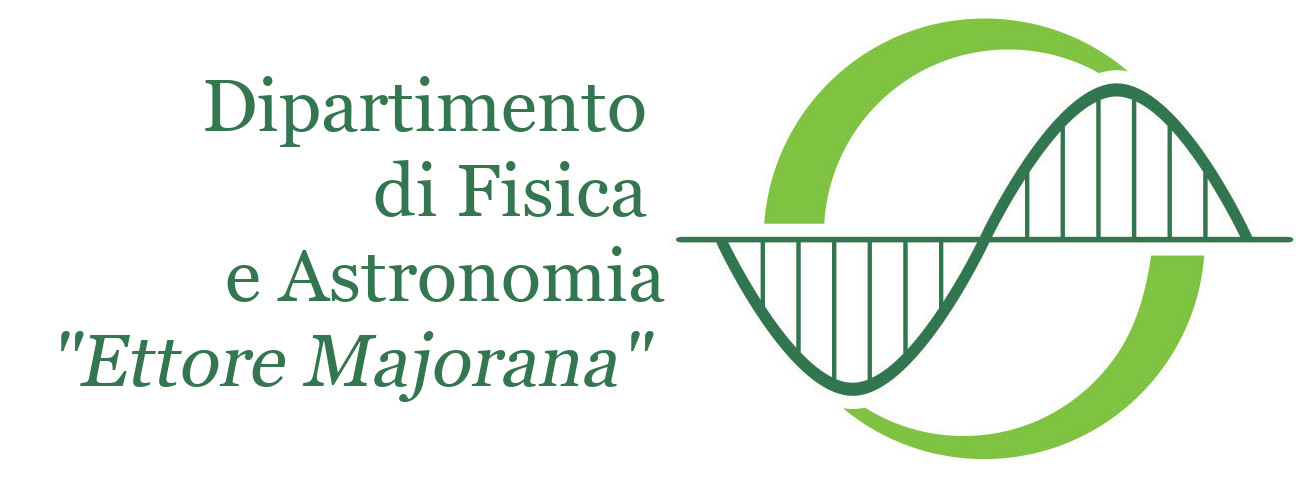}
\end{flushright}
\end{minipage} \\
\medskip
\begin{center}
{\sc PhD Programme in Physics}
\hbox to \textwidth{\hrulefill}
\vspace{1.5truecm}
\end{center}

\begin{center}
\linespread{1.5}
{ \LARGE \bfseries Persistent Currents in Atomtronic Circuits \\
of SU(\textit{N}) Fermions\par} 

\vspace{1.5truecm}

\LARGE Wayne Jordan  \textsc{Chetcuti} 
\vspace{1.5truecm}

{\Large \bfseries January 2023} 
\vspace{2truecm}

\large

\begin{flushright}
\begin{minipage}{0.4\textwidth}
\begin{tabbing}
\emph{{Supervisor:}} Prof. Luigi  \textsc{Amico}\\
\emph{{Co-Supervisor:}} Dr Juan \textsc{Polo}
\end{tabbing}
\end{minipage}
\end{flushright}

\vfill
\hbox to \textwidth{\hrulefill}

\small A thesis submitted in fulfillment of the requirements for the degree of Doctor of Physics at the University of Catania
\end{center}

\vfill
\end{titlepage}

\afterpage{
\newpage~\newpage
}
\clearpage
\rhead{}
\lhead{}
\thispagestyle{plain}
  \begin{dedication}
    In loving memory, \\
    of my little sister Milena Tori 
  \end{dedication}

\chapter{Acknowledgements}
\noindent First and foremost, I would like to express my deepest gratitude to my supervisor Luigi. I admire your attitude, contagious passion for scientific research, and talent in conveying this to others. Thank you for giving me this fantastic opportunity, guiding me, encouraging me, and for your patience. It has been an immense pleasure working with you for the past three years, and I hope we can continue to do so. \\

\noindent On that same note, I extend my heartfelt thanks to my co-supervisor, Juan. From our first interaction, you have always been helpful and eager to share your physics knowledge and intuition. Thank you for taking the time and effort to teach me and entertaining my ideas and suggestions. The time spent with you trying to put out the weekly fires has been one of the highlights of my PhD. \\

\noindent A special thanks go out to the rest of my collaborators, whom I have had the pleasure to interact with throughout my scientific journey: Andreas Osterloh, Tobias Haug, Leong-Chuan Kwek, Tony Apollaro, Mirko Consiglio, Sergi Ramos-Calderer, Carlos Bravo-Prieto, Anna Minguzzi, José Ignacio Latorre and Paolo Castorina. \\

\noindent At the start of my second year, I moved to the Quantum Research Centre at TII in Abu Dhabi. I have encountered many interesting people who created a fun working environment during my time here. Specifically, the Quantum Physics group has made life at the centre quite enjoyable. A special mention goes out to Rico and Vercio for all the fun times yet. \\

\noindent I am very grateful to Dr Libertini of the University of Catania and the staff at QRC for assisting me with the logistics and bureaucratic stuff. \\

\noindent I would also like to thank my friends. To Giampiero, Sergi, Jaideep, Rux and Vicky for helping me acclimate to life here in Abu Dhabi and our fantastic irreplaceable times. I will always remember these moments with a smile on my face. To Rens, Joseph, Steve, and Christian, for the long Zoom calls that helped me keep my sanity during lockdowns. I am grateful to Christian, in particular, for his advice and for sharing his experiences with me. Furthermore, I extend my gratitude to the PHY3569 group for the fun times and all the laughs. Finally, to Andrea and Anna, thank you for always being there and being great and supportive friends. \\

\noindent Most importantly, I am forever grateful to my mother, Mary Rose, for raising me, for her unwavering support, and for sacrifices that she made that I can never equally repay.  \\

\noindent Finally, I would like to thank my dogs, Mika, Quark, Lucky, Brownie, and Bart, for fruitful discussions and comments on this thesis.

\tableofcontents

\chapter{Summary}

\noindent Ultracold atomic systems have emerged as strong contenders amongst the various quantum systems relevant for developing and implementing quantum technologies due to their enhanced control and flexibility of the operating conditions. Recent developments in micro-optics technology paved the way for engineering atomic circuits in various architectures. Ring-shaped geometries, which are a simple case of these circuits, are of particular interest. In such circuits, a guided matter-wave, specifically a persistent current, can be generated by the application of an effective magnetic field. One of the peculiar knobs that can be exploited in cold atoms is the statistics of the quantum fluid flowing in the ring, be they bosons, fermions, or a mixture thereof. Naturally, the persistent current can exhibit specific dependencies and attributes depending on the nature of the quantum matter constituting it. Indeed, such quantum fluids enjoy specific physical properties and quantization rules, which are expected to be harnessed in atomtronic circuital elements with unique features. \\

\noindent In this thesis, we explore persistent currents generated in a ring-shaped quantum gas of strongly interacting $N$-component fermions, the so-called SU($N$) fermions. These  multicomponent fermionic systems, as provided by alkaline earth-like atoms, extend beyond the physics of the typical two-component fermions found in condensed matter systems. We find that the persistent current of $N$-component fermions exhibits a fractional quantization of the angular momentum, with important differences arising on whether the atoms are subject to repulsive or attractive interactions. For repulsive interactions, the \textit{fractional quantization} is manifested by a current whose period is reduced by $1/N_{p}$, with $N_{p}$ being the number of particles in the system. Similarly, the attractive regime also sees a current with a reduced periodicity, albeit with a dependency on the number of components $1/N$. By monitoring these specific properties of the quantization, the persistent current can be used as a diagnostic tool to probe interacting quantum many-particle phenomena.  The fractional quantization of the persistent current can be read out through interference dynamics obtained via homodyne and self-heterodyne protocols. \\

\noindent The systems in physical conditions and parameter ranges discussed in this thesis can be experimentally realized within the current state-of-the-art cold atoms quantum technology. Our results, apart from being a relevant contribution to many-body physics,  provide the `\textit{primum mobile}' for a new concept of matter-wave circuits based on SU($N$) fermionic platforms opening an exciting chapter in the field of atomtronics. Indeed, the specific properties of quantization are expected to provide the core to fabricate quantum devices with enhanced sensitivity like interferometers. At the same time, SU($N$) fermionic circuits show promise in engineering cold atoms quantum simulators with this artificial fermionic matter.

\chapter{Sommario}

\noindent I sistemi atomici ultrafreddi si sono imposti come forti contendenti tra i vari sistemi quantistici rilevanti per lo sviluppo e l'implementazione di tecnologie quantistiche grazie al loro maggiore controllo e alla flessibilità delle condizioni operative. I recenti sviluppi nella tecnologia della micro-ottica hanno aperto la strada all'ingegneria dei circuiti atomici in varie architetture. Le geometrie a forma di anello, che sono un semplice caso di questi circuiti, sono di particolare interesse. In tali circuiti, un'onda-materia guidata, nello specifico una corrente persistente, può essere generata dall'applicazione di un campo magnetico efficace. Uno degli aspetti peculiari dei sistemi atomici ultrafreddi è il controllo della statistica del fluido quantistico che scorre nell'anello, siano essi bosoni, fermioni o una loro miscela. Naturalmente, la corrente persistente può presentare dipendenze e attributi specifici a seconda della natura della materia quantistica che la costituisce. Infatti, tali fluidi quantistici godono di specifiche proprietà fisiche e regole di quantizzazione , che dovrebbero essere sfruttate in elementi circuitali atomtronici con caratteristiche uniche.  \\

\noindent  In questa tesi, esploriamo le correnti persistenti generate in un gas 
quantistico di fermioni fortemente interagenti a $N$-componenti, i cosiddetti fermioni SU($N$), intrappolato in una geometria ad anello. Questi sistemi fermionici multicomponenti, realizzabili con metalli alcalino-terrosi, presentano proprietà fisiche che vanno oltre quelle dei tipici fermioni a due componenti presenti nei sistemi di materia condensata. I nostri risultati mostrano che la corrente persistente dei fermioni a $N$-componenti esibisce un \textit{quantizzazione frazionaria del momento angolare}, con importanti differenze derivanti dal fatto che gli atomi siano soggetti a interazioni repulsive o attrattive. Nel caso di interazioni repulsive, la quantizzazione frazionaria si manifesta con una corrente il cui periodo è ridotto di $1/N_{p}$, ove $N_{p}$ è il numero di particelle nel sistema. Allo stesso modo, anche il regime attrattivo presenta una periodicità ridotta nella corrente, anche se con una dipendenza dal numero di componenti $1/N$. Monitorando queste proprietà specifiche della quantizzazione, la corrente persistente può essere utilizzata come strumento diagnostico per sondare i fenomeni quantistici di molte particelle.  La quantizzazione frazionaria della corrente persistente può essere letta attraverso figure di interferenza ottenute attraverso protocolli cosiddetti ``homodyne'' e ``self-heterodyne''.  \\

\noindent I sistemi fisici discussi in questa tesi possono essere sperimentalmente realizzati mediante le attuali tecnologie quantistiche che utilizzano atomi ultrafreddi. I nostri risultati, oltre a costituire un contributo rilevante alla fisica dei sistemi a molti corpi,  forniscono il `\textit{primum mobile}' per un nuovo concetto di circuiti onda-materia basati su piattaforme fermioniche SU($N$), che aprono un entusiasmante capitolo nel campo dell'atomtronica. Infatti, ci si aspetta che le proprietà specifiche della quantizzazione forniscano le basi per fabbricare dispositivi quantistici con maggiore sensibilità come gli interferometri. Allo stesso tempo, i circuiti di fermioni SU($N$) si dimostrano delle piattaforme promettenti per l'ingegnerizzazione di simulatori quantistici che impiegano atomi freddi di natura fermionica SU($N$).

\chapter{List of publications}

\noindent \underline{The research contained in this thesis is supported by the publications listed below:}

\vspace{0.7em}
\noindent \textbf{Chapter~\ref{chp:repcurr}: Atomtronic circuits with repulsive SU(\textit{N}) matter}

\vspace{0.7em}
\noindent \href{https://scipost.org/SciPostPhys.12.1.033}{1.} \underline{W. J. Chetcuti}, T. Haug, L.-C. Kwek, and L. Amico, ``Persistent current of SU($N$) fermions,'' SciPost Physics, vol. 12 (33), 2022.

\vspace{0.4em}

\noindent \href{https://avs.scitation.org/doi/10.1116/5.0026178}{2.} L. Amico, M. Boshier, G. Birkl, A. Minguzzi, C. Miniatura, L. C. Kwek,
D. Aghamalyan, V. Ahufinger, D. Anderson, N. Andrei, et al., ``Roadmap on Atomtronics: State of the art and perspective,'' AVS Quantum Science, vol. 3, no. 3, p. 039201, 2021.

\vspace{0.4em}
\noindent \href{https://iopscience.iop.org/article/10.1088/1751-8121/ac7016}{3.} M. Consiglio, \underline{W. J. Chetcuti}, C. Bravo-Prieto, S. Ramos-Calderer, A. Minguzzi, J. I. Latorre, L. Amico, and T. J. G. Apollaro, ``Variational quantum eigensolver
for SU($N$) fermions,'' Journal of Physics A: Mathematical and Theoretical, vol. 55,
p. 265301, 2022.

\vspace{0.7em}
\noindent \textbf{Chapter~\ref{chp:probe}: Atomtronic circuits with attractive SU(\textit{N}) matter}

\vspace{0.4em}
\noindent \href{https://www.nature.com/articles/s42005-023-01256-3}{4.} \underline{W. J. Chetcuti}, J. Polo, A. Osterloh, P. Castorina, and L. Amico, ``Probe for bound
states of SU(3) fermions and colour deconfinement,''  Commununications Physics 6 (128), 2023.

\vspace{0.7em}
\noindent \textbf{Chapter~\ref{chp:interfer}: Interference dynamics of matter-waves of SU(\textit{N}) fermions}

\vspace{0.4em}
\noindent \href{https://scipost.org/SciPostPhys.15.4.181}{5.}  \underline{W. J. Chetcuti}, A. Osterloh, L. Amico, and J. Polo, ``Interference dynamics of matter-waves of SU(\textit{N}) fermions,'' SciPost Physics, vol. 15 (181), 2023.

\vspace{0.7em}
\noindent \textbf{Chapter~\ref{chp:ogata}: Exact one-particle density matrix for SU(\textit{N}) fermionic matter-waves in the strong repulsive  limit}

\vspace{0.4em}
\noindent \href{https://scipost.org/SciPostPhys.15.1.006}{6.} A. Osterloh, J. Polo, \underline{W. J. Chetcuti} and L. Amico, ``Exact one-particle density matrix for SU(\textit{N}) fermionic matter-waves in the strong repulsive  limit,'' SciPost Physics, vol. 15 (6), 2023.

\vspace{1.3em}

\noindent \underline{Other publications not contained in this thesis:}
\vspace{0.4em}

\noindent \href{https://www.sciencedirect.com/science/article/abs/pii/S0375960120301055}{7.}  T. J. G. Apollaro, C. Sanavio, \underline{W. J. Chetcuti}, and S. Lorenzo, “Multipartite entanglement transfer in spin chains,” Physics Letters A, vol. 384, no. 15, p. 126306, 2020.

\vspace{0.4em}
\noindent \href{https://www.mdpi.com/1099-4300/23/1/51}{8.} T. J. G. Apollaro and \underline{W. J. Chetcuti}, ``Two-excitation routing via linear quantum channels'', \textit{Entropy}, vol. 23, no. 1, 2021.


\mainmatter 

\fancyhf{}
\fancyhead[RE,LO]{\slshape\nouppercase{\rightmark}}
\fancyfoot[LE,RO]{\thepage}
\pagestyle{fancy} 

\chapter{Introduction}\label{chp:intro}

\noindent Fundamental science and technology are inextricably linked. Basic research furnishes new novel concepts that can be harnessed to engineer devices and instruments with improved specifications. In turn, these technological advancements enable us to investigate fundamental aspects of nature with enhanced precision and sensitivity, prompting further scientific investigation. This virtuous cycle between science and technology, wherein the two continually foster one another, can be credited with ushering epochal changes in human history. The $18^{\mathrm{th}}$ century industrial revolution and, more recently, the first quantum revolution that culminated in the digital era of lasers, electronics, and computers showcase how the symbiosis between science and technology underpins human progress.\\

\noindent Presently, a second quantum revolution is underway. Its primary objective is to fabricate quantum technologies~\cite{dowling}. Quantum technology intertwines basic and applied science to an unprecedented degree: different quantum systems, manipulated and controlled from the macroscopic spatial scale down to the individual or atomic level, can be platforms for quantum devices and simulators with refined capabilities; at the same time, quantum matter constituting the quantum device might exhibit  new fundamental and unexpected physical features due to the specific conditions required for the technology to operate. The defining goal of quantum technology is to harness genuine quantum effects to construct devices with distinctive physical principles that are of practical value in the domains of communication~\cite{Gisin2007}, computation~\cite{Ladd2010}, sensing~\cite{sensingrev}, and simulation~\cite{quantum_simul}. A variety of physical systems have been put forward as suitable candidates to develop and implement these quantum technologies. These platforms range from solid state systems such as arrays of quantum dots~\cite{quantum_dots}, superconducting qubits~\cite{superconducting_quant}, and colour centers (e.g., nitrogen-vacancy centers) in diamonds~\cite{colour_centre} to atomic and molecular ones that include photonic systems~\cite{photonics}, Rydberg atoms~\cite{rydbergtech}, trapped ions~\cite{trapped_ions} and lastly cold atoms~\cite{Bloch2012}, which is the implementation that the present thesis focuses on. The appeal of ultracold atomic platforms consist in their versatility and their enhanced control and flexibility of their operating conditions~\cite{bloch2008many,acinroadmap}. \\

\noindent The field of ultracold atomic physics sees its beginning marked with the milestones of the experimental realization of a Bose-Einstein condensate~\cite{anderson_bec,ketterle_bec} and attaining Fermi degeneracy ~\cite{fermi_degen} as the $20^{\mathrm{th}}$ century drew to a close. After decades of tremendous progress in laser cooling techniques~\cite{chu_nobellecture,cohen_nobel,philips_nobel}, a charge-neutral gas of alkali atoms, placed in a vacuum and spatially confined with suitable electromagnetic fields, was cooled down to temperatures of the order of nanokelvin~\cite{bloch2008many}. At these extremely low temperatures, quantum effects are more pronounced as the thermal de Broglie wavelength is comparable to the average inter-particle distance.  Ultracold atoms platforms are characterized by robust coherences due to the effective shielding of the external environment. They can be realized with fundamentally different quantum statistics of the gas constituent. The atom-atom interactions can be tuned to be attractive or repulsive through Feshbach resonances~\cite{Inouye1998} and can even be enhanced by controlling the potential depth of the optical lattices confining the atoms~\cite{bloch2008many}. Moreover, due to the remarkable progress in micro-optics technology, cold atoms can be trapped in a wide variety of potentials, shapes, and intensities~\cite{Gauthier_faces,halina2016road}. These are some relevant features as to why ultracold atoms provide an important instance of artificial quantum matter that can be used as `hardware' to advance the fabrication of quantum devices~\cite{Bloch2012,georgescu2014quantum,acinroadmap} with practical value such as in sensing~\cite{schmiedmayer_rev_2,sensingrevy}. An interesting application of cold atoms technology is in quantum simulation~\cite{Bloch2012,quantum_simul,georgescu2014quantum}. Originally proposed by Feynman~\cite{fermiyip}, quantum simulators are quantum systems that can be tailored to mimic other physical systems. Accordingly, the physics and dynamics of many-body systems that are not tractable with classical computers can be investigated through quantum simulation. In this regard, cold atoms platforms prove to be a powerful asset due to their adaptability and tuneability.  There are multiple experimental investigations demonstrating that models such as the Bose-Hubbard~\cite{Zoller_tool} and Fermi-Hubbard~\cite{hubbexp,tarruell} can be accurately implemented in ultracold atoms systems. In addition, various 
phenomena have been observed, such as Mott insulators~\cite{Greiner2002,Jrdens2008} and Tonks-Girardeau gases~\cite{Paredes2004}, to name a few. \\

\noindent Atomtronics is an emerging research area in quantum technology exploiting cold atoms trapped with light and magnetic fields to \textit{realize matter-wave circuits} in a variety of different architectures~\cite{amico2005quantum,seaman2007atomtronics,pepino,Amico_2017,amico2021,amico2022}. Atomtronics incorporates the high degree of control and versatility of the cold atoms constituting them. Specifically, the key properties of atomtronic circuits include the robust coherence properties of the quantum fluid flowing through it, the nature of the particles' statistics, tunable atom-atom interactions, and flexible potential landscapes. A natural venue for this research activity has been constructing matter-wave analogues of conventional electronic devices, with the name \textit{atomtronics} being an amalgam of atoms and electronics~\cite{seaman2007atomtronics,progenitor}. Several atomtronic technologies mimicking the functionality of such devices have been realized, ranging from elementary circuits of linear~\cite{ryu2015integrated,Navez_2016,Akatsuka_2017} and annular~\cite{ramanathan2011superflow,eckel2014hysteresis,corman2014quench,wright2022persistent,roati2022imprinting} matter-wave guides to atomic  batteries~\cite{zozubat,caliga2017experimental}, transistors~\cite{transistor_stick,caliga2016principles,anderson2021matter} and diodes~\cite{atomdiode}.  Being characterized by distinctive physical attributes, atomtronics has the potential to realize devices with unprecedented capabilities and novel functionalities compared to their classical counterparts. Recently, atomic components have been fabricated to replicate quantum electronics, such as the atomic counterpart of superconducting quantum interference devices (SQUIDs)~\cite{ramanathan2011superflow,ryu2013experimental,aghamalyan2015coherent,ryu2020quantum}, which are believed to be of paramount importance for guided interferometers~\cite{amico2014superfluid,ryu2015integrated,kim2022one,katy2022matter,Amico2014slm,haug2018readout}. This is just one example illustrating how quantum technologies implemented in other platforms can be studied in a completely new way through cold atoms, especially given the flexibility and dynamical adaptability of the operating conditions.  While the defining goal of the atomtronics field is to fabricate quantum devices of practical value and sensors with enhanced performances, it is also suitable for extending the field of cold atoms quantum simulators. An interesting domain where atomtronics could play a vital role is that of mesoscopic physics, where concepts such as quantum transport and persistent currents can be revisited in a completely new way. In particular, the study of persistent currents in atomtronic circuits is one of the core added values of the field~\cite{amico2021,amico2022}. \\

\noindent The persistent current is one of the purest expressions of mesoscopic behaviour~\cite{imry2002intro}. It is a remarkable effect giving fundamental information on the crossover between the microscopic, purely quantum regime and the macroscopic world: when a mesoscopic ring-shaped physical system is pierced by a magnetic field, it can display quantum coherence by starting a dissipation-less quantized matter-wave current~\cite{gefen1984quantum}. Persistent currents have been extensively investigated in superconductors~\cite{byers1961theoretical,onsager1961magnetic,supercoon} and in normal metallic rings~\cite{trioper,nanobutt,nano2,zvyagin}. Studies of such a phenomenon have been defining a very active research field in physics with significant impacts in technology, culminating with the engineering of several quantum devices of practical value as the aforementioned SQUID~\cite{tinkham2004introduction}.  With the advent of cold atoms quantum technology, persistent currents can be imparted in a variety of systems made of bosonic~\cite{ramanathan2011superflow,kevin2013driving,pandey2019hyper,wolf2021stationary} or fermionic constituents~\cite{wright2022persistent,roati2022imprinting}. Owing to their charge-neutral nature, ultracold atoms cannot be made to flow via magnetic means as charged particles are in electronic and superconducting systems. The simplest way adopted in the first experiments relied on exploiting the equivalence between the Lorentz and Coriolis forces, 
to create an artificial magnetic field through rotation~\cite{bloch2008many,dalibarflux,Goldman_2014}. This can be carried out by stirring the quantum fluid with a barrier~\cite{kevin2013driving,neelystir,wright2022persistent}. Alternatively, one can induce an effective rotation through shaking~\cite{maciejtuna,Struck2013,Atala2014}, whereby the atomic potential is periodically modulated in time. Circulating current states can also be generated by transferring non-zero angular momentum through two-photon Raman transitions~\cite{andersenraman,dalibarflux} and through phase imprinting~\cite{ketterlevortices,Lin2009,helenephaseimprint}. \\

\noindent Angular momentum quantization of the persistent current in $^{87}$Rb atomtronic ring-shaped circuits has been studied both theoretically and experimentally~\cite{ryu2013experimental,ryu2020quantum}. Such studies have been instrumental in defining the atomic counterpart of SQUIDs~\cite{ramanathan2011superflow,ryu2013experimental,aghamalyan2015coherent,ryu2020quantum}. Another facet of persistent currents in atomtronics is as diagnostic tools to probe quantum correlations in many-body systems~\cite{amico2021,amico2022}, defining an instance of \textit{current-based quantum simulators}: In the same spirit as current-voltage characteristics in solid state physics, many-body systems can be probed by monitoring the behaviour of the current flowing through them with respect to changes in the external parameters. For example, the formation of bright solitons in bosonic circuits with attractive interactions is reflected in the persistent current, which displays a fractional quantization of the angular momentum~\cite{polo2020exact,naldesi2020enhancing}. To be specific, when the $N_{p}$-particle bound state is created, the corresponding effective mass leads to having a matter-wave current with a reduced period of $1/N_{p}$ with respect to the free bosonic case. The phenomenon of fractionalization in attracting bosons presents an exciting avenue for current-based simulators in atomtronics and in addition, has been predicted to lead to enhanced performances in rotation sensing~\cite{naldesi2020enhancing,polo2021quantum}.  \\

\noindent Most of the studies carried out so far have been devoted to atomtronics circuits of ultracold bosons, whilst ones comprised of interacting ultracold fermions are still in their infancy. The reason being that it is more challenging to cool~\cite{fermi_degen,pauliblocking}, manipulate~\cite{bloch2008many}, and image~\cite{wright2022persistent,roati2022imprinting} a fermionic gas. Recent experimental advances in cold atoms technology have paved the way for the realization of atomtronic circuits comprised of matter-waves of fermionic natures~\cite{wright2022persistent,roati2022imprinting}. As mentioned previously, persistent currents have been widely explored in solid state systems~\cite{byers1961theoretical,onsager1961magnetic}. However, in cold atoms, fermionic matter-wave currents can be realized in an environment where all system parameters are completely under control. In turn, phenomena such as fermionic superfluidity and the celebrated BCS-BEC crossover can be studied from a different angle~\cite{pecci2021probing,roati2022imprinting}. \\

\noindent This thesis focuses on persistent currents in fermionic atomtronic circuits. However, the quantum fluid flowing in the circuit is not comprised of ordinary two-component fermions but of strongly interacting $N$-component fermions. These so-called SU($N$) symmetric fermions are a new artificial quantum matter that has been recently engineered owing to the developments in the cooling and trapping techniques of alkaline earth-like atoms~\cite{yoshiro_degenerate,taie,Taie2012,Pagano2014,Scazza2014,sonderhouse2020}. \\

\noindent From a theoretical point of view, SU($N$) fermionic systems have been studied since the 1960s. However, models with enlarged SU($N$) symmetry were regarded as idealizations that could not be directly realized in actual physical systems. Accordingly, the experimental realization of SU($N$) fermions has rekindled theoretical interest to conduct quantum simulations with these systems. At the operating conditions of cold atoms, the interactions are dominated by contact $s$-wave scattering, which necessitates that the interacting fermions be in different orientations due to their fermionic statistics. By virtue of their $N$ internal degrees of freedom, the Pauli exclusion principle relaxes for SU($N$) fermions thereby enhancing the number and type of interactions. Such a feature makes alkaline earth-like atoms, especially with lattice confinements, an ideal platform to study exotic quantum matter, including higher spin magnetism~\cite{hermelespin,manmana2011n}, spin liquids~\cite{liquidsun,nataf}, topological matter~\cite{capponi} and Mott insulators~\cite{hermele,Taie2012,scazza2016}. Additionally, SU($N\! >\! 2$) fermions are also relevant to 
areas beyond condensed matter physics. For instance, it can be beneficial to the high energy physics community, where SU(3) symmetry governs the interactions between quarks, to study long-standing problems in the field, such as colour deconfinement in quantum chromodynamics~\cite{Cherng2007superfluidity,Rapp2007color,baym2010bcs}. \\

\noindent In this thesis, we explore the concept of matter-wave circuits based on SU($N$) symmetric degenerate gas platforms. The reasons behind doing so are two-fold: (i) to extend the framework of current-based quantum simulators in atomtronics to fermionic systems; (ii) to utilize the persistent current as a diagnostic tool to probe exotic phenomena in SU($N$) fermionic systems. The thesis is structured as follows. \\

\noindent In \textit{Chapter}~\ref{chp:tools}, we present the main theoretical tools and concepts utilised throughout this thesis. To start, we introduce the Fermi-Hubbard model and provide a description of its exact solution provided by Bethe ansatz. Next, the emergence of SU($N$) symmetry in ultracold atoms systems is explored.  Finally, we round off this chapter by reviewing some experimental considerations concerning different trapping techniques and the realization of SU($N$) fermions. \\

\noindent \textit{Chapter}~\ref{chp:persis} is devoted to the phenomenon of persistent currents. First, we provide a brief overview of persistent currents in condensed matter physics and then focus on the recent developments of these matter-wave currents in ultracold atomic gases. Afterwards, we study from a theoretical viewpoint as to how cold atoms can be used to simulate charged particles moving in a magnetic field and look at some of the experimental schemes that have been implemented to generate persistent currents in bosonic and fermionic systems. Lastly, we discuss the basic properties of persistent currents in the free particle regime and introduce the Leggett theorem, which is an essential theorem for our analysis of persistent currents.  \\

\noindent \textit{Chapter~\ref{chp:repcurr}} focuses on the simple case of an atomtronic circuit provided by a ring-shaped quantum gas of strongly interacting repulsive SU($N$) fermions pierced by an effective magnetic flux~\cite{chetcuti2021persistent}. In particular, we investigate how the persistent current, the response to this applied field, displays specific dependencies on the parameters characterizing the physical conditions of the system. Several surprising effects emerge. Firstly, we find that as a combined effect of spin correlations, interaction, and effective magnetic flux, spin excitations can be created in the ground-state leading to a re-definition of the fundamental flux quantum of the system: fractional values of the angular momentum are dependent on the number of particles in the system. The persistent current landscape is affected dramatically by these changes and displays a universal behaviour.  Moreover, we demonstrate that despite its mesoscopic character, the persistent current is able to detect the onset to a quantum phase transition (from metallic-like to Mott phase).\\

\noindent In \textit{Chapter~\ref{chp:probe}}, our attention shifts to
atomtronic circuits of attracting SU(3) fermions~\cite{chetcuti2021probe}. We find that by utilising the persistent current as our diagnostic tool, one can distinguish between the two types of bound states formed by three-component fermions: trions and colour superfluids (CSFs), which are the analogues of hadrons and mesons in quantum chromodynamics (QCD). Furthermore, we perform a thorough analysis of the persistent current's dependence on interaction and thermal fluctuations, gaining access to a quantitative description of the celebrated colour deconfinement in QCD. Specifically, for small interactions, a crossover is observed from a colourless bound state to coloured multiplets that displays similarities with Quark-Gluon Plasma formation at large temperatures and small baryonic densities.  \\

\noindent \textit{Chapter~\ref{chp:interfer}} deals with the read-out of SU($N$) persistent currents in an experimental setting~\cite{chetcuti_interfer}. Our approach employs both homodyne and self-heterodyne protocols, which are two procedures for interfering ultracold matter-waves that are well-established within the current experimental capabilities. Through interference dynamics, we demonstrate how the fractional values of angular momenta displayed in the persistent current of strongly attractive and repulsive $N$-component fermions, can not only be monitored but also observed in some cases. Additionally, our analysis shows that the study of interference patterns grants us access to both the number of particles and components, two quantities that are notoriously hard to extract experimentally. \\

\noindent In \textit{Chapter~\ref{chp:ogata}}, a theoretical framework to study the exact one-particle density matrix of strongly repulsive SU($N$) fermions in a ring-shaped potential is developed~\cite{osterloh2022exact}. The approach hinges on the fact that at the limit of infinite repulsion, the spin and charge degrees of freedom decouple, simplifying the problem by splitting it into the spinless and SU($N$) Heisenberg models. Then, we consider the specific case of SU($N$) matter flowing in a ring when subjected to an effective magnetic flux. In this context, we show how the developed framework can be used to calculate the interference dynamics of the two read-out protocols introduced in the previous chapter, albeit at system parameters that are not accessible numerically within the current state-of-the-art. \\

\noindent Finally, in \textit{Chapter}~\ref{chp:conc}, we summarize the work presented in this thesis and then provide some future outlooks and perspectives for developments in the field.

\chapter{Basic concepts and models}\label{chp:tools}

In this chapter, we provide an overview of the main theoretical concepts and models used in this thesis to describe the systems under investigation. Section~\ref{sec:sun} is devoted to the main physical systems under consideration in this thesis, which are SU($N$) symmetric fermions. Subsequently, in Section~\ref{sec:sunfhm}, we introduce the SU($N$) Fermi-Hubbard Hamiltonian, an experimentally relevant model describing itinerant, interacting $N$-component fermions on a lattice. Section~\ref{sec:bethesun} details the Bethe ansatz equations for the model. Lastly, we briefly present experimental considerations for ultracold atomic systems, particularly SU($N$) fermions, in Section~\ref{sec:exp}.

\section{SU(\textit{N}) symmetry in ultracold fermionic systems}\label{sec:sun}

\noindent The concept of symmetry is deeply ingrained in nature and the physical laws that describe it~\cite{gross}. Noether's theorem relating continuous symmetries to conservation laws~\cite{noether} 
and the classification of all elementary particles through exchange symmetry are but a few examples that showcase symmetry as a powerful and invaluable tool in physics. Several theories in modern physics were formulated by resorting to symmetry arguments, from Einstein's theories of relativity~\cite{einstein0,einstein} to quantum field theories where the gauge symmetries govern the fundamental interactions~\cite{feynmanqed,Salam1959WeakAE,gross_qcd}. \\

\noindent In this thesis, the SU($N$) symmetry is an essential element. It plays an important role in several branches of physics: SU(3) for the strong interaction in quantum chromodynamics and SU(2) spin symmetry of electrons in solid state physics, being some notable examples. While in condensed matter, SU($N$) symmetry can emerge only effectively and often as a result of finely tuning the parameters~\cite{goerbig,affleck,read,skyrmions,kawashima,regnault,cazalilla_2014,capponi}\footnote{One example is the SU(4) spin-valley symmetry in graphene~\cite{goerbig}, which is associated with the rotational invariance of the Coloumb interaction governing the fractional quantum Hall effect~\cite{goerbig,regnault}.}, ultracold atoms prove to be an ideal platform due to unprecedented degree of control and flexibility of the operating conditions~\cite{bloch2008many,lewenstein_ultracold}. Indeed, the recent milestones in the field paved the way for the experimental realization of SU($N$) symmetric fermionic systems using alkaline earth-like atoms\footnote{The designation alkaline earth-like atoms encompasses not only the group II elements of the periodic table but also those with a similar electronic structure such as zinc and ytterbium, that are elements residing in the $d$ and $f$ blocks, respectively.}~\cite{taie,Taie2012,Pagano2014,rey_spectroscopic,Scazza2014,scazza_parish,scazza2016,Goban2018,not_my_campbell,yoshiro_anti,sonderhouse2020,Tusi2022,Taie2022,miklos}. \\

\noindent The emergent SU($N$) symmetry arises from the decoupling of the nuclear spin $I$ and the electronic angular momentum $J$. Such a phenomenon results in the total internal angular momentum $F$ being provided solely by the nuclear spin degrees of freedom. Consequently, hyperfine interactions are absent, resulting in two-body collisions independent of the nuclear spin with an enlarged SU($N$) symmetry. The strong decoupling between the nuclear and electron angular momentum \textit{without fine-tuning} is an inherent characteristic of fermionic alkaline earth-like atoms ground and metastable excited $^{3}P_{0}$ states as they have zero electronic angular momentum~\cite{cazalilla_2009,gorshkov,cazalilla_2014,capponi}. 
\vspace{1em}
\begin{figure}[h!]
    \centering
    \includegraphics[width=\linewidth]{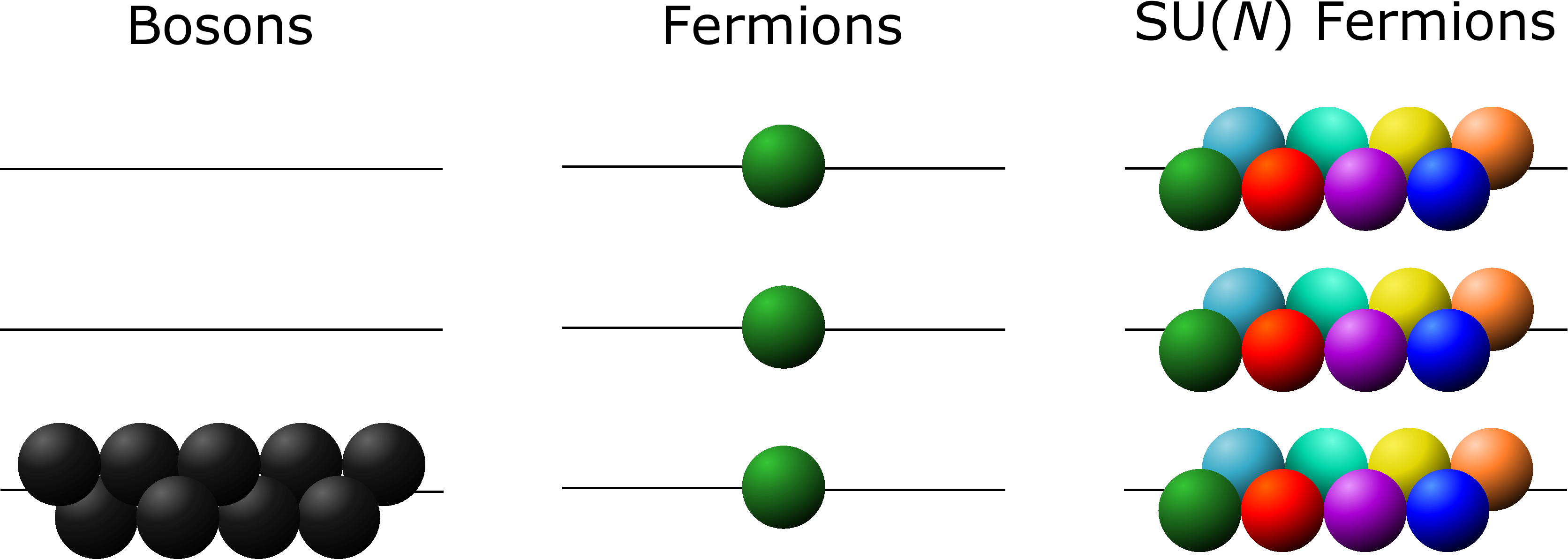}
    \caption{Schematic diagram for the level occupation of bosons, spinless fermions and SU($N$) fermions. Whilst bosons (left) can all occupy the same state, spinless fermions (middle) are restricted by the Pauli exclusion principle and so only one particle can inhabit a given energy level.  On going to SU($N$) fermions (right), the Pauli exclusion principle relaxes, enabling $N$ fermions to occupy in a given level. Accordingly, there are $N\! -\! 1$ interactions for a given particle. Figure reproduced from~\cite{sonderhouse2020}.}
    \label{fig:manyballs}
\end{figure}

\noindent Put simply, SU($N$) symmetric fermions are $N$-component fermions. From Figure~\ref{fig:manyballs}, it is clear that on increasing $N$, the Pauli exclusion principle relaxes, enhancing the number and type of interactions. Eventually, as $N\!\rightarrow\!\infty$, SU($N$) fermions emulate bosons in terms of level occupations. Interactions occurring at the low temperatures of cold atoms experiments are typically characterized by $s$-wave scattering lengths, which result to be vanishing for spinless fermions. Accordingly, SU($N$) fermions present the opportunity to circumvent this constraint by realizing interacting systems of multicomponent fermions, with the increased interplay of interactions leading to exotic physics that extends beyond that of two-component fermions in condensed matter physics~\cite{cazalilla_2014,capponi}. 

\newpage
\noindent Multicomponent fermions are found to possess richer phase diagrams compared to their two-component counterparts. To start, two- and three-dimensional SU($N$) fermionic systems have been theoretically shown to behave like Fermi liquids~\cite{cazalilla_2014,sunliquids}, whose properties display a significant dependence on the number of components~\cite{sunliquids}. On going to one-dimension, it is expected that an SU($N$) Luttinger liquid behaviour emerges~\cite{manmana2011n,capponi}. Recent experiments have shown that the SU($N$) symmetry causes a deviation in the dynamical properties of the spinless and two-component Luttinger liquids~\cite{Pagano2014}. An interesting aspect of SU($N$) fermions is their ability to form bound states with different types and natures when subjected to an attractive interaction~\cite{takahashisu3,Guan2013Fermi}. In this regime, one can revisit established phenomena like BCS pairing~\cite{honerkampbcs,tomoki} and fermionic superfluidity~\cite{Cherng2007superfluidity} from a completely different angle.  On a similar note, repulsive SU($N$) fermions also display interesting phases~\cite{honerkamp2004,capponi}. One such example is that of SU($N$) Mott insulators, which have been experimentally realized and shown to display modified finite temperature properties compared to SU(2) ones on account of their enlarged symmetry~\cite{Taie2012,hazzard_high,Tusi2022}. In addition, the enlarged symmetry of these $N$-component fermions makes them, especially in lattices, an ideal platform to study exotic quantum matter, including higher spin magnetism~\cite{manmana2011n,Cherng2007superfluidity,rey_spectroscopic}, spin liquids~\cite{liquidsun,liquidmila} and topological matter~\cite{hermelespin} and, beyond condensed matter physics, in QCD~\cite{wiese2014towards,Rapp2007color}, lattice gauge theories~\cite{banerjee2012atomic} and the fabrication of synthetic dimensions~\cite{syntheticdim}.

\subsection{Interacting multicomponent fermions}

\noindent At the typical operating conditions of ultracold atoms, i.e., very low temperatures and dilute gases, the relevant contributions to the scattering processes are two-body $s$-wave collisions. Therefore, in this regime, the effective interaction between two atoms can be approximated by the Fermi contact potential~\cite{yang1957quantum,lee_huang_yang}
\begin{equation}\label{eq:pot}
    V(\mathbf{r}-\mathbf{r}') = \frac{4\pi\hbar^{2}}{m}a_{s}\delta (\mathbf{r}-\mathbf{r}'),
\end{equation}
which is a pseudo-potential characterised by the $s$-wave scattering length $a_{s}$~\cite{sakurai}\footnote{Note that this is called pseudo-potential since in the strongly interacting regime, the contact potential is regularised by substituting the Dirac delta with $\delta(\mathbf{r})\frac{\partial}{\partial r}r$~\cite{Busch1998}.}. Due to the particles' statistics, interactions via $s$-wave collisions occur only for fermions with different spin projections and are absent for identical fermions. The Lee-Huang-Yang pseudo-potential introduced in Equation~\eqref{eq:pot} can only be applied to bosons and two-component fermions. As such, it has to be generalised when dealing with spinor bosonic condensates~\cite{spinorho,ohmi,uedaspinor} or fermionic gases with higher spin~\cite{fermiyip}. \\ 

\noindent To obtain the form of interaction occurring in multicomponent systems with SU($N$) symmetry, we start by considering a collision between two spin-$F$ fermionic atoms, with half-odd integer $F$, mediated by a short-range potential. In the absence of symmetry breaking terms, such as an applied magnetic field or non-spherical trapping potentials, this two-body interaction is taken to be rotationally invariant in the ``spinor gas collision'' approximation~\cite{uedaspinor}. In turn, the total angular momentum of the colliding pair, which includes the orbital and internal angular momenta, is conserved. As mentioned previously, in ultracold atom systems, interactions are mainly characterised by $s$-wave collisions, which means that the collision's orbital angular momentum is vanishing. Hence, the total angular momentum of the interaction is given solely by the pair's internal angular momentum $f$, generating a higher dimensional SU(2) representation associated with the rotational invariance of the interatomic potential~\cite{cazalilla_2014,capponi}. Keeping this in mind, we can construct the pseudo-potential between two spin-$F$ fermions as having the form 
\begin{equation}\label{eq:Vf}
    V_{f}(\mathbf{r},\mathbf{r}') = \frac{1}{2}\sum\limits_{f=0,2,...}^{2F-1}g_{f}\mathcal{P}_{f}\delta (\mathbf{r}-\mathbf{r}'),
\end{equation}
where $g_{f} = (4\pi\hbar^{2}/m)a_{f}$ is the coupling constant dependent on the $s$-wave scattering lengths $a_{f}$, $\mathcal{P}_{f}$ corresponds to the projection operator onto states having a total spin-$f$, with $f=0,2,\hdots, 2F-1$. From all the possible values that $f$ can take, we can only consider the even valued ones, since, in $s$-wave collisions it is only the anti-symmetric combinations that are allowed to participate~\cite{uedaspinor,fermiyip}. Indeed, from Equation~\eqref{eq:Vf}, it is evident that there are $(2F+1)/2$ scattering lengths characterizing the collision between the pair of spin-$F$ fermions~\cite{capponi}. In second quantization, the projection operator can be expressed as 
\begin{equation}\label{eq:Pf}
    \mathcal{P}_{f} = \sum\limits_{m_{f} = -f}^{f}A^{\dagger}_{f,m_{f}}A_{f,m_{f}},
\end{equation}
where $A_{f,m_{f}}^{\dagger}$ and $A_{f,m_{f}}$ are the creation and annihilation pairing operators with spin magnetization $m_{f}$~\cite{capponi}, being defined as 
\begin{align}\label{eq:Af}
    A_{f,m_{f}}^{\dagger}(\mathbf{r}) = \sum\limits_{\alpha,\beta = -F}^{F}\bra{F,\alpha;F,\beta}\ket{f,m_{f}}\psi^{\dagger}_{\alpha}(\mathbf{r})\psi^{\dagger}_{\beta}(\mathbf{r}), \\ 
    A_{f,m_{f}}(\mathbf{r}) = \sum\limits_{\alpha,\beta = -F}^{F}\bra{f,m_{f}}\ket{F,\alpha;F,\beta}\psi_{\beta}(\mathbf{r})\psi_{\alpha}(\mathbf{r}),
\end{align}
where $\bra{f,m_{f}}\ket{F,\alpha;F,\beta}$ corresponds to the Clebsch-Gordan coefficient to form a total spin $f$ state from a pair of spin-$F$ particles~\cite{spinorho}. Consequently, the interatomic potential can be expressed in the following form
\begin{align}\label{eq:Vff}
    V_{f}(\mathbf{r}) = \frac{1}{2}&\sum\limits_{f=0,2,...}^{2F-1}g_{f}\sum\limits_{m_{f}}\sum\limits_{\substack{\alpha_{1},\alpha_{2},\\\beta_{1},\beta_{2}}}\int\bra{F,\alpha_{2};F,\beta_{2}}\ket{f,m_{f}}\bra{f,m_{f}}\ket{F,\alpha_{1};F,\beta_{1}} \nonumber\\
    &\times\psi^{\dagger}_{\alpha_{2}}(\mathbf{r})\psi^{\dagger}_{\beta_{2}}(\mathbf{r})\psi_{\beta_{1}}(\mathbf{r})\psi_{\alpha_{1}}(\mathbf{r})\mathrm{d}\mathbf{r}.
\end{align}
A striking consequence of Equation~\eqref{eq:Vff} is the presence of spin-changing processes. The scattering length's dependence on the total spin projection $f$ of the colliding fermions enables the population of different spin states with coupled spin $f$~\cite{krauser}. Due to this, even though the spin projection of the interacting fermions is conserved, typically, their individual one is not~\cite{gorshkov,cazalilla_2014}. \\
 
\noindent There exists a special case where all the scattering lengths, and in turn the coupling constants, are all equal. In such an instance, the pseudo-potential reads
\begin{align}\label{eq:Vfff}
    V_{f}(\mathbf{r}) = \frac{g}{2}&\sum\limits_{f=0,2,...}^{2F-1}\sum\limits_{m_{f}}\sum\limits_{\substack{\alpha_{1},\alpha_{2},\\\beta_{1},\beta_{2}}}\int\bra{F,\alpha_{2};F,\beta_{2}}\ket{f,m_{f}}\bra{f,m_{f}}\ket{F,\alpha_{1};F,\beta_{1}}\nonumber \\
    &\times\psi^{\dagger}_{\alpha_{2}}(\mathbf{r})\psi^{\dagger}_{\beta_{2}}(\mathbf{r})\psi_{\beta_{1}}(\mathbf{r})\psi_{\alpha_{1}}(\mathbf{r})\mathrm{d}\mathbf{r},
\end{align}
and by noting that $\sum_{f}\sum_{m_{f}}\ket{f,m_{f}}\bra{f,m_{f}}$ = 1 forms a complete basis, can be simplified to give
\begin{equation}\label{eq:Veq}
    V_{f}(\mathbf{r}) = \frac{g}{2}\sum\limits_{\alpha\neq\beta}\int\psi^{\dagger}_{\alpha}(\mathbf{r})\psi^{\dagger}_{\beta}(\mathbf{r})\psi_{\beta}(\mathbf{r})\psi_{\alpha}(\mathbf{r})\mathrm{d}\mathbf{r}.
\end{equation}
The equal scattering lengths imply that the interatomic potential is independent of the nuclear spins. During a collision, the nuclear spin typically exerts its influence through its hyperfine coupling with the electron angular momentum. As we alluded to previously, hyperfine interactions are absent in fermionic alkaline earth-like atoms ground-states and vanishing up to leading order in the excited state $^{3}P_{0}$~\cite{ye_optical,ye_nuclear}, owing to the decoupling of the nuclear and electronic spin degrees of freedom~\cite{gorshkov,cazalilla_2014}. Subsequently, the scattering processes are equal and independent of the nuclear spin, with the latter's impact on the collision being to enforce the Pauli exclusion principle, as shown in Equation~\eqref{eq:Veq}. \\ 

\noindent In such a scenario, several appealing properties emerge.  We start by noting that the pseudo-potential in Equation~\eqref{eq:Veq} is in possession of an enhanced symmetry in comparison to the SU(2) symmetry introduced earlier. The symmetry gets enlarged to the SU($N$) group where $N=2F+1$, with the interaction Hamiltonian defined in Equation~\eqref{eq:Veq}, being invariant under all transformations pertaining to this group. Defining the nuclear spin-permutation operators as 
\begin{equation}\label{eq:spinorperm}
    S_{\alpha}^{\beta} = \int \psi^{\dagger}_{\alpha}(\mathbf{r})\psi_{\beta}(\mathbf{r})\mathrm{d}\mathbf{r},
\end{equation}
we have that the interaction Hamiltonian, now denoted by $\mathcal{U}$, commutes with any spin-permutation operator. 
\begin{equation}\label{eq:communist}
    [\mathcal{U},S_{\alpha}^{\beta}] = 0 \hspace{3mm} \forall \alpha,\beta,
\end{equation}
Thus, the interaction is \textit{SU($N$) symmetric} since the spin-permutation operators are the generators of the SU($N$) Lie algebra group satisfying the SU($N$) algebra $[S_{\alpha}^{\beta},S_{\gamma}^{\epsilon}] = \delta_{\beta\gamma} S_{\alpha}^{\epsilon} - \delta_{\alpha\epsilon} S_{\gamma}^{\beta}$~\cite{gorshkov}. An immediate consequence of SU($N$) symmetry is that there are no spin-exchange collisions, with the population of each spin state being conserved. This can be clearly visualized by re-writing interaction potential in the following manner
\begin{equation}\label{eq:Vcons}
    \mathcal{U} = \frac{g}{2}\sum\limits_{\alpha\neq\beta}\int n_{\alpha}(\mathbf{r})n_{\beta}(\mathbf{r})\mathrm{d}\mathbf{r},
\end{equation}
by introducing the density field operator $n_{\alpha}(\mathbf{r}) = \psi^{\dagger}_{\alpha}(\mathbf{r})\psi_{\alpha}(\mathbf{r})$. Additionally, this poses another important implication. An atom with a large nuclear spin can effectively behave like one with a lower spin, by preparing it such that only a subset of the spin states are occupied~\cite{gorshkov,capponi}. Essentially, $N$ can take any value up to $2F+1$.  \\

\noindent The emergent SU($N$) symmetry comes from the inherent properties of fermionic alkaline earth-like atoms. Naturally, it begs the question as to whether such a symmetry can be accessed in others systems by fine-tuning the scattering lengths. Alkali systems with $F\! >\! 1/2$, which already boast enlarged symmetries with independent couplings~\cite{wuexact,capponiconfinement}, can achieve SU($N$) symmetry with fine-tuning, at least in principle. Indeed, it has been shown experimentally that SU(3) symmetry emerges in three-component systems of lithium gases at large magnetic fields but is not sustained at moderate magnetic field strengths~\cite{bartenstein,huckans,jochim_collisional,lompe,lompe2}. As such, it is preferable to utilize alkaline earth-like atoms such as ytterbium and strontium as no fine-tuning is required to attain the enhanced SU($N$) symmetry.

\section{SU(\textit{N}) Fermi-Hubbard model}\label{sec:sunfhm}

\noindent The Hubbard model, originally introduced by John Hubbard to provide an effective description of the electron dynamics in solids~\cite{oghubbard}, is a paradigmatic example addressing the physical properties of strongly interacting quantum many-body systems, ranging from superconductivity to quantum magnetism~\cite{rasetti,montorsi_Hubbard,essler}. Recent successes in the field of cold atoms quantum simulators include the emulation of the Hubbard model~\cite{lewenstein_ultracold,esslinger2010fermi,Mazurenko2017,tarruell} and, very recently, its generalization to $N$-components~\cite{Scazza2014,scazza2016}. Here, we sketch out the derivation for the SU($N$) Fermi-Hubbard model in an ultracold atom setting, which is the central model utilized to describe the physical systems under investigation in this thesis. \\

\noindent From microscopic theory, the many-body Hamiltonian describing a system of $N_{p}$ interacting two-component fermions of mass $m$ can be expressed in second quantization as
\begin{align}\label{eq:firstsec}
    \mathcal{H} = \sum\limits_{\alpha = 1}^{N}\int \Psi^{\dagger}_{\alpha}(\mathbf{r})\mathcal{H}_{0}\Psi_{\alpha} (\mathbf{r})\mathrm{d}\mathbf{r} 
    + \frac{1}{2} \sum\limits_{\alpha,\beta = 1}^{N}\int\Psi^{\dagger}_{\alpha}(\mathbf{r}')\Psi^{\dagger}_{\beta}(\mathbf{r})\mathcal{V}(\mathbf{r}-\mathbf{r}')\Psi_{\beta}(\mathbf{r})\Psi_{\alpha}(\mathbf{r}')\mathrm{d}\mathbf{r}\mathrm{d}\mathbf{r}',
\end{align}
where $\Psi^{\dagger}_{\alpha}(\mathbf{r})$ [$\Psi_{\alpha} (\mathbf{r})$] is the fermionic field operator that creates (annihilates) a fermion with spin projection $\alpha$ at position $\mathbf{r}$, satisfying the anti-commutation relations: $\{\Psi_{\alpha}(\mathbf{r}),\Psi^{\dagger}_{\beta}(\mathbf{r}')\}$ $= \delta_{\alpha,\beta}\delta (\mathbf{r}-\mathbf{r}')$ and $\{\Psi_{\alpha}(\mathbf{r}),\Psi_{\beta}(\mathbf{r}')\} =\{\Psi_{\alpha}^{\dagger}(\mathbf{r}),\Psi_{\beta}^{\dagger}(\mathbf{r}')\}=0 $\footnote{Note that for the sake of convenience, we have re-labelled $\alpha,\beta = -F,\hdots,F\rightarrow \alpha,\beta = 1,\hdots,N$.}. The first term in Equation~\eqref{eq:firstsec} corresponds to the single-particle Hamiltonian $\mathcal{H}_{0}$, which consists of the kinetic energy operator
$(-\hbar^{2}/2m)\nabla^{2}$ and the three-dimensional external potential $V_{\mathrm{ext}}(\mathbf{r})$ providing a lattice structure to the system. The second term $\mathcal{V}(\mathbf{r}-\mathbf{r}')$ denotes the interatomic potential of two-body collisions, which from Equation~\eqref{eq:Veq} reads
\begin{equation}
   \mathcal{V}(\mathbf{r}-\mathbf{r}') = \frac{4\pi\hbar^{2}}{m}a_{I}\delta (\mathbf{r}-\mathbf{r}'), 
\end{equation}
where $a_{I}$ is the scattering length independent of the nuclear spin. The nature and strength of the interactions can be very precisely tuned through Feshbach resonances~\cite{feshbach,fano}, generated via external optical~\cite{fedichev1996influence,julienne,theis2004tuning} or magnetic fields~\cite{Inouye1998,weiner_2003,chin2010feshbach}. The sign of $a_{I}$ be it negative ($a_{I}<0$) or positive ($a_{I}>0$), dictates whether the effective interactions between the atoms are attractive or repulsive respectively.\\

\noindent To write the Hamiltonian in second quantisation requires us to choose a suitable basis. In the presence of a homogenous periodic  potential, such that $V_{\mathrm{ext}}(\mathbf{r}) = V_{0}(\cos^{2}(kx)+\cos^{2}(ky)+\cos^{2}(kz))$, for example~\cite{Zoller_tool}, the eigenfunctions of the one-body Hamiltonian $\mathcal{H}_{0}$ are Bloch functions $\varphi_{a\mathbf{q}}(\mathbf{r})$ with band index $a$ and crystal momentum $q$ running over the first Brillouin zone~\cite{sakurai,ashcroft}. Bloch waves are highly delocalised, spanning the whole lattice, which makes them inappropriate to represent local properties, as is the case with the short-range interaction outlined in Equation~\eqref{eq:pot}. Therefore, we can construct a new basis, one that is complementary to the Bloch basis and readily obtained through its Fourier transform. The Wannier functions are maximally localised functions defined as
\begin{equation}\label{eq:wanndef}
    w_{aj}(\mathbf{r}) \equiv w_{a}(\mathbf{r}-\mathbf{r}_{j}) = \frac{1}{\sqrt{L}}\sum\limits_{\mathbf{q}}e^{-\imath\mathbf{q}\mathbf{r}_{j}}\varphi_{a\mathbf{q}}(\mathbf{r}),
\end{equation}
centered around the lattice potential minimum $r_{j}$ of the $j$-th site, with $L$ denoting the number of lattice sites~\cite{wannier}. Forming an orthogonal basis for different band and site indices, Wannier functions are well suited to handle contact interactions. Expanding the field operators in terms of the Wannier operators $\Psi^{\dagger}_{\alpha}(\mathbf{r}) = \sum_{aj}w_{a}^{*}(\mathbf{r}-\mathbf{r}_{j})c_{aj,\alpha}^{\dagger}$, the Hamiltonian introduced in Equation~\eqref{eq:firstsec} is expressed in the Wannier basis as
\begin{equation}\label{eq:allbands}
    \mathcal{H} = \sum\limits_{i,j,\alpha,a}t_{ij}^{a}c^{\dagger}_{ai,\alpha}c_{aj,\alpha} + \frac{1}{2}\sum\limits_{\substack{i,j,k,l \\ abde}}\sum\limits_{\alpha,\beta}U_{ijkl}^{abde}c^{\dagger}_{ai,\alpha}c^{\dagger}_{bj,\beta}c_{dk,\beta}c_{el,\alpha},
\end{equation}
where $c_{j,\alpha}^{\dagger}$ ($c_{j,\alpha}$) creates (annihilates) an electron with spin $\alpha$ localized at site $j$. These operators obey the canonical fermion algebra $\{c_{i,\alpha},c_{j,\beta}^{\dagger}\} = \delta_{\alpha\beta}\delta_{ij}$ and $\{c_{i,\alpha},c_{j,\beta}\} = \{c_{i,\alpha}^{\dagger},c_{j,\beta}^{\dagger}\}=0$. From this, it is straightforward to see that $(c_{j,\alpha}^{\dagger})^{2} = 0$ thereby ensuring the Pauli exclusion principle. \\

\begin{figure}[h!]
    \hfill\includegraphics[width=0.8\linewidth]{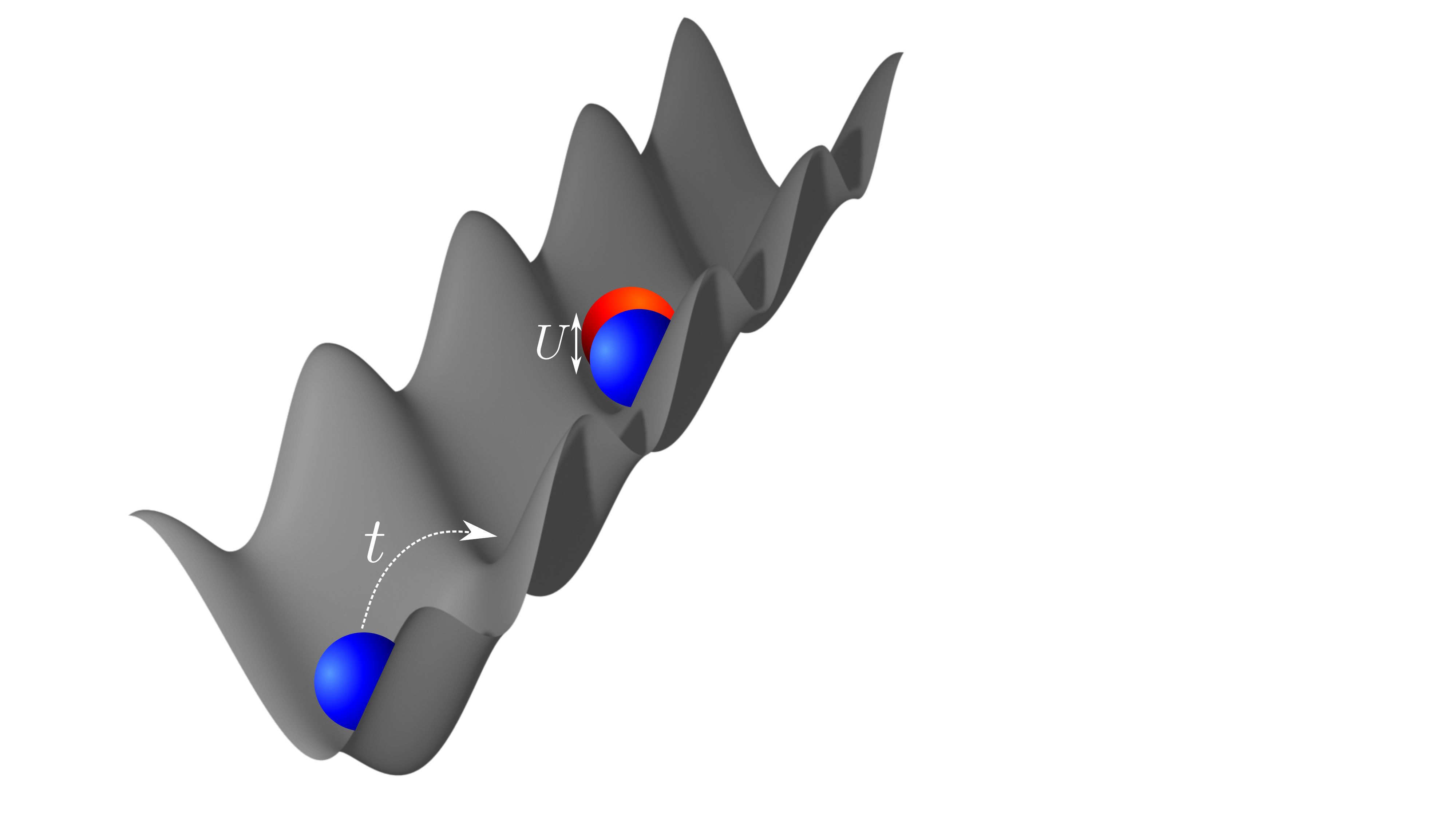}
    \caption{Schematic representation of the SU($N$) Fermi-Hubbard model for $N=2$. The fermions depicted by red and blue spheres are confined in a one-dimensional lattice potential. The fermions tunnel between neighbouring sites with tunneling amplitude $t$ and experience an on-site interaction with strength $U$ on doubly occupied sites.}
    \label{fig:hublat}
\end{figure}

\noindent The tunneling amplitude $t_{ij}^{a}$ with which particles tunnel from site $i$ to site $j$ within a given band $a$, is of the form
\begin{equation}\label{eq:hoppy}
    t_{ij}^{a} = \int w^{*}_{a}(\mathbf{r}-\mathbf{r}_{i})\bigg( -\frac{\hbar^{2}}{2m}\nabla^{2} + V_{\mathrm{ext}}(\mathbf{r}) \bigg)w_{a}(\mathbf{r}-\mathbf{r}_{j}) \mathrm{d}\mathbf{r}.
\end{equation}
Likewise, the interaction strength parameterised by $U_{ijkl}^{abde}$ is given by
\begin{equation}\label{eq:inter}
    U_{ijkl}^{abde} = \int w_{a}^{*}(\mathbf{r}'-\mathbf{r}_{i})w_{b}^{*}(\mathbf{r}-\mathbf{r}_{j})\mathcal{V}(\mathbf{r}-\mathbf{r}')w_{d}(\mathbf{r}-\mathbf{r}_{k})w_{e}(\mathbf{r}'-\mathbf{r}_{l})\mathrm{d}\mathbf{r}\mathrm{d}\mathbf{r}'.
\end{equation}
The Hamiltonian provided in Equation~\eqref{eq:allbands} is the multi-band model. However, in this thesis, we limit ourselves to a more simplified version by considering the effective one-band model. Firstly, we restrict to occupying the lowest Bloch band $a=1$, which is valid when the spectral gap is larger than all energy scales, as is the case at very low temperatures and sufficiently large lattice depths~\cite{sakurai,Zoller_tool}. Next, we consider the tight-binding approximation that maintains the tunneling terms between nearest neighbours~\cite{essler}. The overlap between two non-neighbouring Wannier functions is negligible, especially for deep lattices~\cite{bloch2008many}, and in turn, the tunneling amplitude decays exponentially with distance. With the same logic, we have that the interaction range is small such that the only relevant contribution from Equation~\eqref{eq:inter} is $U_{iiii}^{aaaa}$, fitting in with the narrative of having a contact interaction. Consequently, we may write the full SU($N$) Hubbard Hamiltonian~\cite{cazalilla_2009,gorshkov,cazalilla_2014}, which reads
\begin{equation}\label{eq:fhmsun}
    \mathcal{H}_{\mathrm{SU}(N)} = -t\sum\limits_{\langle i, j\rangle }^{L}\sum\limits_{\alpha = 1}^{N}(c_{i,\alpha}^{\dagger}c_{j,\alpha}+\mathrm{h.c.}) + \frac{U}{2}\sum\limits_{i}^{L}n_{i}(n_{i}-1),
\end{equation}
where $\langle i,j \rangle$ indicates that the summation runs over nearest-neighbours and $n_{i} = \sum_{\alpha}^{N}c_{i,\alpha}^{\dagger}c_{i,\alpha}$ is the local number operator counting the number of multi-occupied sites (sketched in Figure~\ref{fig:hublat} for $N=2$). Note that we have assumed isotropic couplings $t$ in the system and neglected on-site hopping terms $t_{ii}$. In this thesis, all energy scales are measured in units of the tunneling amplitude, which is given by $t=1$. 

\noindent The number of particles per component $N_{\alpha}$ and in turn, the total number of particles $N_{p}$ are conserved in the SU($N$) Hubbard model. This can either be seen from the commutation of the Hamiltonian with the operators, $[\mathcal{H},\hat{N}_{\alpha}] = 0$, defined respectively as $\hat{N}_{\alpha} = \sum_{j=1}^{L}n_{j,\alpha}$
or in a more intuitive way. There are no terms in model~\eqref{eq:fhmsun} that add ($c^{\dagger}_{i,\alpha}c^{\dagger}_{j,\beta}$) or remove ($c_{i,\alpha}c_{j,\beta}$) particles from the system, with ones that enable spin-exchange ($c_{i,\alpha}^{\dagger}c_{j,\beta}$) also being absent. \\

\noindent It is straightforward to show that the full Hamiltonian preserves the SU($N$) symmetry through its commutation with the SU($N$) spin-permutation operators. These operators can be defined as 
\begin{equation}\label{eq:spinmat}
    S^{a} = \frac{1}{2}\sum\limits^{L}_{j=1}\sum\limits_{\alpha,\beta = 1}^{N}c_{j,\alpha}^{\dagger} (\lambda^{a})^{\alpha}_{\beta}c_{j,\beta} \hspace{5mm} a = 1,\hdots, N^{2}-1,
\end{equation}
with $\sigma^{a}$ corresponding to the generators of the SU($N$) generators\footnote{For $N=2$, the generators correspond to the Pauli matrices. In the case of $N\! >\! 2$, the SU($N$) generators are a generalized form of the Gell-Mann matrices (see Appendix~\ref{app:Heisenberg}).}. We see by construction that these operators generate a representation of the SU($N$) Lie algebra. Through its commutation with these spin operators, $[\mathcal{H},S^{a}]=0$, the Hubbard Hamiltonian boasts SU($N$) symmetry. Structurally, the SU($N$) Hamiltonian is the same as the two-component model (setting $N=2$ in Equation~\eqref{eq:fhmsun}). The difference between the two models is subtle in that it lies in the enhanced symmetry of the former due to the $N\! >\! 2$ internal degrees of freedom\footnote{We will refer to the internal degrees of freedom as components, colours or species interchangeably.}, which is quite straightforward to show. Consequently, the SU($N$) Hubbard model displays a richer phase diagram for both repulsive and attractive interactions~\cite{cazalilla_2014,capponi}, which will be explored in this thesis.

\subsection{Integrability of the SU(\textit{N}) Hubbard model}\label{sec:bethesun}

The SU($N$) Fermi-Hubbard model exhibits several rich phenomena that can be attributed to the interplay between the kinetic and interaction terms, having opposing preferences tending to delocalize and localize electrons, respectively. They do not commute, resulting in Hamiltonian~\eqref{eq:fhmsun} not being diagonal in either the Bloch or Wannier basis except in the tight-binding ($U=0$) or atomic limit ($t=0$), respectively. Typically, to solve such a model, one would need to rely on perturbative or numerical methods to understand the underlying many-body physics of the system~\cite{montorsi_Hubbard}. However, this is not the case for the one-dimensional Hubbard model since it belongs to the special class of \textit{quantum integrable systems}~\cite{liebwu}. \\

\noindent Although the notion of integrability in quantum theory is less clear-cut than in classical physics, a well-posed definition can be formulated through scattering. In other words, constraints are imposed on the scattering of the system such that there is no diffraction in any scattering of the particles but a simple exchange of momenta~\cite{beautiful}. Effectively, this corresponds to a closed set of equations that, when solved, yield the spectrum of the model and enable the calculation of several physical properties exactly. Typically, these equations are obtained through the so-called \textit{Bethe ansatz}, originally developed by Bethe to solve the one-dimensional XXX Heisenberg model~\cite{bethe}. Following its introduction, this powerful method has been instrumental in providing the exact solution to a wide variety of systems, ranging from one-dimensional bosonic gases~\cite{liebliniger,lieb_exact}, fermionic gases both in the continuum~\cite{mcguire_rep,mcguire_att,lieb_flicker,gaudin,yangcontinuum,takahashiattractivesu2,sutherland1968,takahashisu3} and lattice~\cite{liebwu,sutherland1975}, to two-dimensional classical spin chains~\cite{sutherland1967,baxter1971} and even in string theory~\cite{Minahan_2003,beisert,stringtheory}.  \\

\noindent The logic behind the Bethe ansatz is common for all systems in that the many-body wavefunction is constructed through an educated guess, which in turn reduces the problem into solving a set of coupled algebraic equations. Generally, diagonalizing a matrix is a transcendental problem\footnote{In accordance with Galois theory, any polynomial of a degree greater than 5 corresponds to a transcendental function leading to a jump in complexity for matrices of this dimensionality~\cite{galois}.}. The route involving the Bethe ansatz equations has a lower complexity, at least in concept, as we can obtain the whole spectrum via algebraic means, enabling us to study physics in situations that can be accessed numerically within the current state-of-the-art.  Remarkably, the Bethe equations can be exactly solved in an explicit way in the thermodynamic limit~\cite{yangyang}.   \\

\noindent The SU(2) Hubbard model was found to be Bethe ansatz integrable in the seminal paper of Lieb and Wu~\cite{liebwu}, where they employed the nested Bethe ansatz approach that was introduced to solve the Gaudin-Yang model~\cite{gaudin,yangcontinuum}.  Unlike its SU(2) counterpart, the SU($N$) Hubbard model is not Bethe ansatz solvable for all system parameters and filling fractions~\cite{beautiful,capponi,frahm1995}. A key step in Bethe ansatz is that the many-particle problem can be factorised into two-body scatterings~\cite{korepin_bogoliubov_izergin_1993}. This concept, which is at the core of the nested Bethe ansatz formulated by Gaudin~\cite{gaudin} and Yang~\cite{yangcontinuum} to handle models with internal degrees of freedom, would go on to serve as the foundation for the Yang-Baxter equation, a sufficient condition for Bethe ansatz integrable systems~\cite{Takhtajan,sklyanin,Faddeev}. In the case of SU(2) fermions, this factorisation is guaranteed by the Pauli exclusion principle. On the other hand, up to $N$ particles can inhabit a given site when dealing with SU($N$) fermions resulting in diffractive scattering, in the sense that the scattering matrix does not obey the Yang-Baxter relations. Indeed, as $N\rightarrow\infty$, the Pauli exclusion principle relaxes, increasing the number of two-body interactions resulting in SU($N$) fermions emulating bosons~\cite{Pagano2014,schlottmann_spin,frahm1995} (see Figure~\ref{fig:manyballs}), which turns out to not be integrable~\cite{haldane,choy,choy_haldane}. \\

\noindent There exist two regimes where two-body interactions are assured, thereby preserving the integrability of model~\eqref{eq:fhmsun}. The \textit{first regime} is attained for filling fractions of one particle per site and very large repulsive values $U\gg t$~\cite{lai,sutherland1975}. In such a setting, which is modeled by the Lai-Sutherland Hamiltonian, the motion of the particles is constrained as they must pay a penalty to traverse to an already occupied site. The \textit{second regime} presents itself in continuum limit with vanishing lattice spacing, described by the Gaudin-Yang-Sutherland model~\cite{sutherland1968,Gu_yang}. An important property of the continuous limit is that the centre of mass dynamics separate from the relative coordinates. Such a property is lost in the lattice theories~\cite{naldesi2020enhancing}.
This is one of the key features explaining the lack of integrability of the lattice regularization of continuous theories. A heuristic way to understand why the system becomes integrable in the continuous limit (for both attractive and repulsive interactions) is to  note that the diluteness condition makes the  probability of more than two particles interacting vanish~\cite{frahm1995}. \\

\begin{figure}[h!]
    \centering
    \includegraphics[width=\linewidth]{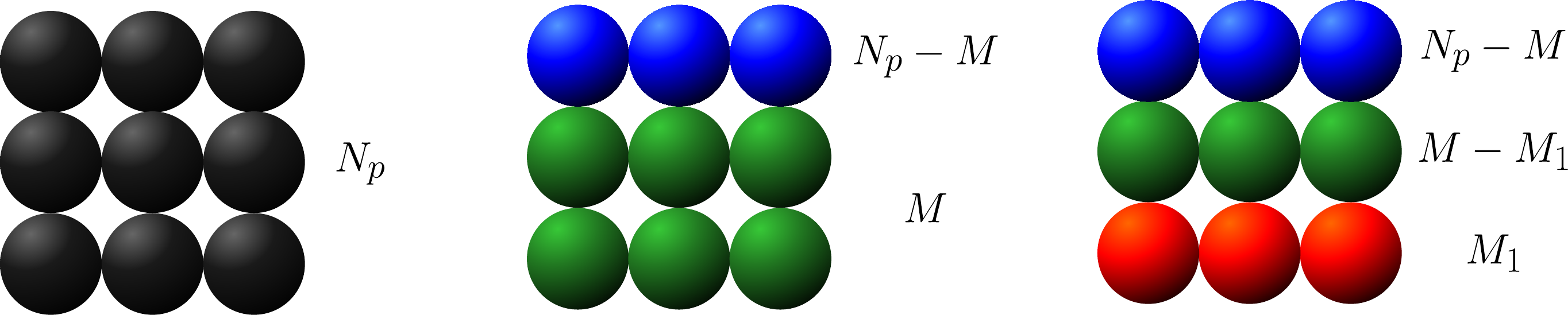}
    \caption{Diagrammatic representation of the logic behind the nested Bethe ansatz approach for SU(3) fermions. Starting out with 9 particles (black), we take 3 of them to be of a given colour (blue) with the rest being green. Subsequently, we take 3 of these to be of another colour(red). The final distribution of particles per colour is that of 3 each corresponding to $N_{p}-M$, $M-M_{1}$ and $M_{1}$.}
    \label{fig:su3visual}
\end{figure}

\noindent The models that govern these two integrable regimes were solved by Sutherland through successive applications of the Bethe-Yang hypothesis on the spin wavefunction coefficients, each time reducing the dimensionality of the problem~\cite{sutherland1968,sutherland1975}\footnote{It is for this reason that this approach is called the \textit{nested Bethe ansatz}.}.
In Appendix~\ref{sec:betheb}, the steps to derive the Bethe ansatz equations for the SU(2) Hubbard model are detailed for a system of $N_{p}$ particles with $M$ flipped spins. Here, we provide a brief overview of the approach for $N$-component fermions. To exemplify this, let us consider $N_{p}$ three-component fermions with $N-M$ particles in colour A, $M-M_{1}$ particles in colour B, and $M_{1}$ particles in colour C as depicted in Figure~\ref{fig:su3visual}. Just as in the two-component case discussed in Appendix~\ref{sec:betheb}, we initially take $M$ particles to be of a different type than the $N_{p}-M$ particles and administer the Bethe-Yang hypothesis for a given sector $Q\in S_{M}$, albeit in a slightly different form to consider the $M$ particles' permutations
\begin{equation}\label{eq:betheyangy}
    \Phi = \sum\limits_{P\in S_{M}} \psi [Q;P]\prod\limits_{n=1}^{M}F(\Lambda_{Pn}^{(1)},y_{Qn}),
\end{equation}
going through all the motions as in the two-component case, the key step being the re-arrangement of the coefficients $\psi[Q;P]$ to acquire the scattering operator (a square matrix of dimensions $M!$). Then, we separate the problem again by taking $M_{1}$ to be of another type different than the $M-M_{1}$ particles, allowing us to apply the Bethe-Yang hypothesis once more for $\psi([Q;P])$ with new spin rapidities $\Lambda_{\beta}^{(2)}$. The last step is the application of periodic boundary conditions giving us the Bethe ansatz equations for SU(3) fermions. This methodology can be straightforwardly generalized to $N$-component fermions and applied until all internal degrees of freedom are eliminated~\cite{schlottmann_spin,lee_schlottmann}. In the case of the Gaudin-Yang-Sutherland model, the Bethe equations for $N_{p}$ SU($N$) symmetric fermions read
\begin{equation}\label{eq:BAksun}
e^{\imath k_{j}L} = \prod\limits_{\alpha =1}^{M_{1}}\frac{k_{j}-\Lambda_{\alpha}^{(1)}+\imath c}{k_{j}-\Lambda_{\alpha}^{(1)}-\imath c},  \hspace{4mm} j=1,\hdots ,N_{p} 
\end{equation}
\vspace{-0.5em}
\begin{equation}\label{eq:BAlsun}
\prod_{\substack{\beta = 1 \\\beta\neq\alpha}}^{M_{r}} \frac{\Lambda_{\alpha}^{(r)} - \Lambda_{\beta}^{(r)} +2\imath c}{\Lambda_{\alpha}^{(r)} - \Lambda_{\beta}^{(r)} -2\imath c}  = \prod\limits_{\beta = 1}^{M_{r-1}} \frac{\Lambda_{\alpha}^{(r)} - \Lambda_{\beta}^{(r-1)} +\imath c}{\Lambda_{\alpha}^{(r)} - \Lambda_{\beta}^{(r-1)} -\imath c}\cdot \prod\limits_{\beta = 1}^{M_{r+1}} \frac{\Lambda_{\alpha}^{(r)} - \Lambda_{\beta}^{(r+1)} +\imath c}{\Lambda_{\alpha}^{(r)} - \Lambda_{\beta}^{(r+1)} -\imath c},\hspace{4mm}\alpha = 1,\hdots , M_{r}
\end{equation}
for $r=1,\hdots ,N\! -\! 1$, where $M_{0}=N_{p}$ and $\Lambda_{\beta}^{0} =k_{\beta}$.  $M_{r}$ corresponds to the number of particles in a given colour, with $k_{j}$ and $\Lambda_{\alpha}^{(r)}$ being the charge and spin rapidities respectively. Note that the Bethe equations for the Lai-Sutherland model are of the same structure, with the added difference that $k_{j}\rightarrow \sin k_{j}$ in the RHS of Equations~\eqref{eq:BAksun} and~\eqref{eq:BAlsun}. Naturally, for $N=2$ we recover the Bethe equations for spin-$\frac{1}{2}$ fermions (see Appendix~\ref{sec:betheb}). 

\newpage
\section{Experimental Aspects of cold SU(\textit{N}) fermions}\label{sec:exp}

In this section, we sketch the logic employed in trapping and controlling systems of cold atoms, with special consideration of SU($N$) fermionic systems. \\

\noindent One of the cornerstones of ultracold atoms experiments is the ability to confine and manipulate atoms in potentials. The realization of smooth, complex, and versatile trapping potentials  has led to the creation of various architectures, prompting an interesting interplay between theory and experiment~\cite{amico2022}. Such confining potentials can be engineered either by magnetic or optical fields~\cite{philips_nobel}.  \\

\noindent \textit{Magnetic trapping techniques} exploit the Zeeman coupling between an external magnetic field $\mathbf{B}$ and the atoms' internal state~\cite{magnets}. The potential corresponding to this interaction is of the form $V_{Z}(\mathbf{r}) = -\mathbf{\mu}\cdot\mathbf{B}(\mathbf{r})$ where $\mathbf{\mu}$ is the magnetic dipole moment of the atoms. In the presence of a homogenous magnetic field, the spins are prone to align themselves parallel ($V_{Z}<0$) or anti-parallel ($V_{Z}>0$) to $\mathbf{B}$~\cite{magtrap}. Therefore, by suitable design, atoms can be trapped in the minima of magnetic fields. Various magnetic trap designs can be crafted: bubbles and sheets via radiofrequency adiabatic potentials~\cite{zob_garraway,Garraway_2016}; stacks of rings (pancakes) and half-moon shapes through time-averaged adiabatic potentials~\cite{pancakes}; complex two-dimensional structures having H-, T- and U-shapes using current-carrying micro-fabricated wires on a substrate (atom chips)~\cite{folmanchips,micromagnets,schmiedmayer_trap}. \\

\noindent \textit{Optical trapping techniques} confine the atoms by an induced dipole interaction. The optical dipole potential generated from this interaction takes the form $U(\mathbf{r}) = -I(\mathbf{r})(3\pi c^{2}\Gamma)/(2w_{0}^{3}\Delta)$ where $I(\mathbf{r})$ corresponds to the intensity of the beam, $c$ is the speed of light in the vacuum and $\Gamma$ is the damping rate of the excited state's population~\cite{grimm}. The sign of the detuning $\Delta = \omega - \omega_{0}$ between the frequency of the laser field $\omega$ and the resonant frequency of the atom $\omega_{0}$ accounts if the dipole force is repulsive or attractive~\cite{miller_res}. Different geometries, such as a three-dimensional cubic grid to triangular lattices~\cite{Zoller_tool}, can be fashioned through spatial variation of the intensity. More complex structures can be crafted by considering more advanced techniques that enable the potential to be tailored in any desired form~\cite{GAUTHIER20211}. Such methods include: (i) ``painted'' time-averaged optical potential~\cite{Houston_2008,Henderson_2009,programmable}; (ii) spatial light modulators~\cite{boyer,Schnelle}; and (iii) digital micro-mirror devices~\cite{Gauthier_faces}.  \\

\noindent In this thesis, we are only concerned with ring-shaped geometries, which have been the subject of intensive investigation in the emerging field of atomtronics~\cite{amico2021,amico2022} and atom interferometry community~\cite{schmiedmayer_rev_2}. Rings traps can be created by both magnetic and optical potentials. Examples of magnetic traps include those constructed by wire structures like atom-chips~\cite{sauer,folman2002}, adiabatic~\cite{Garraway_2016,foot}, and time-averaged adiabatic potentials~\cite{pancakes,Navez_2016,Pandey2019}. In the case of optical traps, for instance, rings can be created by static~\cite{not_kevin,moulder,beattie} or spatial light modulator~\cite{Franke-Arnold,Amico2014slm} generated Laguerre-Gauss beams, painted potentials~\cite{Houston_2008,Henderson_2009,Schnelle}, and through digital micro-mirror devices~\cite{Gauthier_faces}. \\

\noindent Experiment realization of SU($N$) symmetric fermions is mainly carried out with alkaline earth-like atoms.  The absence of hyperfine interactions in the ground-state not only makes them the ideal candidates to investigate SU($N$) symmetry but also as state-of-the-art optical atomic clocks that vastly surpass the current standard of caesium clocks~\cite{cazalilla_2014,ludlow_clocks,Godun2021}. However, this lack of magnetic electronic structure presented some hurdles as some of the conventional methods used to cool, trap and manipulate alkaline atoms in experiments could not be used. \\

\noindent For instance, an all optical cooling setup is required to bring the atoms to quantum degeneracy.  In the standard procedure for cooling atoms, the atoms are loaded into the magneto-optical trap (MOT) where by illumination with red-detuned laser beams, they are cooled to an extent. Subsequently, the atoms are cooled even further by transferring to another magnetic trap, whose depth is changed to allow the ``hot'' atoms to escape the trap in what is known as evaporative cooling~\cite{raab1987,chu_nobellecture}. Seeing as a magnetic trap is reliant on an atom's internal state, it is unfeasible to trap alkaline earth atoms with magnetic means~\cite{yoshiro_degenerate}. As a result, an all optical cooling setup needs to be used, where the evaporative cooling is carried out in an optical trap that typically consists of a crossed optical dipole trap~\cite{Alloptical,not_my_campbell,Scazza2014}. Through this method, fermionic alkaline earth-like atoms have been brought to quantum degeneracy~\cite{yoshiro_degenerate,not_my_campbell,recoil,sonderhouse2020}\footnote{A striking consequence of the enlarged symmetry of SU($N$) fermions is that they can be cooled more efficiently than their SU(2) counterparts. This stems from the enhanced collisions between the internal degrees of freedom during the evaporative cooling, which allows entropy to be removed more efficiently~\cite{Taie2012,sonderhouse2020,hazzard_high,Taie2022}. This phenomenon is called the Pomeranchuk cooling effect~\cite{Taie2012}.}. \\

\noindent Magnetic Feshbach resonances, which are an indispensable tool in cold atoms experiments, cannot be utilised for the same reason outlined above, in that the susceptibility of the nuclear spin degrees of freedom to an external magnetic field is significantly less than their electronic counterparts. However, it is interesting to note that recently orbital Feshbach resonances have been observed~\cite{scazza_parish,fallani_strongly}. These types of resonances, which preserve SU($N$) symmetry, couple the orbital and nuclear spin degrees of freedom through a magnetic field in a similar fashion to the tuning of alkali gases with magnetic Feshbach resonances~\cite{Schfer2020}. Typically, the preferred way to tune the interactions in alkaline earth-like gases is through optical Feshbach resonances~\cite{charlie,takafesh}. The mechanism behind this involves coupling the colliding atomic pair to an excited molecular bound state via lasers. A major drawback of this method is that the SU($N$) symmetry can be broken upon coupling with an excited state possessing a hyperfine structure~\cite{cazalilla_2014}.


\chapter{Persistent currents in ultracold gases}\label{chp:persis}

\noindent Persistent currents are a quantized dissipationless flow of matter reflecting the phase coherence of the system~\cite{imry2002intro}. This purely quantum phenomenon defines a very active research area in fundamental and applied science, especially in the context of quantum sensing and simulation. \\

\noindent The existence of persistent currents was discovered in superconducting circuits at the turn of the 20$^{\mathrm{th}}$ century~\cite{tinkham2004introduction}. Initially associated with superconductivity, it was eventually observed that the two concepts are not intrinsically linked. The matter-wave current arises not due to the zero resistance of the device but from the macroscopic phase coherence in the system~\cite{supercoon,byers1961theoretical,onsager1961magnetic}. Decades later, it was predicted that matter-wave currents should also be observed in normal metals~\cite{trioper,nanobutt,nano2}: when a mesoscopic metal ring is threaded by a magnetic flux, a persistent current can arise in the system~\cite{gefen1984quantum} defining an instance of the Aharonov-Bohm effect in a closed loop~\cite{aharonov}.  In this case, the matter-wave current arises by virtue of the large coherence length of the electrons flowing in the ring with respect to the system length. Such a counter-intuitive phenomenon can only be observed in the quantum regime at very low temperatures where decoherence effects coming from thermal fluctuations are negligible. The magnitude of the persistent current in normal metallic rings is significantly lower than their superconducting counterparts owing to their origin. Moreover, the presence of impurities leads to decoherence, which, even though do not destroy the persistent current completely, reduce its signal dramatically. As a result, it turns out to be quite challenging to detect currents in normal metals due to their very small signals~\cite{magnetcopper,percurrmetobs}. \\

\noindent Soon after the first experiments on bosonic ultracold atoms~\cite{ketterle_bec,anderson_bec}, attention was devoted to the superfluid phenomenon and its consequences. Analogously to the persistent currents discovered in superconducting circuits, dissipationless matter-waves were also observed in superfluids. It is well known that dissipationless matter-waves can occur in superfluids~\cite{pethick_smith_2008,pitaevskii2016bose}, but ultracold atoms provide a platform to realize persistent currents with very specific capabilities that cannot be accessed by  solid state physics implementations. To start with, ultracold atoms have paradigmatically robust coherence properties and control of the physical conditions that can be changed in the course of the experiment `on the fly'~\cite{bloch2008many,Gauthier_faces}. Additionally, their quantum fluid can deal with  different atomic species. Indeed, the first experimental realizations of matter-wave currents of bosonic nature~\cite{phillips_current,ramanathan2011superflow,moulder2012quantized,kevin2013driving,neelystir}. Fermionic persistent currents have also been recently observed experimentally~\cite{wright2022persistent,roati2022imprinting}, enabling the investigation of currents from new angles compared to solid state systems. These new advances have opened new avenues for the investigation of matter-wave currents in exotic platforms such as Rydberg atoms~\cite{perciavalle2022controlled} and SU($N$) fermions~\cite{chetcuti2021persistent,chetcuti2021probe,chetcuti_interfer,osterloh2022exact,vqe}, which have no analogues in condensed matter. \\

\noindent Depending on the fundamental features of the cold atoms matter, persistent currents have been observed to display distinctive properties, reflecting important features of the systems~\cite{polo2020exact,polo2021quantum,naldesi2019rise,naldesi2020enhancing,pecci2021probing}. This fact, combined with the aforementioned enhanced control and flexibility, makes persistent currents a valuable tool for diagnosing many-body systems. Finally, after a decade of basic research activity, the field is now at a stage where technology based on this mesoscopic phenomenon is starting to surface. In particular, persistent currents are the core added value of Atomtronics, the newly emerged field of guided ultracold atoms technology~\cite{amico2021,amico2022}. A fruitful starting point in this context has been to consider devices in an analogy of quantum electronics~\cite{seaman2007atomtronics,progenitor}. The atomic analog of the SQUID~\cite{ryu2013experimental,aghamalyan2015coherent,haug2018readout,ryu2020quantum}, Sagnac interferometry~\cite{sagnaccorn,kim2022one}, or gyroscopes~\cite{kasevichgyro,katy2022matter}. Quantum devices based on persistent currents of cold atoms have the potential to define radically new quantum devices depending on the particular features of the system. Examples in this direction are the bright soliton interferometers~\cite{polosoliton,sagnaccorn}. \\

\noindent The rest of the chapter is organized in the following manner. Section~\ref{sec:coldmo} details the introduction of an artificial gauge field in a Hamiltonian, allowing matter to be put in motion. Section~\ref{sec:basicspec} is devoted to the basic properties of persistent currents. 

\section{Artificial gauge fields in cold atoms systems}\label{sec:coldmo}

\noindent A natural question that emerges is how to set these cold atoms in motion. Being charge-neutral in nature, ultracold atoms are not affected by Lorentz forces when subjected to a magnetic field. Nevertheless, cold atoms can emulate the behaviour of charged particles in a magnetic field through the application of a synthetic gauge field~\cite{dalibarflux,Goldman_2014}, which can be implemented through various techniques. One method of inducing rotation in the system is to stir the quantum fluid, typically carried out by a moving barrier~\cite{fetter_rev,kevin2013driving,neelystir,wright2022persistent}. The Coriolis force, which is the response to the applied rotation, mimics the action of the Lorentz force on a charged particle. Another approach is to utilize phase imprinting, where an arbitrary phase is imparted to the system through tailored time-dependent laser potentials~\cite{moulder,helenephaseimprint,roati2022imprinting}. Circulating current states in lattices can be prepared through Floquet engineering~\cite{Goldman2016}: in such a scheme, the confining potential is modulated periodically in time. By choosing a suitable modulation in the limit of a large driving frequency, the system is described by an effective time-independent target Hamiltonian. As such, one can engineer complex tunneling terms that correspond to a synthetic magnetic flux~\cite{maciejtuna,Atala2014}. Lastly, we point out that it was recently proposed to engineer persistent currents through machine-learning~\cite{haugmachine}. Through these methods, matter-wave currents in ultracold atomic gases have been experimentally realized in both bosonic~\cite{phillips_current,ramanathan2011superflow,wolf2021stationary} and, very recently, in fermionic systems~\cite{wright2022persistent,roati2022imprinting}. Below, we illustrate how synthetic gauge fields are introduced in the Hamiltonian, describing the system by providing a specific example of inducing a rotation by stirring the quantum fluid with a moving barrier. \\

\noindent Consider $N_{p}$ particles of mass $m$, that can be either bosonic or fermionic in nature, residing on a ring of radius $R$ interacting by a contact potential such that the system is described by the following Hamiltonian
\begin{equation}\label{eq:genhamm}
    \mathcal{H}_{0} = -\sum\limits_{j}^{N_{p}}\frac{\hbar^{2}}{2m}\frac{\partial^{2}}{\partial x_{j}^{2}} + g\sum\limits_{i<j}^{N_{p}}\delta(x_{i}-x_{j}),
\end{equation}
where $g$ corresponds to the interaction strength. To rotate the condensate, we introduce a time-dependent potential barrier denoted by $V(x-\Omega R t)$ moving at an angular velocity $\Omega$, such that 
\begin{equation}\label{eq:genhamm1}
    \mathcal{H}(\Omega,t) = H_{0} + V(x-\Omega Rt).
\end{equation}
By switching over to the rotating reference frame having the same frequency as the potential barrier, such that $\mathcal{H}_{\mathrm{rot}} = U^{\dagger}(t)\mathcal{H}(\Omega,t)U(t)$, the time dependency of the Hamiltonian is removed 
\begin{equation}\label{eq:hamrot}
    \mathcal{H}_{\mathrm{rot}} = \mathcal{H}_{0} + V(x) - \Omega\hat{L}_{z},
\end{equation}
with $U = \exp(\imath\hat{L}_{z}\Omega t/\hbar)$. The ring is taken to lie in the $x$-$y$ plane such that the $z$-component of the angular momentum denoted by $\hat{L}_{z}$ is perpendicular to it. Accordingly, the position of the $j$-th particle on the ring is given by the arc $R\varphi_{j}$ with $\varphi_{j}$ being the azimuthal angle on the ring. Translating the system from Cartesian to polar coordinates, $x = R\cos(\varphi)$ and $y=R\sin(\varphi)$, the angular momentum can be expressed as
\begin{equation}
    -\Omega \hat{L}_{z} = \imath\hbar \Omega \frac{\partial}{\partial\varphi},
\end{equation}
and in turn the Hamiltonian can be recast into the following form
\begin{equation}\label{eq:rothamm}
    \mathcal{H}_{\mathrm{rot}}= \sum\limits_{j=1}^{N_{p}}\frac{1}{2m}\bigg(\frac{-\imath\hbar}{R}\frac{\partial}{\partial \varphi_{j}} - m\Omega R\bigg)^{2} + \frac{g}{R}\sum\limits_{i<j}^{N_{p}}\delta (\varphi_{i}-\varphi_{j}) - \frac{1}{2}m\Omega^{2}R^{2}.
\end{equation}
The action of the induced rotation produces a shift in the momentum operator and the total energy of the system. At the single particle level, we can compare $\mathcal{H}_{\mathrm{rot}}$ with the Hamiltonian $\mathcal{H}_{c}$ describing a particle with charge $q$ and momentum $p=-\imath\hbar\partial_{x}$ subjected to a gauge field $A$
\begin{equation}\label{eq:hamlor}
    \mathcal{H} = \frac{(p-qA)^{2}}{2m}.  
\end{equation}
It becomes immediately apparent that the two expressions have a similar structure to one another with $qA = m\Omega R$. As such, the analogy that an artificial gauge field $m\Omega R$ generated through rotation mimics the action of an actual magnetic field holds. The angular velocity $\Omega$ that induces the rotation can be attributed to the Coriolis force and as such, is typically called the Coriolis flux. \\

\noindent Next, we showcase how these artificial gauge fields give rise to the persistent currents, which are the main subject of investigation in this thesis following the derivation for electrons in metallic rings carried out in~\cite{VIEFERS20041}. For simplicity, the single-particle case is considered. The time-independent Schr\"{o}dinger equation in the presence of a gauge field takes the following form
\begin{equation}\label{eq:hammm}
    -\frac{\hbar^{2}}{2m}\bigg( \frac{1}{R}\frac{\partial}{\partial\varphi} - \frac{\imath}{\hbar}m\Omega R\bigg)^{2}\psi_{n} (\varphi) = E_{n}\psi_{n}(\varphi).
\end{equation}
where $E_{n}$ denotes the $n$-th energy level and $\psi_{n}$ is the corresponding single-particle wavefunction. Multiplying both sides by $\psi_{n}^{*}$ and integrating over the whole space, we arrive to
\begin{equation}
    E_{n} = -\frac{\hbar^{2}}{2m}\int\limits_{0}^{2\pi}R\psi_{n}^{*}K^{2}(\varphi)\psi_{n},
\end{equation}
with $K$ is the kinetic term of the Hamiltonian in Equation~\eqref{eq:hammm}. Performing the derivative with respect to the flux and subsequently integrating by parts we find that~\cite{VIEFERS20041}
\begin{equation}
    \frac{\partial E_{n}}{\partial \Omega} = \frac{1}{2\pi R}\frac{\imath \hbar}{2m} \int\limits^{2\pi}_{0}R[\psi_{n}^{*}K\psi_{n} - \psi_{n}K^{*}\psi_{n}^{*}]\mathrm{d}\varphi = -\frac{1}{2\pi}\int\limits_{0}^{2\pi}j_{n}(\varphi)\mathrm{d}\varphi . 
\end{equation}
Note that the last term in the above expression $j_{n}(\varphi)$ corresponds to the textbook definition of the probability current in quantum mechanics~\cite{sakurai}, which is the analogue to an electric current in electromagnetism. This shows that persistent currents can be defined as the derivative of a thermodynamic potential with respect to the flux. In the canonical ensemble, the persistent current $I(\Omega)$ corresponds to 
\begin{equation}\label{eq:helmenergy}
    I(\Omega) = -\frac{\partial F(\Omega)}{\partial \Omega},
\end{equation}
where $F$ denotes the Helmholtz free energy~\cite{zvyagin}. At zero temperatures, only the ground-state level of the system is occupied and the current can be defined as 
\begin{equation}\label{eq:groundper}
    I(\Omega) = -\frac{\partial E_{0}}{\partial\Omega},
\end{equation}
with $E_{0}$ being the ground-state energy of the system. It must be stressed that even though the derivation for the persistent current was performed for the free particle case, one obtains the same result in the presence of interactions since the flux dependence in the Hamiltonian manifests itself in the kinetic operator of the many-body Hamiltonian. \\

\noindent So far, the flux dependence of the system was explicitly introduced in the Hamiltonian. However, one can perform a unitary transformation on the system by choosing a suitable gauge in order to obtain a Hamiltonian that is field free and the effect of the flux becomes encoded in the periodic boundary conditions, that now become ``twisted''. The unitary transformation under consideration is of the form 
\begin{equation}\label{eq:gauag}
    \mathcal{U}(2\pi) = \exp\bigg(-\frac{\imath}{\hbar}\int m\Omega R\, R\mathrm{d}\varphi\bigg).
\end{equation}
Upon applying this transformation, the Hamiltonian goes back to its field free form as in Equation~\eqref{eq:genhamm} and the wavefunction acquires a phase as it is makes a turn around the closed loop: $\psi(0) = \psi (2\pi)\mathcal{U}(2\pi)$. Such a phenomenon is due to the Aharonov-Bohm effect~\cite{aharonov}, a mesoscopic phenomenon for which a particle acquires a phase shift as it travels in a closed path. For our case, of a particle moving around a closed loop, we find that the Aharonov-Bohm phase is
\begin{equation}\label{eq:phaseaha}
    -\frac{\imath}{\hbar}\Delta\theta = -\frac{\imath}{\hbar} \oint\limits_{2\pi} \tilde{A}\, R\mathrm{d}\varphi = 2\imath\pi\frac{\Omega}{\Omega_{0}}
\end{equation}
with $\Delta\theta$ being the phase difference in the wavefunction at the boundary condition and $\tilde{A} = m\Omega R$. The quantity $\Omega_{0} = \hbar/(mR^{2})$ is the elementary (bare flux quantum). In this thesis, we work in the rotating reference frame unless explicitly stated. Additionally, the flux threading the system is denoted by $\phi$ and the corresponding bare flux quantum by $\phi_{0}$, which is taken be be equal to 1. 

\subsection{Peierls substitution}

\noindent The main system considered in this thesis is described by the SU($N$) Hubbard Hamiltonian, which is a lattice model. Therefore, in the following we will describe how to incorporate an effective magnetic flux into a lattice description. Neglecting interactions, the one-body Hamiltonian subjected an artificial gauge field $\tilde{A}$ reads
\begin{equation}
    \mathcal{H} = \sum\limits_{j=1}^{N_{p}} \frac{(p_{j}-\tilde{A})^{2}}{2m},
\end{equation}
In the presence of a lattice, the wavefunction can be expressed in terms of $\Psi(x) = \sum_{j}w_{j}(x)a_{j}$ where $a_{j}$ corresponds to the annihilation operator on site $j$ and $w_{j}(x)$ is the Wannier function discussed in Chapter~\ref{chp:tools}. The flux dependence of the Hamiltonian can be gauged away by introducing the following transformation into the Wannier function
\begin{equation}
    \tilde{w}_{j}(x) = \exp\Bigg(\frac{\imath}{\hbar}\int\limits_{x_{j}}^{x}\tilde{A}(x')\mathrm{d}x'\Bigg)w_{j}(x),
\end{equation}
where $x_{j}$ is the position of lattice site $j$. Therefore, we start by calculating
\begin{equation}
\mathcal{H}\tilde{w}_{n}(x) = \Bigg (\sum\limits_{j=1}^{N_{p}} \frac{(p_{j}-\tilde{A}(x))^{2}}{2m}\Bigg)\exp\Bigg(\frac{\imath}{\hbar}\int\limits_{x_{n}}^{x}\tilde{A}(x')\mathrm{d}x'\Bigg)w_{n}(x),
\end{equation}
and making use of the fact that $p_{j} = -\imath\hbar\partial_{x_{j}}$, we find that 
\begin{equation}
\mathcal{H}\tilde{w}_{n}(x) = \exp\Bigg(\frac{\imath}{\hbar}\int\limits_{x_{n}}^{x}\tilde{A}(x')\mathrm{d}x'\Bigg)\Bigg (\sum\limits_{j=1}^{N_{p}} \frac{(p_{j}-\tilde{A}(x) +\tilde{A}(x))^{2}}{2m}\Bigg)w_{n}(x),
\end{equation}
such that the flux dependence of the Hamiltonian is gauged away. The effective hopping amplitudes of the lattice are given by 

\begin{align}
\int \tilde{w}_{l}^{*}(x)\mathcal{H}\tilde{w}_{n}(x)\mathrm{d}x =\int w_{l}^{*}(x)\exp\Bigg(   -\frac{\imath}{\hbar}\int\limits_{x_{l}}^{x}\tilde{A}(x')\mathrm{d}x' + \frac{\imath}{\hbar}\int\limits_{x_{n}}^{x}\tilde{A}(x')\mathrm{d}x'\Bigg)\mathcal{H}_{0}w_{n}(x)\mathrm{d}x, \nonumber \\
=\exp\Bigg(\frac{\imath}{\hbar}\int\limits_{x_{n}}^{x_{l}}\tilde{A}(x')\mathrm{d}x'\Bigg)\int w_{l}^{*}(x) \exp\Bigg( \frac{\imath}{\hbar}\oint\limits_{x_{n}\rightarrow x\rightarrow x_{j} \rightarrow x_{n}}A(x')\mathrm{d}x\Bigg)\mathcal{H}_{0}w_{n}(x)\mathrm{d}x,
\end{align}
with $\mathcal{H}_{0} = \sum_{j}\frac{p_{j}^{2}}{2m}$. In the case where the flux threading the system is sufficiently uniform at the atomic scale, the closed path integral is approximately zero.  Consequently, we have that 
\begin{align}\label{eq:hoppeier}
\int \tilde{w}_{l}^{*}(x)\mathcal{H}\tilde{w}_{n}(x)\mathrm{d}x 
=\exp\Bigg(\frac{\imath}{\hbar}\int\limits_{x_{n}}^{x_{l}}\tilde{A}(x')\mathrm{d}x'\Bigg)\int w_{l}^{*}(x) \mathcal{H}_{0}w_{n}(x)\mathrm{d}x .
\end{align}
The above expression corresponds to that of the hopping amplitude introduced in Chapter~\ref{chp:tools} albeit with an extra complex phase factor
\begin{equation}
    t_{ij} = t_{ij}\exp\Bigg(\frac{\imath}{\hbar}\int\limits_{x_{i}}^{x_{j}}A(x')\mathrm{d}x'\Bigg).
\end{equation}
This phase factor is called the Peierls phase factor and the technique employed to get it is known as the Peierls substitution~\cite{Peierls1933,essler}. The tunneling amplitudes considered here are for hopping between arbitrary lattice sites. In this thesis, we only consider isotropic nearest-neighbour hopping amplitudes such that $x_{j} = x_{i}+1$. \\

\noindent For the SU($N$) Hubbard model under consideration describing a system of $N_p$ fermions with SU($N$) symmetry residing in a ring-shaped lattice composed of $L$ sites threaded with a magnetic flux $\phi$ 
\begin{equation}\label{eq:fluxHuba}
\mathcal{H}_{\textrm{SU}(N)} = -t\sum_{j=1}^L\sum\limits_{\alpha = 1}^{N}\big (e^{\imath \frac{2\pi\phi}{L}}c^{\dagger}_{\alpha ,j}c_{\alpha ,j+1} + \textrm{h.c.}\big ) + U\sum\limits_{j}^{L}\sum\limits_{\alpha <\beta}^{N}n_{j,\alpha}n_{j,\beta},
\end{equation}
the effective magnetic field is realized by performing the Peierls substitution $t\rightarrow t e^{\imath \frac{2\pi\phi}{L}}$. For lattice models, one can compute the persistent current either through the numerical derivative of Equation~\eqref{eq:groundper} or through the current operator obtained through the Hellmann-Feynmann theorem\footnote{$\mathrm{d}E/\mathrm{d}\phi = \langle \psi(\phi)| \mathrm{d}\mathcal{H}(\phi)/\mathrm{d}\phi|\psi(\phi)\rangle$} and evaluated on the ground-state of Equation~\eqref{eq:fluxHuba}.
\begin{equation}\label{eq:newcurr}
    \hat{I}(\phi) = \frac{2\imath\pi t}{L}\sum\limits_{j=1}^{L}\sum\limits_{\alpha=1}^{N}(e^{2\imath\pi\phi /L}c_{j,\alpha}^{\dagger}c_{j+1,\alpha} - \mathrm{h.c.}),
\end{equation}
with the factor $1/L$ accounting for the persistent current's mesoscopic nature. Note that in lattice systems, the center of mass and relative coordinates are no longer decoupled as opposed to their continuous counterparts leading to particular lattice effects~\cite{polo2020exact,naldesi2020enhancing}.

\section{Properties of persistent currents in the non-interacting regime}\label{sec:basicspec}

\noindent Here we discuss the basic properties of persistent currents in non-interacting fermionic systems. This will set the stage for the rest of this thesis, where we will be considering the effects of particle interactions both in the attractive and repulsive regimes. \\

\noindent As discussed in this chapter, persistent currents can be understood in terms of the energy landscape as a function of the effective magnetic flux. By considering the gauge transformation described in Equation~\eqref{eq:gauag}, one can readily observe that the eigenstates of the Hamiltonian are given by plane waves with the added constraint imposed by the twisted boundary conditions. The momentum associated to these eigenstates is given by 
\begin{equation}
    k_{n} = \frac{2\pi}{L_{R}}(n-\phi),
\end{equation}
and therefore the corresponding energy reads
\begin{equation}\label{eq:enflux}
    E = \sum\limits_{\{n\}}\frac{\hbar^{2}}{2m}\bigg[\frac{2\pi}{L_{R}}(n-\phi)\bigg]^{2},
\end{equation}
with $L_{R}$ denoting the length of the ring and $\{n\}$ being the set of quantum numbers for spinless fermions~\cite{takahashi2005thermodynamics}. The quantum numbers $n$ are related to the charge quantum numbers $I_{j}$ (see Appendix~\ref{sec:betheb}) in the following manner: $I_{j}=n$ and $I_{j}=n+\frac{1}{2}$ for systems with an odd and even number of particles respectively. The sum of these quantum numbers, which is always integer by construction, is related to the total angular momentum perpendicular to the ring's plane denoted by $\ell$ per particle such that $N\frac{\sum_{j}I_{j}}{N_{p}}=\ell$. 
\begin{figure}[h!]
    \centering
    \includegraphics[width=\linewidth]{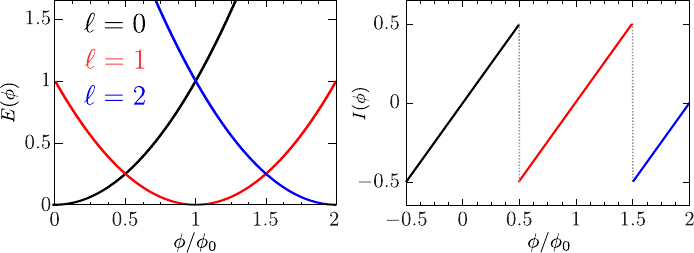}
    \put(-250,140){(\textbf{a})}
    \put(-30,140){(\textbf{b})}
    \caption{(\textbf{a}) Schematic of the single-particle energy $E(\phi)$ as a function of the effective magnetic flux $\phi$. As one traverses from one parabola to the next with increasing $\phi$, the angular momentum quantum number $\ell$ of the system increases to minimize the ground-state energy. (\textbf{b}) The corresponding persistent current $I(\phi)$ as a function of flux. The jumps in the current coincide with the crossings between the different parabolas.}
    \label{fig:eneryU0exact}
\end{figure}

\noindent The energy spectrum of the system as a function of the effective magnetic flux is presented in Figure~\ref{fig:eneryU0exact}. The spectrum is periodic with the flux $\phi$ having a period fixed by the elementary flux quantum $\phi_{0}$, analogous to the Bloch theorem of particles in a periodic lattice. Such a theorem is originally due to Leggett, stating that the persistent current is dictated by the effective flux quantum and is independent of disorder~\cite{leggett1991}. As the flux threading the system increases, the set of quantum numbers $\{n\}$ shifts such that the energy is minimized. This shift in the quantum numbers occurs precisely at the level crossings between parabolas with different $\ell$. Such a change is also reflected as jumps in the persistent current, which displays a characteristic sawtooth behaviour --Figure~\ref{fig:eneryU0exact}(\textbf{b}). 
\begin{figure}[h!]
    \centering
    \includegraphics[width=\linewidth]{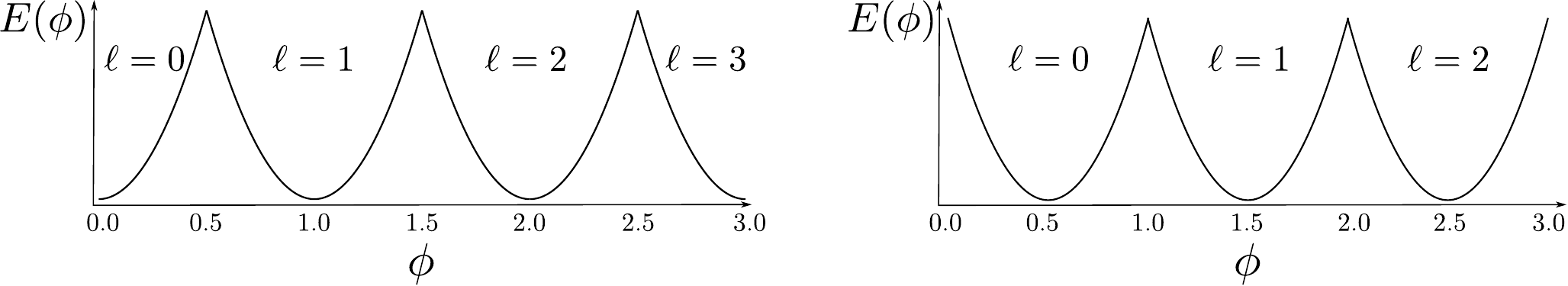}
    \caption{Energy as a function of the effective magnetic flux, denoted by $E$ and $\phi$ respectively, for systems with $N_{p} = (2n+1)$ (left) and $N_{p} = (2n)$ (right) for integer $N$. The difference between the left and right panels stems from the parity of the system, which is diamagnetic and paramagnetic respectively depending on whether the ground-state energy increases or decreases with the flux $\phi$. The degeneracy point $\phi_{d}$, which is the point where two parabolas cross, is at (half-odd) integer values for (diamagnetic) paramagnetic systems. Figure adapted from~\cite{chetcuti_interfer}. }
    \label{fig:barb}
\end{figure}

\begin{figure}[h!]
    \centering
    \includegraphics[width=0.6\linewidth]{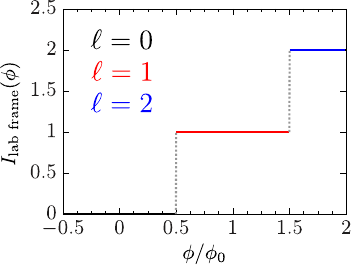}
    \caption{Persistent current $I(\phi)$ as a function of the effective magnetic flux in the lab frame.  Jumps in $I(\phi)$ correspond to the change of angular momentum in the system. }
    \label{fig:eneryU0exactlab}
\end{figure}

\noindent In the case of spinless fermions, the degeneracy point $\phi_{d}$ corresponding to these level crossings can occur either at integer or half-odd integer values of the flux depending on the even/odd parity of the system respectively. This is another facet of the Leggett theorem, which states that the parity of the energy and in turn the persistent current, is diamagnetic [paramagnetic] for systems $N_{p} = (2n+1)$ [$N_{p} = (2n)$] particles for integer $n$. For bosons, the parity of the current is always diamagnetic~\cite{naldesi2019rise}. Diamagnetic (paramagnetic) systems are characterized by whether the ground-state energy increases (decreases) on increasing the flux --Figure~\ref{fig:barb}. We note that the energy spectrum in the presence of the lattice behaves in the same manner and can be straightforwardly checked through discrete Fourier transformation into momentum space, where any translationally invariant one-body Hamiltonian is diagonal in the Bloch basis~\cite{essler}. \\

\noindent When going to the lab frame, the quantized nature of the persistent current is displayed by a characteristic step-like behaviour shown in Figure~\ref{fig:eneryU0exactlab}. Such a behaviour has been reported in ultracold atoms experiments with remarkable precision~\cite{kevin2013driving}.  Additionally, it is important to point out the difference in the energy landscapes between the rotating and lab frame. Whilst in the former the energy is periodic, in the latter the energy increases on going to larger angular momenta presenting local minimas at integer/half-odd integer flux values depending on the parity of the system~\cite{VIEFERS20041}. Such a feature reflects that the condensate becomes less stable at larger values of the angular momentum, which will be tackled briefly in Chapter~\ref{chp:interfer}.


\chapter{Atomtronic circuits with repulsive SU(\textit{N}) fermionic matter}\label{chp:repcurr}

\noindent Atomtronics is the technology of matter-wave circuits of ultracold atoms~\cite{amico2021,amico2022}. Most of the studies so far have been devoted to atomtronic circuits of ultracold bosons, while ones comprised of ultracold fermions still require extensive investigation. However, recent advances in the field have broken new ground with the experimental realization of atomtronic circuits of two-component fermions~\cite{wright2022persistent,roati2022imprinting}. This thesis aims to expand the scope of fermionic atomtronic circuits, with quantum fluids comprised of the SU($N$) fermionic matter discussed in Chapter~\ref{chp:tools}. These strongly interacting $N$-component fermions, as provided by alkaline earth-like gases, are very relevant both for high-precision measurement~\cite{ludlow_clocks,poliprecision,poli2013optical} and to enlarge the area of cold atoms quantum simulators of many-body systems~\cite{livi2016synthetic,kolkowitz2017spin,scazza2016}, which is in line with the recent research activity in atomtronics~\cite{amico2021,amico2022}. \\

\noindent In this chapter, we initiate this venture by focusing on SU($N$) fermions with repulsive interactions trapped in a ring-shaped circuit of mesoscopic size~\cite{imry2002intro} and pierced by an artificial gauge field. The persistent current, which is the response to this applied field, provides a standard avenue to probe the coherence of the system~\cite{gefen1984quantum}. Here, we analyze the specific dependence of the persistent current's quantization properties on the parameters characterizing the physical conditions of the system. Different regimes depending on the filling fractions are explored. Specifically, we consider the incommensurate and commensurate fillings since, in these regimes, the numerical results of the system can be monitored with the exact Bethe ansatz analysis~\cite{sutherland1968,sutherland1975}. \\

\noindent The chapter is structured as follows. In Section~\ref{sec:mod3}, the model and methods are introduced. Section~\ref{sec:betherep} tackles the Bethe ansatz of the model in the strong repulsive regime, where it becomes markedly simplified. Sections~\ref{sec:incomm} and~\ref{sec:comm} are devoted to the results achieved for the incommensurate and commensurate filling regimes. Specific parity effects of the current are handled in Section~\ref{sec:parity3}. Conclusions are presented in closing Section~\ref{sec:conc3}.

\newpage

\section{Model and methods}\label{sec:mod3}
A system of $N_p$ fermions with SU($N$) symmetry residing in a ring-shaped lattice composed of $L$ sites threaded with a magnetic flux $\phi$ can be modeled using the SU($N$) Hubbard model
\begin{equation}\label{eq:fluxHub}
\mathcal{H}_{\textrm{SU}(N)} = -t\sum_{j=1}^L\sum\limits_{\alpha = 1}^{N}\big (e^{\imath \frac{2\pi\phi}{L}}c^{\dagger}_{\alpha ,j}c_{\alpha ,j+1} + \textrm{h.c.}\big ) + U\sum\limits_{j}^{L}\sum\limits_{\alpha <\beta}^{N}n_{j,\alpha}n_{j,\beta}.
\end{equation}
In what follows, the energy scale is given by setting the hopping strength $t=1$ and the on-site interaction to $U>0$ to account for the repulsive interactions. The effective magnetic field is realized by performing the Peierls substitution $t\rightarrow t e^{\imath \frac{2\pi\phi}{L}}$ outlined in Chapter~\ref{chp:persis} into the SU($N$) Hubbard Hamiltonian defined in Equation~\eqref{eq:fhmsun}. \\

\noindent For $N =2$, the Hubbard model describing spin-$\frac{1}{2}$ fermions is obtained. In this case, the Hamiltonian (\ref{eq:fluxHub}) is integrable by Bethe Ansatz for any $U/t$ and filling fractions $\nu=N_{p}/L$~\cite{liebwu}. On the other hand, as we discussed in Chapter~\ref{chp:tools}, the Bethe ansatz integrability for $N\! >\! 2$ is preserved only in two regimes. The first is in the continuous limit of vanishing lattice spacing governed by the Gaudin-Yang-Sutherland model describing SU($N$) fermions with delta interactions~\cite{sutherland1968,takahashisu3,Guan2013Fermi}, with such a regime achieved by model~\eqref{eq:fluxHub} in the dilute limit of small $\nu\ll 1$. The other integrable regime of model~\eqref{eq:fluxHub} is obtained for filling fractions $\nu = 1$ of one particle per site and large repulsive values of $U\gg t$, for which the system is modeled by the Lai-Sutherland Hamiltonian~\cite{sutherland1975,lai,capponi}. \\

\noindent According to the general theory of Bethe ansatz solvable models, the spectrum is obtained through the solution of coupled transcendental equations, parameterised by a specific set of numbers called the quantum numbers~\cite{takahashi2005thermodynamics}. For the specific case of the integrable SU($N$) Hubbard model, the many-body excitations are customarily labeled by the  quantum numbers: $I_{j}, \; j = 1,\hdots, N_{p}$ and $J_{\beta_{l}}, \; \beta_{l}=1,\hdots, M_{l}\hspace{2mm}\mathrm{for}\hspace{2mm} l = 1,\hdots, N\! -\! 1$, where $I_{j}$ and $J_{\beta_{l}}$ are the charge and spin quantum numbers respectively, and $M_{l}$ refers to the number of particles with a given component~\cite{essler,takahashisu3,sutherland1968} (see Section~\ref{sec:betherep}). At zero flux, the ground-state is found to be characterized by quantum numbers with consecutive sequences: $I_{j} = I_{1}, I_{1}+1,I_{1}+2,\hdots I_{1}+N_{p}$ and $J_{\beta_{l}} = J_{\beta_{l}} , J_{\beta_{l}} +1,J_{\beta_{l}} +2,\hdots J_{\beta_{l}} +M_{l}$. Instead, for non-vanishing flux, we shall see that the quantum numbers configurations $\{ I_{j}, J_{\beta_{l}} \}$ can change. Any variations of the quantum numbers from their ground-state configurations results in the creation of excitations: sequences of $I_{j} = I_{1}-1,{\atop\vee},I_{1}+1,I_{1}+2,\hdots I_{1}+N_{p}$ and $J_{\beta_{l}} = J_{\beta_{l}} -1,{\atop \vee}, J_{\beta_{l}} +1,J_{\beta_{l}} +2,\hdots J_{\beta_{l}} +M_{l}$ with `holes' corresponding  to excitations; in particular holes in $\{J_\beta\}$  characterize the so-called spinon excitations~\cite{andrei1995integrable}. For SU($N$) fermions, there can be $N\!-\! 1$ different types of such spinon states~\cite{frahm1995,lee_schlottmann} as opposed to one type for SU(2). Indeed, the Hubbard models for SU($2$) and SU($N$) fermions enjoy very different physics. For incommensurate fillings, a metallic behaviour is found with characteristic oscillations of the spin-spin and charge correlation functions that, for $N\! >\! 2$ can be coupled to each other.  At integer filling fractions $\nu =1$, fermions may be in a Mott phase. Such a phase is suppressed only exponentially for $N=2$~\cite{liebwu}; in striking contrast, for $N\! > \! 2$ the system displays a Mott transition for a finite value of $U/t$~\cite{cazalilla_2014,capponi,xu2018interaction}.

\noindent At mesoscopic size, the properties discussed above are displayed as specific traits~\cite{imry2002intro}. By mesoscopic effects, we refer to those  arising on length scales that are comparable with the particles’ coherence length. In this regime, even though the application of the magnetic flux does not change the nature of the possible excitations, as we shall see that the latter may be indeed promoted to ground-states. Our diagnostic tool is the persistent current, providing access to the particles' phase coherence~\cite{gefen1984quantum, shastry1990twisted}. At zero temperature, the persistent current of the system is given by $I(\phi ) = -\frac{\partial E_{0}}{\partial\phi}$ where $E_0$ is the ground-state energy (see Chapter~\ref{chp:persis}). The persistent current is of mesoscopic nature in that it corresponds to  $1/L$ corrections of the ground-state energy~\cite{gefen1984quantum}. For a quantum system in a ring, the angular momentum is quantized. Accordingly, the persistent current displays a characteristic sawtooth behaviour, with a periodicity that Leggett proved to be fixed by the effective flux quantum of the system~\cite{leggett1991}. This theorem is important in our approach since the periodicity of the persistent current reflects the structure of the ground-state. In the case of a gas of non-interacting particles, the effective flux quantum is the bare flux quantum $\phi_{0}$; while for a BCS ground-state, the period is halved due to the formation of Cooper pairs~\cite{byers1961theoretical,onsager1961magnetic}. Similarly, for bosonic systems, a persistent current with a period of $1/N_{p}$, has been found, indicating the formation of a bound state of $N_{p}$ particles~\cite{polo2020exact,naldesi2020enhancing,polo2021quantum}. \\

\noindent In our approach, we combine exact diagonalization or DMRG~\cite{whitedmrg,itensor} results with, whenever possible, those obtained from Bethe ansatz. Specifically, in the integrable regimes of dilute systems and a filling of one particle per site and large interactions, the Bethe ansatz results (through the Bethe quantum numbers introduced above) are exploited as bookkeeping to monitor the eigenstates provided by the numerical results. This way, the nature and physical content of the system's ground-state can be established as functions of the parameters. Here, systems with an equal number of particles per species are considered. 

\subsection{Bethe ansatz in the limit of large repulsive interactions}\label{sec:betherep}

\noindent In the continuous limit, the SU($N$) Hubbard model tends to the Gaudin-Yang-Sutherland Hamiltonian describing $N$-component fermions with a delta interaction of strength $c$~\cite{sutherland1968,decamp2016high}, which reads
\begin{equation}~\label{eq:GYSham}
\mathcal{H}_{GYS} = \sum\limits_{m =1}^{N}\sum\limits_{i=1}^{N_{m}}\bigg (-i\frac{\partial}{\partial x_{i ,m}} - \frac{2\pi}{L_{R}}\phi\bigg )^{2} + 4c\sum\limits_{m<n}^{N}\sum\limits_{i,j}\delta(x_{i ,m} -x_{j ,n}),
\end{equation}
where $N_{m}$ is the number of electrons with colour $\alpha$ of  with $m = 1,\dots N$, and $L_{R}$ being the size of the ring. The model is integrable by Bethe ansatz through Equations~\eqref{eq:BAksun} and~\eqref{eq:BAlsun}, which in the presence of an effective magnetic flux read:
\begin{equation}\label{eq:BAksunflux}
e^{\imath (k_{j}L_{R}-\phi)} = \prod\limits_{\alpha =1}^{M_{1}}\frac{k_{j}-\Lambda_{\alpha}^{(1)}+\imath c}{k_{j}-\Lambda_{\alpha}^{(1)}-\imath c}  \hspace{4mm} j=1,\hdots ,N_{p},
\end{equation}
\begin{equation}\label{eq:BAlsunflux}
\prod_{\substack{\beta = 1 \\\beta\neq\alpha}}^{M_{r}} \frac{\Lambda_{\alpha}^{(r)} - \Lambda_{\beta}^{(r)} +2\imath c}{\Lambda_{\alpha}^{(r)} - \Lambda_{\beta}^{(r)} -2\imath c}  = \prod\limits_{\beta = 1}^{M_{r-1}} \frac{\Lambda_{\alpha}^{(r)} - \Lambda_{\beta}^{(r-1)} +\imath c}{\Lambda_{\alpha}^{(r)} - \Lambda_{\beta}^{(r-1)} -\imath c}\cdot \prod\limits_{\beta = 1}^{M_{r+1}} \frac{\Lambda_{\alpha}^{(r)} - \Lambda_{\beta}^{(r+1)} +\imath c}{\Lambda_{\alpha}^{(r)} - \Lambda_{\beta}^{(r+1)} -\imath c} \hspace{4mm}\alpha = 1,\hdots , M_{r},
\end{equation}
for $r = 1,\hdots ,N\! -\! 1$ where $M_{0} = N_{p}$, $M_{N}=0$ and $\lambda_{\beta}^{(0)} = k_{\beta}$. $M_{r}$ corresponds to the colour with $k_{j}$ and $\Lambda_{\alpha}^{(r)}$ being the charge and spin momenta respectively. The energy of the system is given by $E = \sum_{j}^{N_{p}}k_{j}^{2}$. The flux dependence in the Bethe equations, is obtained by applying the twisted boundary conditions~\cite{shastry1990twisted} discussed in Chapter~\ref{chp:persis} in place of the typical periodic boundary conditions illustrated in Appendix~\ref{sec:betheb}\footnote{It must be stressed that the flux is taken to be independent of the particles' colour and so it only enters in the first equation. If the twist in the wavefunction is colour dependent, the solvability of the model would not be assured~\cite{luigitwist}.}. Taking the SU(3) case as an example, one obtains a set consisting of three non-linear equations\footnote{For the sake of conveniency, we changed $\Lambda_{\alpha}^{(1)}$ and $\Lambda_{\alpha}^{(2)}$ to $\Lambda_{\alpha}$ and $\lambda_{a}$.}
\begin{equation}\label{eq:chargesu3rep}
e^{\imath( k_{j}L_{R} - 2\pi\phi)} = \prod\limits_{\alpha = 1}^{M_{1}} \frac{4( k_{j}-\Lambda_{\alpha}) +\imath U}{4(k_{j}-\Lambda_{\alpha}) -\imath U} \hspace{3mm} j=1,\hdots, N_{p},
\end{equation}
\vspace{-1em}
\begin{equation}\label{eq:spin1su3rep}
\prod\limits_{\beta \neq\alpha}^{M_{1}} \frac{2(\Lambda_{\alpha}-\Lambda_{\beta}) +\imath U}{2(\Lambda_{\alpha}-\Lambda_{\beta}) -\imath U} = \prod\limits_{j =1}^{N_{p}} \frac{4(\Lambda_{\alpha}-k_{j}) +\imath U}{4(\Lambda_{\alpha}- k_{j}) -\imath U} \prod\limits_{a =1}^{M_{2}} \frac{4(\Lambda_{\alpha}-\lambda_{a}) +\imath U}{4(\Lambda_{\alpha}-\lambda_{a}) -\imath U} \hspace{3mm} \alpha = 1,\hdots, M_{1},
\end{equation}
\begin{equation}\label{eq:spin2su3rep}
\prod\limits_{b \neq a}^{M_{2}} \frac{2(\lambda_{a}-\lambda_{b}) +\imath U}{2(\lambda_{a}-\lambda_{b}) -\imath U} = \prod\limits_{\alpha =1}^{M_{1}} \frac{4(\lambda_{a}-\Lambda_{\alpha}) +\imath U}{4(\lambda_{a}-\Lambda_{\alpha}) -\imath U} \hspace{3mm} a = 1,\hdots,M_{2}.
\end{equation}
Note that we opted to take $c=U/4t$ in the Bethe equations above, such that the results could be matched with the numerical ones. The relation between the interaction strengths in the two limits is straightforwardly obtained during the mapping (see Appendix~\ref{sec:mappinggys}). Subsequently by taking the logarithm of Equations~\eqref{eq:chargesu3rep} through~\eqref{eq:spin2su3rep} and noting that 
\begin{equation}
  2 i \arctan x= \pm\pi + \ln  \frac{x-i}{x+i},
\end{equation}
the Bethe equations can be recast into the following form
\begin{equation}\label{eq:logrep1}
 k_{j}L_{R} +2 \sum\limits_{\alpha=1}^{M_{1}} \arctan \bigg[\frac{4( k_{j}-\Lambda_{\alpha})}{U}\bigg] = 2\pi (I_{j}+\phi),
\end{equation}
\vspace{-1.5em}
\begin{equation}\label{eq:logrep2}
2\sum\limits_{j=1}^{N_{p}}\arctan \bigg[\frac{4(\Lambda_{\alpha}- k_{j})}{U}\bigg] +2 \sum\limits_{a=1}^{M_{2}}\arctan \bigg[\frac{4(\Lambda_{\alpha}-\lambda_{a})}{U}\bigg] -2 \sum\limits_{\beta=1}^{M_{1}} \arctan\bigg[\frac{2(\Lambda_{\alpha}-\Lambda_{\beta})}{U}\bigg]= 2\pi J_{\alpha},
\end{equation}
\vspace{-1em}
\begin{equation}\label{eq:logrep3}
2 \sum\limits_{\beta=1}^{M_{1}}\arctan \bigg[\frac{4(\lambda_{a}-\Lambda_{\beta})}{U}\bigg]- 2 \sum\limits_{b =1}^{M_{2}}\arctan \bigg[\frac{2(\lambda_{a}-\lambda_{b})}{U}\bigg]  = 2\pi L_{a}, 
\end{equation}
where $I_{j}$, $J_{\alpha}$ and $L_{a}$ are the previously mentioned Bethe quantum numbers, the first being associated with the charge quasimomenta and the other two for the spin momenta.\\

\noindent In the limit $\frac{U}{N_{p}}\rightarrow\infty$~\cite{yu1992persistent,OgataShiba90}, we observe that the $k_{j}/U$ terms will tend to zero, and so they can be neglected since they are significantly smaller in magnitude compared to the spin momenta. Accordingly, the Bethe equations simplify and making use of the anti-symmetry of the arctangent function $\arctan(-x)=-\arctan(x)$, it can be shown that the quasimomenta $k_{j}$ can be expressed as 
\begin{equation}\label{eq:kincomm}
 k_{j}L_{R} = 2\pi \Bigg [ I_{j}  +\frac{1}{N_{p}} \Bigg(\sum\limits_{\alpha=1}^{M_{1}}J_{\alpha} +\sum\limits_{a=1}^{M_{2}}L_{a}\Bigg )+\phi \Bigg ].
\end{equation}
Consequently, the system's energy can be recast into 
\begin{equation}\label{eq:energconti}
E =  \bigg (\frac{2\pi}{L}\bigg )^{2} \Bigg [\sum\limits_{j}^{N_{p}}I_{j}^{2} + 2\sum\limits_{j}^{N_{p}}I_{j}\Bigg(\frac{X}{N_{p}}+\phi \Bigg) + N_{p}\bigg( \frac{X}{N_{p}}\bigg )^{2} + N_{p}\Bigg (2\phi\frac{X}{N_{p}}+ \phi^{2} \Bigg)\Bigg ],
\end{equation}
where $X = \big(\sum_{\alpha=1}^{M_{1}}J_{\alpha} +\sum_{a=1}^{M_{2}}L_{a}\big )$. In the case of SU($N$) fermions, one would still have the same expression for the energy, with the added difference that $X = \sum_{l}^{N\! -\! 1}\sum^{M_{l}}_{\alpha_{l}}J_{\beta_{l}}$ accounting for the $N\! -\! 1$ spin rapidities. \\

\noindent For the other integrable regime of the SU($N$) Hubbard model, i.e., commensurate filling fractions in the presence of a lattice, we find that $k_{j}$ has the same form as Equation~\eqref{eq:kincomm}. The corresponding energy in the lattice model is $E = -2\sum_{j}^{N_{p}}\cos k_{j}$, which comes out from lattice regularisation as discussed in Chapter~\ref{chp:tools}. Therefore, by assuming that the charge quantum numbers $I_{j}$ are a consecutive integer/half-integer set, the ground-state energy is expressed as:
\begin{equation}\label{eq:energlat}
E_{0}(\phi ) = -E_{m}\cos \Bigg[\frac{2\pi}{L}\bigg ( D+ \frac{X}{N_{p}}+ \phi \bigg  ) \Bigg ],
\end{equation}
where $E_{m} = 2\sin \big(\frac{N_{p}\pi}{L}\big )/\sin \big (\frac{\pi}{L}\big )$ and $D = \frac{I_{max}+I_{min}}{2}$. The above expression is a generalization of the ground-state energy of SU(2) fermions obtained in~\cite{kusmar,yu1992persistent}. As we stated previously, the continuous limit can be accessed by considering model~\eqref{eq:fluxHub} with dilute filling fractions. Such a statement can be confirmed through the small-angle approximation, with the energies of the two models being related in the following manner: $E_{GYS} = E_{\mathrm{SU}(N)}+2N_{p}$. \\

\noindent The energy expressions provided in Equations~\eqref{eq:energconti} and~\eqref{eq:energlat} hold in the limit of infinite repulsion. At large but finite interactions, we need to introduce a correction to the energy that will be called the energy spin correction $E_{s}$. When $U$ is infinite, the charge momenta $k_{j}$ are of order unity, whilst the spin momenta $\Lambda_{\beta}$ are of order $U$. As such, by defining the scaled variables $x_{\alpha}$ as,
 \begin{equation}\label{eq:xacorr}
 x_{\alpha} =\displaystyle{\lim_{U \to \infty}} \bigg(\frac{2\lambda_{\alpha}}{U}\bigg ),
 \end{equation}
and performing a Taylor expansion of the arctangent function in Equation~\eqref{eq:chargesu3rep} to leading order in $k_{j}/U$, we find the following correction to the quasimomenta 
\begin{equation}\label{eq:kcorr}
\Delta k_{j} = - 2\frac{ \sin k_{j}}{UL}\sum\limits_{\alpha}^{M}\frac{1}{x^{2}_{\alpha}+\frac{1}{4}}. 
\end{equation}
Correspondingly, the energy correction takes the form
\begin{equation}\label{eq:heisse}
E_{s} = -\frac{4}{UL}\Bigg (\sum\limits_{j=1}^{N_{p}}\sin^{2}k_{j}\Bigg )\sum\limits_{\alpha}^{M}\frac{1}{x^{2}_{\alpha}+\frac{1}{4}} = J_{eff}\sum\limits_{\alpha}^{M}\frac{1}{x^{2}_{\alpha}+\frac{1}{4}},
\end{equation} 
having the same form in the continuous limit albeit that $\sin k_{j}\rightarrow k_{j}$. The energy $E_{s}$ corresponds to that of an SU($N$) Heisenberg anti-ferromagnetic spin chain with $N_{p}$ spins, $M$ of which are flipped, having an exchange coupling $J_{eff}$~\cite{yu1992persistent}. This connection between the Hubbard and Heisenberg models will be explored in more detail in Chapter~\ref{chp:ogata}.

\section{Incommensurate fillings}\label{sec:incomm}
\begin{figure}[h!]
	\centering
	\includegraphics[width=0.495\textwidth]{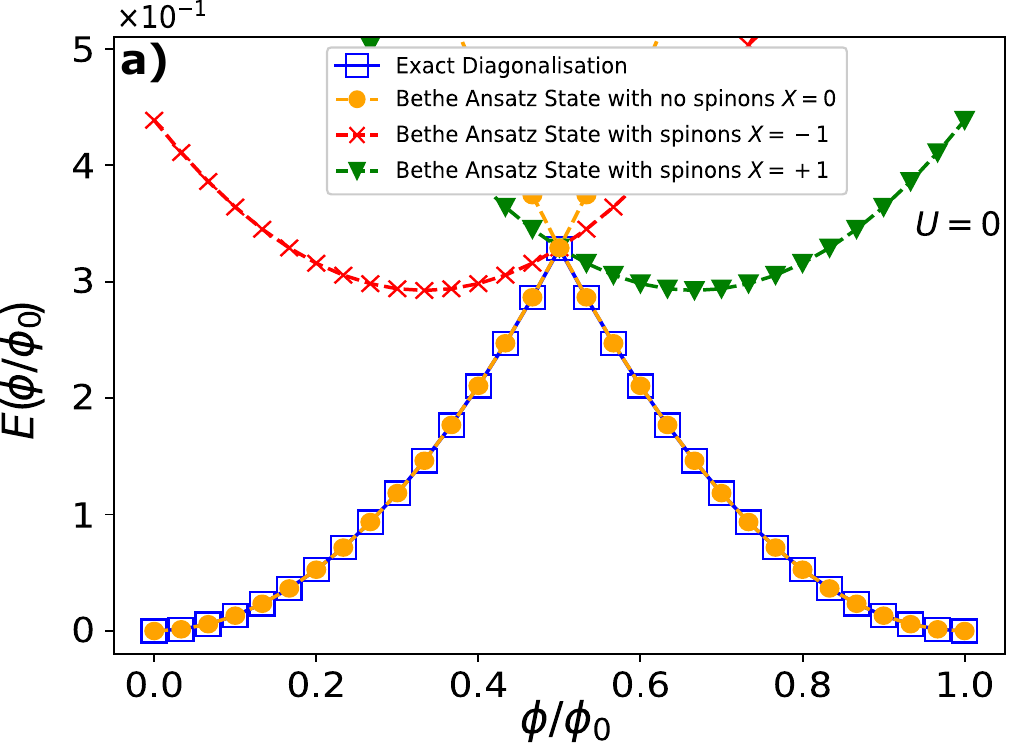}
	\includegraphics[width=0.495\textwidth]{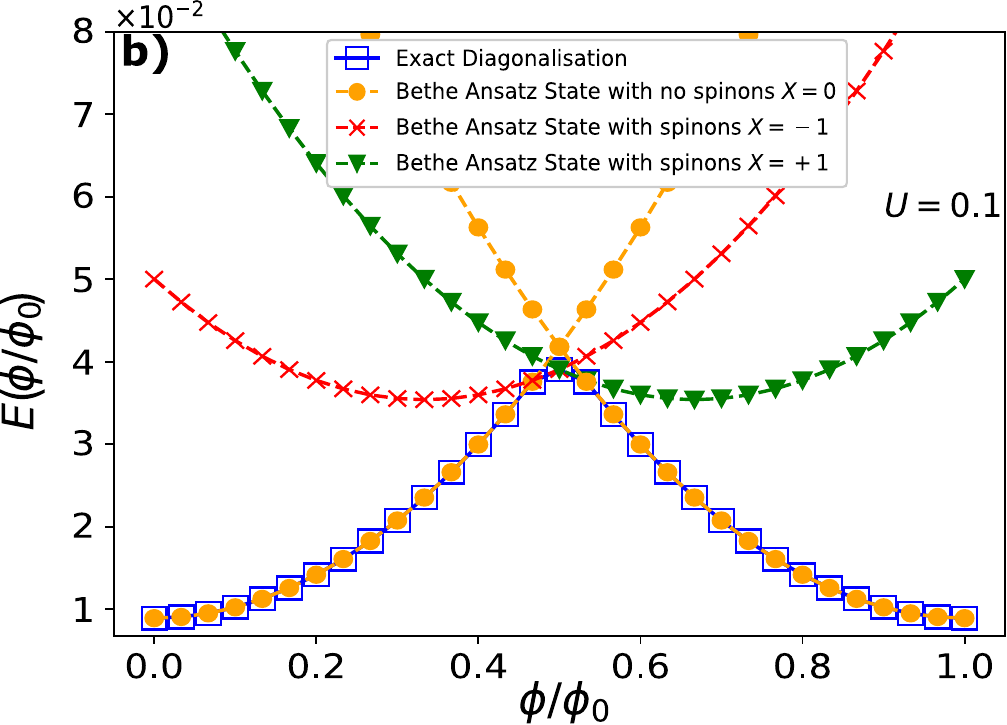}
	\includegraphics[width=0.495\textwidth]{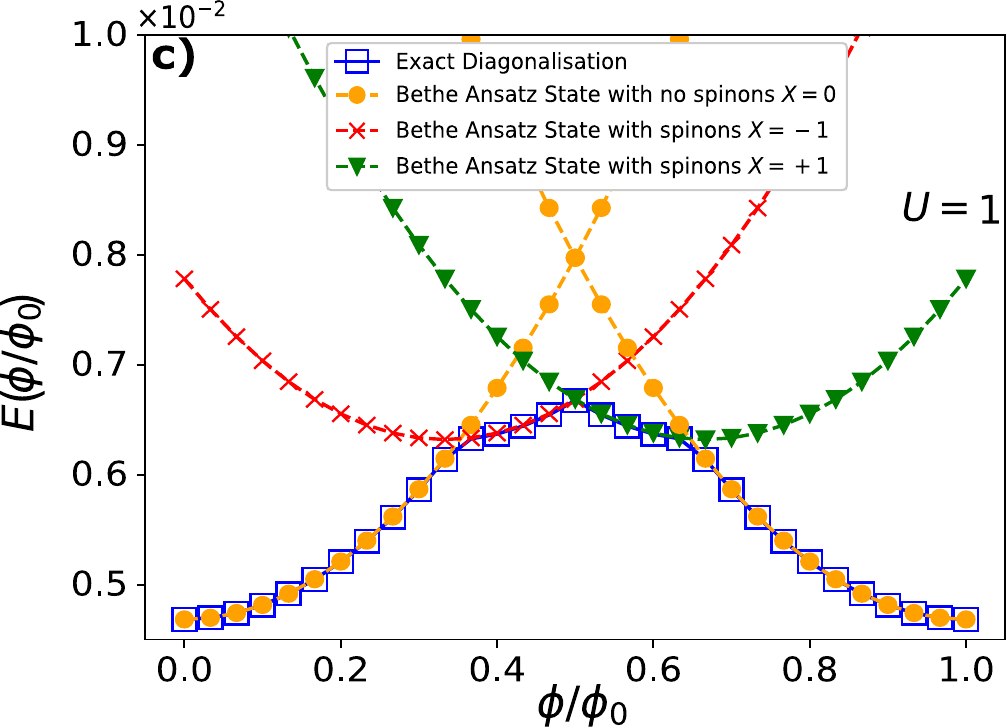}
	\includegraphics[width=0.495\textwidth]{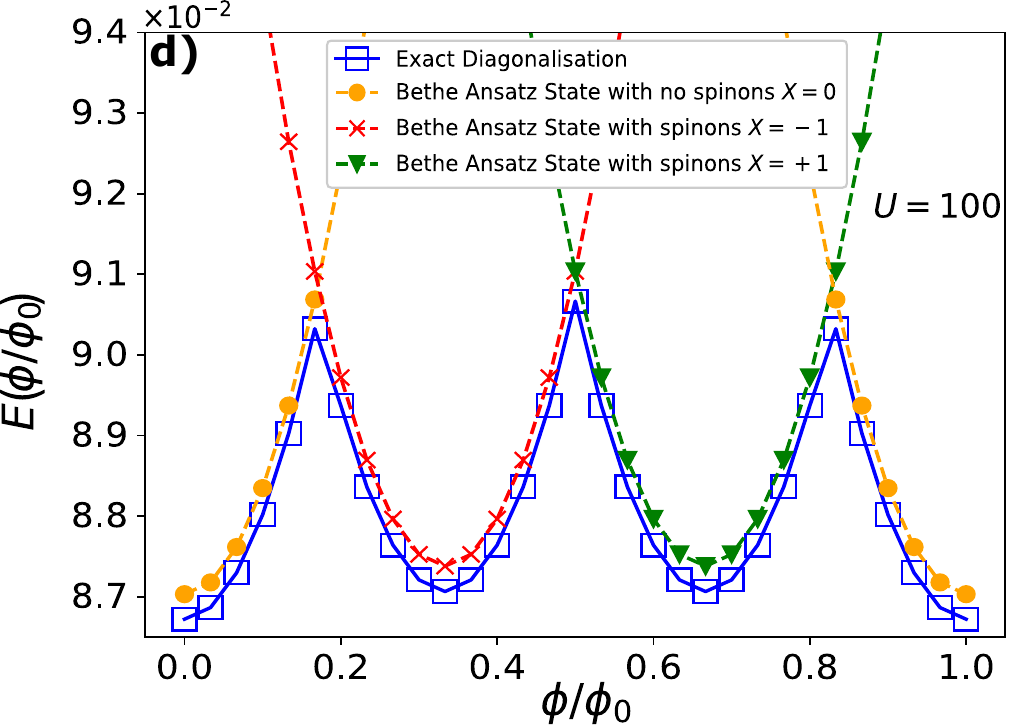}
	\caption{Spinon creation in incommensurate SU($N$) fermionic systems. The case of $N=3$ is considered for $N_{p}=3$ fermions residing on a ring of  $L=30$ sites. The above figures show how the Bethe ansatz energies need to be characterized by spinon quantum numbers in order to have the actual ground-state for various values of the interaction $U$. Results calculated with the Bethe ansatz of the Gaudin-Yang-Sutherland model and exact diagonalization. Figure adapted from~\cite{chetcuti2021persistent}.}
	\label{fig:spccomm}
\end{figure}

Our analysis begins in the continuous limit of low filling fractions wherein we can rely on exact results of the Gaudin-Yang-Sutherland model Bethe ansatz. The numerical analysis shows that, by increasing the effective magnetic flux, specific energy level crossings occur in the ground-state of the system. The Bethe ansatz analysis allows us to recognize such level crossings as ground-state transitions between no-spinons and spinons states --Figure~\ref{fig:spccomm}. 

\begin{figure}[h!]
	\centering
	\includegraphics[width=0.7\textwidth]{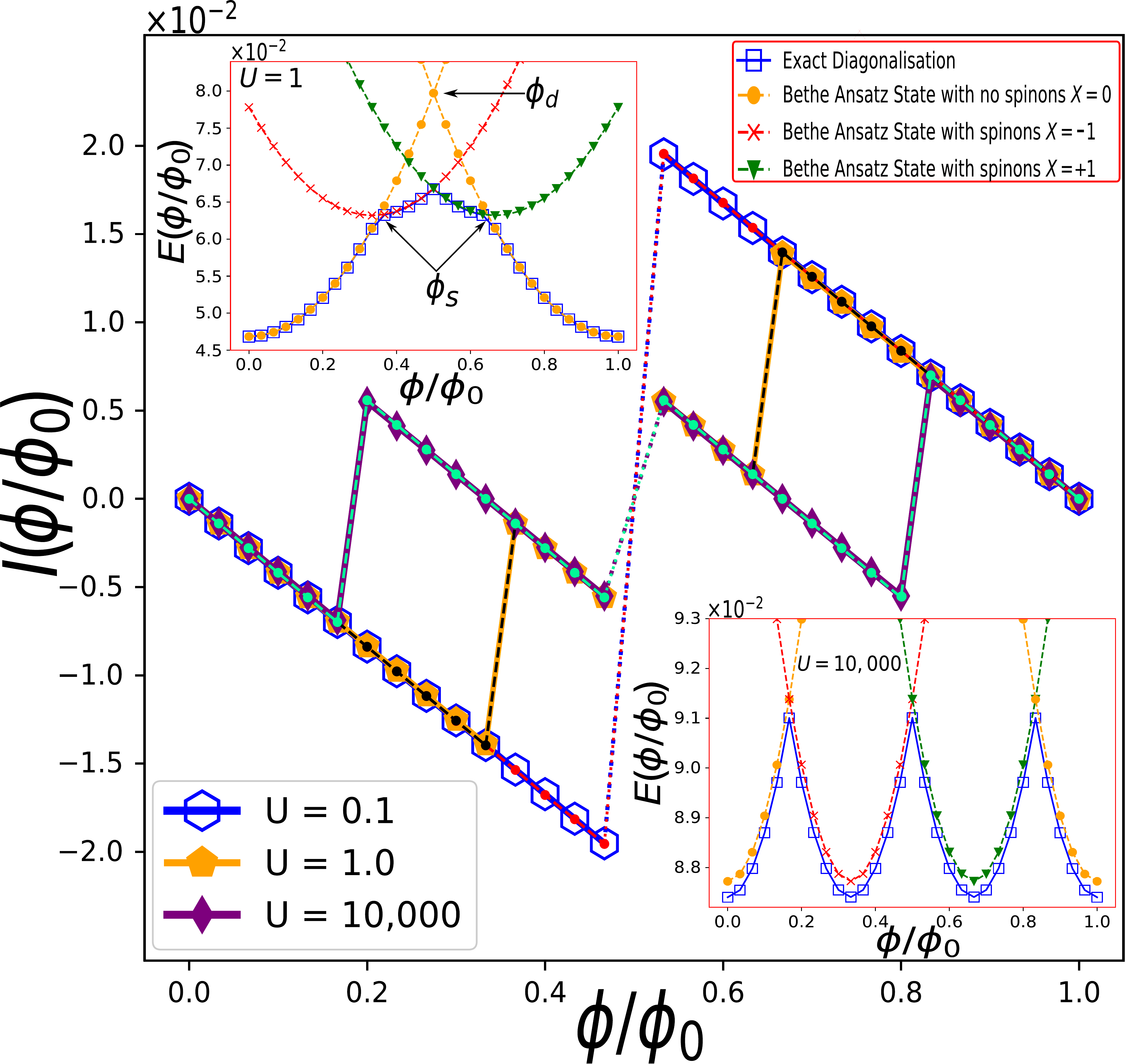}
	\caption{Persistent current $I(\phi )$ for SU($3$) fermions with different interaction strengths $U$ in the dilute filling regime  of the Hubbard model. The exact diagonalization for $N_{p}=3$ SU(3) fermions in a ring of $L=30$ sites is monitored with the Bethe ansatz of the Gaudin-Yang-Sutherland model. The red, black and green dots in the main figure depict the Bethe ansatz results for the persistent current for $U$ = 0.1, 1.0 and 10,000 respectively. These dots are meant to be a guide to the eye, to aid in perceiving the fractionalization of the persistent current. Insets show how the Bethe ansatz energies need to be characterized by $X\neq 0$, to be the actual ground-state. At $U=0$, the ground-state energy is a periodic sequence of parabolas meeting at degeneracy points $\phi_d$ ($\phi_d=1/2$ for the case in this figure). $\phi_{s}$ corresponds to the flux at which spinons are created. Figure taken from~\cite{chetcuti2021persistent}.}
	\label{fig:dilute}
\end{figure}

\noindent To obtain the minimum energy, for a given value of the flux $\phi$, one requires that the summation over the spin rapidities $X$ satisfies the degeneracy point equation having the form~\cite{kusmar, yu1992persistent}
\begin{equation}\label{eq:degeneq}
\frac{2w - 1}{2N_{p}}\leq \phi +D \leq \frac{2w+1}{2N_{p}}\hspace{5mm} \textrm{where} \hspace{5mm} X = -w,
\end{equation}
with $w$ only being allowed to have integer or half-integer values due to the nature of the spin rapidities. Consequently, to reach the target value of $X$, we find that spinons can be created in the ground-state as the effective magnetic flux increases. Such a phenomenon occurs as a specific `screening' of the external flux, which being a continuously adjustable quantity, can be compensated by the spin excitations (quantized in nature) only partially. This in turn results in an imbalance and causes the persistent current to display characteristic oscillations with a period of ${1}/{N_{p}}$ in the ground-state as the flux is varied. Therefore, a curve with $N_{p}$ cusps/parabolic-wise segments per flux quantum emerges. Correspondingly, each parabola is characterized by fractional quantized values of the angular momentum. In the limit of infinite interaction, the persistent current is analytically obtained through the derivative of Equation~\eqref{eq:energconti} as
\begin{equation}\label{eq:4b}
I(\phi ) = -2\bigg (\frac{2\pi}{L}\bigg )^{2}\sum\limits_{j}^{N_{p}}\bigg[I_{j} + \frac{X}{N_{p}}+\phi\bigg ]. 
\end{equation}
This expression shows that, in this regime, the persistent current displays a reduced periodicity $1/N_{p}$ shorter than the bare flux quantum $\phi_{0}$ --Figure~\ref{fig:dilute}. Therefore, in this regime, the bare flux quantum of the system is evenly shared among all the particles. Such a feature was evidenced for two-component fermions in the large interaction regime~\cite{kusmar,yu1992persistent}. 

\newpage
\begin{table}[h!]
\centering
\begin{tabular}{|c| c|c|c|c|c|c|c|}
\hline
 & \multicolumn{2}{c|}{Configuration 1}&\multicolumn{2}{c|}{Configuration 2}&\\
\hline
Magnetic flux & $J_{\alpha_{1}}$  &  $J_{\alpha_{2}}$  & $J_{\alpha_{1}}$  &  $J_{\alpha_{2}}$  &$X$\\
\hline
$0.0-0.1$&$\{-0.5,0.5\}$ &$\{0\}$&$\{-0.5,0.5\}$ &$\{0\}$     &$0$ \\
$0.2-0.5$&$\{-1.5,0.5\}$&$\{0\}$&$\{-0.5,0.5\}$&$\{-1\}$      &$-1$ \\
$0.6-0.8$&$\{-0.5,1.5\}$&$\{0\}$&$\{-0.5,0.5\}$&$\{+1\}$      &$+1$  \\
$0.9-1.0$&$\{-0.5,0.5\}$ &$\{0\}$&$\{-0.5,0.5\}$ &$\{0\}$     & $0$ \\
\hline
\end{tabular}
\caption{\label{T:totalrep}Spin quantum number configurations as functions of the flux $\phi$ for $N_{p} = 3$ fermions with SU(3) symmetry obtained by Equation~\eqref{eq:degeneq}. On increasing $\phi$, the ground-state configuration sees the introduction of `holes' and is no longer consecutive. As a characteristic property of SU($N\!>\!2$), the target value $X$ can be reached via two different configurations of $J_{\alpha_{1}}$ and $J_{\alpha_{2}}$ leading to degenerate states.}
\end{table}
\noindent In our analysis, we find that spinon creation defines a phenomenon occurring for any value of interaction. Specifically, it is observed for $UL/N_{p}\gg 1$ for any number of components $N$. While the fractionalization is mainly dependent on the number of particles, the spinon creation mechanism displays a non-trivial dependence on $N$. Indeed, the different $N\! -\! 1$ spinon configurations are found to play a relevant role in the fractionalization. The quantity $X=\sum_{j}^{N\! -\! 1}\sum_{\beta_j}^{M_{j}}J_{\beta_j}$ can be exploited to characterize the properties of the specific spinon excitations that are created in the ground-state.

\begin{figure}[h!]
	\centering
	\includegraphics[width = 0.65\textwidth]{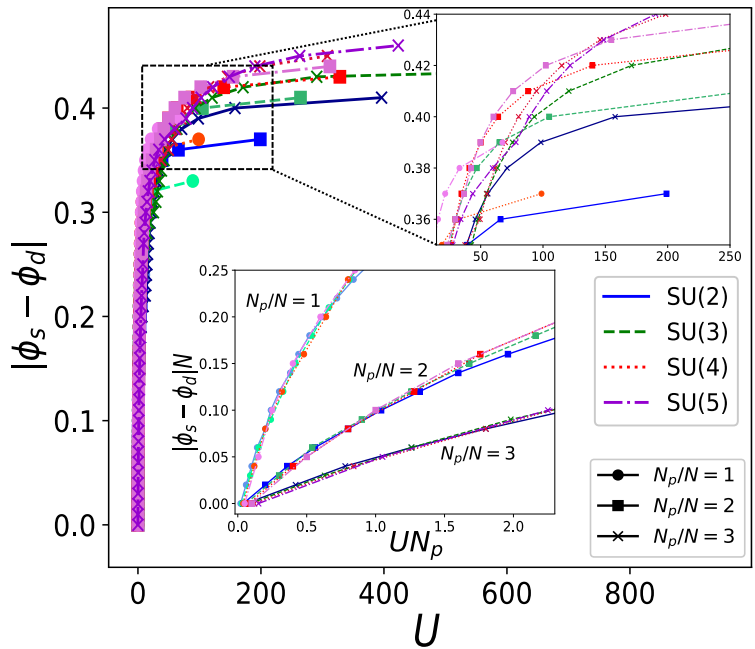}
	\caption{Spinon creation flux distance $|\phi_{s}-\phi_{d}|$ against interaction $U$. Minimum value of $U$ required for spinons to be created in the ground-state for a given $\phi$ is recorded. All the distances $|\phi_{s}-\phi_{d}|$ at which the state with no spinons crosses states with any spinon states are monitored, where $\phi_{s}$ is the flux at which spinons are created and $\phi_{d}$ is the degeneracy point (see Figure~\ref{fig:dilute}). Bottom inset depicts $|\phi_{s}-\phi_{d}|$ against $U$, rescaled by $N$ and $N_{p}$ respectively, in the limit of low $UN_{p}$. All the presented results are obtained by the Bethe ansatz of model~\eqref{eq:GYSham} for $L_{R}=40$.}
	\label{fig:spincre1}
\end{figure}

\noindent Specifically, for small and intermediate interactions, while the system's ground-state with no spinons is found to be non-degenerate, the one with spinons can be made of degenerate multiplets corresponding to Bethe states with distinct configurations of the quantum numbers $J_{\beta_j}$. This is only applicable for SU($N\!>\!2$) fermions as they have different sets of spin quantum numbers (see Table~\ref{T:totalrep}). By further increasing interactions, the spinon states organize themselves in multiplets of increasing degeneracy on a wider interval of the flux. At large but finite interactions, the exact Bethe ansatz analysis shows that the spectrum can be reproduced by a suitable continuous limit of a SU($N$) $t-J_{eff}$ model with $J_{eff}$ defined in Equation~\eqref{eq:heisse}. We remark that the specific features of the SU($N$) fermions enter the entire energy spectrum of the system through the SU($N$) quantum numbers $\{ I_{a}, J_{\beta_{j}} \} $.  We note that, in the infinite $U$ regime, the  ground-state reaches the highest degeneracy --see lower inset of Figure~\ref{fig:dilute}. 

\begin{figure}[h!]
	\centering
    \includegraphics[width = 0.68\textwidth]{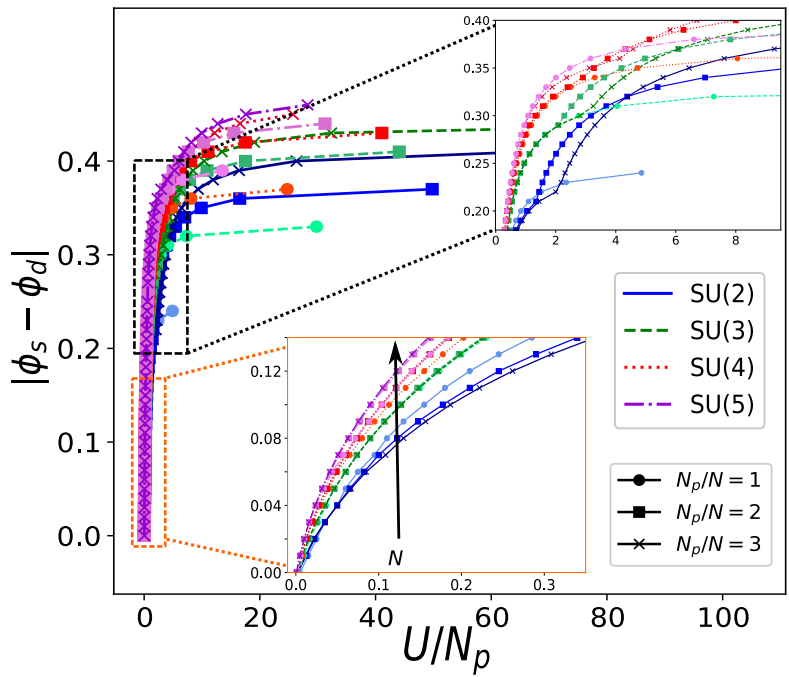}
	\caption{Spinon creation flux distance $|\phi_{s}-\phi_{d}|$ against the interaction per particle $U/N_{p}$. Bottom (top) inset contains data in the low (intermediate) $U/N_{p}$ regime. The discontinuities observed in the intermediate $U/N_{p}$ regime when $N_{p}/N \! >\! 1$, are more pronounced for larger values of $N_{p}/N$ for a system with the same $N_{p}$ but different $N$. All the presented results are obtained by  Bethe ansatz of model~\eqref{eq:GYSham} for $L_{R}=40$.}
	\label{fig:spincre2}
\end{figure}

\noindent As a global view of spinon creation in the ground-state, we monitor, for different values of $U$, $N$, and $N_p$, the values of the flux  $\phi_{s}$ at which the ground-state energy in the system is no longer given by a state with no spinons --Figure~\ref{fig:spincre1}. Such values provide the number of spinons that can be present in the ground-state at a given interaction.  At moderate interactions, spinon production is found to be a universal function of $N_{p}/N$ --see lower inset of Figure~\ref{fig:spincre1}: for systems with lower $N_{p}$, spinons are generated at a lower value of interaction. On going to larger interactions, spinon production is dictated by $N_{p}$, with a fine structure that is determined by $N$:  Systems with higher $N_{p}$ produce spinons at a lower value of $U$; for fixed $N_p$, systems with a lower value of $N_{p}/N$ generate spinons at a lower $U$ --see upper inset of Figure~\ref{fig:spincre1}. Such a phenomenon depends on the specific degeneracies of the system discussed previously, that facilitate spinon creation by increasing $N$ (see Table~\ref{T:totalrep}). This feature emerges also by analysing the dependence of the phenomenon on the interaction per particle $U/N_{p}$ --Figure~\ref{fig:spincre2}. We observe that $N$ enhances spinon production. While the number of spinons decreases with $N_{p}$ for two-component fermions, such a trend appears to be reversed for $N>2$. For intermediate interaction values, discontinuities arise in the curves in cases where $N_{p}/N \!> \!1$ -- see upper inset of Figure~\ref{fig:spincre2}. These discontinuities  correspond to jumps $\Delta X$ in the spinon character $X$. By comparing systems with the same $N_p$ but different $N$, we note that the discontinuities tend to be smoothed out by increasing $N$ and $L$ --Figure~\ref{fig:spincre2}. The value of $\Delta X$ results to be parity dependent. 
\begin{figure}[h!]
	\centering
	\includegraphics[width=0.49\textwidth]{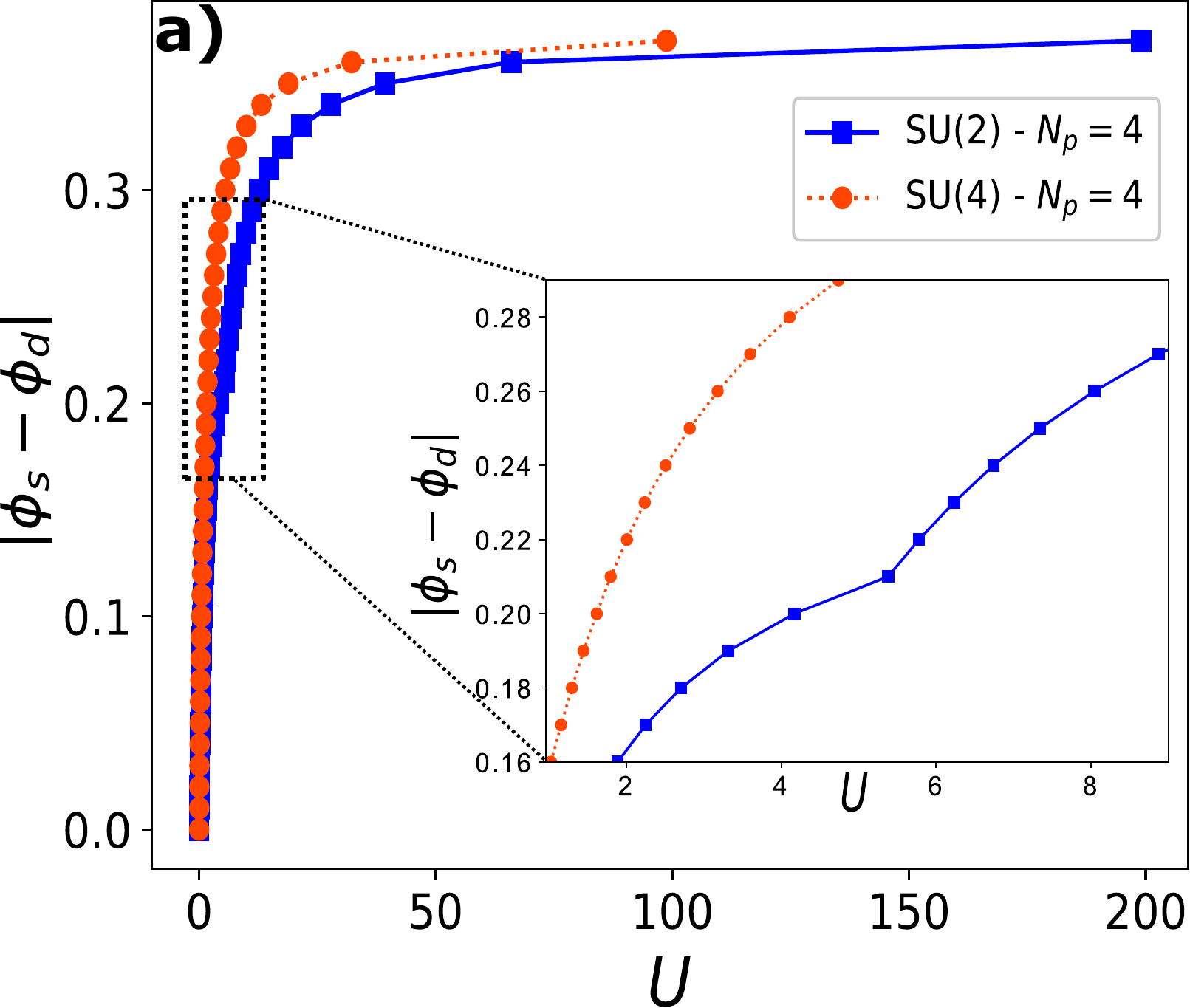}
	\includegraphics[width=0.49\textwidth]{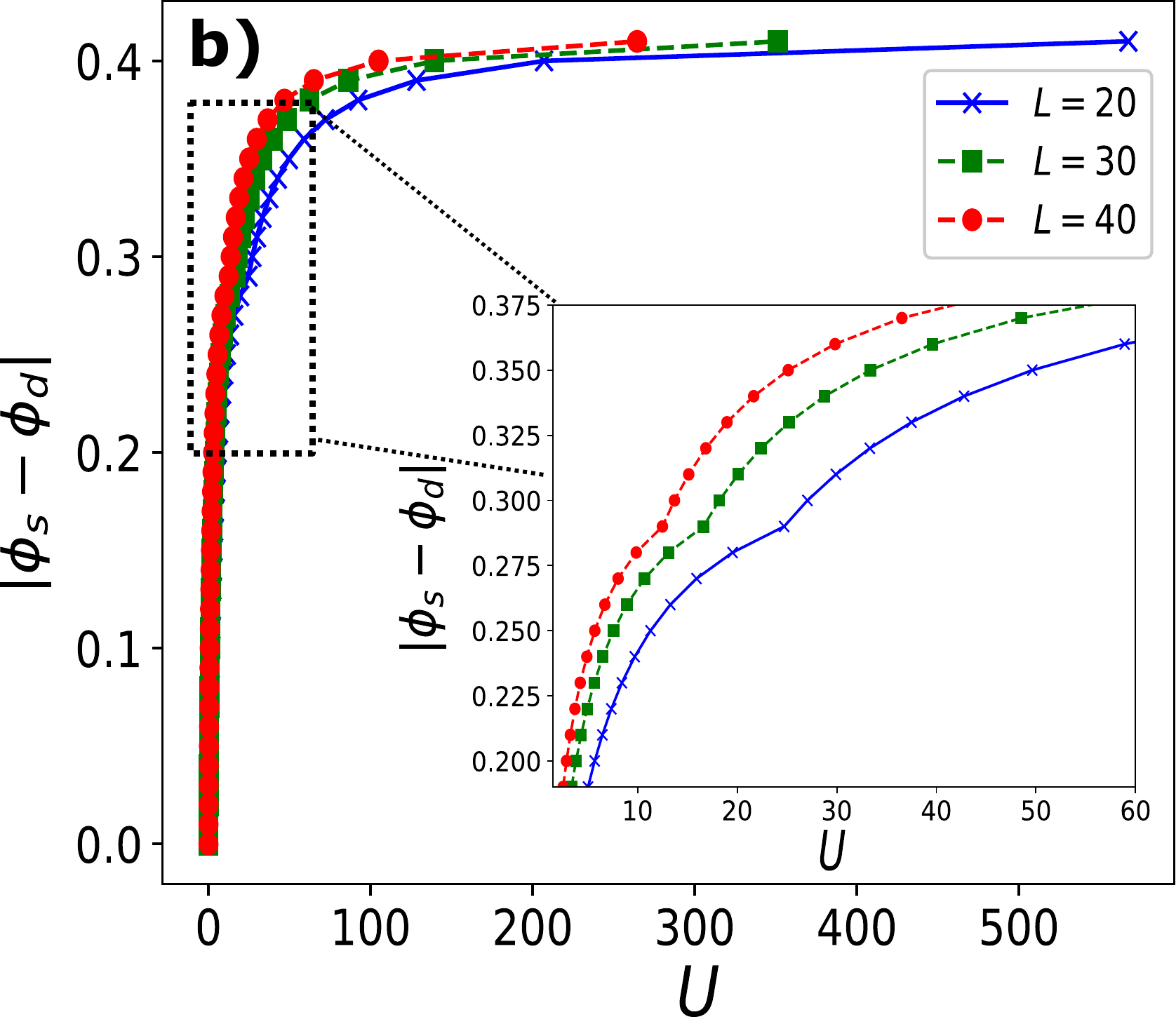}
	\caption{(\textbf{a}) Spinon creation flux distance $|\phi_{s}-\phi_{d}|$ against interaction $U$ is considered for a ring of $L=40$ sites with $N_{p} = 4$ fermions with $N=2$ and $N=4$ components, where $\phi_{s}$ is the flux at which spinons are created and $\phi_{d}$ is the degeneracy point. The intermediate $U$ regime (inset) highlights the discontinuity present in the SU(2) case that disappears for SU(4). (\textbf{b}) Spinon creation flux distance for a system of $N_{p}=6$ particles with SU(3) symmetry for various system sizes, $L_{R}=\{20,30,40\}$. The discontinuity becomes less pronounced with increasing ring size. All the presented results are obtained with Bethe ansatz of the Gaudin-Yang-Sutherland model. Figure reprinted from~\cite{chetcuti2021persistent}.}
	\label{ww}
\end{figure}

\section{Commensurate fillings}\label{sec:comm}

At integer fillings of one particle per site $\nu=1$, the system can become a Mott insulator. In the Mott phase, the motion of particles is constrained as they need to pay an energy penalty $U$ to move around\footnote{Mott insulators are different than band insulators in that the latter, which occurs for $\nu = N$, the particles are not able to move because of the Pauli exclusion principle as the lowest Bloch band is completely filled.}. As such, the hopping process occurs virtually and the system can be effectively described by an antiferromagnetic SU($N$) Heisenberg model that captures the low energy physics of model~\ref{eq:fluxHub}. Through second order perturbation theory one obtains~\cite{essler,cazalilla_2014}
\begin{equation}\label{eq:hesinse}
    \mathcal{H}_{H} = \frac{2t^{2}}{U}\sum\limits_{\langle i,j\rangle}\sum\limits_{\alpha,\beta}S_{\alpha}^{\beta}(i)S_{\beta}^{\alpha}(j),
\end{equation}
where the spin-ladder operators $S_{\alpha}^{\beta}(i)=c_{i,\alpha}^{\dagger}c_{i,\beta}$ obey the SU($N$) algebra discussed in Chapter~\ref{chp:tools}. An alternative way to confirm that model~\eqref{eq:fluxHub} tends to the Heisenberg model is by looking at the analytical expression for the energy derived in Section~\ref{sec:betherep}. For $N_{p}=L$, it is clear that the energy $E_{0}$ in Equation~\eqref{eq:energlat} is vanishing and thus the only contribution would come from $E_{s}$ in Equation~\eqref{eq:heisse} corresponding to the energy of a Heisenberg spin chain. 
\begin{figure}[h!]
	\centering
	\includegraphics[width=0.32\textwidth]{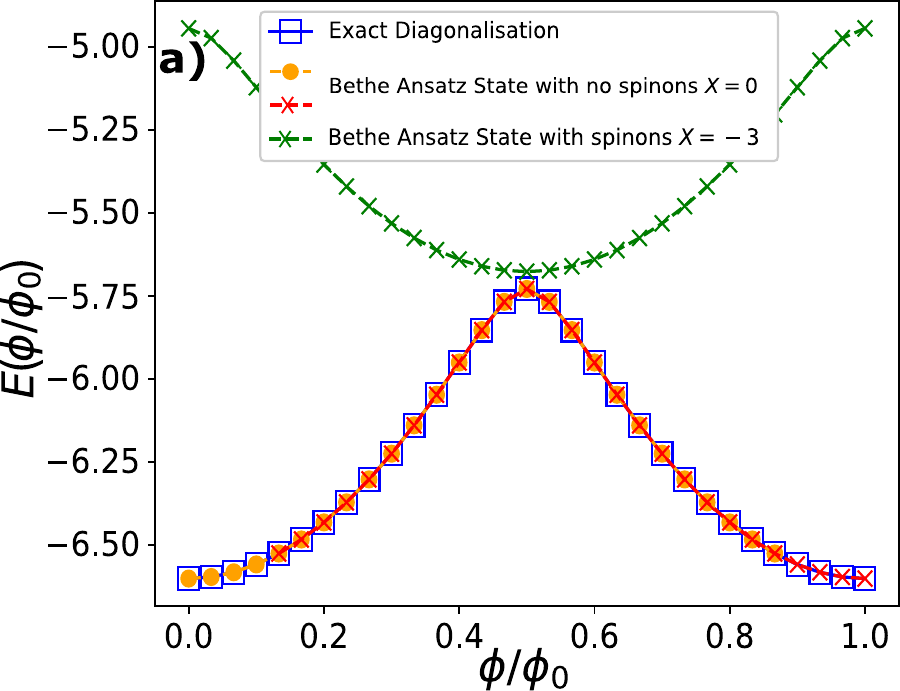}
	\includegraphics[width=0.32\textwidth]{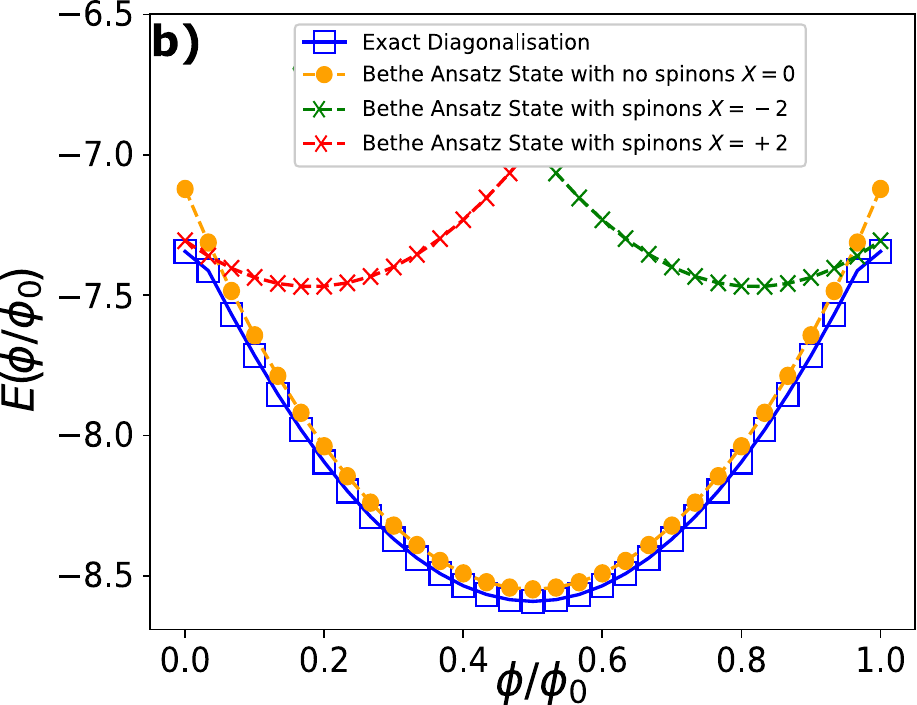}
	\includegraphics[width=0.32\textwidth]{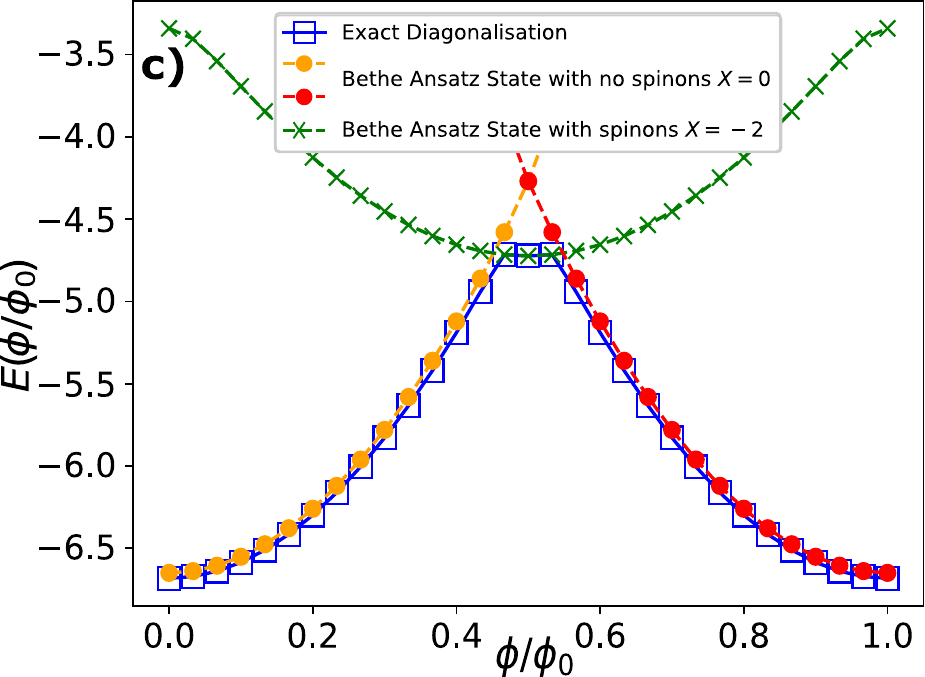}
    \includegraphics[width=0.32\textwidth]{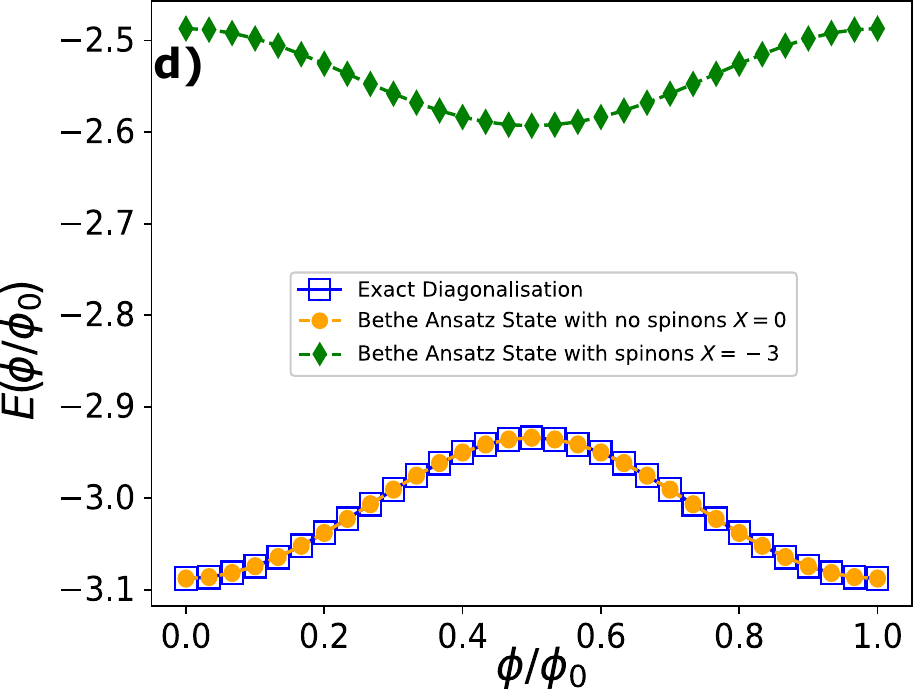}
	\includegraphics[width=0.32\textwidth]{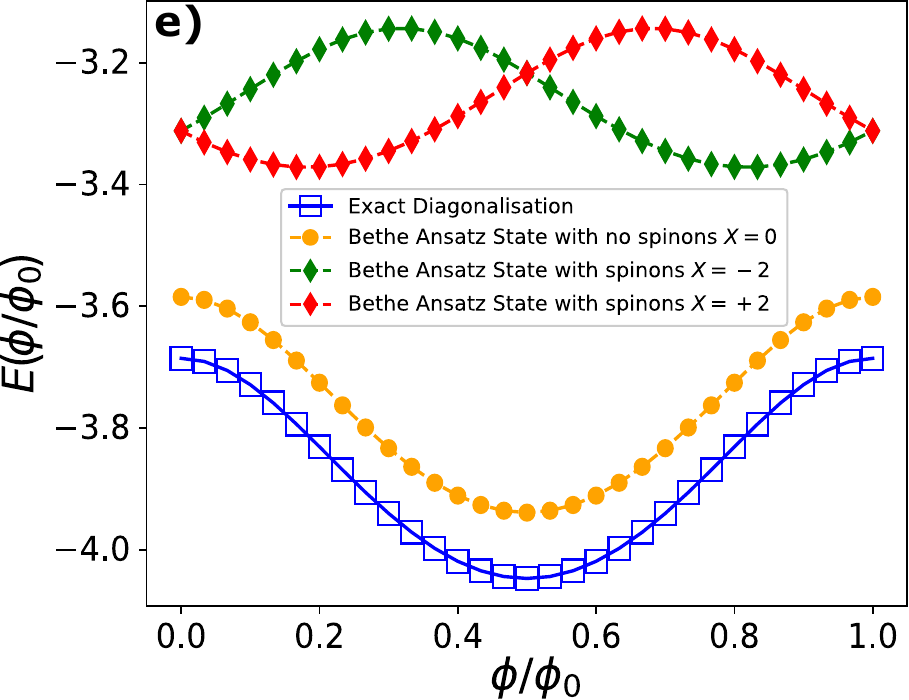}
	\includegraphics[width=0.32\textwidth]{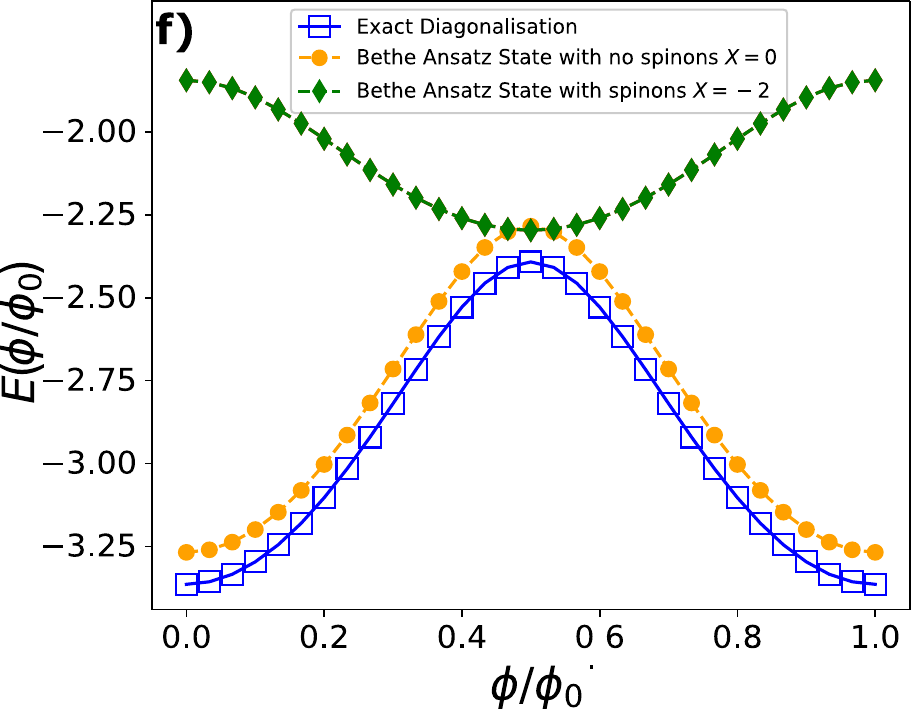}
	\caption{Spinon creation in commensurate SU($N$) fermionic systems for $N_{p}=6$, $N=2$ (left column), $N_{p}=6$, $N=3$ (center column) and $N_{p}=4$, $N=4$ (right column). The Bethe ansatz energies characterized by different spinon configurations needed to make up the ground-state of the system, are considered as functions of the effective flux $\phi$ at $U=1$ (top row) and $U=5$ (bottom row). The Bethe ansatz states with $X=0$ are depicted by orange and red to indicate that the $I_{j}$ quantum numbers are shifted due to being in parabolas with different angular momenta. All the presented results are obtained with Bethe ansatz of the Lai-Sutherland model for $N_{p}=L$ and exact diagonalization. Figure adapted from~\cite{chetcuti2021probe}.}
	\label{fig:pex}
\end{figure}

\noindent The effective Heisenberg model is exactly solvable by Bethe ansatz compared to the fermionic model~\eqref{eq:fluxHub}, as the former meets the necessary criteria for integrability outlined in Chapter~\ref{chp:tools}. We note that the Lai-Sutherland Bethe ansatz can reproduce the qualitative features of the low-lying states of the model obtained by numerics even for intermediate interactions, and as expected, for large interactions, Bethe ansatz and numerics match exactly --Figure~\ref{fig:pex}. In our analysis, we observe that for $N=2$, spinon states have energies larger than the ground-state for any value of interaction. Conversely, for $N\!>\!2$ spinons can be created below a given interaction threshold $U^{*}$; above $U^{*}$ spinon energies result in being well separated from the ground-state energy. The spinon creation process is affected substantially by the onset to a gapped phase in the charge sector that opens up at $U^{*}$, marking the transition from a superfluid phase to the Mott insulating phase. Accordingly, level crossings between no-spinons and spinons states are suppressed --Figure~\ref{fig:pex}.
\begin{figure}[h!]
\centering
\includegraphics[width=0.5\textwidth]{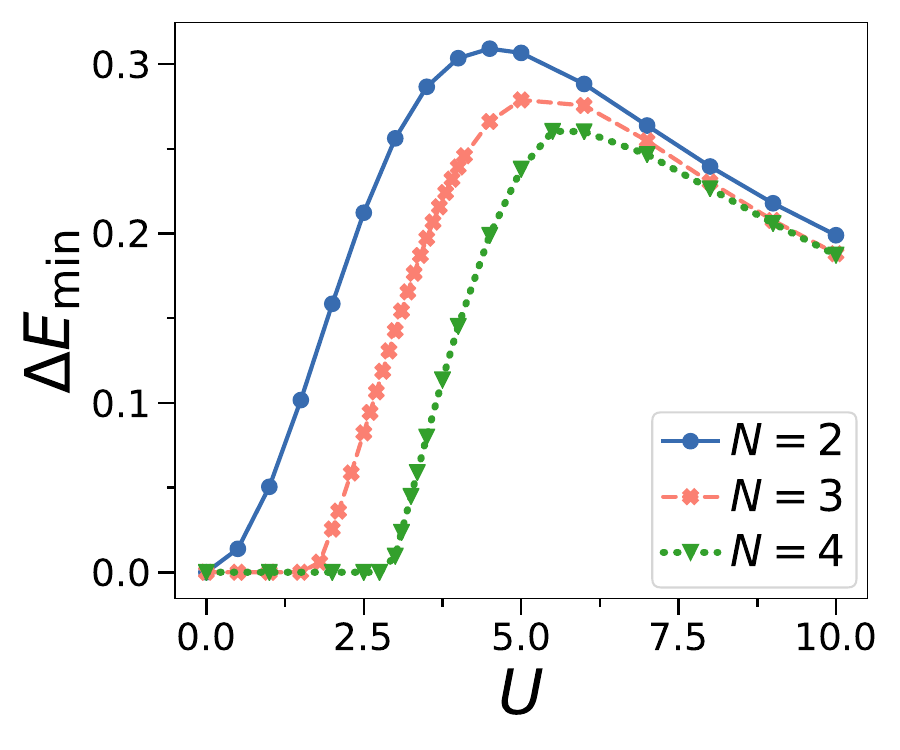}
\caption{SU($N$) energy gap at integer filling. Minimal energy gap $E_{\textrm{min}} = \textrm{min}_{\phi}(\Delta E)$ for different $N$ against $U$ at comparable system sizes ($N=2$ and $N=4$ with $L=8$ and $N=3$ with $L=9$). All curves were obtained by exact diagonalization. Figure adapted from~\cite{chetcuti2021persistent}.}
\label{fig:gap}
\end{figure}

\noindent While for $N=2$, the spectral gap opens up for any repulsive interactions as reported in the seminal paper of Lieb and Wu~\cite{liebwu}, for $N\!>\!2$ it opens up at a finite value of the interaction~\cite{assaraf1999metal,hermele,capponi}. Consequently, the system enters a Mott phase for $U>U_\text{c}$ in the thermodynamic limit~\cite{manmana2011n,cazalilla_2014,capponi}. For our mesoscopic system, we 
observe the gap indicating the onset to the Mott phase transition --Figure~\ref{fig:gap}. The opening of the spectral gap gets delayed with increasing $N$ as the umklapp processes responsible for bringing it about become less relevant at small interaction values~\cite{assaraf1999metal}. \\

\begin{figure}[h!]
\centering
\includegraphics[width=0.9\textwidth]{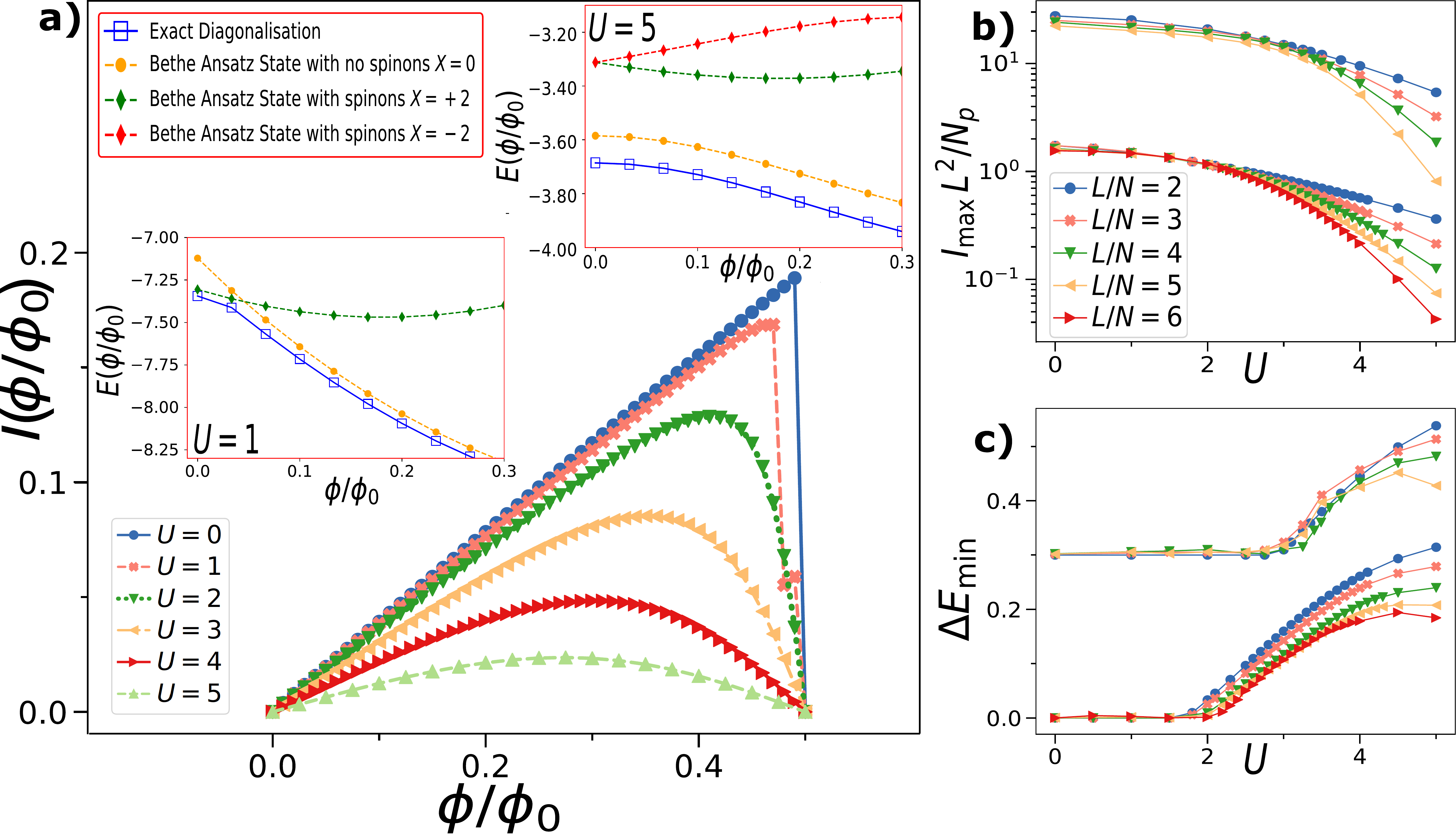}
\caption{(\textbf{a}) SU($3$) persistent current $I(\phi)$ at integer fillings  against flux $\phi$ for $N_{p}=L=9$. Insets display Bethe ansatz results of the Lai-Sutherland model for different spinon configurations $X$ for $N=3$ and $N_{p}=L=6$ compared with exact diagonalization. (\textbf{b}) Maximal current $I_\text{max}=\max_\phi(I)$ for $N=3$ (lower curves) and $N=4$ (upper curves, shifted by factor 20) plotted against interaction $U$. (\textbf{c}) Minimal energy gap $E_\text{min}$ against $U$ for $N=3$ (lower curves) and $N=4$ (upper curves, shifted by 0.3). $\Delta E_{min}$  is around the same specific value of interaction for larger system sizes ($L\ge 8$), which depends on $N$ ($U\approx2$ for $N=3$, $U\approx3$ for $N=4$).  All curves with $L>9$ were calculated with DMRG, and the rest with exact diagonalization. Figure adapted from~\cite{chetcuti2021persistent}.}
\label{fig:SUNAll}
\end{figure}

\begin{figure}[h!]
	\centering
	\includegraphics[width=0.495\textwidth]{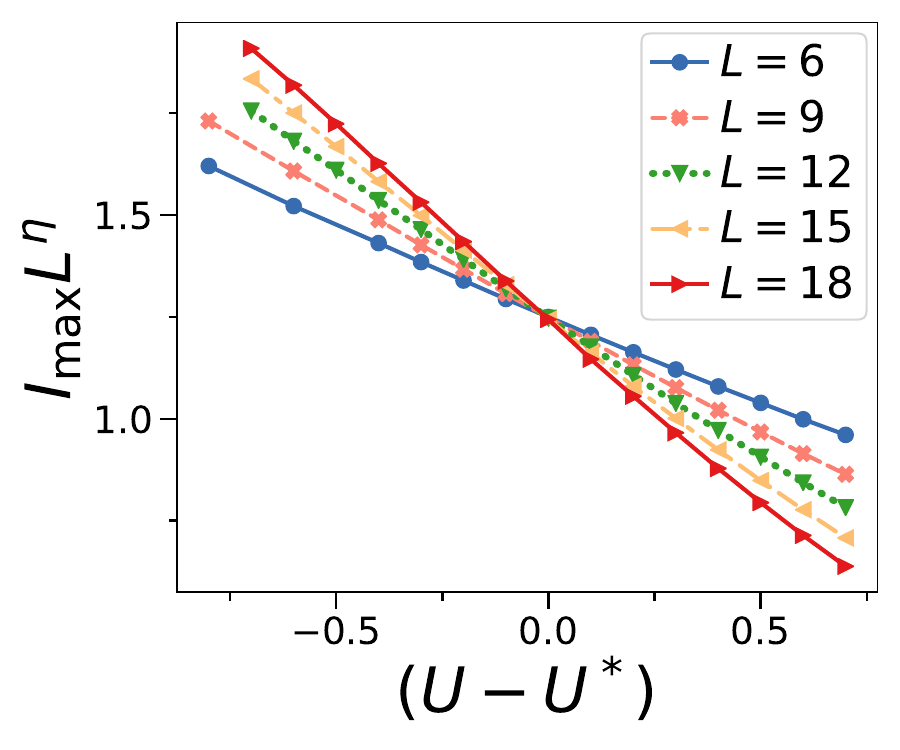}
	\includegraphics[width=0.495\textwidth]{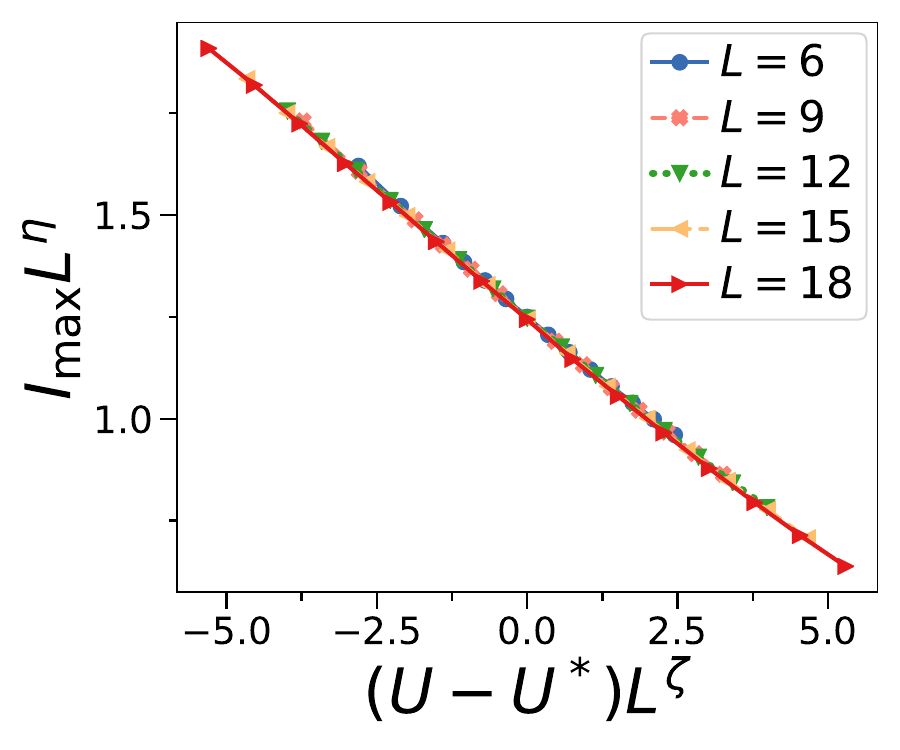}
	\caption{Finite-size scaling of the persistent current of SU(3) fermions.  (\textbf{a}) finite-size critical crossing of maximum current $I_\text{max}$ at $U^*=2.9$ (\textbf{b}) Data collapse. The critical indices for the scaling are $\eta \approx 0.2$ and $\zeta \approx 0.7$. The results were calculated with exact diagonalization for $L=6,9$, with larger $L$ obtained with DMRG. Figure reprinted from~\cite{chetcuti2021persistent}.}
	\label{fig:MottScaling}
\end{figure}

\noindent In spite of its mesoscopic nature, the persistent current is found to be able to indicate the onset to the Mott phase transitions. Firstly, at small interactions, the current is a nearly perfect sawtooth that is eventually smoothed out for stronger interactions --Figure~\ref{fig:SUNAll}(\textbf{a}). Such behavior is found to hold for all $N$, reflecting the opening of the spectral gap. Furthermore, the maximum amplitude of the persistent current $I_\text{max}=\max_\phi(I)$ and $\Delta E_{\min}$ is suppressed exponentially for large $U$ since the motion of the particles is restricted --Figure~\ref{fig:SUNAll}(\textbf{b}). To get a better understanding, we carry out a finite-size scaling analysis~\cite{barber1983finite} of the persistent current for values of the interaction around the Mott instability. In Figure~\ref{fig:MottScaling}(\textbf{a}), the persistent current displays a crossing point at a particular value $U^{*}\approx2.9$ (see also~\cite{schlottmann1992metal,assaraf1999metal,buchta2007mott,manmana2011n}). A clear data collapse is obtained in Figure~\ref{fig:MottScaling}(\textbf{b}) for critical indices $\eta\approx 0.2$ and $\zeta \approx 0.7$.

\newpage
\section{Parity effects}\label{sec:parity3}

\begin{figure}[h!]
\centering
\includegraphics[width=0.76\textwidth]{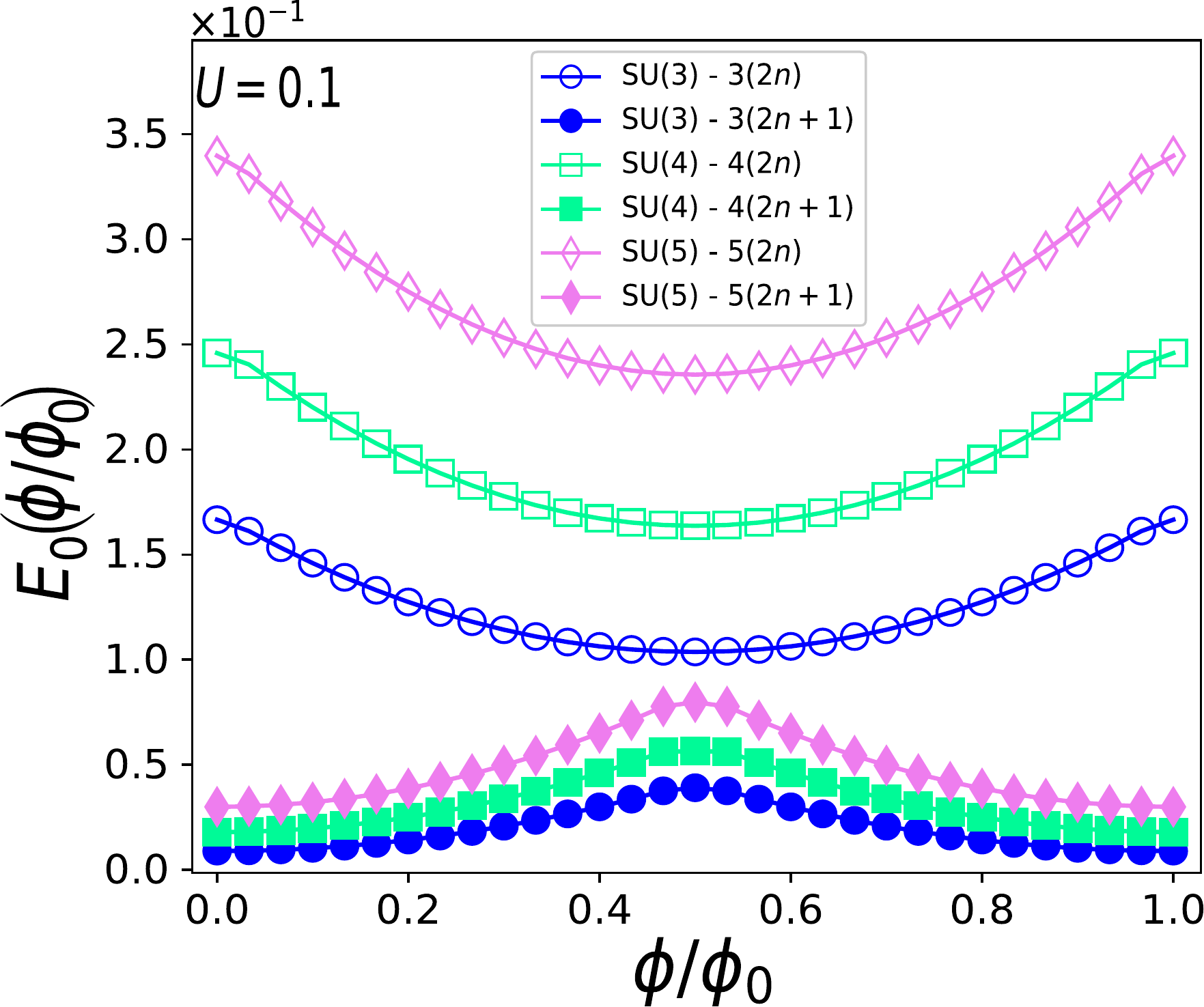}
\caption{Parity effect for SU($N$) fermions. Ground-state energy $E_{0}(\phi )$ as a function of the effective magnetic flux $\phi$ for different $N$ ranging from 3 (circles) to 5 (diamonds). Since the energy is suppressed (increased) by the effective magnetic field, systems with an even (odd) number of particles per component are  paramagnetic (diamagnetic). All the presented results are obtained by Bethe ansatz of Gaudin-Yang-Sutherland model for $L_{R}=30$, with $N_{p}$ taken to be 1 particle and 2 particles per species for each $N$ corresponding to $n=0,1$ respectively. Figure adapted from~\cite{chetcuti2021persistent}.}
\label{fig:parinc}
\end{figure}

\noindent Specific parity effects are observed for SU($N$) fermions. Both for commensurate and incommensurate fillings, the persistent current is found to be diamagnetic (paramagnetic) for ring systems containing $N_p=(2n+1) N $ [$N_p=(2n) N$] fermions, with  $n$ being an integer. The nature of the current can be deduced by looking at the ground-state energy of the system, whereby if the system has a minimum (maximum) at zero flux, then it is diamagnetic (paramagnetic) --Figure~\ref{fig:parinc}. Such phenomena generalize the parity effects of $2n$/$2n+1$ spinless fermions discussed in Chapter~\ref{chp:persis} and of $4n$/$4n+2$ two-component fermions~\cite{waintal2008persistent}.

\begin{figure}[h!]
	\centering
	\includegraphics[width=0.495\textwidth]{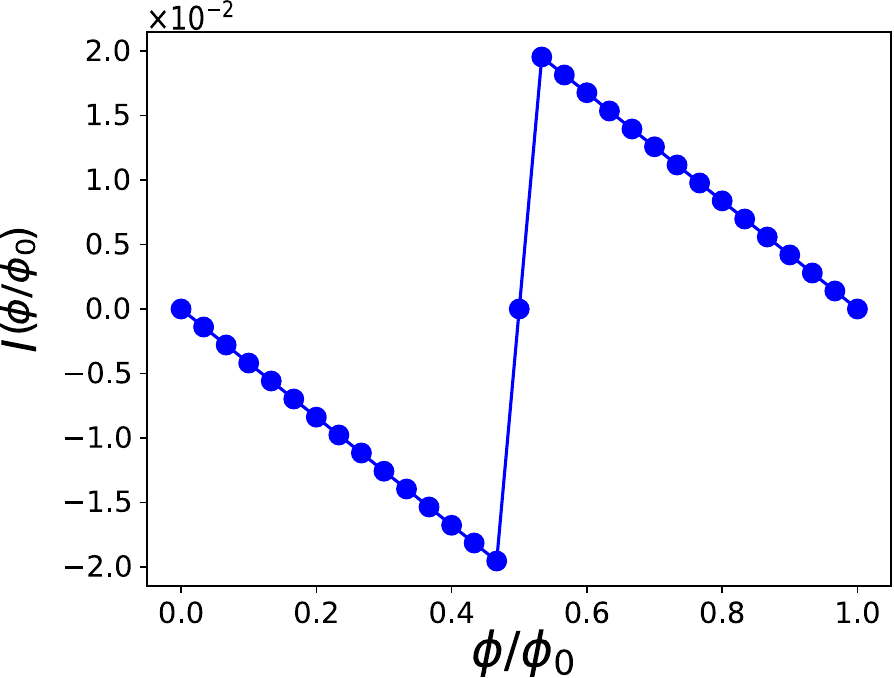}
	\includegraphics[width=0.495\textwidth]{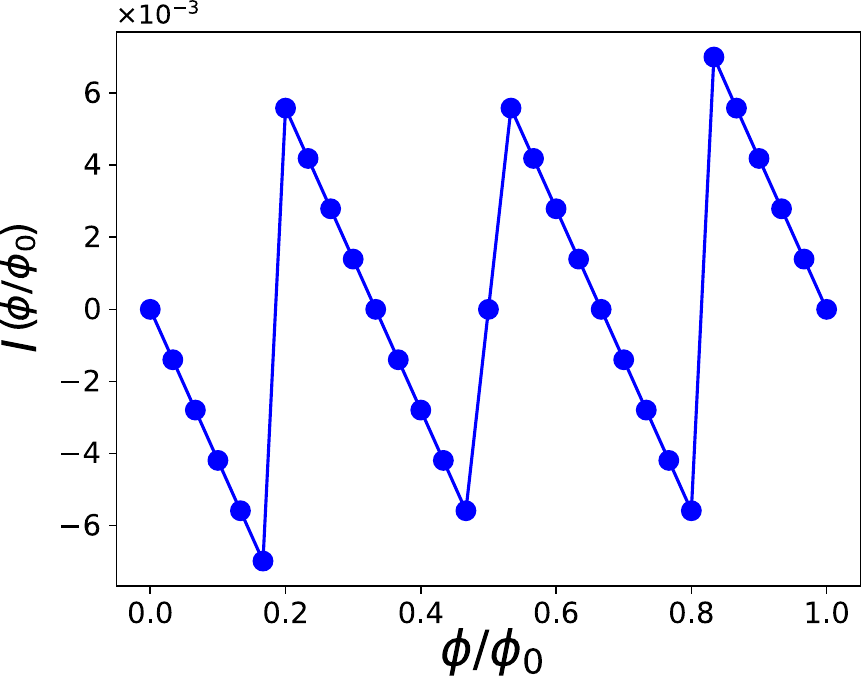}\\
	\includegraphics[width=0.495\textwidth]{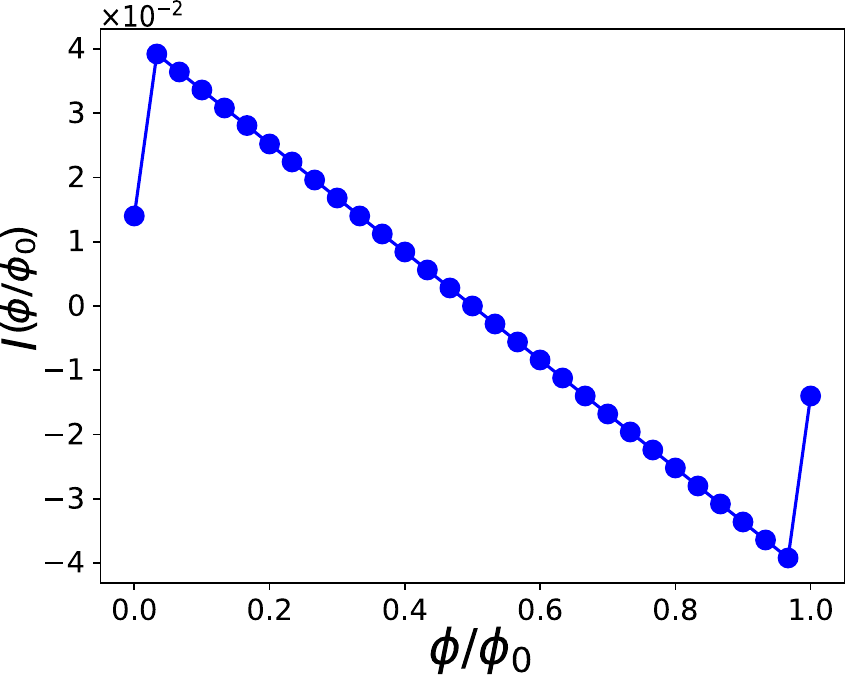}
	\includegraphics[width=0.495\textwidth]{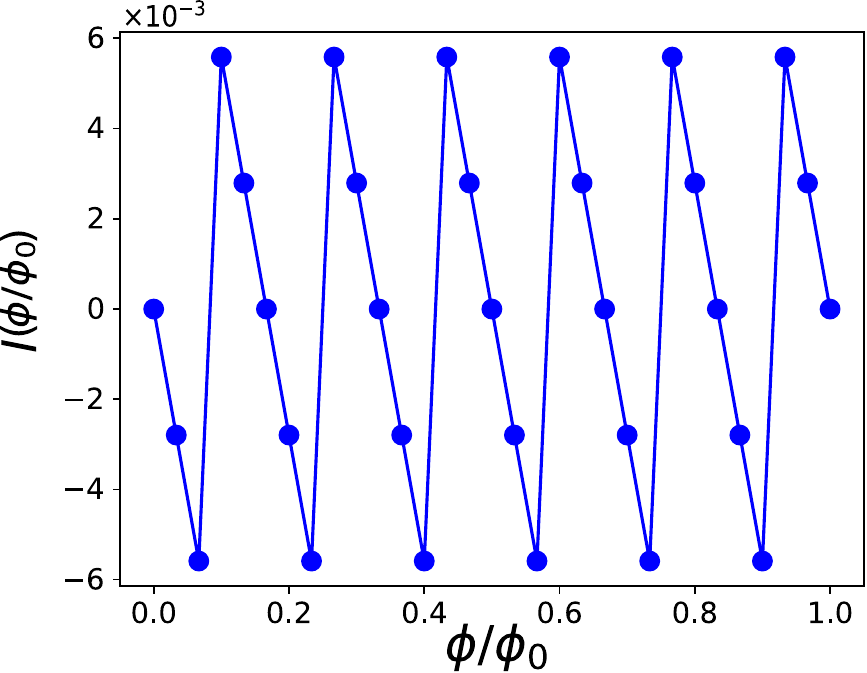}
	\caption{SU($3$) persistent current $I(\phi)$ at incommensurate fillings for different interaction strengths $U$. Left (right) panels correspond to the persistent current for weak (strong) interactions for $N_{p}=3$ ($N_{p}=6$) particles in the top (bottom) row. On going to large repulsive interactions, the persistent current goes from a paramagnetic to a diamagnetic nature for $N_{p}=6$ as it fractionalizes. On the other hand, the parity of the current for $N_{p}=3$ remains diamagnetic. All results are calculated with Bethe ansatz for the Gaudin-Yang-Sutherland model for a ring size of $L_{R}=30$ with $U=0.1$ and $U=10,000$. Figure reprinted from~\cite{chetcuti2021persistent}.}
	\label{fig:pex1}
\end{figure}

\noindent For incommensurate fillings, the behaviour of this parity effect holds for small and intermediate interactions but is washed out on going to the large repulsive limit. Indeed, the character of the current is always diamagnetic since the fractionalization of the bare flux quantum causes the ground-state energy to always be at a minimum at zero flux. The washing out of the parity effect can be clearly observed in Figure~\ref{fig:pex1}, whereby a comparison of SU(3) systems with $N_{p}=3$ and $N_{p}=6$ clearly shows the stark difference in the nature of the persistent current for the latter case between the different interaction regimes. For incommensurate fillings, the parity effect is still observed at large interactions since the opening of the spectral gap prohibits energy level crossings, thereby suppressing fractionalization. 

\section{Conclusions}\label{sec:conc3}

In this chapter, the coherence of a quantum gas of SU($N$) interacting fermions as quantified by the  persistent current, is explored. The analysis is carried out both for incommensurate and commensurate filling $\nu$ fractions. The nature of the ground-state of the system is highlighted by corroborating the numerical analysis (exact diagonalization and DMRG) with Bethe ansatz, which allows access to the specific physical nature of the system's states. 
\vspace{1em}

\noindent For both incommensurate and commensurate fillings, the ground-state can have a spinon nature.  Such a phenomenon implies that the spin correlations can lead to a re-definition of the system's effective flux quantum and, in the regime of incommensurate fillings, yields the $1/N_p$ fractional periodicity for the persistent current observed at large interactions --see insets of Figure~\ref{fig:dilute}.  The reduction of the effective flux quantum  indicates that a form of `attraction from repulsion' can occur in the system, with this feature being consistent with superconducting pairing mediated by pure repulsive electronic interactions~\cite{chakra2001electronic}. Despite the similarities, such a phenomenon follows a very different route from the flux quantum fractionalization  occurring for electrons with pairing force interaction (that could be compared to our study for $N=2$ only)~\cite{byers1961theoretical,onsager1961magnetic} and for bosons with attractive interaction  (occurring as a consequence of quantum bright solitons formation)~\cite{naldesi2019rise, naldesi2020enhancing}: For SU($N$) fermions, the persistent current and the aforementioned re-definition of the flux quantum  reflect the coupling between the spin and matter degrees of freedom.  Recently, it was demonstrated how the flux fractionalization could allow one to approach the Heisenberg quantum limit for rotation sensing~\cite{naldesi2020enhancing}. Our study indicates that SU($N$) systems can provide a platform for high-precision sensors. 

\vspace{1em}
\noindent The ground-state spinon creation displays a marked dependence on the number of components $N$ with distinctions between the $N=2$ and $N>2$ cases --Figures~\ref{fig:spincre1} and~\ref{fig:spincre2}. At moderate interactions, spinon production is found to be a universal function of $N_{p}/N$ --see lower inset Figure~\ref{fig:spincre1}. At integer fillings, spinon creation is suppressed by increasing interactions --Figure~\ref{fig:pex}. The characteristic sawtooth shape of the current is smoothed out --Figure~\ref{fig:SUNAll}. This feature arises since the Mott gap that opens up hinders both the motion of the particles and the creation of spinons in the ground-state.  Remarkably, a clear finite-size scaling behaviour is observed for $N>2$, albeit the persistent current is a mesoscopic quantity --Figure~\ref{fig:MottScaling}. Such a result provides an operative route for the detection of the Mott phase transition in SU($N$) systems, a notoriously challenging problem in the field.  

\vspace{1em}
\noindent A specific parity effect is demonstrated to occur whereby the current is of diamagnetic (paramagnetic) nature for systems comprised of $N_p=(2n+1) N $ [$N_p=(2n) N$] fermions --Figure~\ref{fig:parinc}.  Indeed, for both non-integer and integer fillings fractions, we demonstrate how the results of Byers-Yang, Onsager, and Leggett  on the landscape of the system persistent current can be generalized to SU($N$) fermions~\cite{byers1961theoretical,onsager1961magnetic,leggett1991}. This parity effect is found to be washed out for incommensurate filling fractions in the large interaction regime on account of spinon creation in the ground-state. On the other hand, in the commensurate regime, the parity effect still holds. 

\vspace{1em}
\noindent Lastly, we point out that the persistent current of repulsive SU($N$) fermions has been recently studied by us through a variational quantum eigensolver (VQE)~\cite{vqe}, a hybrid quantum-classical algorithm~\cite{vqerev}, apt for execution on noisy intermediate-scale (NISQ) computers~\cite{tobrev}. By extending the Jordan-Wigner fermion-to-qubit mapping (see Appendix~\ref{sec:vqejw}), a parameterized quantum circuit for the SU($N$) Hubbard model is devised to determine its ground-state energy and, in turn, the persistent current\footnote{\parbox{\linewidth}{For the sake of homogeneity of this thesis, we did not go into too much detail of the machinery employed in the VQE protocol. A detailed explanation can be found in~\cite{vqe}.}}. The distinctive features of the persistent current discussed in this chapter are captured when considering circuits with suitable depth. Thus, the employed approach highlights the highly correlated nature of the fractionalized persistent current.


\chapter{Atomtronic circuits with attractive SU(\textit{N}) fermionic matter}\label{chp:probe}

Mutually attracting quantum many-body systems can form bound states, with their nature depending on the particles' quantum statistics. For bosons, all particles can be bound together, giving rise to the formation of `bright solitons'~\cite{strecker2002formation,Kanamoto2005symmetry,calabrese2007dynamics,naldesi2019rise}. Conversely, such states are hindered for fermions as they are subjected to the Pauli exclusion principle: two-component fermions with opposite spins can form bound pairs~\cite{leggett2006quantum}. However, such restrictions are loosened when considering SU($N$) fermions that have been recently engineered in cold atoms systems~\cite{gorshkov,yoshiro_degenerate,Scazza2014,sonderhouse2020}. Due to them having $N$ different internal states, these multicomponent fermions satisfy the requirement necessitated by the dominant $s$-wave interactions in cold atoms, resulting in the formation of bound states of different types and natures. \\

\noindent Fueled by the recent aforementioned research activity in quantum technology, a considerable interest has been devoted to three-component fermions~\cite{honerkampbcs,Cherng2007superfluidity,Rapp2007color,catelani2008phase,Capponi2008molecular,Klingschat2010exact,Batchelor2010exact,kuhn2012phase,pohlmann2013trion,Guan2013Fermi}. On one hand, this can provide paradigmatic features of the bound states that can be formed for the general cases of $N\!>\!2$. On the other, three-component fermions are of special interest because of their  potential to mimic  quarks and specific aspects of  quantum chromodynamics (QCD)~\cite{Greensite2011,Cherng2007superfluidity,Rapp2007color,tomoki,Klingschat2010exact}
for which it can be advantageous to explore ``low-energy" quantum analogues~\cite{baym2010bcs,rico2018so,banerjee2012atomic,tajima2021cooper,dalmonte2016lattice}.
Specifically, SU($3$) fermions can form two types of bound states: a colour superfluid (CSF) wherein  two colours are paired,
and the other  is unpaired; and a  trion where all colours are involved in the bound state. Trions and CSFs are the analogues of hadrons and mesons (quark-quark pairs) in quantum chromodynamics (QCD). As such, important aspects of the QCD phase diagram like colour deconfinement and resonance formation in nuclear matter can be analysed in cold atom platforms. Despite these bound states being thoroughly analysed in the literature,
devising physical observables paving the way to explore the nature of the SU($3$) bound states in cold atoms systems remains a challenging problem.  \\

\noindent In this chapter, we demonstrate how the persistent current, an experimentally accessible quantity~\cite{phillips_current,ramanathan2011superflow,wright2022persistent,roati2022imprinting}, generated in a ring-shaped gas of strongly attracting three-component fermions, can provide the sought-after observable to study the problem. Specifically, we explore how trion and CSF bound states correspond to specific ways in which the persistent current frequency responds to the effective magnetic flux. By monitoring the persistent current for different regimes of interactions, we demonstrate how thermal fluctuations can lead to a specific deconfinement of the bound states. As an experimental probe in the cold atoms quantum technology, we analyse the time-of-flight imaging (TOF) of the momentum distribution~\cite{amico2022,amico2021,ramanathan2011superflow,beattie,lewenstein2012ultracold}. \\

\noindent The chapter is organized as follows. Section~\ref{sec:methods4} is devoted to the model, with the details for the Bethe ansatz of attractive systems provided in Section~\ref{sec:bethegys4}. Section~\ref{sec:pcp4}, results for the persistent current as a probe for SU(3) bound states are presented. Finite temperature effects are dealt with in Section~\ref{sec:fintemp}. Conclusions in Section~\ref{sec:conc4} close out this chapter.

\section{Model and Methods}\label{sec:methods4}%

\noindent To model $N_{p}$ strongly interacting fermions trapped in an $L$-site ring-shaped lattice pierced by an effective magnetic flux $\phi$, we employ the SU($N$) Hubbard model introduced in Chapter~\ref{chp:repcurr}, albeit in a slightly different form:
\begin{equation}\label{eq:Ham4}
\mathcal{H} = \sum\limits_{j=1}^{L}\sum\limits_{\alpha=1}^{3}\bigg[-t( e^{\imath\frac{2\pi\phi}{L}}c_{j,\alpha}^{\dagger}c_{j+1,\alpha}+\textrm{h.c.})+\sum_{\alpha<\beta}U_{\alpha\beta}n_{j,\alpha}n_{j,\beta}\bigg],
\end{equation}
where the interaction strength is now parameterised by $U_{\alpha\beta}$, which now exhibits a dependence on the colours that are interacting. By choosing asymmetric values of the interaction $U_{\alpha\beta}$, the SU(3) symmetry of the Hamiltonian is broken. This can be straightforwardly checked through the commutation relation outlined in Equation~\eqref{eq:communist}. In what follows, we consider attractive interactions such that $U_{\alpha\beta}\!<\!0$ and utilize $U$ to refer to symmetric interactions between all colours. \\

\noindent Just like in the repulsive case, model~\eqref{eq:Ham4} with attractive \textit{symmetric} interactions is exactly solvable. However, to account for the creation of bound states in the ground-state, the Bethe equations admit complex solutions for the quasimomenta $k_{j}$ via the Takahashi string hypothesis (see Section~\ref{sec:bethegys4}). Furthermore, the integrability of the model is restricted to the dilute lattice limit $\nu=N_{p}/L\!\ll\!1 $, where the physics of the system is captured by the Gaudin-Yang-Sutherland model~\cite{sutherland1968,takahashisu3}. The reason being that the strongly attractive regime of model~\eqref{eq:Ham4} cannot be recast into a Lai-Sutherland anti-ferromagnet as the condition of one particle per site cannot be met~\cite{sutherland1975,capponi}. This, in turn, spoils the integrability of the model, as the system becomes `diffractive' in that the scattering matrix no longer obeys the Yang-Baxter relation. \\

\noindent Bound states of different natures can arise in systems described by model~\eqref{eq:Ham4}: CSF bound states, wherein two colours form a bound pair with the other colour remaining unpaired; trions, wherein all the three colours are bounded. One way of achieving CSFs is by lifting the SU(3) degeneracy through unequally spaced Zeeman splittings between the colours in the grand-canonical ensemble, leading to different phase transitions between the bound states~\cite{Batchelor2010exact,kuhn2012phase,Guan2013Fermi}. Another possibility is to break the SU(3) symmetry~\cite{honerkampbcs,catelani2008phase,Klingschat2010exact,Guan2013Fermi}. Here, SU(3) symmetry is broken explicitly in the canonical ensemble by choosing asymmetric interactions between the components, as in Hamiltonian~\eqref{eq:Ham4}. \\

\noindent Recently, the SU(3) bound states of the Hubbard model have been investigated through correlation functions~\cite{Capponi2008molecular,Klingschat2010exact,pohlmann2013trion}. In addition to correlation functions, we employ the persistent currents $I(\phi )$ given by Equation~\eqref{eq:newcurr}, as the main diagnostic tool to study our system. 
The reduced periodicity it experiences upon the formation of a bound state~\cite{byers1961theoretical,polo2020exact,pecci2021probing} makes for an ideal probe to characterize the different bound states. Relying on the experimental capability of addressing fermions of different colours separately~\cite{sonderhouse2020}, especially to analyse the broken SU($3$) cases, we utilize the species-wise persistent current. At finite temperature $T$, this can be defined through thermodynamics potentials as  $I_{\alpha}(\phi) = -\partial F_{\alpha}(\phi)/\partial\phi$, with $F$ being the system's Helmholtz free energy in the canonical ensemble~\cite{byers1961theoretical}, or through the persistent current operator $\hat{I}$ defined in Equation~\eqref{eq:newcurr}:
\begin{equation}\label{eq:tempcurr}
     I(\phi) = \frac{1}{\mathcal{Z}}\textrm{tr}\{ \hat{I} e^{-\beta \mathcal{H}}\},
\end{equation}
where $\mathcal{Z} = \mathrm{tr}(e^{-\beta\mathcal{H}})$ is the partition function and $\beta = 1/(k_{B}T)$ with $k_{B}$ being the Boltzmann constant.  Note that the calculation of the species-wise persistent current, which requires two-point correlations, is very challenging to implement in Bethe ansatz. \\

\noindent Our approach relies on utilising a combination of numerical methods such as exact diagonalization and DMRG~\cite{whitedmrg,itensor}, as well as Bethe ansatz results whenever possible, in order to identify and characterize the bound states of SU($3$) fermions, for systems with 
an equal number of particles $N_{p}$ per colour. In Chapter~\ref{chp:repcurr}, we explored in great detail how Bethe ansatz results (through the quantum numbers) can be exploited as bookkeeping to monitor the eigenstates provided by the numerical results. For attractive interactions, one is not able to compare numerical results with the exact solution. Despite the formal mapping between lattice and continuous SU($N$) theories works for arbitrary interaction (see Appendix~\ref{sec:mappinggys}), the case of attractive interactions requires extra care. The catch lies in the formation of bound states that can have a correlation length comparable to or smaller than the lattice spacing $\Delta$. Specifically, for $\Delta$ larger than the coherence length, one will not be able to observe the correlation functions' decay as the latter is overshadowed by the former. This is particularly evident for small-sized systems. The previously outlined condition can be satisfied for large-sized systems and small values of the interaction.  In order to obtain a meaningful continuous limit to be quantitatively comparable with the lattice theory at small fillings, the vanishing lattice spacing $\Delta$ should come with a suitable rescaling of $U$. While such an aspect has been analysed for bosonic theories (see f.i.~\cite{oelkers2007ground,naldesi2019rise}), for fermions it still requires investigation. Even though one needs to rescale $U$ for SU(2) fermions, this problem can be circumvented due to the Lieb-Wu Bethe ansatz, which allows one to validate the Bethe ansatz results with numerical methods. However, the SU($N\!>\!2$) Hubbard model is not Bethe ansatz integrable for lattice systems. As such, one is not able to monitor the numerical results with Bethe ansatz.  Nonetheless, the continuous limit is Bethe ansatz integrable by the Gaudin-Yang-Sutherland model, and it is vital in understanding the underlying physics of the model.

\subsection{String hypothesis for the Gaudin-Yang-Sutherland model}\label{sec:bethegys4}

\noindent Typically, the Bethe ansatz equations admit a complex solution for the spectral parameters $k_{j}$ and $\Lambda_{\alpha}$. Whilst for real solutions, it is clear that the wavefunction corresponds to the superposition of plane waves, complex solutions are synonymous with the formation of bound states: the wavefunction experiences an exponential decay with respect to its relative coordinates~\cite{essler,takahashi2005thermodynamics}. It is worth mentioning that complex solutions are present for both repulsive and attractive interactions. However, in the former, 
these correspond to highly excited states as it is more energetically favourable for particles to scatter than forming a pair. On the other hand, for attractive interactions bound states appear in the ground and low-lying excited states satisfying the symmetry that $\mathcal{H}(U)=-\mathcal{H}(-U)$~\cite{essler}.
\begin{figure}[h!]
    \center
    \includegraphics[width=0.6\textwidth]{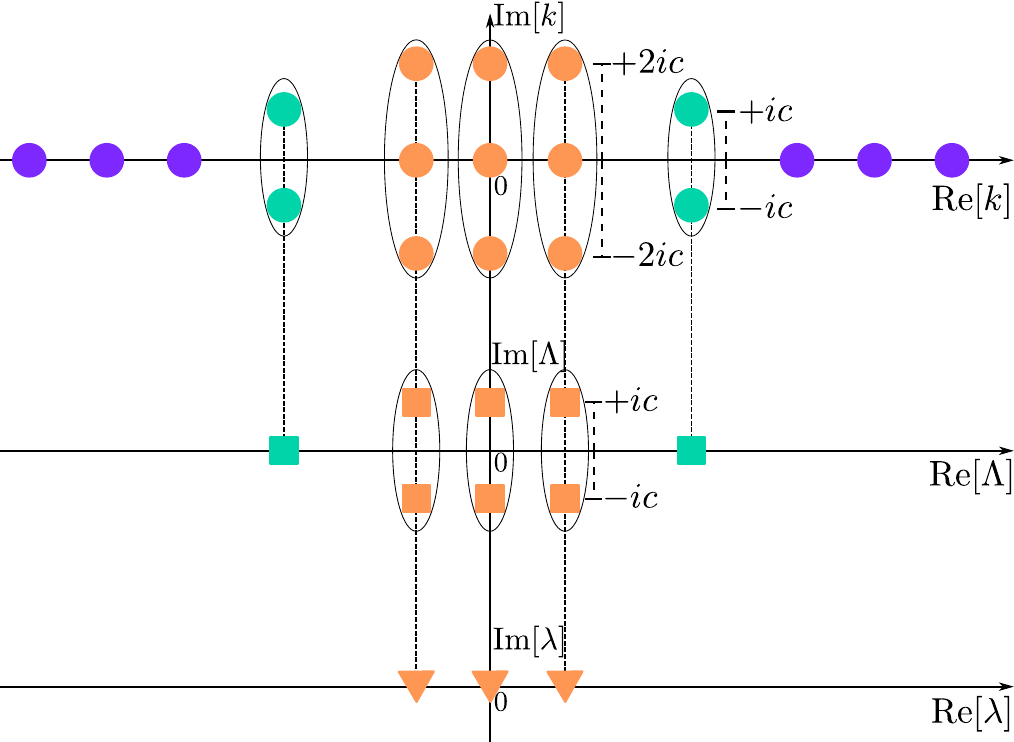}%
    \caption{Figure of merit for the quasimomenta string configuration in the ground-state for $N=3$, with $n_{1}=6$ (purple), $n_{2} = 2$ (green) and $n_{3} = 3$ (orange). The bound states in the system are represented by an oval structure, and they are present both for the charge (circles) and spin rapidities (squares and triangles). Accompanying a charge bound state of length $m$, are spin bound states of decreasing length, with $m=2$ being the minimum length one needs to form a bound state. The real part of the charge and spin bound states of length $m$ is given by the spin rapidity corresponding to $\lambda^{(m-1)}$. The figure above is based on the one in~\cite{takahashisu3}.}
    \label{fig:taka}
\end{figure}

\noindent To account for complex solutions for the roots of the Bethe ansatz equations, we employ the \textit{string hypothesis}~\cite{takahashi2005thermodynamics,deguchi2000thermodynamics}. The string hypothesis assumes that the rapidities, denoted by $p$, take the form  $p^{\pm} = \xi \pm \imath |\epsilon|$ for real $\xi$ and $\epsilon$. These string solutions, so-called due to their arrangement in the complex plane (see Figure~\ref{fig:taka}), are pairwise complex conjugates to ensure that observables like energy and momenta are real valued. The procedure to determine the structure of the string is pretty straightforward~\cite{essler}: one of the rapidities is taken to be complex and inserted into the product form Bethe equations~\eqref{eq:BAksun} and~\eqref{eq:BAlsun}. For large $L_{R}$, the left-hand side of the equation corresponding to this rapidity becomes exponentially large or goes to zero, depending on the sign of the imaginary part. Consequently, to fulfil the Bethe equations, we require that the right-hand side corresponds to either a pole or a zero, which is satisfied when $\xi = \lambda_{\alpha}$ and $\epsilon = c$, with $c$ corresponding to the interaction in the Gaudin-Yang-Sutherland model~\eqref{eq:GYSham}. In principle, $q$ can either correspond to the quasimomenta resulting in what is known as a $k\mathrm{-}\Lambda$, where two $k$ combined with one spin rapidity $\Lambda$; or it can correspond to $\Lambda$ such that the strings involve only spin rapidities giving rise to $\Lambda$ strings. \\

\noindent In the case of SU($N$) fermions with strongly attractive interactions such that $L_R|c|\!\gg\! 1$, there are both $k\mathrm{-}\Lambda$ and $\Lambda$ strings. Let us start by considering an equal number of particles per colour such that the quasimomenta $k_{j}$ form charge bound states of size $N$ as permitted by the Pauli exclusion principle, with the following string configuration~\cite{yin2011effective}
\begin{equation}\label{eq:kstrings}
    k_{q,j}^{N} = \Lambda_{q}^{(N\!-\! 1)} + \imath(N+1-2j)c + \mathcal{O}(\imath\delta |c|),
\end{equation}
where $j=1,\hdots, N$ and $q = 1,\hdots, N_{p}/N $. Accompanying each charge bound state are $N-2$ sets of spin strings, whose form in the ground-state reads
\begin{equation}\label{eq:lambdastrings}
    \Lambda_{q,\alpha}^{(r)} = \Lambda_{q}^{(N\! -\! 1)}+ \imath (m-r+1-2\alpha)c + \mathcal{O}(\imath\delta |c|),
\end{equation}
with $\alpha = 1,\hdots, N-r$ for $r=1,\hdots, N-2$. The term $\mathcal{O}(\imath\delta |c|)$ is included in every string and it accounts for any deviations for the ideal string hypothesis, with $\delta$ being a very small quantity whose order is that of $\exp (-L_R|c|)$.  As $L\rightarrow\infty$, $\mathcal{O}(\imath\delta |c|)$ vanishes and one is able to utilise the ideal strings. Note that it is justified to take ideal strings when working in the limit of $L_R|c|\!\gg\! 1$, which what we consider unless explicitly stated. \\

\noindent If we consider unequal populations per species, several types of charge bound states composed of $m$ particles can arise, with the number of particles ranging from 2 to $N$. For the SU(3) case, we can have $n_{1}$ unpaired fermions ($m=1$), $n_{2}$ pairs of two-body bound states ($m=2$) and $n_{3}$ three-body bound states $m=3$ such that $N_{p}=n_{1}+2n_{2}+3n_{3}$, $M_{1}=n_{2}+2n_{3}$ and $M_{2}=2n_{3}$. Consequently through Equations~\eqref{eq:kstrings} and~\eqref{eq:lambdastrings}, the quasimomenta and corresponding spin strings for all the different phases of SU(3) fermions, can be written as~\cite{Guan2013Fermi}
\begin{align}
\centering
\label{eq:strings}
    & k^{1}_{\alpha} = k_{\alpha} &
    &k^{2}_{\beta}=  
\begin{cases}
    \Lambda_{\beta}+\imath c\\
    \Lambda_{\beta}-\imath c             
\end{cases}  &
&k^{3}_{a}= 
\begin{cases}
    \lambda_{a}+2\imath c\\
    \lambda_{a} \\
    \lambda_{a}-2\imath c             
\end{cases}\nonumber & \\
     &\color{white}0 & 
     &\Lambda_{\beta} = \Lambda_{\beta} &
     &\Lambda_{a}= 
\begin{cases}
    \lambda_{a}+\imath c\\
    \lambda_{a}-\imath c             
\end{cases}\nonumber & \\
    &\color{white}0 &
    &\color{white}0&
    &\lambda_{a}=\lambda_{a}&
\end{align}
for each case respectively where $\alpha = 1,\hdots ,n_{1}$, $\beta = 1\hdots, n_{2}$ and $a = 1,\hdots, n_{3}$. Substituting the ideal strings into the product form Bethe ansatz equations~\eqref{eq:chargesu3rep},~\eqref{eq:spin1su3rep} and~\eqref{eq:spin2su3rep}, and subsequently taking the logarithm, we obtain
\begin{equation}\label{eq:l1 string}
k_{j}L_R = - 2\sum\limits_{\alpha = 1}^{n_{2}}\arctan \bigg(\frac{k_{j}-\Lambda_{\alpha}}{c}\bigg) - 2\sum\limits_{a = 1}^{n_{3}}\arctan \bigg(\frac{k_{j}-\lambda_{a}}{2c} \bigg) +2\pi I_{j} +\Phi, 
\end{equation}
\begin{align}\label{eq:l2string}
 2\Lambda_{\alpha}L_R =  &-2\sum\limits_{j = 1}^{n_{1}}\arctan \bigg(\frac{\Lambda_{\alpha}-k_{j}}{c}\bigg)  - 2\sum\limits_{a = 1}^{n_{3}}\bigg[ \arctan \bigg(\frac{\Lambda_{\alpha}-\lambda_{a}}{c}\bigg) +\arctan \bigg(\frac{\Lambda_{\alpha}-\lambda_{a}}{3c}\bigg)\bigg]\nonumber \\
 &- 2\sum\limits_{\substack{\beta = 1 \\ \beta\neq\alpha}}^{n_{2}}\arctan \bigg(\frac{\Lambda_{\alpha}-\Lambda_{\beta}}{2c}\bigg) +2\pi J_{\alpha} +2\Phi, 
\end{align}
\begin{align}\label{eq:l3string}
3\lambda_{a}L_R =  &-2\sum\limits_{j = 1}^{n_{1}}\arctan \bigg(\frac{\lambda_{a}-k_{j}}{2c}\bigg)-  2\sum\limits_{\alpha = 1}^{n_{2}}\bigg[\arctan \bigg(\frac{\lambda_{a}-\Lambda_{\alpha}}{c}\bigg)+\arctan \bigg(\frac{\lambda_{a}-\Lambda_{\alpha}}{3c}\bigg) \bigg]\nonumber\\
&-2\sum\limits_{\substack{b = 1 \\ b\neq a}}^{n_{3}}\bigg[\arctan \bigg(\frac{\lambda_{a}-\lambda_{b}}{2c}\bigg)+\arctan \bigg(\frac{\lambda_{a}-\lambda_{b}}{4c}\bigg)\bigg]+2\pi K_{a}+3\Phi, 
\end{align}
for $j=1,\hdots, n_{1}$, $\alpha = 1,\hdots , n_{2}$ and $a = 1,\hdots , n_{3}$ with $I_{j}$, $J_{\alpha}$ and $K_{a}$ being the quantum numbers associated to the charge, first and second spin rapidities respectively. These are called Takahashi's equations for SU(3) fermions with attractive delta interaction~\cite{takahashisu3}. The total energy is given by
\begin{equation}\label{eq:4app15}
E  = \sum\limits_{j=1}^{n_{1}}k_{j}^{2} + \sum\limits_{\alpha=1}^{n_{2}}(2\Lambda_{\alpha}^{2}-2c^{2}) + \sum\limits_{a=1}^{n_{3}}(3\lambda_{a}^{2}-8c^{2}).
\end{equation}
\noindent From the outset, it becomes clear from the construction of the Takahashi equations that a system with an equal number of particles per species, will consist solely of trions: the population of the third hyperfine level given by $M_{2}$, corresponds to the number of trions $n_{3}$. It immediately follows that trions are formed in the system whenever possible, with this statement being confirmed by the expression for the energy. At any value of the interaction $|c|$, a trion is more stable (lower energy) than that of a CSF ($n_{1}=n_{2}=1$). Therefore, to form an unpaired particle or a CSF in the continuous limit for $L_R|c|\!\gg\!1$, the SU(3) symmetry needs to be broken. Alternatively, as mentioned previously, one could consider the system in the grand-canonical ensemble with adjustable chemical potentials allowing for multiple phase transitions between unpaired, paired, trions and a mixture thereof through colour selective magnetic fields~\cite{Batchelor2010exact,kuhn2012phase}.  

\section{Probe for bound states of three-component fermions}\label{sec:pcp4}

At small interactions, the persistent current is found to be a function with a period given by the elementary flux quantum $\phi_{0} = \hbar/mR^{2}$ with $m$ and $R$ denoting the atoms' mass and ring radius respectively. However, on going to stronger interactions, the persistent current displays fractionalization reducing its period. 
\begin{figure}[h!]
\centering
    \includegraphics[width=\linewidth]{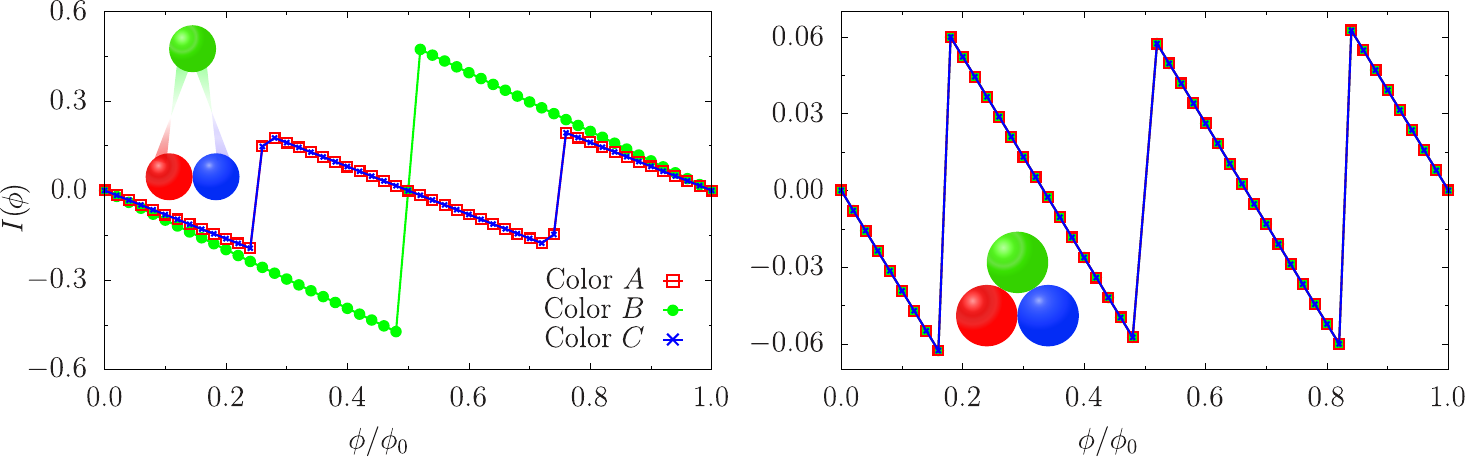}
    \put(-237,115){(\textbf{a})}
    \put(-22,115){(\textbf{b})}
    \caption{Persistent current $I(\phi)$ of the three colours against the effective magnetic flux $\phi/\phi_{0}$. The left (right) figure depicts the persistent current of a CSF (trion). The interactions for the CSF are $|U_{AB}|=|U_{BC}| = 0.01$ and $|U_{AC}|=3$. For a trion, $|U| = 3$ for all colours. All presented results are obtained for $N_{p} = 9$ and  $L = 15$ using DMRG. The lines are meant to be a guide to the eye for the reader, to aid in perceiving the fractionalization.  Figure adapted from~\cite{chetcuti2021probe}.}
    \label{fig:main}
\end{figure}

\noindent Contrary to attractive bosons~\cite{naldesi2019rise,polo2020exact,polo2021quantum} or repulsive $N$-component fermions~\cite{yu1992persistent,chetcuti2021persistent}, the reduction of the period in attracting $N$-component fermions does not depend on the total number of particles present in the system. Instead, the effective mass of the bound state, which amounts to the number of particles constituting it, dictates the period of the current. Starting from the CSF case, we observe a halved period $\phi_{0}/2$ for the paired particles, whilst the unpaired particles maintain the bare periodicity $\phi_{0}$ --Figure~\ref{fig:main}(\textbf{a}). On the other hand, in the SU(3) symmetric case, the persistent current exhibits a tri-partite periodicity reflecting the presence of trions --Figure~\ref{fig:main}(\textbf{b}).
\begin{figure}[h!]
    \centering
    \includegraphics[width=0.495\linewidth]{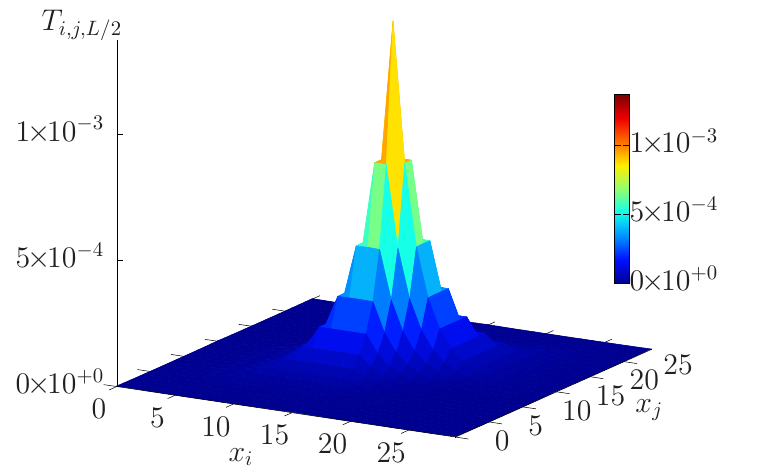}
    \includegraphics[width=0.495\linewidth]{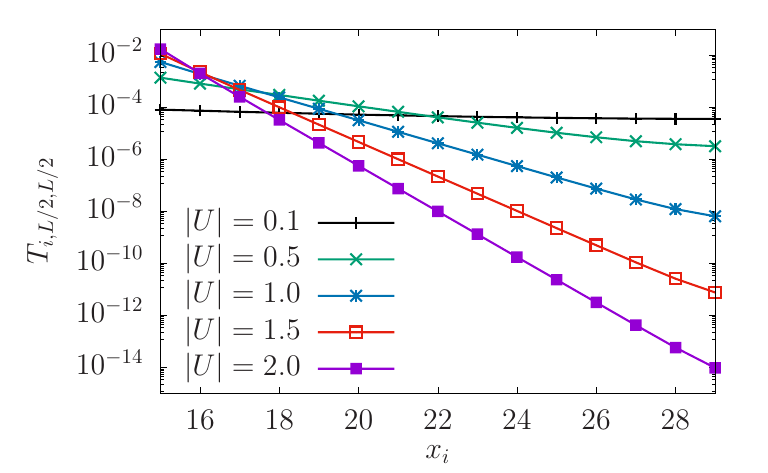} 
    \put(-435,120){(\textbf{a})}
    \put(-210,120){(\textbf{b})} 
    \caption{(\textbf{a}) Three-body correlation function, $T_{i,j,L/2}$, and its decay (\textbf{b}) for $L = 30$ and $N_{p}$ = 3 with a trion configuration for $|U_{AB}| = |U_{AC}| = |U_{BC}|= 0.5$. All results were obtained with exact diagonalization for flux $\phi = 0$. Figure adapted from~\cite{chetcuti2021probe}.}
    \label{fig:densdensdenstrion}
\end{figure}

\noindent The nature of the bound states can be corroborated through analysis of the three-body correlation function defined as:
\begin{equation}\label{eq:three}
T_{i,j,k} = \langle c^{\dagger}_{i,A}c^{\dagger}_{j,B}c^{\dagger}_{k,C}c_{k,C}c_{j,B}c_{i,A}\rangle ,
\end{equation}
where $i,j,k$ denote the lattice sites, $A,B,C$ correspond to the different species and $c_{j\alpha}^{\dagger}$ is the typical fermionic creation operator for colour $\alpha$ at site $j$. Similar correlation functions have been studied in~~\cite{Capponi2008molecular,Klingschat2010exact,pohlmann2013trion}. Analysing the decay of this correlation function for both trions and CSFs highlights the nature of the bound states in our lattice system.
\begin{figure}[h!]
\centering
    \includegraphics[width=0.495\linewidth]{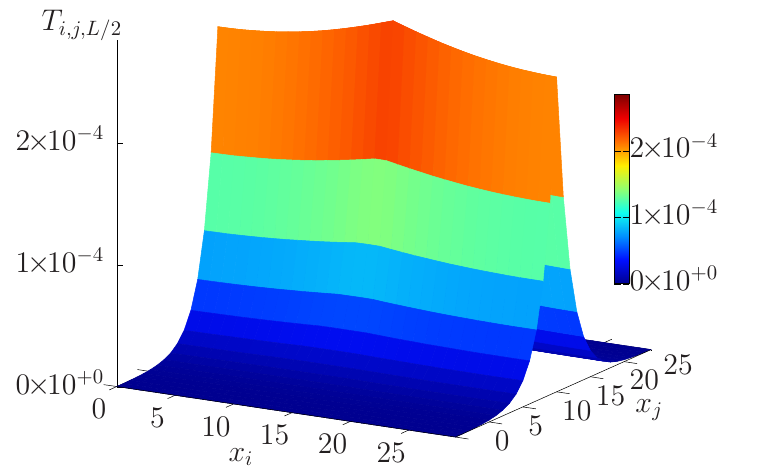}
    \includegraphics[width=0.495\linewidth]{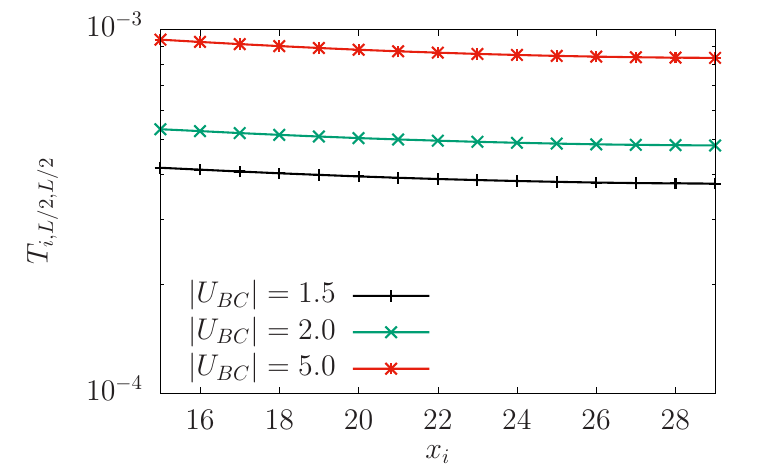}
    \put(-435,120){(\textbf{a})}
    \put(-210,120){(\textbf{b})}    
    \caption{
    (\textbf{a}) Three-body correlation function, $T_{i,j,L/2}$, for $L = 30$ and $N_{p}$ = 3 with a CSF configuration for $|U_{AB}|=|U_{AC}|= 0.01$ and $|U_{BC}|= 5$. (\textbf{b}) The decay of the correlation function for different values of $|U_{BC}|$ is depicted. All results were obtained with exact diagonalization for flux $\phi = 0$. Figure adapted from~\cite{chetcuti2021probe}.}
        \label{fig:d3csf}
\end{figure}

\noindent Here, we consider two instances of the three-body correlator~\eqref{eq:three} for a CSF and a trion presented in Figures~\ref{fig:densdensdenstrion} and~\ref{fig:d3csf} respectively, by fixing one of the axes such that $T_{i,j,L/2}$. In panel (\textbf{b}), we calculate the corresponding decay for different values of $U$. For a trionic case, the decay shows us that the correlation length associated to the bound state decays exponentially in $\{i,j,k\}$, meaning that the localization of the effective molecules increases. This should be contrasted with the results of the CSF, which only displays an exponential decay on $j$ as the bound states are of a lower degree, i.e., consist of two particles. \\

\noindent Finally, we point out that demonstrating that the long-distance three-body correlation length dominates over any other correlator, gives us a clear signature of the nature of the bound state, and shows that in our exact diagonalization simulations (in the three particle sector), trions are formed as soon as attractive interactions are present in the system. 

\subsection{Bethe ansatz analysis of SU(3) symmetric bound states}

\noindent Further insight on the persistent current fractionalization of trions can be gained through the Bethe ansatz analysis of the Gaudin-Yang-Sutherland model. Due to SU(3) symmetry breaking, the CSF case is out of reach of Bethe ansatz. \\

\noindent As we discussed in Section~\ref{sec:bethegys4}, the main technical feature of attracting fermions is that the ground-state is composed of bound states. In integrable theories, this corresponds to complex Bethe rapidities, which we have already sketched out how such complex solutions in the limit of $L_R|c|\!\gg\! 1$ are arranged in the Takahashi string solutions. For small systems with weak interactions, Takahashi's equations assuming ideal string configurations of the solution, cannot be used to solve the Bethe ansatz equations and, in turn, access the energy. To this end, we obtain the exact solution of the original Bethe equations, i.e., in their product form, without imposing any constraints on $c$ and $L_{R}$. Our approach is a numerical iterative method, wherein we start close to $c\lesssim 0$ that can be straightforwardly calculated analytically and determine the solution at every step with increasing interactions. Note that for a solution to be valid: it must be continuous in energy, does not display jumps in the logarithmic form, and consists of distinct rapidities such that the solution of the Bethe equations is well defined.
\begin{figure}[ht!]
    \centering
    \includegraphics[width=0.6\textwidth]{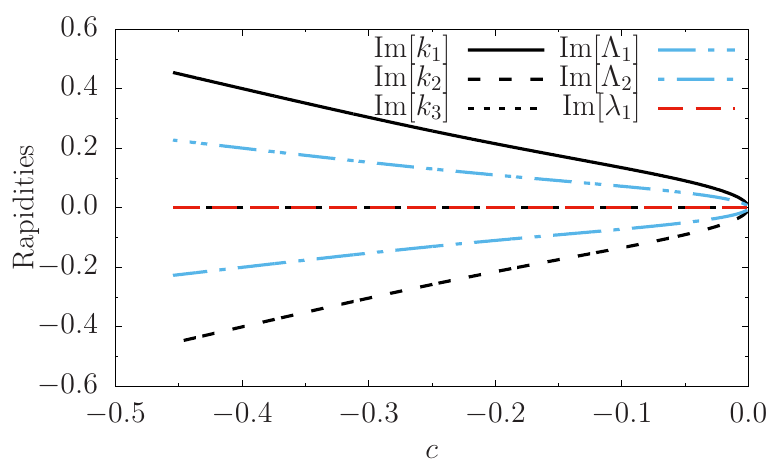}%
    \caption{The complex part of the rapidities against interaction $c$. As the interaction increases, the imaginary part of the rapidities goes to the ideal string limit as outlined in Equations~\eqref{eq:strings}. Note that in this case, the real part of the rapidities $k_j$, $\Lambda_\alpha$ and $\lambda_a$ is always zero. Results were obtained using the product form of the Bethe ansatz equations of the Gaudin-Yang-Sutherland model~\eqref{eq:chargesu3rep},~\eqref{eq:spin1su3rep} and~\eqref{eq:spin2su3rep}.}
    \label{fig:prodbethea}
\end{figure}

\noindent Trions are observed to be formed for any arbitrary small attraction in the three-particle sector (see~\cite{pohlmann2013trion,capponi}), corroborating the results obtained from the correlation functions. Through Figure~\ref{fig:prodbethea}, we can confirm that for the three-particle case considered here, a trion appears immediately. The reason being that both $k_{j}$ and $\Lambda_{\alpha}$ admit complex solutions for any $c\!<\! 0$.  If it were the case that a CSF was formed as an intermediate step, then there would have been a critical value of the interaction up to which its imaginary part would be null, as per the structure of the string solutions in Equation~\eqref{eq:strings}. 
\begin{figure}[ht!]
    \centering
    \includegraphics[width=0.5\textwidth]{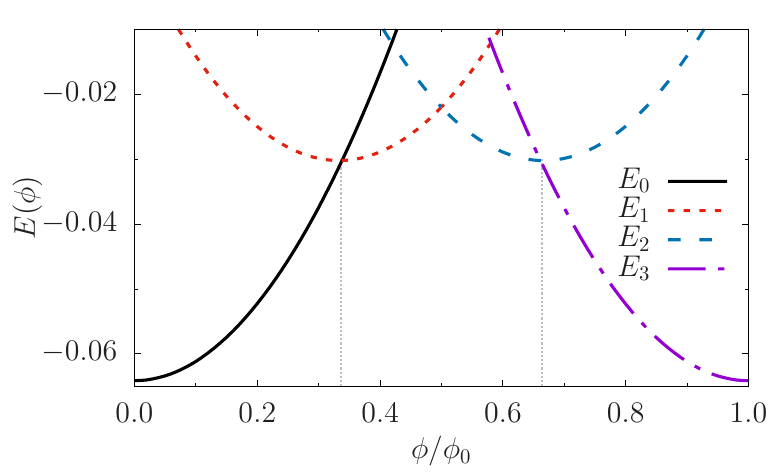}%
    \includegraphics[width=0.5\textwidth]{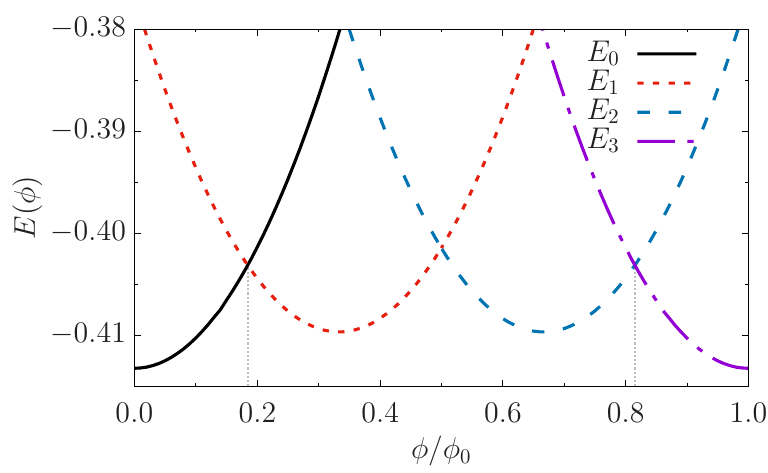}
    \put(-388,45){(\textbf{a})}
    \put(-175,45){(\textbf{b})}
    \caption{Results obtained from solving the product form Bethe ansatz equations of the Gaudin-Yang-Sutherland model~\eqref{eq:chargesu3rep},~\eqref{eq:spin1su3rep} and~\eqref{eq:spin2su3rep} for (\textbf{a}) $|c|=0.155$ and $|c|=0.45$ with $N_p=3$ and $N=3$. In (\textbf{a}), the energy $E(\phi)$ starts to fractionalize, whilst in 
    (\textbf{b}) the fractionalization is more pronounced and the tri-partite periodicity is almost reached. The size is fixed to $L_R=20$ for all curves. }
    \label{fig:prodbethe}
\end{figure}

\begin{figure}[h!]
    \centering
    \includegraphics[width=\linewidth]{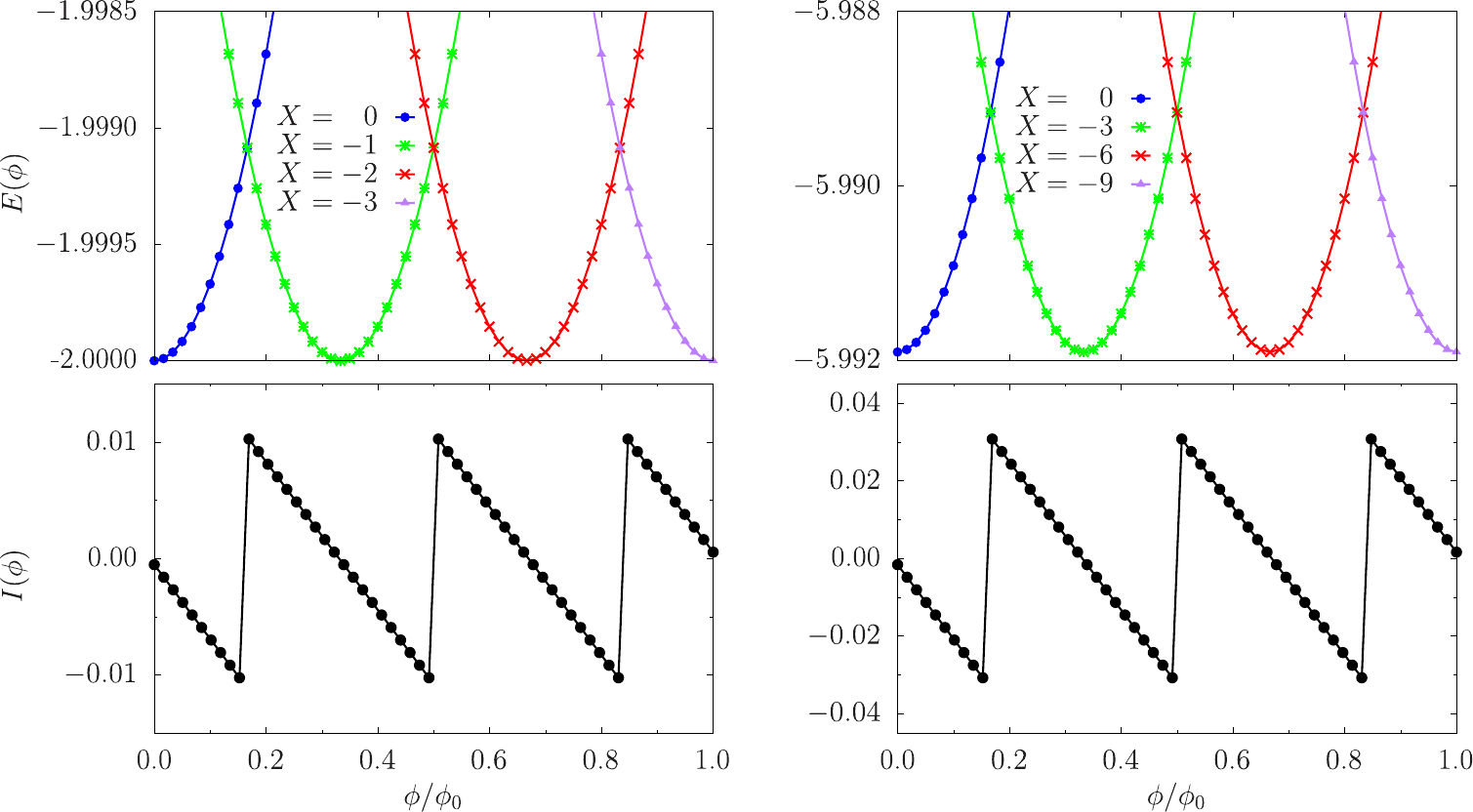}%
     \put(-238,220){(\textbf{a})}
     \put(-23,220){(\textbf{b})}
    \put(-238,110){(\textbf{c})}
     \put(-23,110){(\textbf{d})}
    \caption{The top row (\textbf{a}) and (\textbf{b}) depict how the Bethe ansatz energies $E(\phi)$ need to be characterized by the quantum numbers ($X = \sum_{a}K_{a}$) in order to have the actual ground-state for increasing flux $\phi/\phi_{0}$.  The bottom row (\textbf{c}) and (\textbf{d}) show the corresponding persistent current $I(\phi)$ for a system consisting solely of trions with $N_{p}=3,9$ respectively. All the presented results were obtained with the Takahashi equations of the Gaudin-Yang-Sutherland model for $L_{R}=60$ and $|c|=2$. }
    \label{fig:bet}
\end{figure}
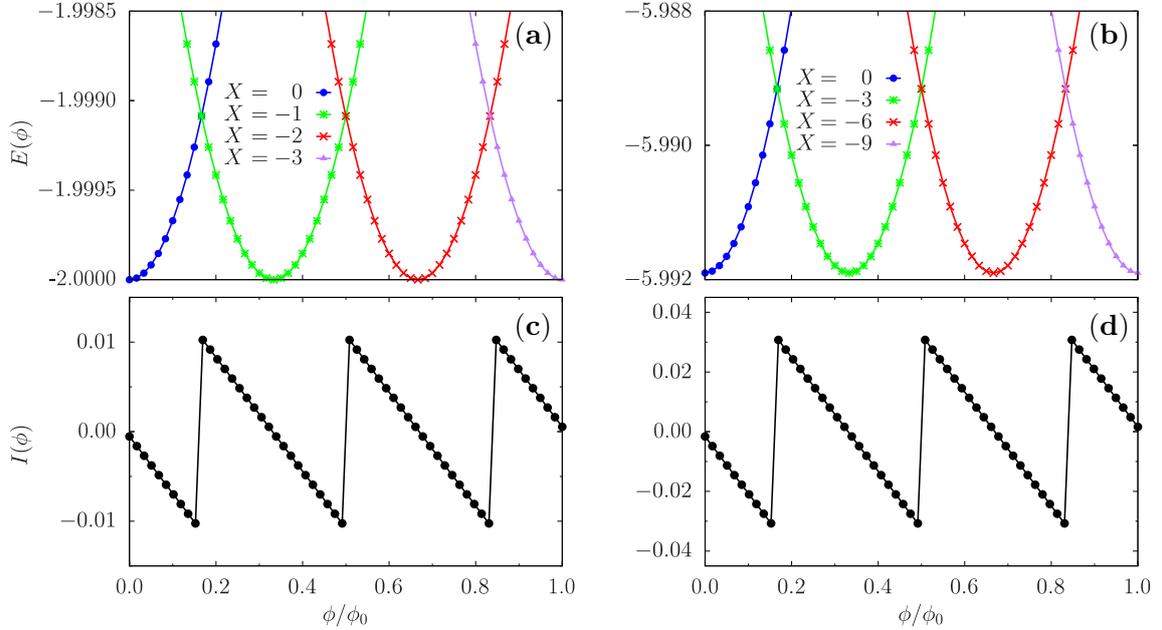

\noindent The energy obtained from the solution of the product form displayed in Figure~\ref{fig:prodbethe}, albeit exhibiting fractionalization, does not demonstrate the perfect tri-partition of the period $\phi_{0}/3$. This implies that there are deviations from the ideal string hypothesis. However, on increasing interactions and accessing the regime of large $L_{R}c/t$, the results coincide with the analysis based on Takahashi's equations corresponding to the formation of a three-body bound state --Figure~\ref{fig:bet}.  

\noindent The mechanism leading to fractionalization in the attractive regime can be properly understood by studying the Bethe ansatz equations as $|c|\rightarrow\infty$. In this limit, for a system comprised solely of trions ($n_{1}=n_{2}=0$), the Bethe equations are reduced to the following form 
\begin{equation}\label{eq:inf1}
\lambda_{a} = \frac{2\pi}{L_R}\bigg(\frac{1}{N}K_{a} + \phi\bigg),
\end{equation}
where $\lambda_{a}$ is the spin rapidity associated to the three-body bound states and $K_{a}$ being the aforementioned quantum numbers. Consequently, we find an exact expression for the zero temperature persistent current, which reads
\begin{equation}\label{eq:inf3}
 I(\phi) = -6\bigg(\frac{2\pi}{L_R}\bigg)^{2}\sum\limits_{a}^{n_{3}}\bigg[\frac{K_{a}}{N}+\phi\bigg].
\end{equation}
This expression implies that as the flux, which is taken to be positive, is increased, the quantum numbers $K_{a}$ need to shift to take on a negative values so as to counteract the increase in flux. Here, a change in the quantum numbers creates excitations in the ground-state by having energy level crossings between the ground and excited states, that in turn cause the fractionalization --Figures~\ref{fig:prodbethe} and~\ref{fig:bet}. These excitations, being discrete in nature, can only partially compensate for the increase in flux resulting in oscillations with a reduced period of $1/N$, thereby accounting for the `size' of the bound state. \\

\noindent Although the fractionalization in both attractive and repulsive regimes shares several qualitative features, the mechanism facilitating it, is markedly different. Remarkably, for the trionic system, the Bethe equations are decoupled as in Equation~\eqref{eq:inf1} such that $\lambda_{a}$ is only dependent on its own quantum number. However, this does not occur for repulsive interactions. Although a similar expression is obtained in the latter (see Equation~\eqref{eq:kincomm}), it is still dependent on the spin quantum numbers of the other particles. Thus, whilst in the repulsive case, a single spin excitation aims to counteract the flux for the whole system, the same does not hold true for its attractive counterpart. Indeed, the decoupling between the ``different bound'' states (trions), implies that all the trions corresponding to a rapidity $\lambda_{a}$, need to shift their quantum number $K_{a}$ to counteract the increase in flux and minimize the energy. An example on how to change the quantum numbers with increasing flux for a system with attractive interactions is presented in Table~\eqref{tab:taa}. 
\begin{table}[ht!]
\centering
\begin{tabular}{|c|c|}
    \hline
     Magnetic Flux & $K_{a}$ \\
     \hline
     0.0 -- 0.1 &  \{$-$1,0,1\} \\
     0.2 -- 0.5 &  \{$-$2,$-$1,0\} \\
     0.6 -- 0.8 & \{$-$3,$-$2,$-$1\} \\
     0.9 -- 1.0 & \{$-$4,$-$3,$-$2\} \\
     \hline
\end{tabular}
     \caption{Quantum number configurations with the flux for SU(3) fermions with $N_{p}=9$ for a system containing only three trions. }\label{tab:taa}
\end{table} 

\noindent Immediately, one can observe that all the quantum numbers are shifted with increasing flux as opposed to the repulsive case outlined in Table~\eqref{T:totalrep}. Additionally, unlike its repulsive counterpart, there is only one set of quantum numbers implying that the fractionalized parabolas are not degenerate.

\section{Parity effects}\label{sec:par4}
One consequence of Leggett's theorem is related to the parity of the persistent current, which, as we explored in Chapter~\ref{chp:repcurr} is diamagnetic [paramagnetic] for (2$n$+1)$N$ [(2$n$)$N$] fermions with $n$ being an integer number~\cite{leggett1991,chetcuti2021persistent}. These parity effects are demonstrated to be washed out upon fractionalization for two-component fermions with attractive interactions~\cite{pecci2021probing} and for strongly repulsive $N$-component fermions~\cite{chetcuti2021persistent}, with the persistent current exhibiting diamagnetic behaviour irrespective of the number of particles in the system. In the case of trions, the persistent current maintains its parity despite undergoing fractionalization. \\

\begin{figure}[h!]
    \includegraphics[width=0.5\textwidth]{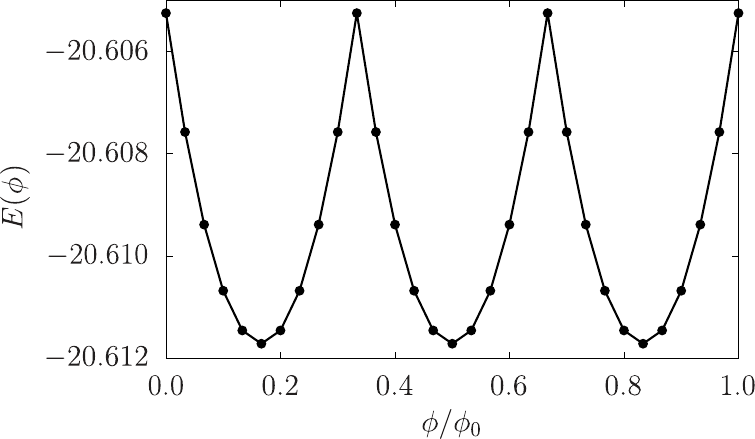}%
    \includegraphics[width=0.5\textwidth]{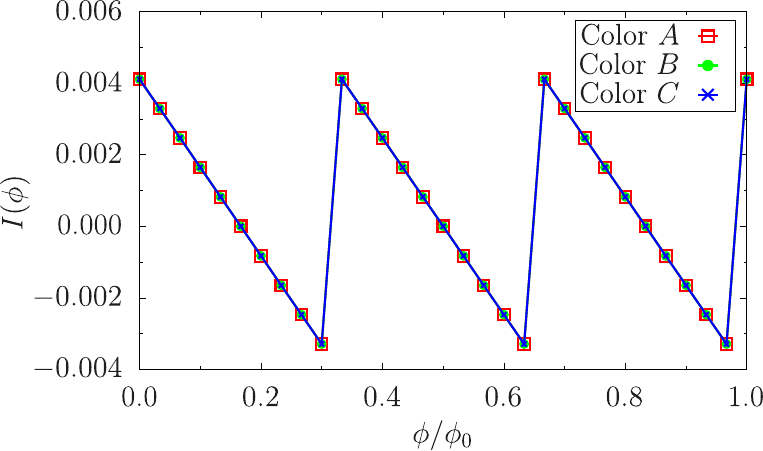}
    \put(-378,33){(\textbf{\textbf{a}})}
    \put(-172,33){(\textbf{\textbf{b}})}
    \caption{(\textbf{a}) Ground-state energy $E(\phi)$ and (\textbf{b}) the corresponding persistent current $I(\phi)$ as a function the flux $\phi/\phi_{0}$ for a system of $N_{p}=6$ with $L=20$ in a trion configuration for $|U_{AB}| = |U_{AC}|=|U_{BC}|=3$. All results were obtained with exact diagonalization. Figure adapted from~\cite{chetcuti2021probe}.}
    \label{fig:partrion}
\end{figure}

\begin{figure}[h!]
    \includegraphics[width=0.5\textwidth]{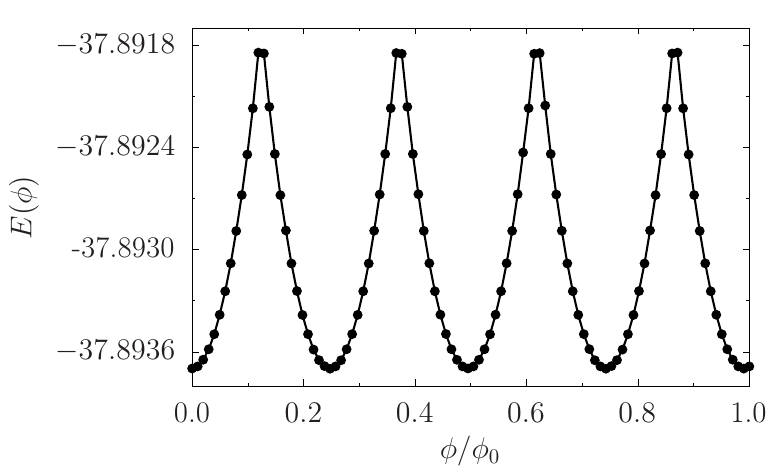}%
    \includegraphics[width=0.5\textwidth]{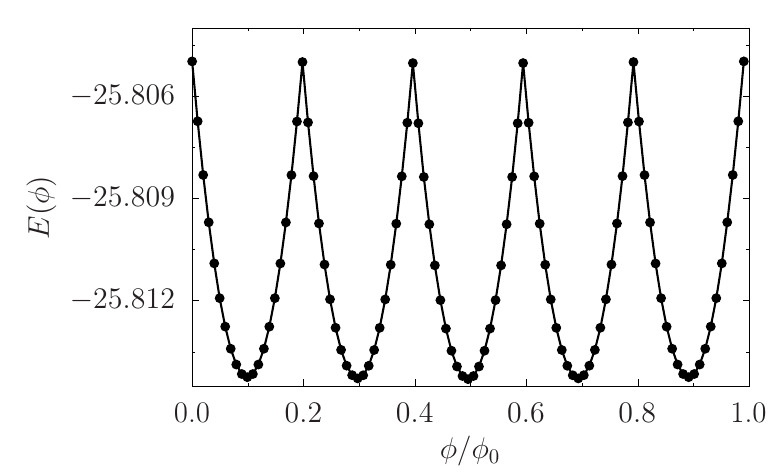}
    \put(-372,113){(\textbf{\textbf{a}})}
    \put(-157,113){(\textbf{\textbf{b}})}
    \caption{Ground-state energy $E(\phi)$ as a function the flux $\phi/\phi_{0}$ for a system of (\textbf{a}) $N_{p}=8$ and (\textbf{b}) $N_{p}=10$ fermions with SU(4) and SU(5) symmetric fermions in $N$-body bound state configuration with $L=15$. Results were obtained with DMRG with $|U|=3$ ($|U|=1$) for the $N=4$ ($N=5$) case. }
    \label{fig:parsun}
\end{figure}

\noindent Taking a look at a system of six SU(3) symmetric fermions with equal interactions, as depicted in Figure~\ref{fig:partrion}, we find that for larger interactions, not only does the system fractionalize, but it also retains its paramagnetic nature. Such a feature can be attributed to the fact the bound states composed of an odd (even) number of fermions have an (anti-)symmetric wavefunction~\cite{yin2011effective}. Essentially, one can effectively treat an $N$-body bound state of attractive multicomponent fermions as composite fermions or bosons, described by spinless fermions with $p$-wave interactions or super Tonks Girardeau phase\footnote{
The super Tonks-Girardeau phase arises upon abruptly switching from strongly repulsive coupling in the Tonks-Girardeau phase to strong attractive interactions. The attractive phase is more strongly correlated than its repulsive counterpart, earning the former its moniker of super Tonks-Girardeau gas~\cite{batchelor_defsuper}.} of attracting bosons for odd and even $N$ respectively~\cite{yin2011effective}. This is in line with the fact that persistent currents of bosons are always diamagnetic for any number of particles~\cite{polo2020exact}. On going to systems with $N=4$ ($N=5$), we observe that the parity effect is lost (kept), thereby reaffirming the previous statements --Figure~\ref{fig:parsun}.

\begin{figure}[h]
    \includegraphics[width=0.5\textwidth]{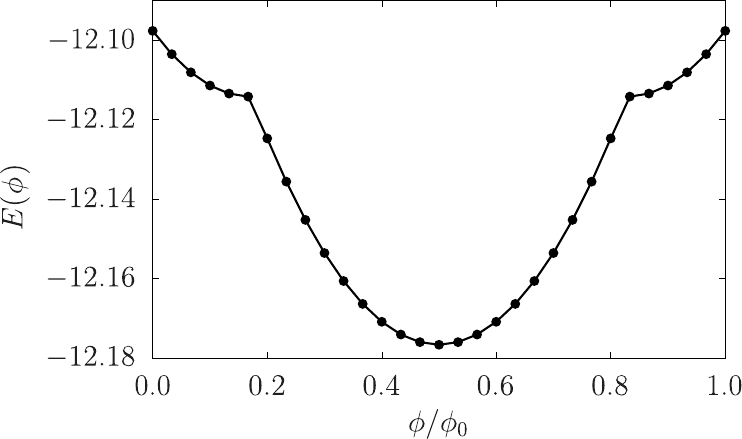}%
    \includegraphics[width=0.5\textwidth]{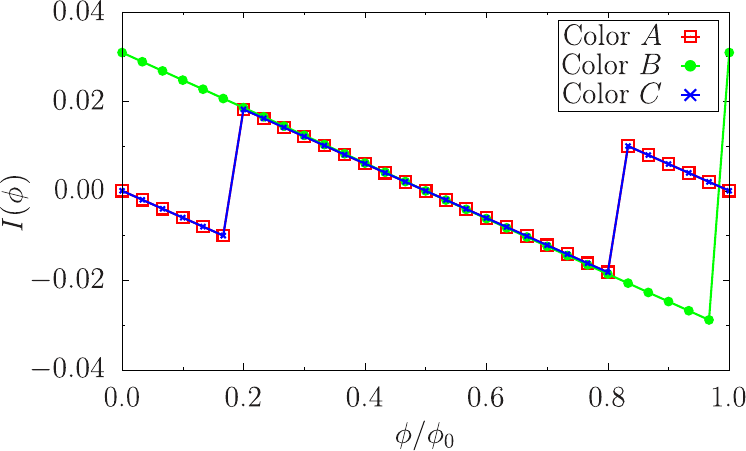}
    \put(-378,33){(\textbf{\textbf{a}})}
    \put(-172,33){(\textbf{\textbf{b}})}
    \caption{(\textbf{a}) Ground-state energy $E(\phi)$ and (\textbf{b}) the corresponding persistent current $I(\phi)$ as a function the flux $\phi/\phi_{0}$  for a system  of $N_{p}=6$ with $L=20$ in a CSF configuration for $|U_{AB}| = |U_{BC}|=0.01$ and $|U_{AC}|=1$. All results were obtained with exact diagonalization. The lines in the right panel are meant as a guide to the eye for the reader. Figure adapted from~\cite{chetcuti2021probe}.}
    \label{fig:parcsf}
\end{figure}

\noindent In contrast with the trion case, a system composed solely of CSFs has a different behaviour --Figure~\ref{fig:parcsf}. For the paired colours, the persistent current is diamagnetic reflecting the formation of composite bosons with double the mass~\cite{leggett1991,pecci2021probing}. However, the remaining unpaired particles display diamagnetic (paramagnetic) behaviour if they are odd (even), indicating that they are nearly free.

\section{Finite temperature effects}\label{sec:fintemp}

Having established the persistent current as an effective diagnostic tool to probe the various bound states of SU(3) fermions, we turn our attention to monitor its behaviour at finite temperature, in view of the recent experimental realization of fermionic currents~\cite{wright2022persistent,roati2022imprinting}.
\begin{figure}[h!]
    \includegraphics[width=\linewidth]{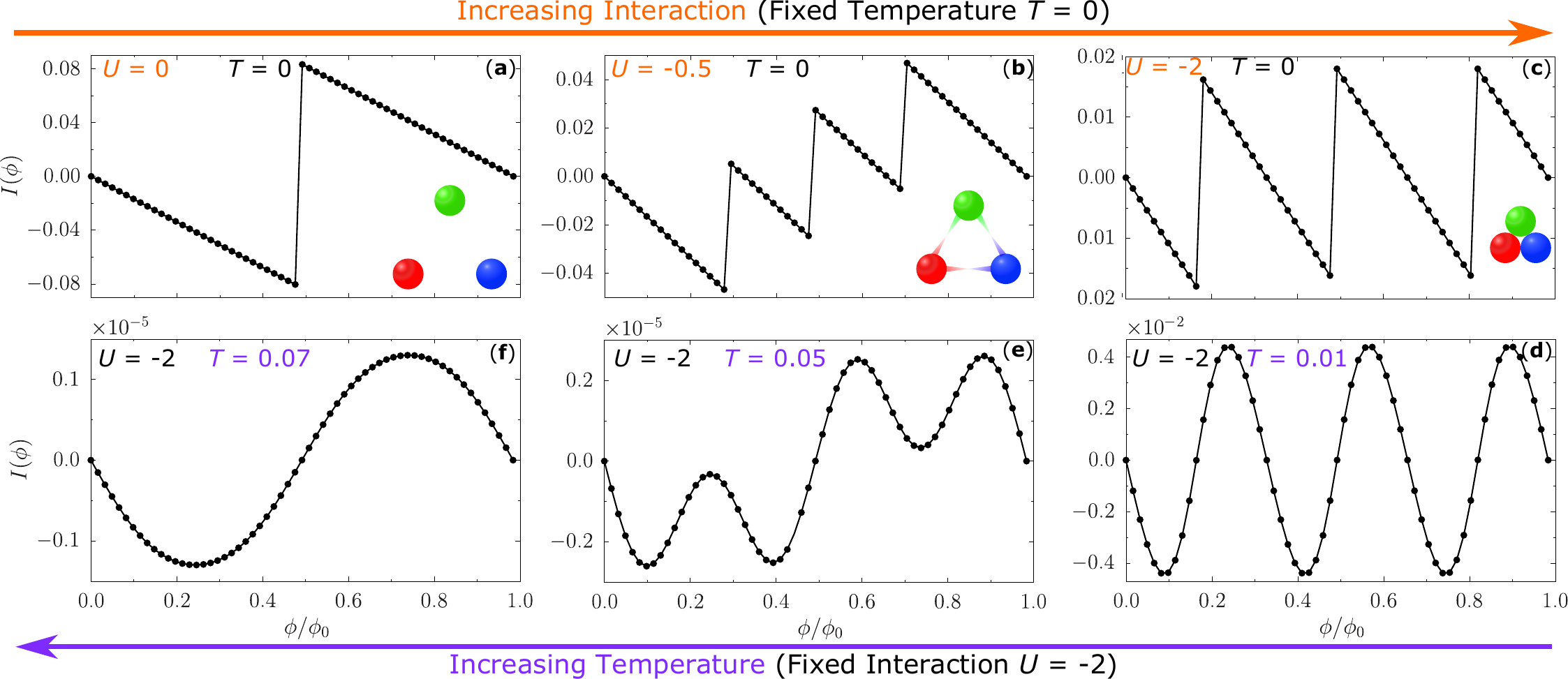}%
    \caption{Persistent current $I(\phi)$ of SU(3) symmetric fermions, defined in Equation~\eqref{eq:tempcurr}, for various interactions $U$ (temperatures $T$) in the upper (lower) panel. Top panels (\textbf{a})-(\textbf{c}) for fixed $T=0$ and varying $U$: persistent current fractionalizes with increasing $U$ with the bare period $\phi_{0}$ being reduced $\phi_0/N$ at strong interactions. Bottom panels (\textbf{d})-(\textbf{f}) for fixed $U=-2$ and increasing $T$: persistent current regains the period $\phi_0$ upon increasing $T$. The results were obtained by exact diagonalization with $N_{p} =3$ and $L=15$. Lines are a guide to the eye for the reader.}
    \label{fig:tempcurr}
\end{figure}

\noindent The interplay between thermal fluctuations and attractive interactions leads to remarkable effects in the persistent current. Besides the generic smoothening of the saw-tooth behaviour, finite temperature causes specific changes in the frequency of the persistent current. From Figure~\ref{fig:tempcurr} (bottom panel), we observe that qualitatively the effect of thermal fluctuations mimics a reduction in interaction strength: the persistent current regains its original single particle frequency on increasing temperature. Such a phenomenon is consistent with the thermal effects on two-component fermions with repulsive interactions~\cite{patu}, which also holds for attractive interactions, irrespective of the number of particles in the system (see Appendix of~\cite{chetcuti2021probe}). Additionally, we point out that the amplitude of the finite temperature persistent current is exponentially reduced on comparison with its zero temperature counterpart. One possible explanation is that temperature can activate transitions between different angular momentum states, thereby causing a decay in the current's amplitude~\cite{joantemp}. Alternatively, the populated levels give rise to persistent currents that are essentially equal in magnitude but opposite in sign, i.e., different directionalities, thereby suppressing the current. \\

\noindent So far, the mechanism behind the `de-fractionalization' brought about by the increase in temperature has not been addressed. Turns out that the persistent current exhibits specific dependencies on the nature of the interactions governing present in the systems, ones that are not subtle. Such behaviour can be properly understood by carrying out a quantitative description of the persistent current frequency, which is presented in Section~\ref{sec:quanti} for trions.

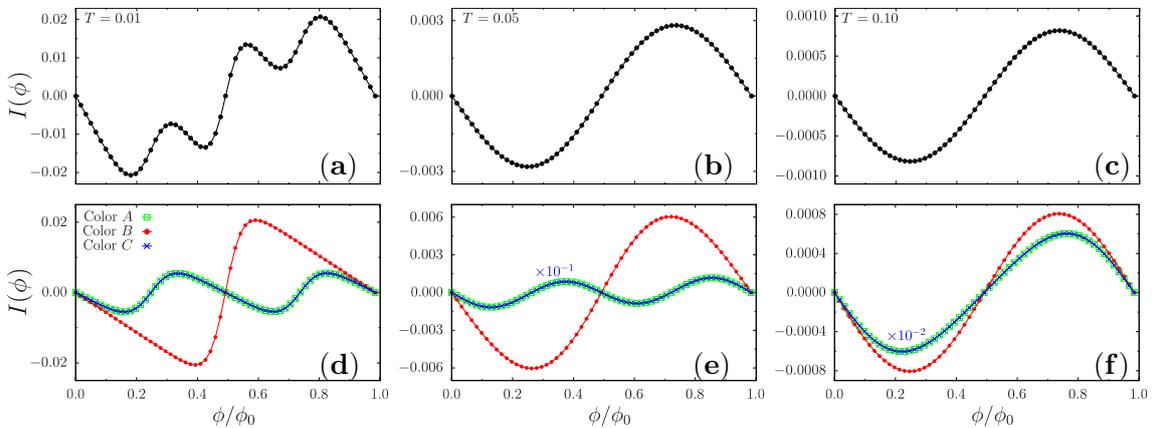
\begin{figure}[h!]
    \includegraphics[width=\linewidth]{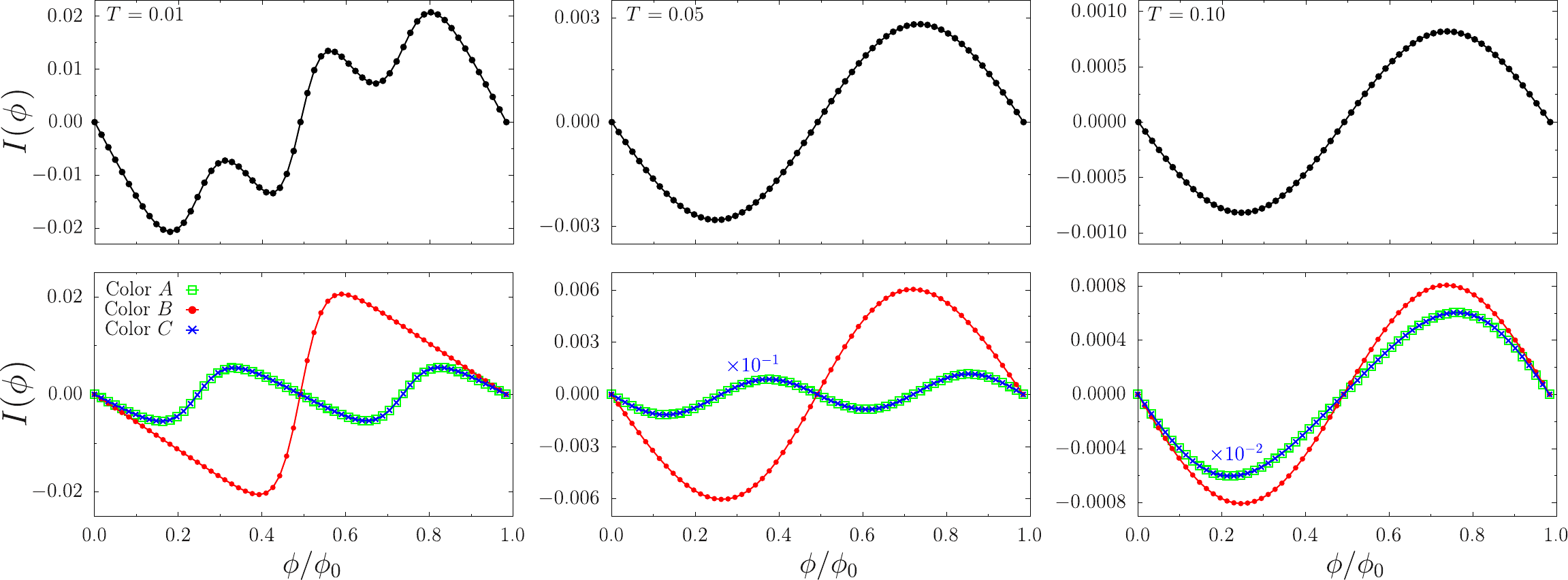}%
    \put(-310,95){(\textbf{\textbf{a}})}
    \put(-170,95){(\textbf{\textbf{b}})}
    \put(-20,95){(\textbf{\textbf{c}})}
    \put(-310,20){(\textbf{\textbf{d}})}
    \put(-170,20){(\textbf{\textbf{e}})}
    \put(-20,20){(\textbf{\textbf{f}})}
    \caption{Top (bottom) panels depict the persistent current $I(\phi)$ for all (each) colours against the flux $\phi/\phi_{0}$ for different values of the temperature $T$. All the results were obtained with exact diagonalization for a system of $N_{p}=3$ with $L=20$ in a CSF configuration for $|U_{AB}| = |U_{BC}|=0.01$ and $|U_{AC}|=3$. The lines  are meant as a guide to the eye for the reader. Figure adapted from~\cite{chetcuti2021probe}.}
    \label{fig:csftemp}
\end{figure}

\noindent In the case of a CSF configuration, finite temperature affects the current in a similar manner. At a glance, the current's frequency indicates that, compared with trions, CSFs are less robust  to thermal fluctuations. However, looking at the persistent current per species paints a more interesting picture. The total current has a period given by the bare flux quantum purely because the amplitude of the two colours in the pair is smaller than that of the unpaired colour. Interestingly enough, the temperature required to break the interaction between the pair is higher than that required for a system of symmetric trions with the same interaction.

\subsection{Quantitative description of thermal effects in SU(3) symmetric systems}\label{sec:quanti}
To study the specific dependence of the persistent current on temperature and interaction, we analyse its power spectrum. This is achieved by taking the Fourier transform such that \begin{equation}\label{eq:four1}
   C_{n} = \frac{1}{P}\int\limits_{P}X(x)\cdot e^{-\frac{2\imath\pi}{P} n x}\textrm{d}x,
\end{equation}
where $C_{n}$ denotes the Fourier coefficient, $P$ is the period of the function and $X$ is the periodic function, which in our case is the finite temperature persistent current as expressed in Equation~\eqref{eq:tempcurr}. \\

\begin{figure}[h!]
    \includegraphics[width=0.5\textwidth]{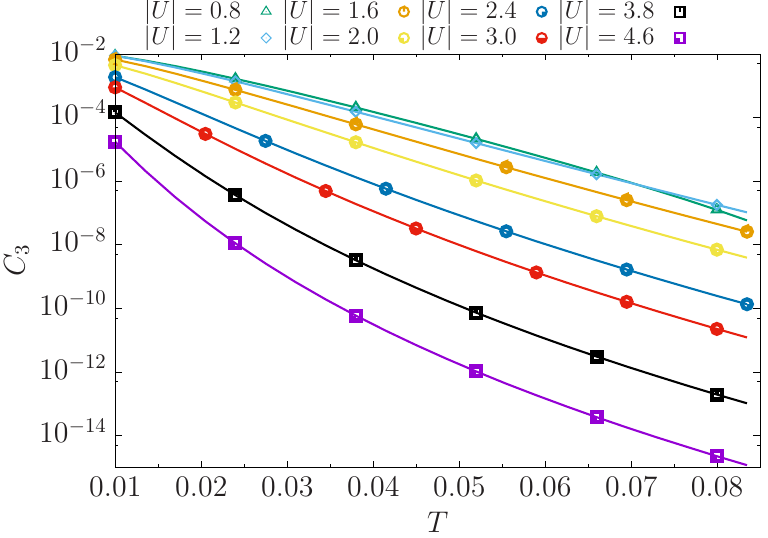}%
    \includegraphics[width=0.5\textwidth]{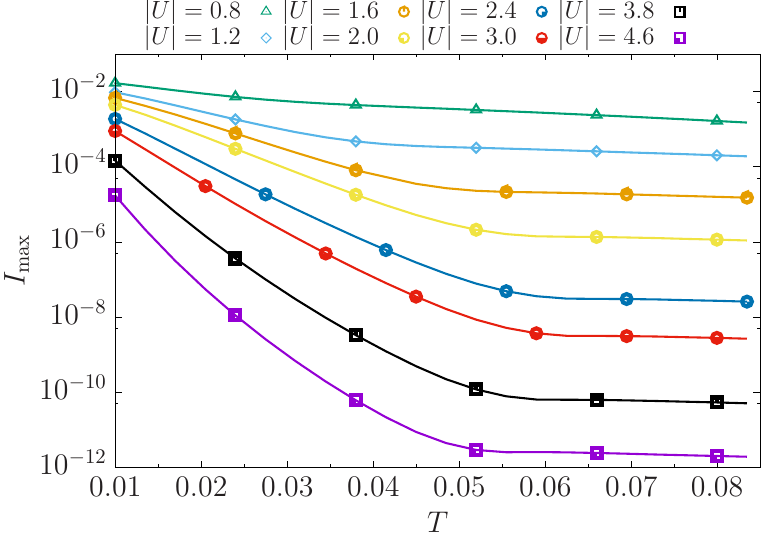}
    \put(-392,33){(\textbf{\textbf{a}})}
    \put(-175,33){(\textbf{\textbf{b}})}
    \caption{(\textbf{a}) Fourier weight $C_{3}$ as a function of temperature $T$ for different values of the interaction $U$. (\textbf{b}) depicts the corresponding maximum amplitude of the persistent current $I_{\mathrm{max}}$ against temperature $T$ for different $U$. The presented results were obtained with exact diagonalization for $N_{p} = 3$ and $L=15$. Figure adapted from~\cite{chetcuti2021probe}. }
    \label{fig:fourier}
\end{figure}
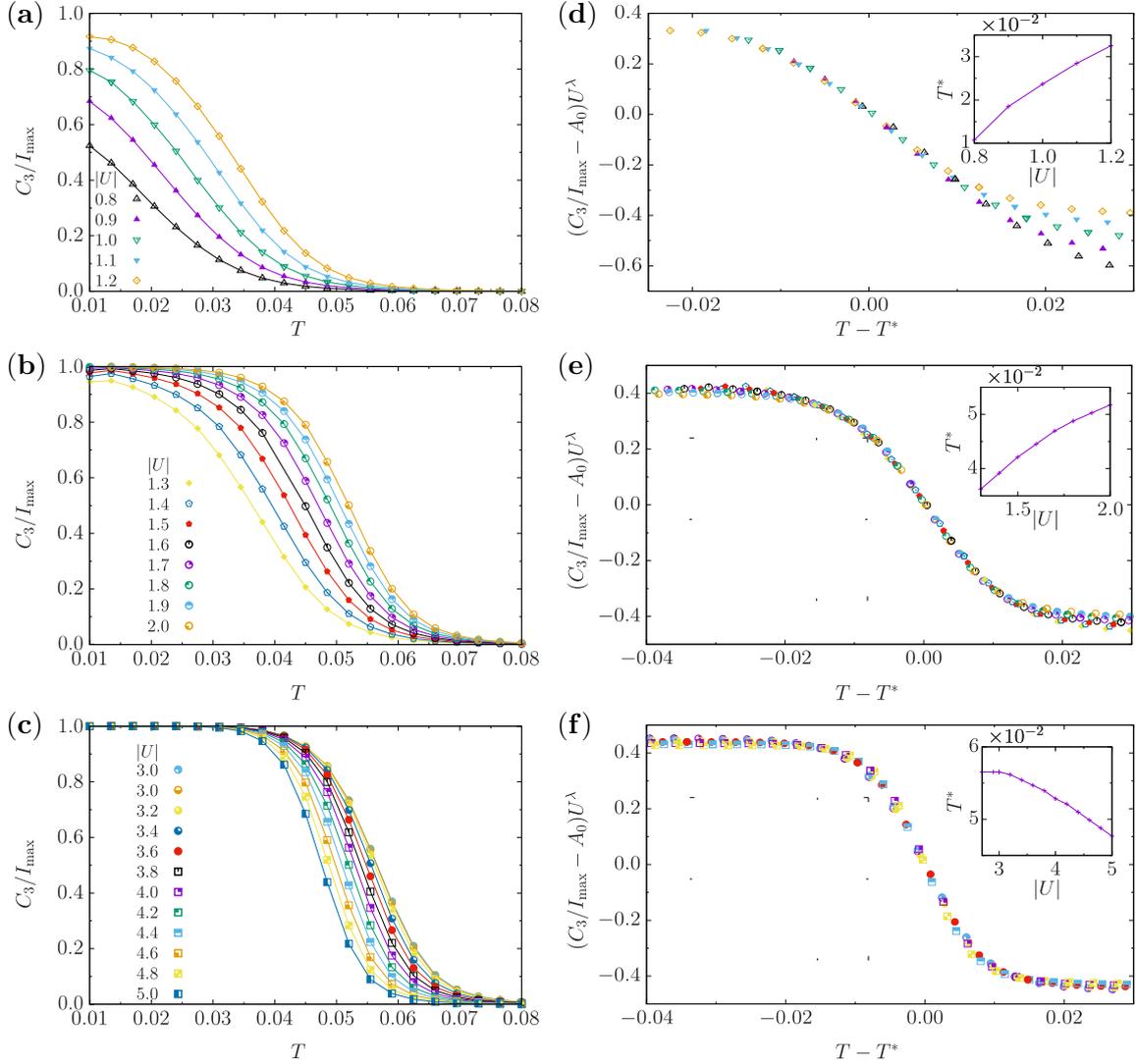
\begin{figure}[h!]
    \centering
    \includegraphics[width=\linewidth]{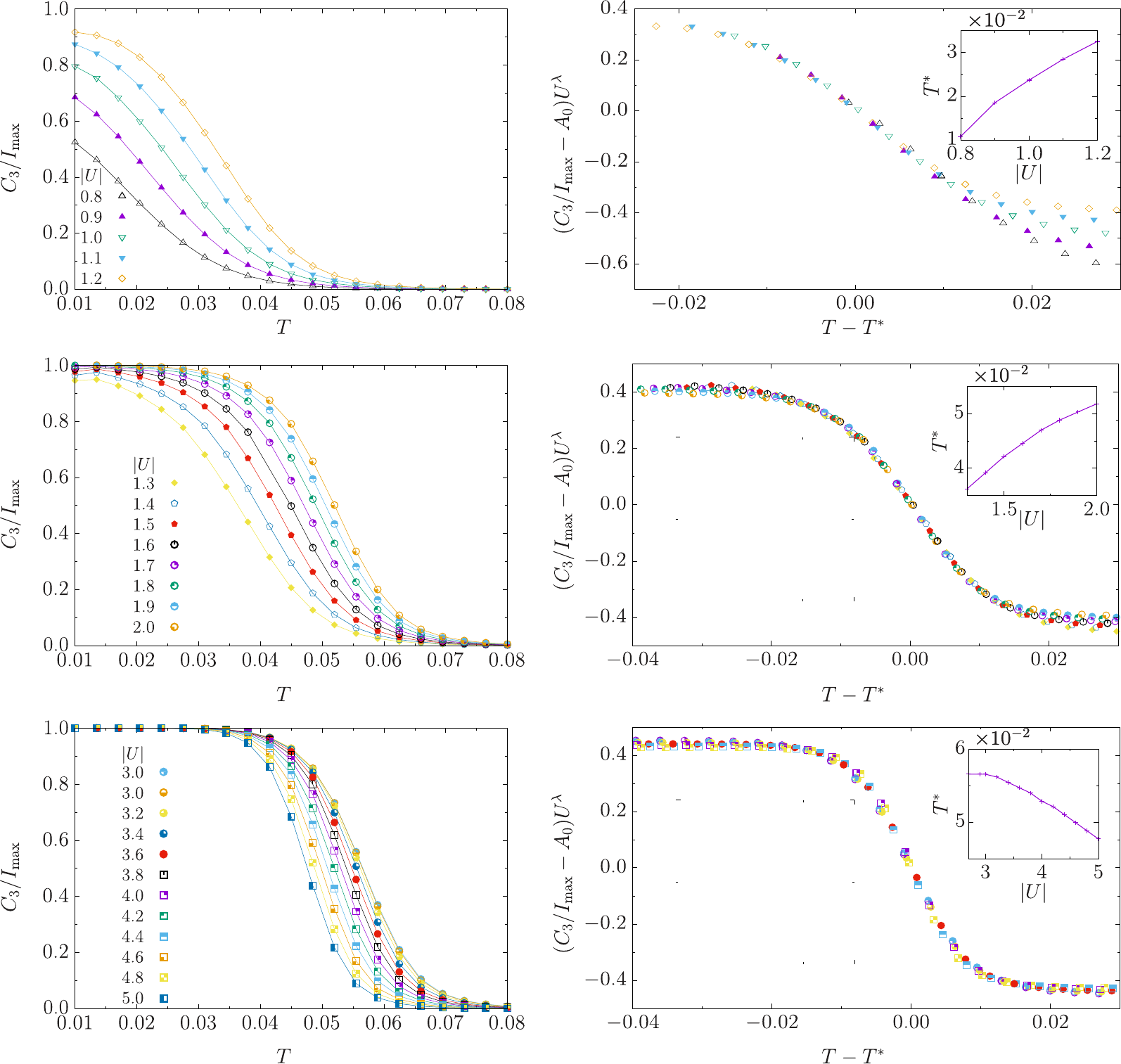}%
    \put(-430,398){(\textbf{a})}
    \put(-430,263){(\textbf{b})}
    \put(-430,125){(\textbf{c})}
    \put(-220,398){(\textbf{d})}
    \put(-220,263){(\textbf{e})}
    \put(-220,125){(\textbf{f})}
    \caption{Interplay between temperature $T$ and interaction $U$ for the persistent current $I(\phi)$.  The three plots demonstrate that the normalized Fourier weight $C_3(T^*)/I_{max}(T^*)$ as a function of temperature (left panels) obeys distinct laws in the different regimes of interaction (right panels): weak interactions (top); intermediate interactions (middle); and strong interactions (bottom).  The constant shift in $C_3(T^*)/I_{max}(T^*)$, is fixed for all curves by $A_{0}=0.5$. Top right insets display the temperature displacement $T^{*}$ as a function of interaction.  $T^*$ is defined by $C_3 (T^*)/I_{max}(T^*)=1/2$. In the regime of weak $U$ displayed in (\textbf{a})  $\lambda=-1.25$ and $T^*$ is an increasing function of $|U|$. For intermediate $U$ depicted in (\textbf{b}) $\lambda=-0.33$ and $T^*$ is still increasing, but with a different algebraic law. For the strong $U$ regime in (\textbf{c}), $\lambda=-0.1$ and $T^*$ is decreasing with $|U|$. All results were obtained with exact diagonalization for $N_{p}=3$ and $L=15$ with $T$ ranging from 0.01 to 0.08. Figure adapted from~\cite{chetcuti2021probe}.}
    \label{fig:scales}
\end{figure}
\begin{figure}[h!]
    \includegraphics[width=0.5\textwidth]{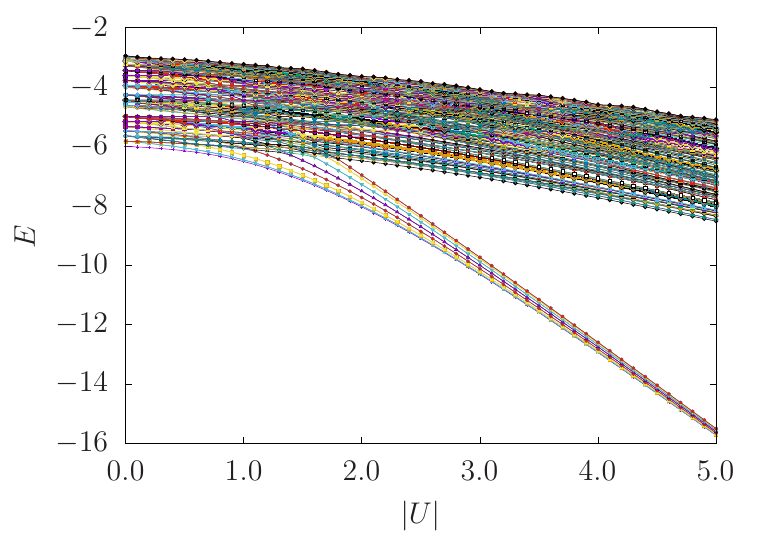}%
    \includegraphics[width=0.5\textwidth]{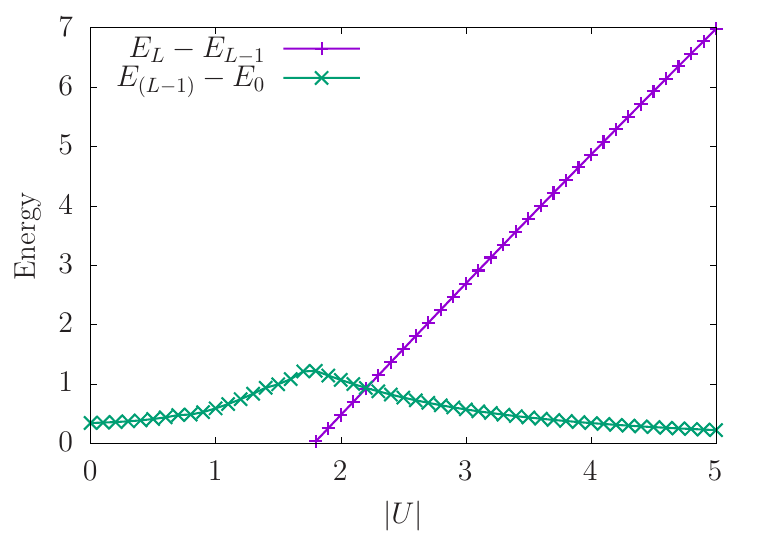}%
    \caption{(\textbf{a}) Energy spectrum $E$ as a function of the interaction $U$. (\textbf{b}) The green line depicts the energy spacing in the bound states sub-band denoted by $E_{L-1}-E_{0}$ where $E_{0}$ is the ground-state energy. Energy gap between the two sub-bands $E_{L}-E_{L-1}$, represented by the purple line. Results were obtained with exact diagonalization for $N_{p} = 3$ and $L=15$. Figure adapted from~\cite{chetcuti2021probe}. }
    \label{fig:spec}
\end{figure}

\noindent Specifically, we consider the Fourier weight $C_{3}$, which corresponds to the formation of trions --Figure~\ref{fig:fourier}(\textbf{a}), and follows its decay with increasing temperature. The coefficient is re-scaled by the maximum amplitude $|I_{\mathrm{max}}|$ of the persistent current displayed in Figure~\ref{fig:fourier}(\textbf{b}). Upon plotting the normalized Fourier weight $C_3/|I_{\mathrm{max}}|$, we observe three markedly distinct regimes: regime (I) at weak values of the interaction where $C_{3}$ becomes more robust to thermal effects on increasing $U$ --Figure~\ref{fig:scales}(\textbf{a}); regime (III) with $C_{3}$ being more susceptible to finite temperature for strong interactions--Figure~\ref{fig:scales}(\textbf{c}) lastly regime (II) for intermediate interactions --Figure~\ref{fig:scales}(\textbf{b}).  The existence of these three regimes is corroborated by the energy spectrum --Figure~\ref{fig:spec}. For small and intermediate $U$, the system is characterized by a continuous band, in which, beyond a certain energy threshold, the bound and scattering states are interwoven. Upon increasing $U$, a gap opens up, splitting the states in two distinct sub-bands by an increasing energy gap (linearly);  the energy levels within the states sub-band result to be separated by a level spacing that is suppressed by the interaction --Figure~\ref{fig:spec}(\textbf{b}). 

\noindent Our analysis shows that at small and moderate interactions, the persistent current and its frequency in particular, arise from thermal fluctuations populating the scattering states (for the band structure of the system, see~\cite{pohlmann2013trion} and Figure~\ref{fig:spec}). Here, the relevant parameter is the relative size between interaction $U$ and thermal fluctuations $T$ (measured in units of $t/k_B$). At moderate $U$ the bound states can remain well-defined for large $U/T$, whilst  for smaller values of  $U/T$, the bound states' deconfinement occurs because the temperature makes scattering states accessible. At stronger $U$, the relevant contributions to the persistent current come from the bound states' sub-band only. For such a `gas of bound states', the periodicity of $I(\phi)$ changes because the temperature allows the different frequencies  of the excited states to contribute to the current.  In this regime, since the level spacing between the bound states' energy levels decreases, the thermal effects are increasingly relevant by increasing interaction. 

\begin{figure}[h!]
    \centering
    \includegraphics[width=0.6\linewidth]{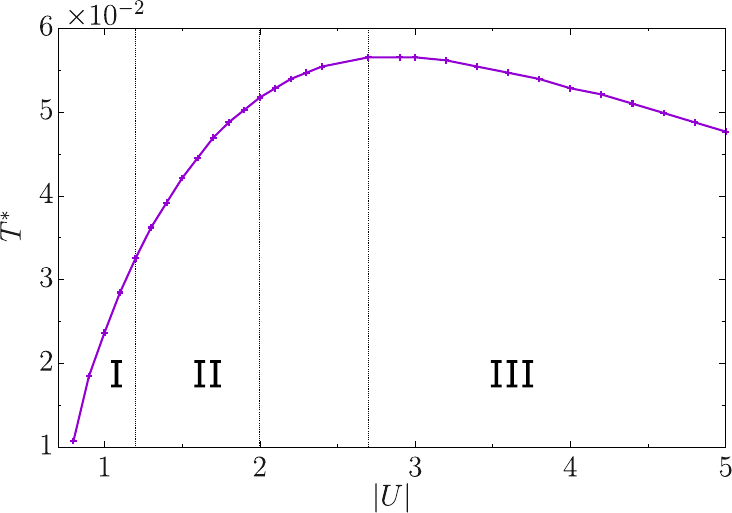}
    \caption{Crossover temperature $T^{*}$ against interaction $U$ for  the three interaction regimes for: weak, intermediate and strong denoted as I, II and III respectively. Figure adapted from~\cite{chetcuti2021probe}.}
    \label{fig:tempc}
\end{figure}

\noindent By employing the logic of the finite-size scaling machinery~\cite{barber1983finite}, we are able to identify the functional dependence of $W=C_{3}/I_{\mathrm{max}}$ on the parameter $T$ and $U$ in the aforementioned different interaction regimes. Our ansatz for $W$ is of the following form:
\begin{equation}\label{eq:scal1}
    (W-A_0)U^{\lambda} = G(T-T^{*}),
\end{equation}
where  $T^{*}$ is a  crossover temperature defined by $W(T^*)=1/2$. The value of  $\lambda$  is determined in such a way that a single functional law of the combination of  $U$ and $T$ is obtained. We observe that  $T^{*}$ depends non-monotonically on the interaction parameter $|U|$, indicating 
that the change $W$ undergoes is distinct in all three regimes --Figure~\ref{fig:tempc}.

\noindent By inspection, we adopt the following expression for $T^{*}$
\begin{equation}\label{eq:regime1&2}
T^*\approx T_{0,R} - a_{R}(U-U_{0,R})^2 ,
\end{equation}
with $T_{0,R}$, $U_{0,R}$, $a_{R}$ and $b_{R}$ being fitting parameters in the three regimes labelled by $R=\text{I, II, III}$, whose values are provided in Table~\ref{tab:scaling stuffa}. The parameter $U_{0,R}$ was chosen such that it could clearly distinguish between the three regimes: $U_{0,\mathrm{I}}$ lies in the region where the spectrum starts to splits such that nearly all excited states are scattering states --Figure~\ref{fig:spec}; $U_{0,\mathrm{II}}$ corresponds to a clear distinction between the bound and scattering states bands with the end of regime II being taken such that all avoided level crossings between the two are resolved with the two bands becoming fully separate; and $U_{0,\mathrm{III}}$ was taken to be the same as $U_{0,\mathrm{II}}$ since a clear distinction of the bound states from the scattering states is already achieved. Note that despite the small temperatures considered here, all bound states belonging to the lowest band are involved in the dynamics.

\noindent Inserting the respective values for all regimes into a Taylor expansion for the function $G$, demonstrates 
that in the vicinity of the crossover $|U-U_{0,\mathrm{II}}|<2.5$,
no impact is made on final form of the function G, especially near $T^*$.
Upon re-arrangement we have that
\begin{equation}\label{eq:scal2}
    W-A_0 = g_{R}(U) G_R\bigg(\frac{T-T_{0,R}}{(U-U_{0,R})^{\mu_{R}}}\bigg),
\end{equation}
where $g_{R}=U^{\lambda_{R}}(U-U_{0,R})^{\mu_{R}}$, with the values $\mu_{R}$ and $\lambda_{R}$ are listed in Table~\ref{tab:scaling stuffa}. Therefore, we found a functional form for the decay of the persistent current periodicity with temperature and interaction in the aforementioned three different interaction regimes. Remarkably, in the crossover region between the colour deconfinement region and the bound states gas, the two regimes result to be indistinguishable --Figure~\ref{fig:scales} (\textbf{b}). \\

\begin{table}[ht!]
    \centering
    \begin{tabular}{|c|c|c|c|c|c|c|}
        \hline
         & $T_{0,R}$ &$U_{0,R}$ & $a_{R}$ & $b_{R}$&$\mu_{R}$ & $\lambda_{R}$ \\
          \hline
         $R=\mathrm{I}$ & 0.040 & 1.61&0.044 &2& 2&1.25\\
         \hline
         $R=\mathrm{II}$ &0.056 &2.56 &0.013&2&2 &0.33 \\
         \hline
         $R=\mathrm{III}$ &0.056 &2.56&0.0027&1.459&1.459&0.1 \\
         \hline
    \end{tabular}
         \caption{Table containing the fitting parameters $T_{0,R}$, $U_{0,R}$, $a_{R}$, $b_{R}$, $\mu_{r}$ and $\lambda_{R}$ in the three regimes for Equations~\eqref{eq:regime1&2} and ~\eqref{eq:scal2}. }\label{tab:scaling stuffa}
\end{table}

\newpage

\section{Time-of-flight measurement}\label{sec:tof4}

\noindent In cold atom systems, it has been demonstrated that most of the features of the persistent current can be precisely probed  through time-of-flight (TOF) imaging~\cite{amico2005quantum}. TOF expansion entails the calculation of the particle density pattern, which for long expansion times corresponds to the momentum distribution defined as
\begin{equation}\label{eq:tof4}
 n_{\alpha}(\textbf{k}) = |w(\textbf{k})|^{2}\sum_{j,l}e^{\imath \textbf{k}\cdot(\textbf{x}_{j}-\textbf{x}_{l})}\langle c_{j,\alpha}^{\dagger}c_{l,\alpha}\rangle,
\end{equation}
where $w(\textbf{k})$ is the Fourier transform of the Wannier function and $\textbf{x}_{j}$ denotes the position of the lattice sites in the plane of the ring. 

\vspace{3mm}

\begin{figure}[h!]
    \includegraphics[width=\linewidth]{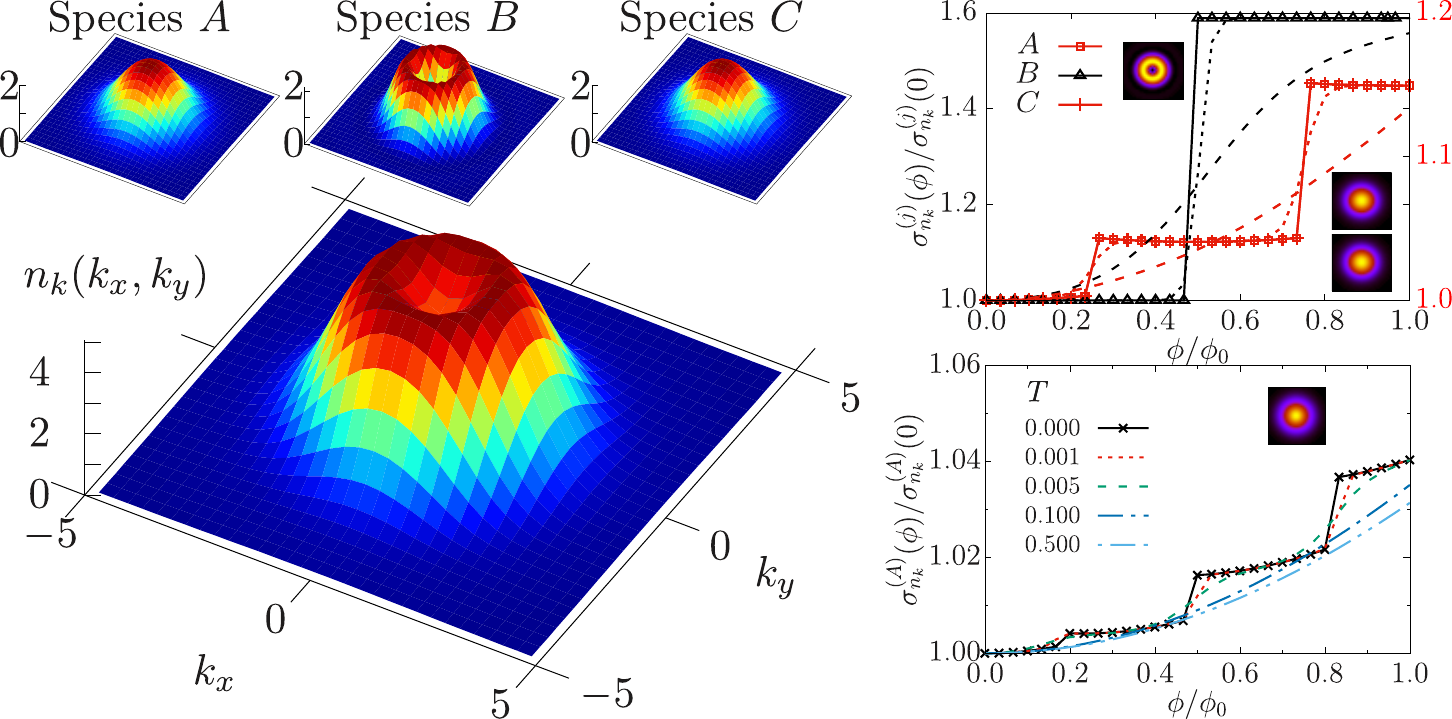}
    \put(-430,200){(\textbf{a})}
    \put(-170,200){(\textbf{b})}
    \put(-170,100){(\textbf{c})}
    \caption{(\textbf{a}) TOF expansion of the CSF configuration. Main (top) panel displays the TOF expansion, $n(k)$, for all (each) colours. Interactions are set to $|U_{AB}|=|U_{BC}|=0.01$ and $|U_{AC}|=5$. 
    (\textbf{b}-\textbf{c})Variance of the TOF expansion, $\sigma_{n_k}(\phi)$, against the effective magnetic flux $\phi$. Panel (\textbf{b}) shows the CSF configuration with the same interaction strengths described in the main panel for $T= \{0,0.01,0.1\}$ displayed solid, dotted and dashed lines respectively. Panel (\textbf{c}) corresponds to the trionic configuration at $|U| = 5$ for different $T$. Insets next to the curves in (\textbf{b}) show the momentum distribution $n_k(k_x,k_y)$ of each component at $\phi=1$, while in (\textbf{c}) we only show one colour due to SU(3) symmetry. The presented results are done for $N_{p}=3$ and $L=10$ using exact diagonalization. Figure adapted from~\cite{chetcuti2021probe}.}
    \label{fig:TOF}
\end{figure}

\noindent To read out the nature of the states in our system, it is important that such images arise as an interference pattern of the gas wavefunctions. For the specific case of coherent neutral matter circulating with a given angular momentum, a characteristic hole is displayed. Due to the reduced coherence in the system, characterized by the exponential decay in the three-body correlation functions (see Figures~\ref{fig:densdensdenstrion} and~\ref{fig:d3csf}), no holes have been found in TOF of bound states~\cite{naldesi2020enhancing,pecci2021probing,pecci2021phase}. Nevertheless, current states and the corresponding angular momentum quantization emerge as discrete steps in the variance of the width of the momentum distribution~\cite{moulder2012quantized,kevin2013driving,Ryu_2014},  
\begin{equation}
    \sigma^{(\alpha)}_{n_k}=\sqrt{\langle \hat{n}_\alpha^2\rangle-\langle \hat{n}_\alpha\rangle^2}.
\end{equation}
Here, trions display three steps in the variance reflecting the tri-partite periodicity of the current. For CSF states, we find a characteristic TOF with decreased density in the center of the interference pattern. By analysing, the different colour contributions to the momentum distribution, we figured out that the images arise as a sum of the hole corresponding to the delocalised weakly coupled species and the smeared peak corresponding to the bound state of the paired particles. Such bound states are found to be characterized in the TOF by just two steps in $\sigma$ (reflecting the particle pairing) --Fig.~\ref{fig:TOF}. On increasing temperature, the steps in the variance for both trions and CSFs vanish since the single particle frequency is reinstated.  Moreover, at finite temperature, the persistent current gets suppressed owing to the occupation of levels close to the Fermi level~\cite{patu}. The positive and negative contributions of these occupations to the current are quite similar, which is reflected by a broadening of the momentum distribution and, in turn, this smoothens out the hole corresponding to species $B$.

\section{Conclusions}\label{sec:conc4} 

\noindent In this chapter, we studied the bound states of attracting three-component fermions through the frequency of the persistent current both at zero and finite temperature. To this end, we apply a combination of Bethe ansatz and numerical methods that, especially for the finite temperature results, are among the very few non-perturbative approaches that can be applied to our system.  Our analysis hinges on the fact that the  {\it{effective}} flux quantum, defined by the frequency of the persistent current, provides information on the nature of the particles involved~\cite{leggett1991,byers1961theoretical,naldesi2019rise,chetcuti2021persistent}.  For our specific system of attractive SU(3) fermions, such a frequency  {\it  indicates that three-colour bound states are formed, irrespective of the number of particles}. This $N=3$ case is the general feature we find for SU($N$) attracting fermions whose bound states are formed by $N$ particles; in contrast to repulsive fermions and attractive bosons in which the frequency is fixed by the number of particles. Our analysis can clearly distinguish between trions and CSFs: the first are characterized by the persistent currents of the three species displaying a periodicity that is increasingly reduced by interaction until reaching $1/3$ of the original periodicity (for large interaction); CSFs, instead, result in persistent currents having two different periodicities for the different species --Figure~\ref{fig:main}. \\

\noindent Finite temperature induces specific changes in the persistent current frequency. We analysed the interplay between the interaction and thermal fluctuations quantitatively and {obtained specific laws describing it}. For  mild interactions, the frequency of the persistent current changes as result of the population of the scattering states.  Indeed, we observe that the phenomenon occurs as a  {\it crossover  from a colourless bound state to coloured multiplets, governed by the ratio $U/T$ without an explicit SU(3) symmetry breaking} --Figure~\ref{fig:tempcurr} and Figure~\ref{fig:scales}(\textbf{a}). Although specific non-perturbative effects near the QCD transition are missed by our analogue system (such as string breaking and colour charge screening), the  bound states' deconfinement in this regime displays similarities  with  the Quark-Gluon plasma formation at large temperatures and small baryonic density~\cite{satz2013probing}.
For stronger attraction, the system defines a gas of bound states separated from the scattering states by a finite energy gap.  {\it In this regime, a `single particle' thermal persistent current arises from the  combination of the frequencies characterizing the different energy levels in the bound state sub-band} --Figure~\ref{fig:scales}(\textbf{c}). 
In the crossover region between the ones dominated by the scattering-states and gas of bound states, the change of the frequency of persistent current  takes place with an identical functional dependence on interaction and temperature --Figure~\ref{fig:scales}(\textbf{b}). 
On increasing the interaction, the bound states' sub-band gets tighter, and so, the temperature is increasingly  relevant to wash out the fractionalization of the persistent current's periodicity.  \\

\noindent The suggested implementation of our work is provided by cold atoms. Thus, we studied the time-of-flight images of the system obtained by releasing the cold atoms from the trap --Figure~\ref{fig:TOF}. Although the characteristic hole does not open up in the momentum distribution, one can still gain insight of the persistent current periodicity through the variance of the width of the momentum distribution: fractional values of the angular momentum correspond to discrete steps. Naturally, on increasing temperature, the steps disappear on account of the reinstatement of the single particle frequency.


\chapter{Interference dynamics of matter-waves of SU(\textit{N}) fermions}\label{chp:interfer}

\noindent The physics of the SU($N$) Hubbard model has been explored in great detail in both the repulsive (Chapter~\ref{chp:repcurr}) and attractive (Chapter~\ref{chp:probe}) regimes by utilizing the persistent current as a diagnostic tool. In the same spirit as current-voltage characteristics, we have observed how the fractional quantization of the angular momentum gives crucial information about the physical nature of the system. Such results reaffirm the notion that persistent currents can be used to define an instance of current-based quantum simulators to probe the correlations of interacting quantum many-particle systems~\cite{amico2021,amico2022,naldesi2019rise,polo2020exact,pecci2021probing}. \\

\noindent In the cold atoms infrastructure, the experimental read-out of persistent currents is done through interference dynamics. There are two commonly applied approaches, homodyne and heterodyne protocols. For homodyne protocols, the system of interest interferes with itself. Such logic has been widely employed in ultracold atoms experiments in time-of-flight (TOF) images of the atoms' density for both bosons and fermions~\cite{amico2021,amico2022,ramanathan2011superflow,lewenstein2012ultracold,beattie,eckel2014hysteresis,wright2022persistent}. Through this measurement technique, the angular momentum quantization of a circulating current state can be monitored~\cite{kevin2013driving,amico2005quantum}. With heterodyne phase detection protocols, the phase portrait of the system flowing along the ring is acquired through its additional interference with a non-rotating quantum degenerate system placed at the center of the ring. Such a protocol has been experimentally realized both for bosons~\cite{andrews1997,eckel2014interferometric,corman2014quench,mathew2015self} and very recently for fermions~\cite{roati2022imprinting}. The emerging fringe pattern is a spiral interferogram whose topological features (number of arms and dislocations) reflect the properties of a circulating current state~\cite{corman2014quench,haug2018readout,pecci2021phase,roati2022imprinting}. \\

\noindent Here, we investigate the fractionalization of the persistent current flowing in an SU($N$) atomtronic circuit, modeled by the Hubbard model in Chapter~\ref{chp:repcurr}, by analyzing the interference dynamics of matter-waves of SU($N$) fermions generated through homodyne and heterodyne protocols. Additionally, we demonstrate how the resulting interference patterns reflect important features of the system, including the specific angular momentum fractionalization and parity effects characterizing  the system. Particularly, we highlight how our approach may be utilized to detect the number of particles $N_{p}$ and components $N$, both of which are notoriously hard to extract from an experimental setting~\cite{zhao2021heuristic}.

\section{Homodyne protocol for SU(\textit{N}) fermions} \label{sec:mom}

In homodyne protocols, characteristic phase portraits arise when the atoms residing in the ring interfere during time-of-flight expansion. The basic idea behind this commonly employed measurement technique is as follows: the confinement potential is switched off, releasing the trapped gas, enabling it to expand freely in space owing to its internal momentum. After some time, generally 10-20ms after release, the spatial distribution of the gas $n(\mathbf{r},t)$ is measured through absorption imaging techniques such as CCD contrast images. For sufficiently long times, the atom density distribution corresponds to the momentum distribution. This statement holds as long as the expansion is free, i.e., the interaction between the particles is negligible, which is typically the case for dilute gases. The momentum distribution $n(\mathbf{k})$ is defined as the Fourier transform of the one-body correlator,
\begin{equation}\label{eq:momentumm}
    \langle n(\textbf{k})\rangle = \int e^{\imath \textbf{k}(\textbf{r}-\textbf{r}')} \langle \Psi^{\dagger}(\textbf{r})\Psi (\textbf{r}') \rangle \textrm{d}\textbf{r}\textrm{d}\textbf{r}',
\end{equation}
where $\textbf{k}$ is the momentum and $\mathbf{r}$, $\mathbf{r}'$ correspond to two positions on the ring's circumference. Expanding the field operators in the basis set of the single-band Wannier functions $w(\textbf{r}-\textbf{r}_{j})$ localised at the $j$-th lattice site, we have that $\Psi (\textbf{r}) = \sum_{j}^{L}w(\textbf{r}-\textbf{r}_{j})c_{j}$ with $L$ being the number of lattice sites and $\mathbf{r}_{j}$ denoting their position in the ring's plane. In turn, the expression for $n(\textbf{k})$ in Equation~\eqref{eq:momentumm} can be adapted to a lattice system and re-cast into
\begin{equation}\label{eq:TOF}
    n_{\alpha}(\textbf{k}) = |w(\textbf{k})|^{2}\sum\limits_{j,l}e^{\imath\textbf{k}(\mathbf{r}_{l}-\mathbf{r}_{j})}\langle c_{l,\alpha}^{\dagger}c_{j,\alpha}\rangle ,
\end{equation}
where $w(\textbf{k})$ is the Fourier transform of Wannier function and $\alpha$ corresponds to the colour of the fermionic particle. Note that we are taking the harmonic approximation for the lattice sites corresponding to a adopting a Gaussian form for the Wannier functions. \\

\noindent The momentum distribution of particles on a ring is one of the few observables related to momenta that can be experimentally probed~\cite{iskin2009momentum} and is of particular interest in the field of atomtronics since the persistent current is visible in experiments by studying the particles' momentum distribution~\cite{amico2005quantum,phillips_current,ramanathan2011superflow,kevin2013driving,amico2021,amico2022}. In the case of coherent neutral matter circulating in a ring with a given angular momentum quantization, a characteristic hole is observed in the momentum distribution~\cite{amico2005quantum,perez2021coherent}. On the other hand, no hole is observed when there is a reduced coherence, e.g., for attractive interactions~\cite{naldesi2020enhancing,pecci2021probing,pecci2021phase,chetcuti2021probe}. Nonetheless, looking at the variance of the momentum distribution width $\sigma^{(\alpha)}_{n_{k}}$, given by $ \sigma^{(\alpha)}_{n_{k}} = \sqrt{\langle n_{\alpha}^{2}\rangle  -\langle n_{\alpha}\rangle^{2} }$, one is still able to observe the corresponding angular momentum quantization through the appearance of discrete steps~\cite{kevin2013driving,naldesi2020enhancing}. In what follows, we make an in-depth analysis of the momentum distribution of SU($N$) fermions for both attractive and repulsive regimes. As we shall see, the distinct physical features and characteristics of these two regimes can be aptly captured through homodyne interference images.\\
\begin{figure}[h!]
    \centering
    \includegraphics[width=0.85\linewidth]{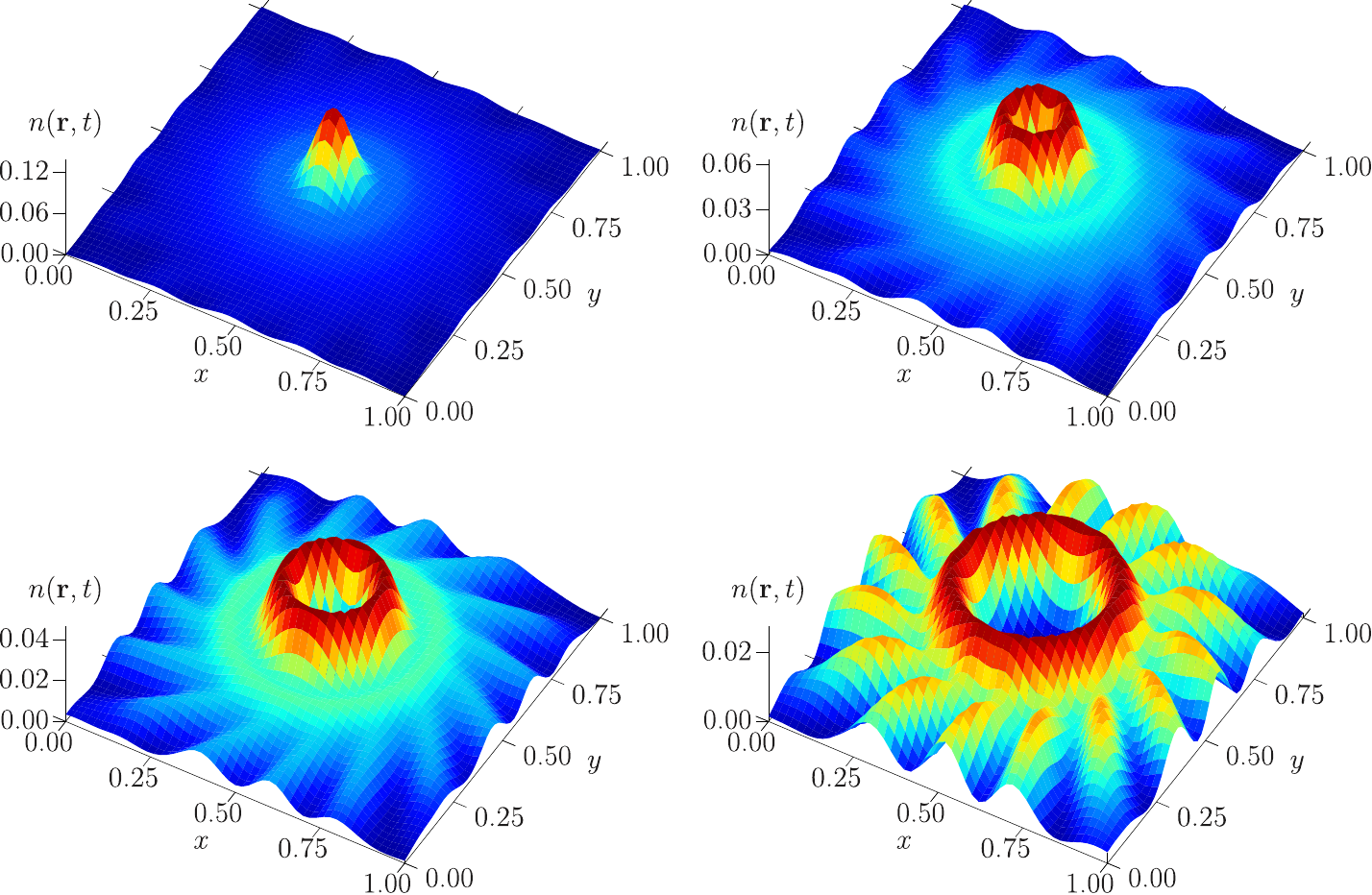}
    \put(-375,210){(\textbf{a})}
    \put(-185,210){(\textbf{b})}
    \put(-375,90){(\textbf{c})}
    \put(-185,90){(\textbf{d})}
    \caption{Density distribution $n(\textbf{r},t)$ at an intermediate time expansion $t=3$, for various flux values $\phi$. On increasing $\phi$, the sharply peaked Gaussian (top left) goes a characteristic hole with spirals radiating from it. As the size of the hole increases, so does the intensity of the spirals. Results were calculated with exact diagonalization for $N_{p}=4$ with $N=2$ and $L=15$ sites at interaction $U=0$ for $\phi/\phi_{0} = 0,1,2,4$. Figure reprinted from~\cite{chetcuti_interfer}.}
    \label{fig:shuriken}
\end{figure}

\begin{figure}[h!]
    \centering
    \includegraphics[width=\linewidth]{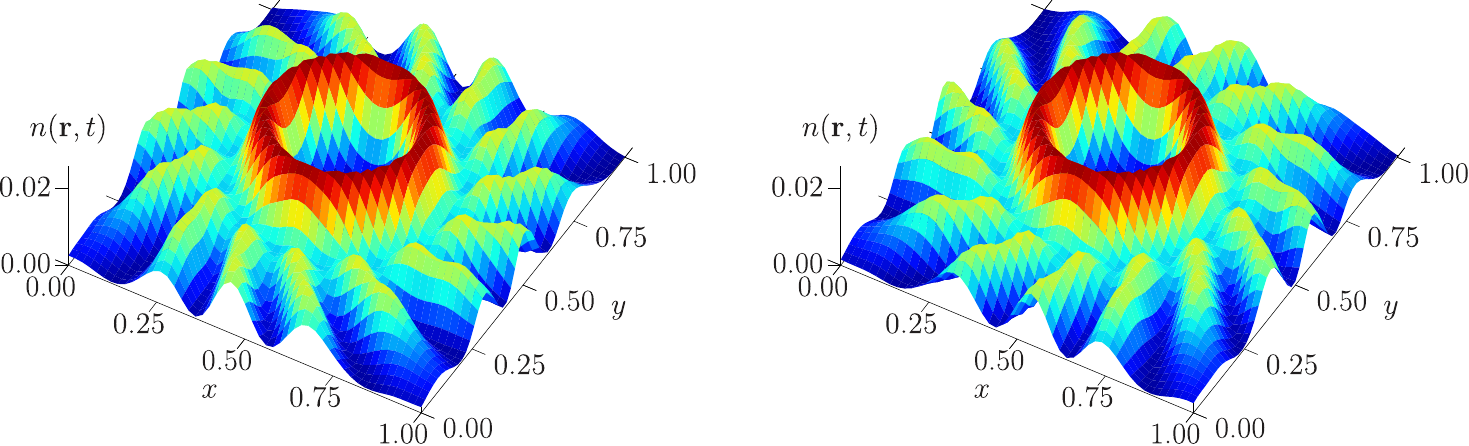}
    \put(-430,110){(\textbf{a})}
    \put(-210,110){(\textbf{b})}
    \caption{Density distribution $n(\textbf{r},t)$ at an intermediate time $t=3$. The right (left) panel corresponds to $\phi/\phi_{0}=4$ ($\phi/\phi_{0} = -4$) with the direction of the protruding spirals being clockwise (anti-clockwise). Consequently, the direction of the rotation by the artificial gauge field is reflected in the quantum shuriken.The results were calculated with exact diagonalization for $N_{p}=3$ with $N=3$ and $L=15$ at $U=0$. Figure adapted from~\cite{chetcuti_interfer}.}
    \label{fig:leftright}
\end{figure}
\noindent Before doing this, we remark that the intermediate time expansion of the atomic density distribution $n(\mathbf{r},t)$ is particularly interesting. Initially when the trap is opened at time $t=0$, one observes the Wannier functions localized on each site. Upon release and expanding for intermediate times, the ring displays a characteristic hole with protruding spirals, giving rise to a peculiar shape resembling a `shuriken'--Figure~\ref{fig:shuriken}. The direction of the spirals, be they clockwise or anti-clockwise, gives an indication of the directional flow of the current --Figure~\ref{fig:leftright}. Eventually, at longer times, one recovers the momentum distribution.

\subsection{Free particles}%
\begin{figure}[h!]
    \centering
    \includegraphics[width=0.68\linewidth]{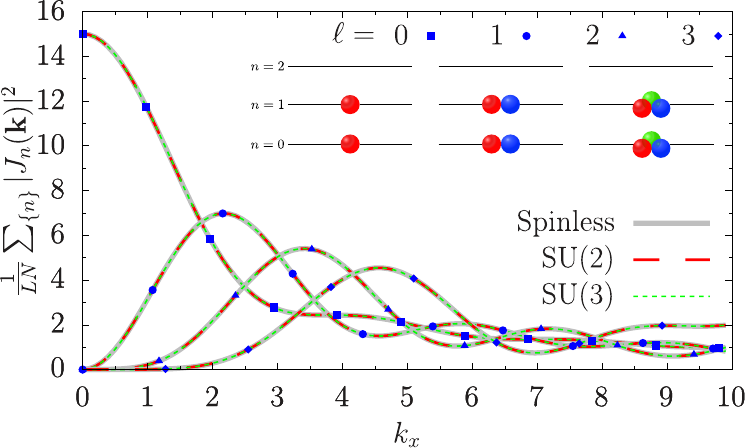}
    \caption{Main figure depicts the momentum distribution $\sum_{\{n\}}|J_{n}(\textbf{k})|^{2}$ re-scaled by the number of components $N$ against $k_{x}$. For the $\ell=0$ configuration, the momentum distribution has a finite value at the origin. However for $\ell >0$, the function collapses to zero at the origin. Insets depict the ground-state configuration $\ell=0$ for spinless, SU(2) and SU(3) fermions containing 2,4,6 particles respectively at $U=0$. Figure taken from~\cite{chetcuti_interfer}.}
    \label{fig:mom}
\end{figure}

\begin{figure}[h!]
    \centering
    \includegraphics[width=0.83\linewidth]{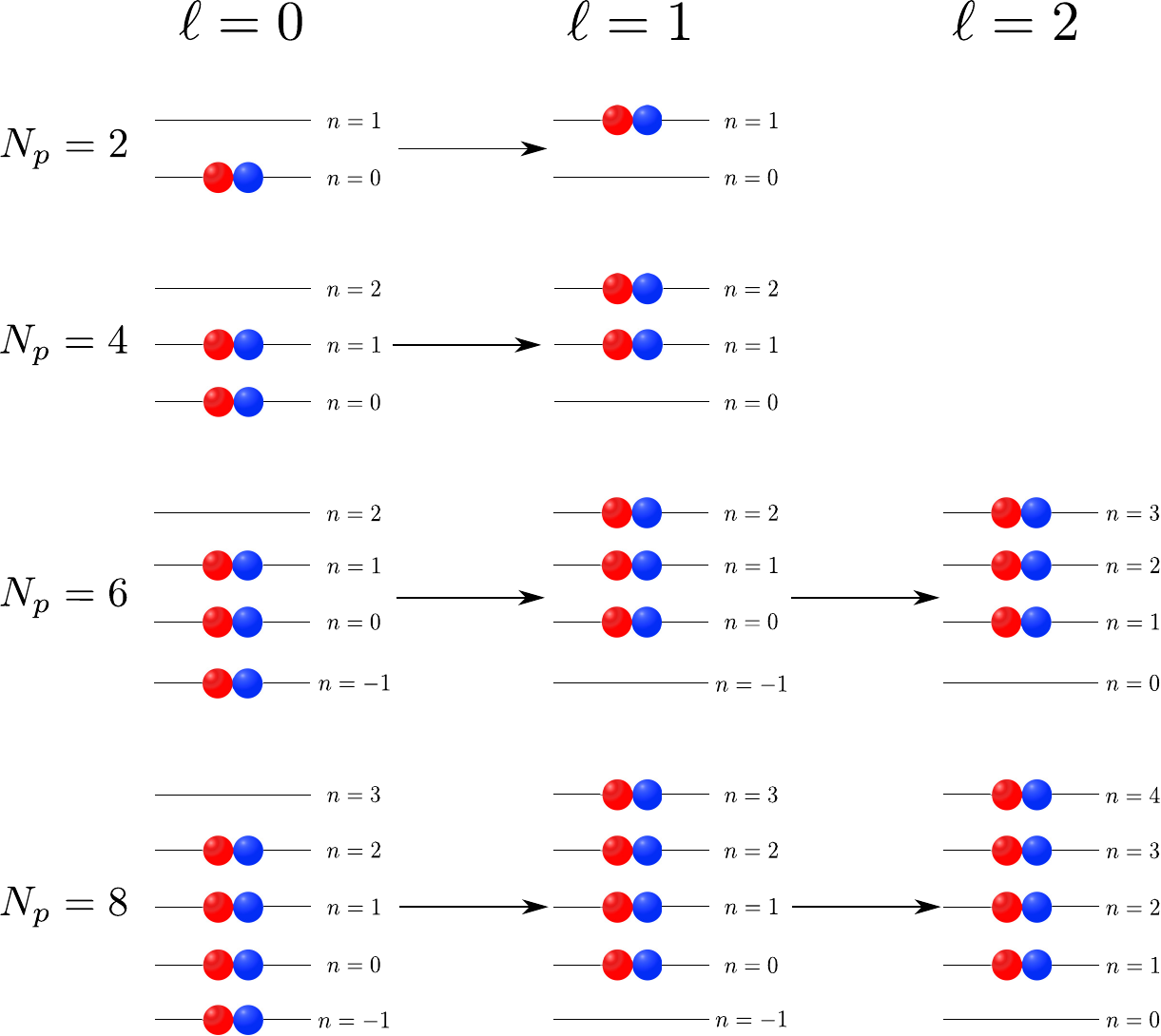}
    \caption{SU(2) particle occupation energy levels for $N_{p}=2,4,6,8$ particles and their displacement with angular momentum quantization $\ell$ at zero interaction. It is clear that with increasing $N_{p}$ a higher value of $\ell$ is required for the particles to vacate the $n=0$ level. As the effective magnetic flux increases, we pass from the first to the second parabola with $\ell=0$ and $\ell=1$ respectively (see Figure~\ref{fig:barb}). The Fermi sphere is displaced accordingly to counteract the increase in flux. In the case of $N_{p}=2$ ($N_{p}=4$), 
    no particles occupy the $n=0$ level such that the angular momentum $\ell_{H}$ for a hole to open up corresponds to $\ell=1$, with $\phi_{H} = 0.5$ ($\phi_{H}=1.0$) respectively. The value of $\phi_{H}$ denotes the flux value where we traverse to the energy parabola with angular momentum $\ell_{H}$. In contrast to the two previously discussed cases, one needs to go to $\ell=3$ for systems with $N_{p}=6,8$ to clear the $n=0$ level. As such, the hole opening in the momentum distribution is delayed to a larger value of $\phi$.  Figure adapted from~\cite{chetcuti_interfer}. }
    \label{fig:leve}
\end{figure}

The momentum distribution of spinless fermions at zero interactions is a sum of discrete Bessel functions $\sum_{\{n\}} J_{n}(\textbf{k})$ of order $n$, with $\mathbf{k}$ denoting the momentum and $n$ being the quantum numbers of the levels the particles occupy~\cite{aghamalyan2015coherent,pecci2021phase}. From this expression, it is clear that the momentum distribution is dependent on the sets of $n$, with the ground-state configuration being such that they are distributed symmetrically around zero. Apart from the zeroth order Bessel function, all other orders are zero-valued at $\mathbf{k}=0$ --see Figure~\ref{fig:mom}. Consequently, when the particles inhabit the $n=0$ level corresponding to the zeroth order Bessel function, the momentum distribution is always peaked at the origin. Such is the case for $\ell=0$ in Figure~\ref{fig:mom}. When threaded by an effective magnetic flux, the ground-state energy displays periodic oscillations characterized by a given angular momentum $\ell$. As the flux increases and we move from one energy parabola with a given $\ell$ to the next, the quantum numbers $n$ need to be changed to counteract the increase in flux and minimize the energy (see Figure~\ref{fig:leve}). 

\noindent Eventually, the set of $\{n\}$ is such that no spinless particles inhabit the $n=0$ level at a given value of $\ell$. Being a sum of discrete Bessel functions, the momentum distribution becomes zero-valued at the origin and a hole opens up~\cite{pecci2021phase}. The value of the angular momentum $\ell$ needs to be such that Fermi sphere of spinless fermions is displaced by the ceiling function $\lceil \frac{N_{p}}{2}\rceil$~\cite{amico2005quantum,pecci2021phase} (see Figure~\ref{fig:leve} for a schematic diagram). Therefore, there is a `delay' in the flux values to observe the hole for increasing $N_{p}$. This needs to be contrasted with bosons in a Bose-Einstein condensate, which due to the different statistics all reside in the $n=0$ level at $\ell =0$. In turn, there is no `delay' for the characteristic hole, which opens up at $\ell=1$ irregardless of $N_{p}$. 

\noindent The same logic used can be applied to SU($N$) fermions. With increasing $N$, the restriction imposed on the system by the Pauli exclusion principle relaxes: $N$ particles can occupy a given level --see Figure~\ref{fig:mom}. Accordingly, the Fermi sphere needs to be displaced less with increasing $N$. We find that for a system with $N_{p}$ SU($N$)-symmetric fermions, the momentum distribution collapses at the origin when the Fermi sphere is displaced by $\lceil\frac{N_{p}}{2N}\rceil$. In other words, the angular momentum required for a hole to open up is $\ell_{H}=\lceil\frac{N_{p}}{2N}\rceil$ and $\phi_{H}$ is the flux at which one transitions to the energy parabola with this corresponding angular momentum. For $N_{p}<N$, all particles will reside in the $n=0$ level. Indeed, as $N\rightarrow\infty$, SU($N$) fermions behave as bosons with regards to the level occupation. \\

\noindent Systems with an equal and commensurate value of $W = \frac{N_{p}}{N}$ display similar features. The persistent current's parity is one such feature whereby it is diamagnetic [paramagnetic] if $N_{p} = (2m+1)N$ [$N_{p} = (2m)N$] with $m$ being an integer~\cite{leggett1991,chetcuti2021persistent}. Likewise, we have that the momentum distribution re-scaled by $N$, is the same for equal and commensurate $W$ --Figure~\ref{fig:mom}. Clearly, this is not the case when $W$ is commensurate but not equal for different SU($N$). Owing to the different particle occupations, the momentum distribution and consequently the shifts in the sets of $\{n\}$ and angular momentum $\ell_{H}$ for a hole to appear is different. 

\subsection{Interacting particles}
Having established the basis for the momentum distribution of SU($N$) fermions at zero interactions, we turn our attention to the repulsive and attractive regimes. At small values of the interaction, the same features as in the free fermion case are observed (see also~\cite{pecci2021phase} for SU(2)). Here, we focus on the regimes of intermediate and infinite interactions.

\subsubsection{Repulsive interactions}
\begin{figure}[h!]
    \centering
    \includegraphics[width=0.93\linewidth]{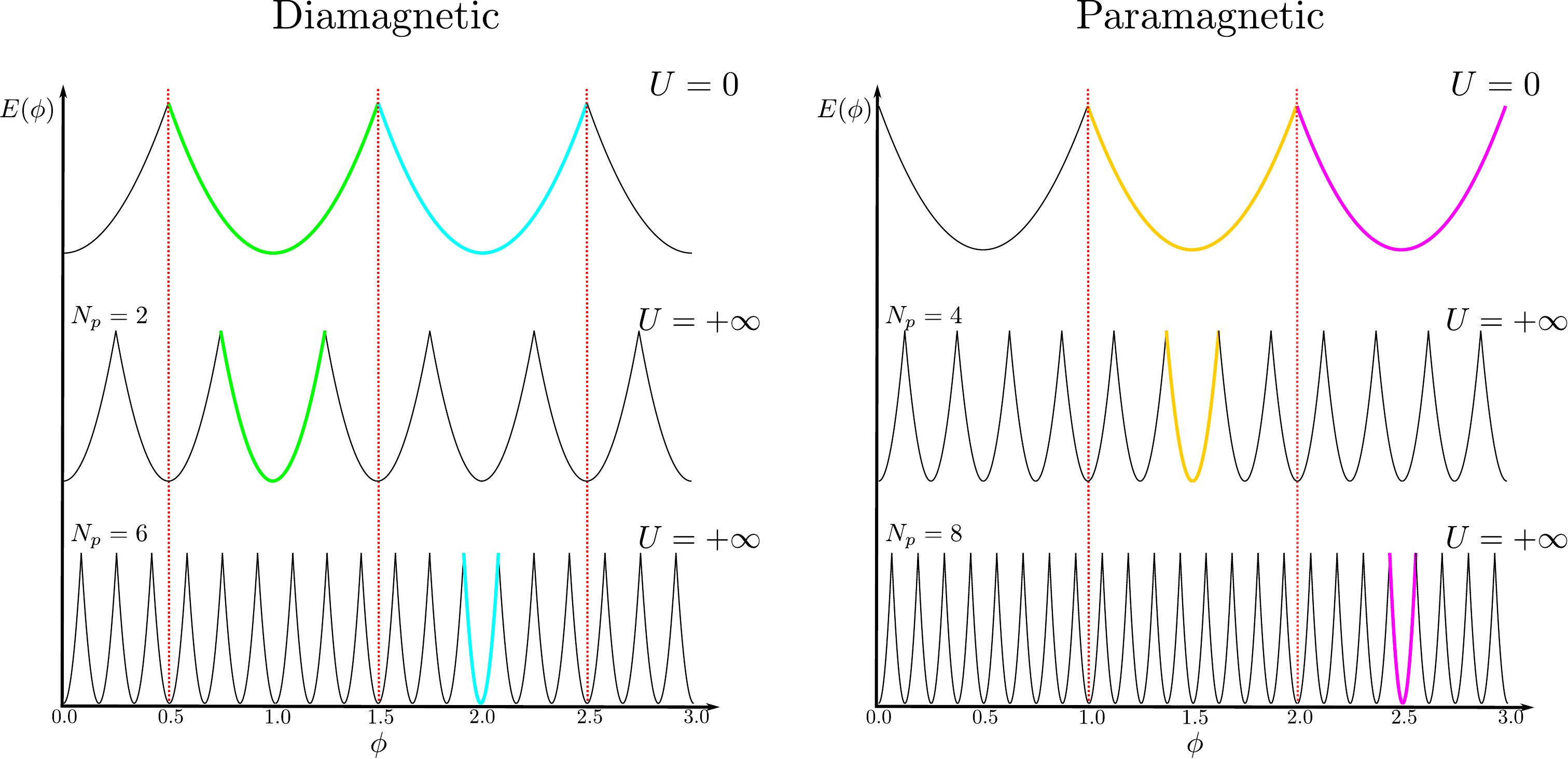}
    \caption{Schematic figure of the energy $E$ against effective magnetic flux $\phi$ for diamagnetic (left) and paramagnetic (right) cases. Top panel depicts the cases at zero interaction for SU(2) fermions (holds for any $N$). Middle and bottom panels correspond to having infinite repulsive interactions $U$ for different number of particles. In all cases, we observe that there is an additional delay to the one at $U=0$. Figure reprinted from~\cite{chetcuti_interfer}.}
    \label{fig:mess}
\end{figure}

For strong repulsive interactions, the persistent current fractionalizes with a reduced period dependent on the number of particles $N_{p}$ (see Chapter~\ref{chp:repcurr}). Originating from level crossings between the ground and excited states to counterbalance the increase in flux, the initial parabola observed at $U=0$ evolves into $N_{p}$ piece-wise parabolas/peaks --Figure~\ref{fig:mess}.

\begin{figure}[h!]
    \centering
    \includegraphics[width=0.49\linewidth]{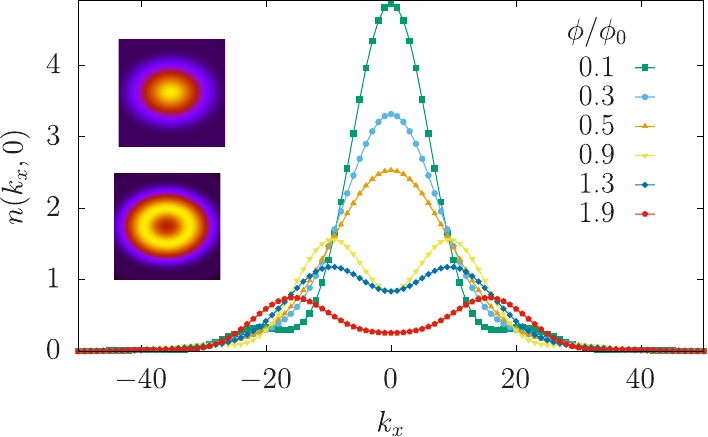}
    \includegraphics[width=0.49\linewidth]{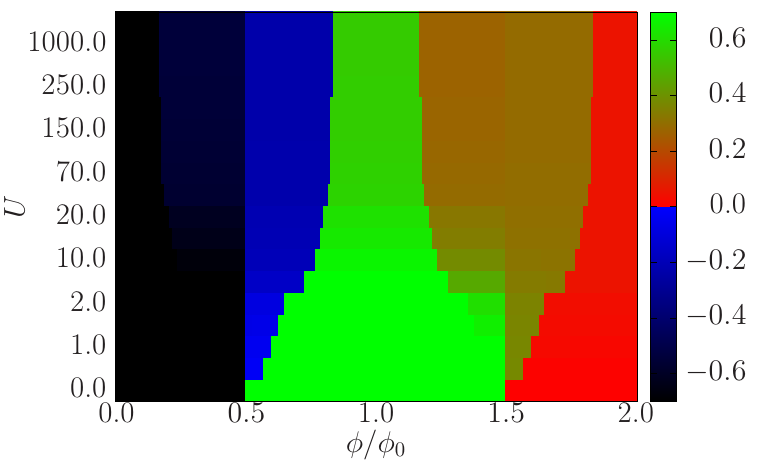}
    \put(-420,125){(\textbf{a})}
    \put(-210,125){(\textbf{b})}
    \caption{(\textbf{a}) Cross-section of the momentum distribution $n(k_{x},0)$ at strong repulsive interactions $U=10,000$ for various values of the effective magnetic flux $\phi$. When the threshold imposed by the fractionalization is surpassed, the momentum distribution collapses at $k_{x} =0$ and a depression is observed. For larger values of $\phi$, the depth of the depression increases. (\textbf{b}) Second derivative of the momentum distribution $n(k_{x},k_{y})$ evaluated at $k_{x}=k_{y}=0$, defined as $\frac{\partial^{2}n(k_{x},0)}{\partial k_{x}^{2}}\big|_{k_{x}=0}$, as a function of $\phi$ and $U$. A change in the sign of the second derivative denotes going from a peak (negative) to a depression (positive) that corresponds to a change in colour from blue to green respectively. At $U=0$ the hole opens up at $\phi = 0.5$. On increasing the interaction, the depression appears at larger $\phi$ thereby reflecting the fractionalization in the system. As $U\rightarrow\infty$, the system achieves complete fractionalization and the flux at which a depression is displayed corresponds to $\phi_{D}$. Note that the $y$-axis is not linear in values of the interaction. Results were obtained with exact diagonalization for $N_{p}=3$ SU(3) symmetric in $L=15$ sites. Figure taken from~\cite{chetcuti_interfer}.}
    \label{fig:TOFrp}
\end{figure}

\noindent The fractionalization phenomenon results in a momentum distribution depression at infinite repulsion. In contrast with the characteristic hole observed at zero and weak interactions, the depression is not zero-valued at the origin but a local minimum: i.e., a dip in the momentum distribution --Figure~\ref{fig:TOFrp}(\textbf{a}). Apart from cases like the one depicted in
Figure~\ref{fig:TOFrp}(\textbf{a}), we generally find a non-monotonous behaviour in the peak of the momentum distribution. 

\noindent Interestingly enough, the depression appears at fluxes $\phi_{D}$ that are found to be significantly larger than the flux values corresponding to the cases in which the angular momentum is quantized to integer values. In other words, the fractionalization in the system causes a specific {`delay'} in observing the momentum distribution depression --see Figure~\ref{fig:TOFrp}(\textbf{a}).  We remark that this `delay' is an add-on to the one observed at zero interactions. It is solely dependent on $N_{p}$ and independent of $N$ due to the nature of the fractionalization. The depression in the momentum distribution occurs at $\phi_{D} = \phi_{H}+\frac{N_{p}-1}{2N_{p}}$ where $\phi_{H}$ is the flux at which a hole appears at zero interaction. \\

\begin{figure}[h!]
    \centering
    \includegraphics[width=0.49\linewidth]{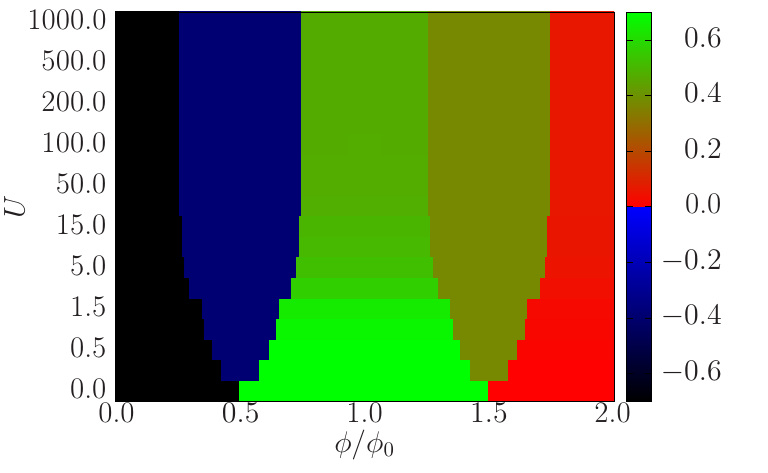}
    \includegraphics[width=0.49\linewidth]{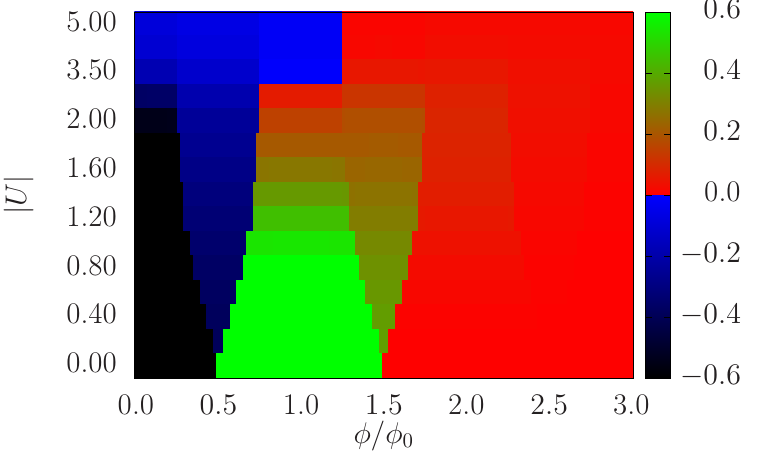}
    \put(-435,110){(\textbf{a})}%
    \put(-210,110){(\textbf{b})}%
    
    \includegraphics[width=0.49\linewidth]{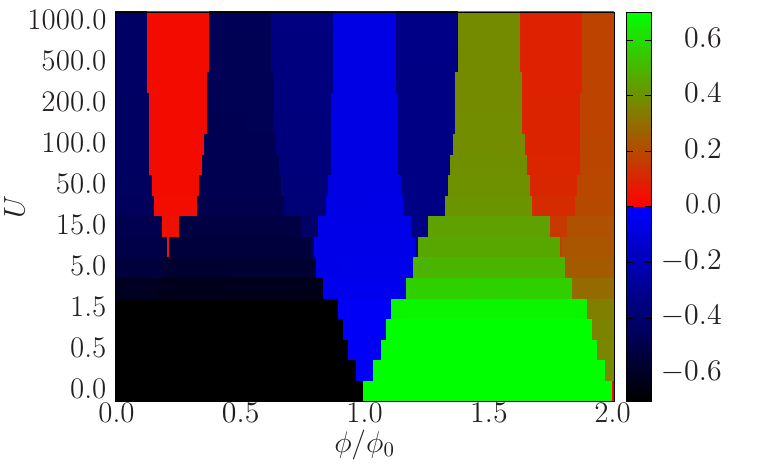}
    \includegraphics[width=0.49\linewidth]{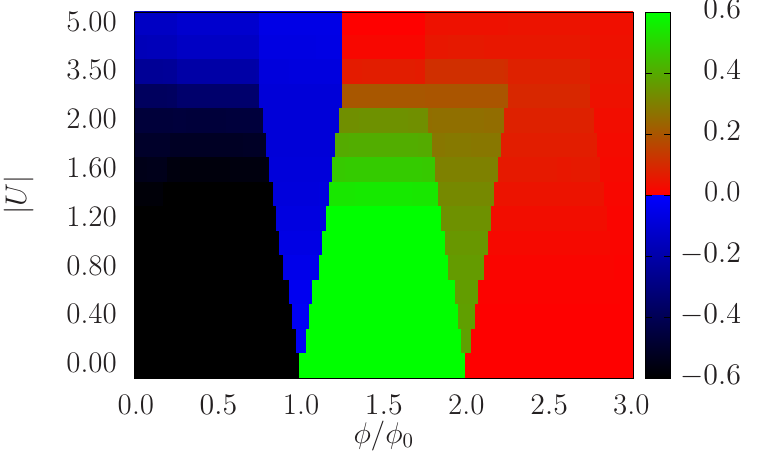}
    \put(-435,110){(\textbf{c})}%
    \put(-210,110){(\textbf{d})}%
    
    \includegraphics[width=0.49\linewidth]{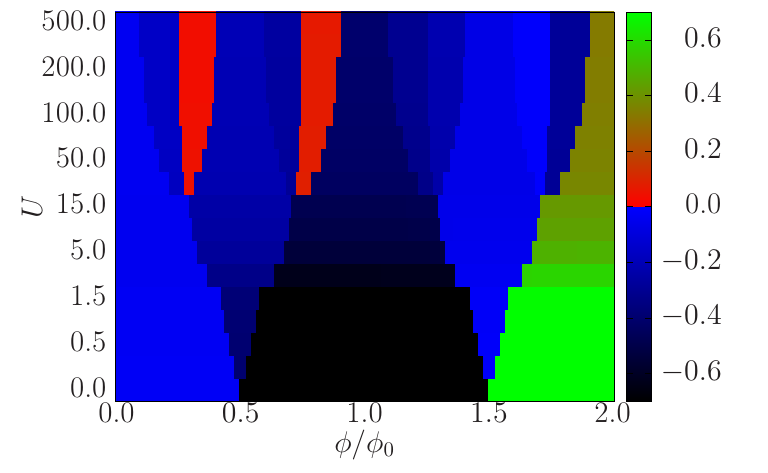}
    \includegraphics[width=0.49\linewidth]{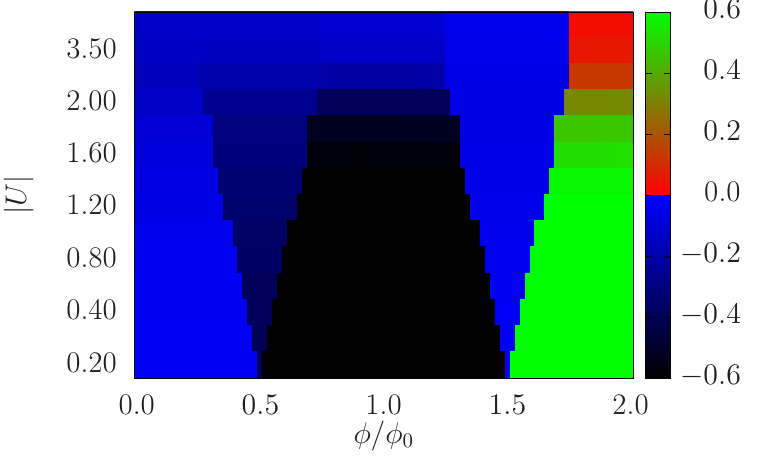}
    \put(-435,110){(\textbf{e})}%
    \put(-210,110){(\textbf{f})}%
    \caption{Second derivative of the momentum distribution
    $\frac{\partial^{2}n(k_{x},0))}{\partial k_{x}^{2}}\big|_{k_{x}=0}$ evaluated at $k_{x}=k_{y}=0$, as a function of the effective magnetic flux $\phi$ and interaction $U$. The left and right panels correspond to the repulsive and attractive regimes respectively for $N_{p}=2$ (top), $N_{p}=4$ (middle) and $N_{p}=6$ (bottom) SU(2) fermions. In both regimes, we observe that (i) as the number of particles increase the depression is delayed to higher flux values; (ii) there is an extra delay for stronger interactions. However, a notable difference is that in the attractive case the depression is smoothed out with increasing interactions. Results were obtained with exact diagonalization for $L=15$ sites. Note that the $y$-axis is not linear in the values of the interaction. Figure taken from~\cite{chetcuti_interfer}.}
    \label{fig:maprp}
\end{figure}

\noindent For intermediate interactions, the depression is monitored through the second derivative of the momentum distribution as a function of the effective magnetic flux and interaction. Indeed, a change in the derivative's sign from negative to positive corresponds to a peak and a depression respectively. By keeping track of the flux values at which this occurs, we gain insight into the degree of fractionalization in the system --Figure~\ref{fig:TOFrp}(\textbf{b}). Lastly, we remark that for cases where $W\neq 1$, there is the alternation between peak (blue) and depression (red) at flux values preceding $\phi_{H}$ as is observed in Figures~\ref{fig:maprp}(\textbf{c}) and (\textbf{e}). These red areas are small in value and correspond to plateaus in an experimental setting.

\subsubsection{Attractive interactions}

\begin{figure}[h!]
    \centering
    \includegraphics[width=0.49\linewidth]{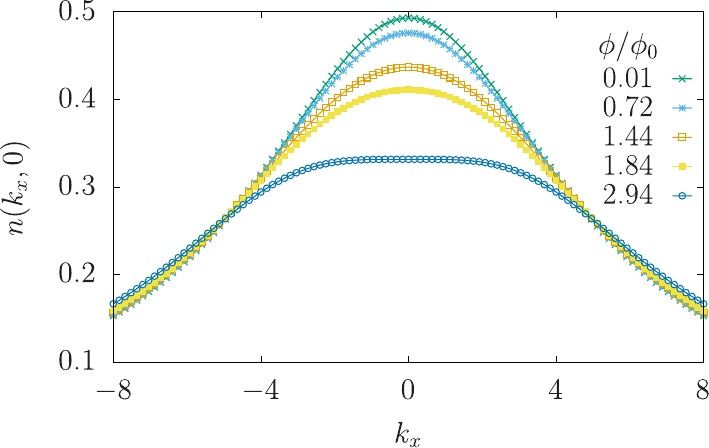}
    \includegraphics[width=0.49\linewidth]{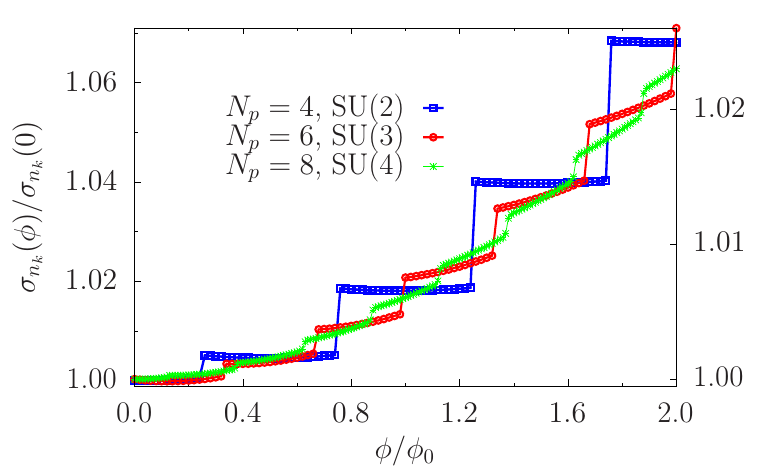}
    \put(-420,110){(\textbf{a})}
    \put(-210,110){(\textbf{b})}
    \caption{(\textbf{a}) Cross-section of the momentum distribution $n(k_{x},0)$ for $N_{p}=6$ fermions with SU(3) symmetry for various values of the effective magnetic flux $\phi$ at $U=-5$. (\textbf{b}) Variance of the momentum distribution $\sigma_{n_{k}} (\phi)$, against the effective magnetic flux $\phi$ for $N_p = 4$ (blue), $N_p = 6$ (red) and $N_{p}=8$ (green) fermions with SU(2), SU(3) and SU(4) symmetry respectively for attractive interactions. Results were obtained for $L=15$ with exact diagonalization for the SU(2) and SU(3) case, with DMRG being employed for the SU(4) case for interactions strengths of $U=-5$ and $U=-3$ respectively. Figure adapted from~\cite{chetcuti_interfer}.}
    \label{fig:TOFat}
\end{figure}

As we saw in Chapter~\ref{chp:probe}, the formation of bound states is reflected by a fractionalized current with a reduced period $\phi_{0}/m$, where $m$ denotes the number of particles in the bound state. Even though the state acquires current, the reduced coherence from the particles' localization, drastically reduces the visibility of the hole in the momentum distribution --Figure~\ref{fig:TOFat}(\textbf{a}). 

\noindent At intermediate interactions, where bound states are not yet fully formed, a momentum distribution depression appears --Figure~\ref{fig:maprp}(\textbf{b}), (\textbf{d}) and (\textbf{f}). 
Just like in the repulsive case, the depression experiences a two-fold `delay' that for sufficiently strong interactions appears at $\phi_{D} = \phi_{H} + \frac{N\! -\! 1}{2N}$. However, in this case the `delay' depends on both the particle number (through $\phi_{H}$) and the number of components (second term). Eventually, on increasing interactions, the visibility of the depression diminishes as it flattens out. For larger flux values, beyond $\phi_{D}$, a depression emerges again. However, one should keep in mind that in an experimental setting there is a trade-off between the winding number (magnitude of the current) and stability of the condensate~\cite{kevin2013driving,fetter_stability}. \\

\noindent It has already been established that at large attractive interactions, the momentum distribution does not provide us with direct information on the persistent current fractionalization.  Nevertheless, the variance of the width of the momentum distribution $\sigma_{n_{k}}$ has been demonstrated to be a figure of merit for fractional currents~\cite{kevin2013driving,Ryu_2014,naldesi2020enhancing,chetcuti2021probe} as displayed in Figure~\ref{fig:TOFat}(\textbf{b}). For one flux quantum, we observe two, three and four discrete steps in one flux quantum ($\phi/\phi_{0}=1$) corresponding to SU(2), SU(3) and SU(4) fermions respectively. \\

\begin{figure}[h!]
    \centering
    \includegraphics[width=0.5\linewidth]{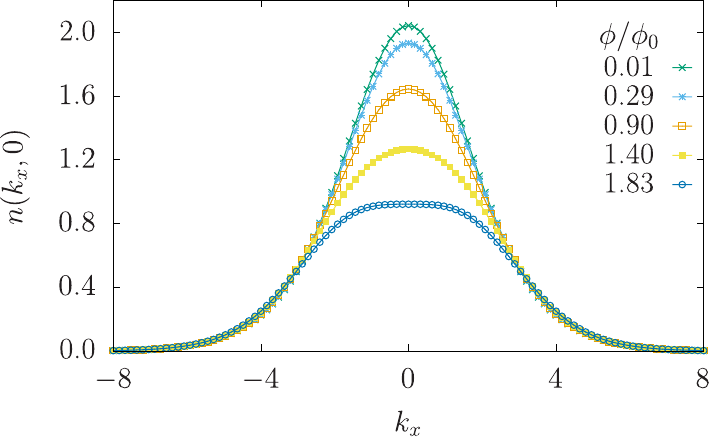}%
    \includegraphics[width=0.5\linewidth]{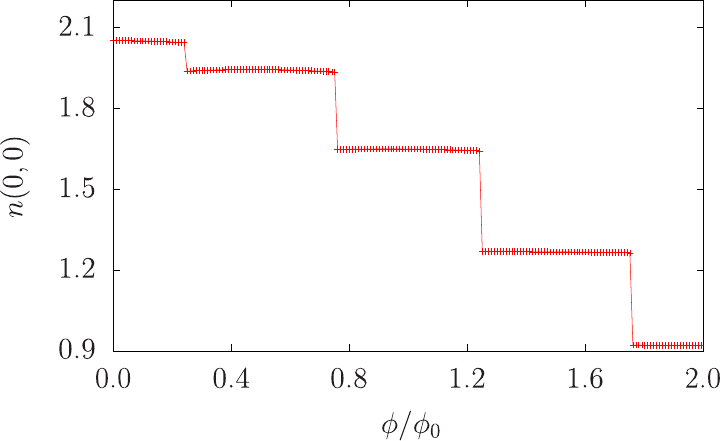}%
    \put(-425,110){(\textbf{a})}%
    \put(-211,110){(\textbf{b})}%
    
    \includegraphics[width=0.5\linewidth]{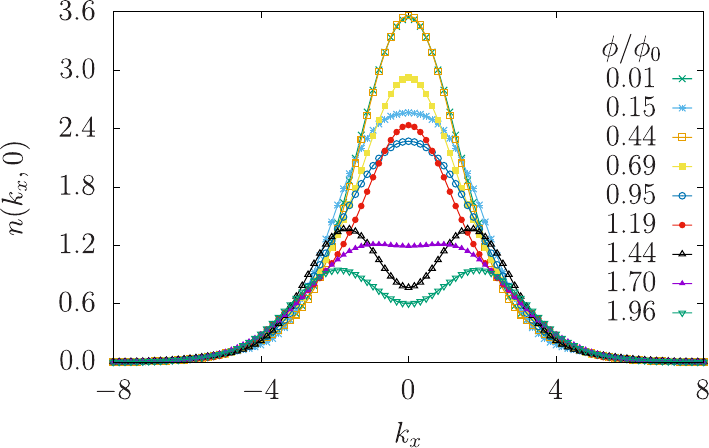}%
    \includegraphics[width=0.5\linewidth]{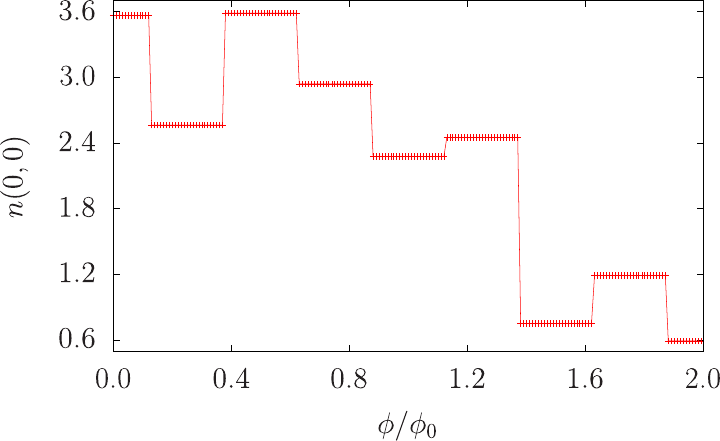}%
    \put(-425,110){(\textbf{c})}
    \put(-211,110){(\textbf{d})}
    \caption{Left panels depict the cross-section of the momentum distribution $n(k_{x},0)$ for $U=-5$ (top) and $U=10,000$ (bottom) as a function of the effective magnetic flux $\phi$. On the right panels, there is the corresponding momentum distribution $n(0,0)$ for $k_{x}=k_{y}=0$ against flux. Results were obtained with exact diagonalization for $N_{p}=4$, $N=2$ for $L=15$. Figure adapted from~\cite{chetcuti_interfer}.}
    \label{fig:alty}
\end{figure}

\noindent In the repulsive case, the variance is not a good measure, which can be attributed to the peculiar behaviour observed in Figures~\ref{fig:maprp}(\textbf{c}) and (\textbf{e}). To understand this properly, we examine the momentum distribution cross-section depicted in Figure~\ref{fig:alty}(\textbf{c}) and (\textbf{d}), where it is observed that the height of the peak/depression varies non-monotonically with $\phi$. This in turn affects variance of the width, which displays instances of expansions and contractions. Such behaviour is in sharp contrast with its attractive counterpart, where the behaviour of the peak is monotonically decreasing --Figures~\ref{fig:alty}(\textbf{a}) and (\textbf{b}). In view of this fact, it proves to be a good diagnostic tool for the persistent current fractionalization.

\section{Self-heterodyne interferograms of SU(\textit{N}) fermions} \label{sec:intfer}

Properties of circulating current states can also be detected through self-heterodyne phase detection protocols~\cite{eckel2014interferometric,corman2014quench,mathew2015self,haug2018readout,pecci2021phase,roati2022imprinting}. The mechanism behind this procedure is the same as the homodyne protocol, with the added difference that the annular ring is made to co-expand with an additional condensate at its center, whose role is to act as a phase reference. Initially separated, the ring and center undergo time-of-flight expansion, during which their combined wavefunction evolves in time and interferes with itself, giving rise to characteristic spiral interferograms. Topological features of the spiral pattern reveal information on the current's orientation (clockwise or anti-clockwise) and intensity (phase winding number). \\

\noindent In an experimental setting, the spiral interference patterns are obtained through in-situ measurements of a single co-expansion. A single-shot experiment such as this one could be simulated from a theoretical point of view. It would require combining the eigenfunctions of the Hamiltonian modeling the ring with the wavefunction of the center and then replicating the detection protocol for all the particles involved in the expansion~\cite{castin1997relative}. However, such an approach requires a large number of particles in order for spirals to be produced, which is not viable in our case. As such, we turn to reconstructing the particle density $n(\mathbf{r},t)$ through expectation values, which amounts to taking the average over several experimental realizations of the co-expansion. Each instance of the co-expansion is characterized by a randomly distributed relative phase between the two condensates that averages out over multiple experimental runs. As such, no interference effects appear when calculating the one-body density matrix, and any information about the current's configuration is washed out. \\

\noindent On account of the ring and center being initially decoupled, there are no uncertainties about the particle's origin, be it from the former or the latter. For this reason, the density operator corresponding to the measurement of a single particle does not yield any information on the relative phase pattern between the two systems~\cite{haug2018readout}. According to the particle number-phase uncertainty relation, we can acquire information about the phase by erasing any with regards to the particles' origin. This can be achieved by resorting to higher-order correlators and measuring two or more particles, such that the uncertainty about their origin, which can either be from the center or the ring, increases~\cite{castin1997relative,haug2018readout}. As such, in line with the previous protocols for bosons~\cite{haug2018readout} and quite recently for fermions~\cite{pecci2021phase}, we focus on the density-density correlator for an expanding ring and an additional site in the center at a fixed time $t$. The two-body correlator is defined in the following way
\begin{equation}\label{eq:cov}
    G(\textbf{r},\textbf{r}',t) = \sum\limits_{\alpha,\beta}^{N}\langle n_{\alpha}(\textbf{r},t)n_{\beta}(\textbf{r}',t)\rangle.
\end{equation}
The density operator is expressed as $n_{\alpha}(\textbf{r},t) = \psi^{\dagger}_{\alpha} (\textbf{r},t)\psi_{\alpha} (\textbf{r},t)$ where $\psi^{\dagger}_{\alpha} = (\psi^{\dagger}_{R,\alpha} + \psi^{\dagger}_{C,\alpha})$ being the field operator of the whole system, ring and the center denoted by $R$ and $C$ respectively. Seeing as the ring and center are intially decoupled prior to their release from their confinement potential, at time $t=0$ the ground-state can be seen as a product state $|\phi\rangle = |\phi\rangle_{R}\otimes|\phi\rangle_{C}$. \\

\noindent Assuming free expansion for $t\geq 0$, the number of terms in $G(\textbf{r},\textbf{r}',t)$ can be significantly reduced. Firstly, terms consisting of an odd number of creation or annihilation operators have a null expectation value due to particle conservation. Likewise, terms where either both creation or annihilation operators act on one system also vanish. As such, the only surviving terms are those comprised of an equal number of creation-annihilation pairs, one acting on the ring and another on the center. From the remaining terms, we focus only on the cross-terms involving contributions from both the center and the ring~\cite{pecci2021phase}, seeing as they are the ones giving rise to the interference:
\begin{equation}\label{eq:interx}
    G_{R,C} = \sum\limits_{\alpha,\beta}^{N}\sum\limits_{j,l=1}^{L}I_{jl}(\textbf{r},\textbf{r}',t)\big[ N_{0,\alpha} \delta_{\alpha,\beta}(\delta_{jl} - \langle\phi_{R}|c_{l,\alpha}^{\dagger}c_{j,\alpha} |\phi_{R}\rangle ) + \delta_{\alpha,\beta}(1-N_{0,\alpha})\langle\phi_{R}|c_{l,\alpha}^{\dagger}c_{j,\alpha} |\phi_{R}\rangle  \big ].
\end{equation}
In the above expression, we defined $I_{jl}(\textbf{r},\textbf{r}',t) = w_{c}(\textbf{r}',t) w_{c}^{*}(\textbf{r},t) w_{l}^{*}(\textbf{r}'-\textbf{r}_{l}',t) w_{j}(\textbf{r}-\textbf{r}_{j},t)$ as the Wannier functions of the interfering terms and $N_{0} = \langle\phi_{C}|c_{0,\beta}^{\dagger}c_{0,\beta} |\phi_{C}\rangle$ corresponds to the expectation value of the number operator center, which in the current protocol is always equal to one. Consequently, the second term in Equation~\eqref{eq:interx} does not contribute to the interference pattern such that 
\begin{equation}\label{eq:interf}
    G_{R,C} = \sum\limits_{\alpha}\sum\limits_{j,l}I_{jl}(\textbf{r},\textbf{r}',t) \langle c_{l,\alpha}^{\dagger}c_{j,\alpha}\rangle ,
\end{equation}
where the $\delta_{ij}$ in the first term of Equation~\eqref{eq:interx} is neglected in order to enhance the visibility of the spirals. \\

\noindent Just like in the momentum distribution, the particles' statistics are reflected in the self-heterodyne interference patterns. As was previously mentioned, the fermionic particles occupy distinct levels associated with different momenta owing to their anti-symmetric nature. Hence, when the fermions start to circulate, the imparted phase gradient of the wavefunction couples to the various momenta. Each of these momenta acquire a different phase and correspond to individual particle orbitals. At intermediate time expansions, the multiple phases recombine, giving rise to the spiral-like interference pattern, as well as accounting for the emergence of dislocations (radially segmenting lines) in the interferograms. Such a phenomenon can be properly visualized in the free fermionic case. In this regime, it is straightforward to show that the interference pattern in Equation~\eqref{eq:inter} can be re-cast into the following form~\cite{pecci2021phase}
\begin{equation}\label{eq:disloca}
    G_{R,C} = -\frac{N}{L}\sum\limits_{\{n\}}I_{n}(\mathbf{r})I_{n}^{*}(\mathbf{r'}),
\end{equation}
with $I_{n}(\mathbf{r})I_{n}^{*}(\mathbf{r'})$ being a spiral in the $x$-$y$ plane having $n$ arms, where $\{n\}$ is the set of quantum numbers corresponding to the levels occupation. From this expression, it is clear that at fixed $\mathbf{r}'$ the interference pattern arises as a superposition of the different contributions $I_{n}(\mathbf{r})I_{n}^{*}(\mathbf{r}')$. \\

\noindent In what follows, we build up on the recent work carried out on interferograms of SU(2) fermions at zero and weak interactions~\cite{pecci2021phase} by generalizing to SU($N$) fermions and extending the analysis to the intermediate and strongly interacting regimes. Such regimes cannot be accessed with DMRG. Indeed, it is well known that DMRG has issues with convergence for large interactions. Furthermore, the problem is exacerbated by multi-degenerate ground-states present in strongly repulsive SU($N\!> \! 2$) systems. As a result, we have opted to carry out the self-heterodyne analysis by considering two-component systems that can be addressed with exact diagonalization. Nonetheless, this does not hinder our analysis since, as we shall see, interferograms with equal $W=N_{p}/N$ display similar characteristics as was the case with the momentum distribution. Seeing as the features of the interferogram depend on the particle distribution, then one would expect that system with commensurate and equal $W$ displays interferograms with the same qualitative characteristics.

\subsection{Free Particles}
\begin{figure}[h!]
    \centering
    \includegraphics[width=\linewidth]{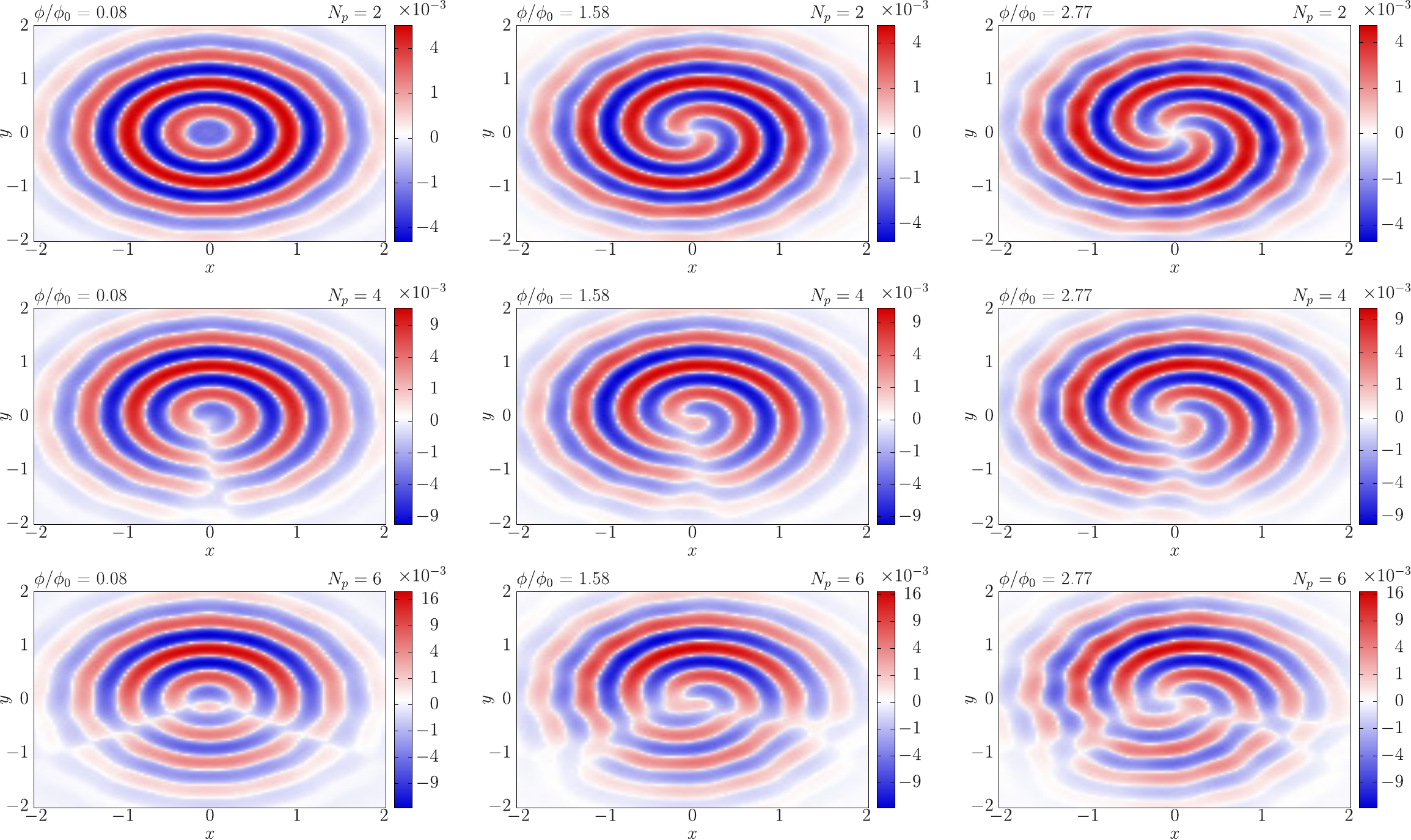}
    \caption{The interference $G_{\mathrm{R,C}}$ between ring and center for $N_{p}=2,4,6$ particles with SU(2) symmetry, is shown as a function of the effective magnetic flux $\phi$ at zero interaction and short time $t=0.033$. For $N_{p}=2$ (top) and $N_{p}=4$ (middle), the number of spirals corresponds to the angular momentum quantization $\ell$. Note that at $\phi=1.58$, the difference in spirals arises due to the different parity of the two systems, that correspond to a degeneracy point of $\phi = s+1/2$ and $\phi = s$ respectively, with $s$ being an integer. In contrast to the $N_{p}=2$ case, we observe dislocations (segmenting lines) in the interference patterns for $N_{p}=4,6$. Lastly, by comparing $N_{p}=2$ and $N_{p}=6$ at $\phi=1.58$ we observe that the latter displays only one spiral instead of two like its counterpart. The reason being that at $U=0$ for $N_{p}=6$, the hole opens up at $\phi=1.5$. All correlators are evaluated with exact diagonalization for $L=15$ by setting $\textbf{r}' = (0,R)$ and radius $R=1$. The color bar is non-linear by setting $\mathrm{sgn}(G_{\mathrm{R,C}})\sqrt{ |G_{\mathrm{R,C}}|}$. Figure taken from~\cite{chetcuti_interfer}.}
    \label{fig:interU0}
\end{figure}

At zero and weak interactions for short time expansions, one observes spiral-like patterns, along with dislocations in these interferograms~\cite{pecci2021phase}. These dislocations arise from the various momenta components that contribute to the expansion and indicate the suppression of the particle density at these positions in space. The amount of dislocations, given to $W-1$, grants us access to the number of particles present --Figure~\ref{fig:interU0}. Furthermore, these dislocations do not depend on the flux threading the system.

\noindent For bosons, the order of the spirals gives an indication of the number of rotations, or rather the angular momentum quantization $\ell$ of the current~\cite{corman2014quench,haug2018readout,roati2022imprinting}. However, it is not as straightforward when it comes to fermions. Owing to their different statistics, the level occupation of fermions is broader than that of bosons. As such, in order for spirals to emerge in the interferograms, the given system of fermions needs to displace its Fermi sphere by
$\big\lceil \frac{N_{p}}{2N}\big\rceil$, 
which corresponds to the characteristic hole in the momentum distribution --Figure~\ref{fig:interU0}.
After clearing this threshold, the number of spirals accrues with increasing angular momentum. 
\begin{figure}[h!]
    \includegraphics[width=\linewidth]{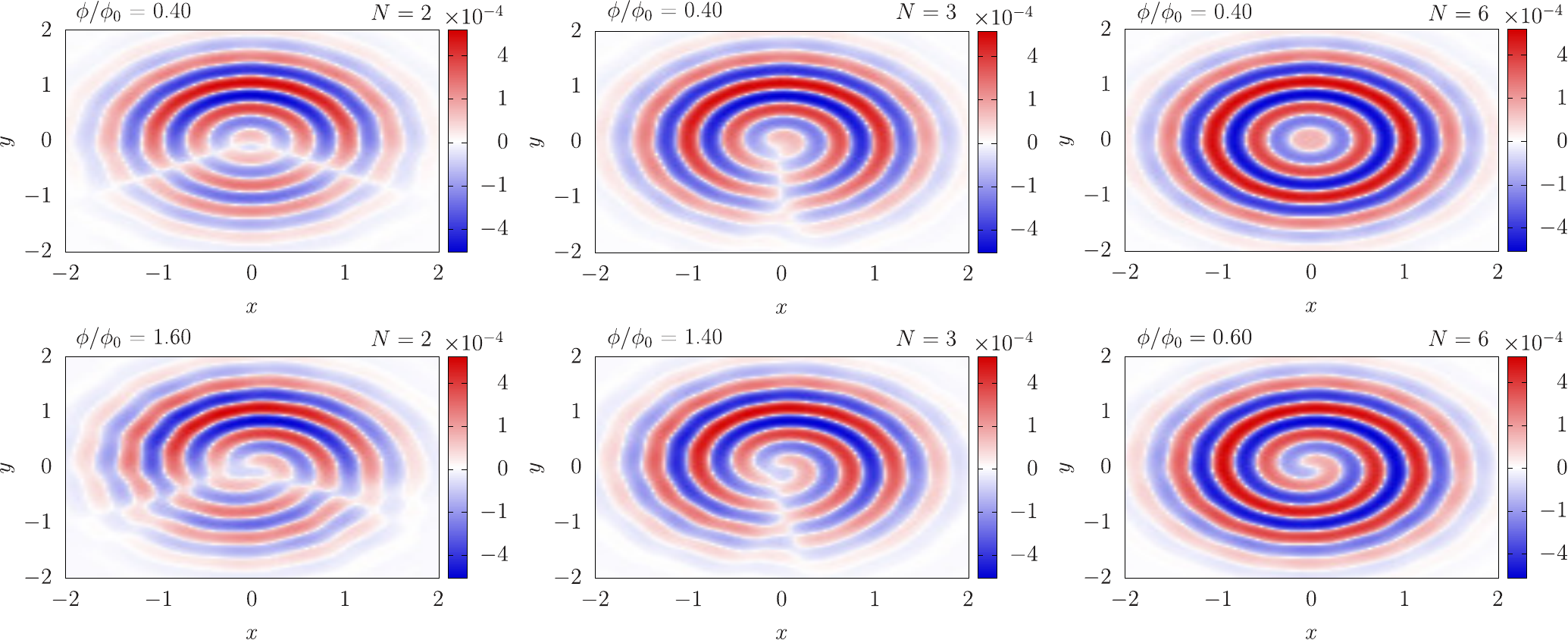}
    \caption{The interference $G_{\mathrm{R,C}}$ between ring and center for $N_{p}=6$ particles with $N=2,3,6$, as a function of the effective magnetic flux $\phi$ at $U=1$ and short time $t=0.025$. 
    In the top panel, the number of dislocations in the interferograms is 2,1,0 for $N=2$ (left), $N=3$ (middle) and $N=6$ (right) respectively. The bottom panel depicts the value of $\phi$ where a spiral is observed. On increasing $N$, the spiral appears at lower $\phi$ since the displacement of the Fermi sphere is inversely proportional to $N$. All correlators are evaluated with DMRG for $L=15$ by setting $\textbf{r}' = (0,R)$ and radius $R=1$. The color bar is non-linear by setting $\mathrm{sgn}(G_{\mathrm{R,C}})\sqrt{ |G_{\mathrm{R,C}}|}$. Figure adapted from~\cite{chetcuti_interfer}.} 
    \label{fig:sunint}
\end{figure}

\noindent When $W=1,2$, the number of spirals reflect the angular momentum quantization (since these cases only require one Fermi sphere `shift') --see upper and middle panels of Figure~\ref{fig:interU0}. For $W>2$, the number of observed spirals is not indicative of the angular momentum quantization --see lower panel in Figure~\ref{fig:interU0}. Fixing the number of particles and increasing the number of components, enables more particles to inhabit a given level. Indeed, for $N_{p}>N$, SU($N$) fermions behave as bosons with regards to level occupations --Figure~\ref{fig:sunint}. Lastly, we point out that for equal and commensurate $W$, the interference patterns display the same qualitative features --see Figures~\ref{fig:interU0} and~\ref{fig:sunint}.

\subsection{Interacting particles}

\subsubsection{Repulsive interactions}

\begin{figure}[h!]
    \centering
    \includegraphics[width=\linewidth]{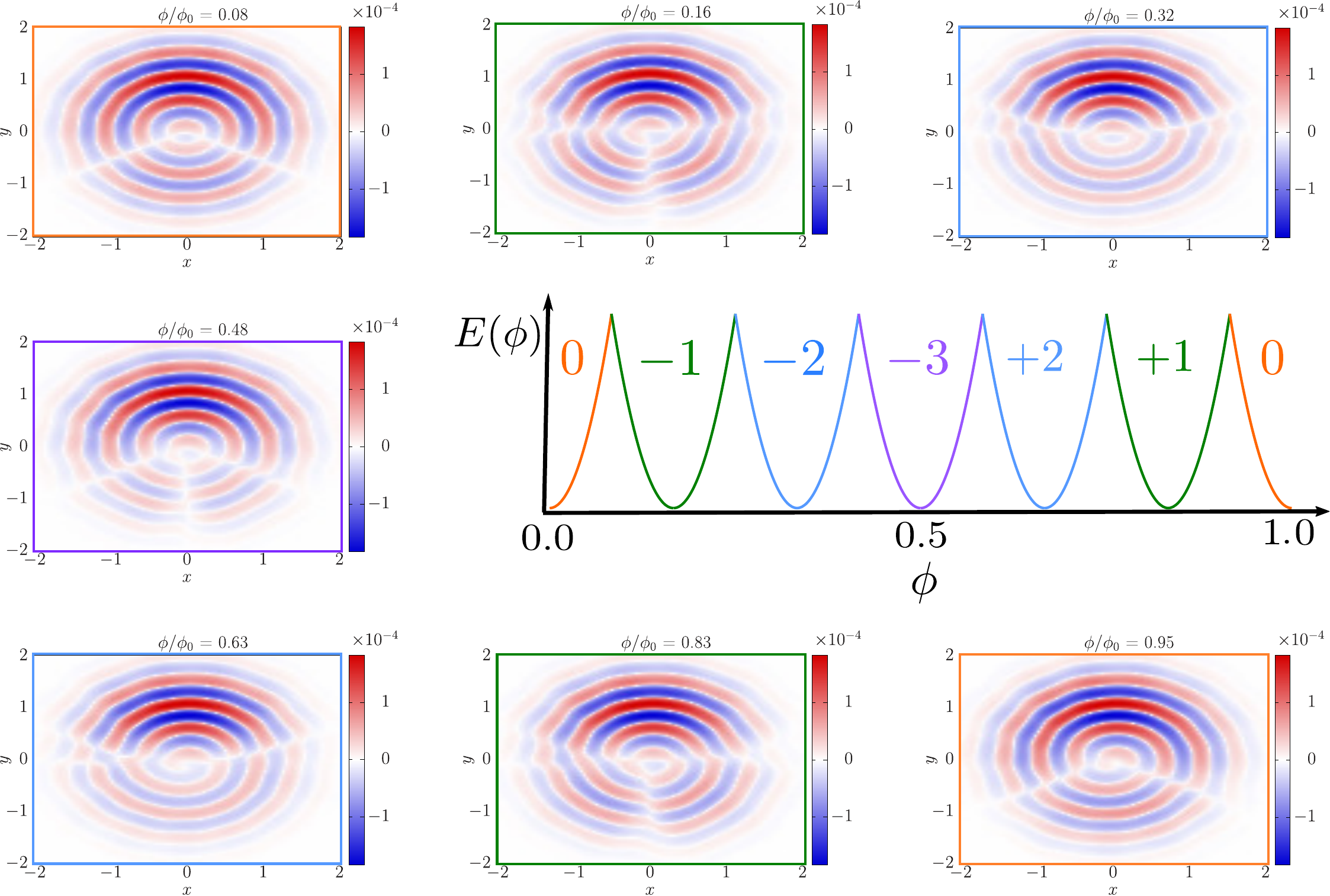}
    \caption{Interference $G_{\mathrm{R,C}}$ between ring and center for $N_{p}=6$ particles with SU(2) symmetry against the effective magnetic flux $\phi$ at $U=1000$ and short time $t=0.025$. Middle right panel is the schematic for the fractionalized energy parabolas at infinite repulsive interactions for the corresponding system (numbers on parabola correspond to the spin quantum numbers $X$ in the Bethe ansatz solution introduced in Chapter~\ref{chp:repcurr}).
    The interference pattern corresponding to the first parabola (orange), displays two dislocations as in the zero interaction case. In the next parabola (green), where there is a generation of a spin excitation, we observe three dislocations. Traversing the other parabolas, we find that apart from their number, the orientation of the dislocations changes as well --see upper left (right) panel correspond to a downward (upward) V-shape. Additionally, we note that parabolas with the equal $X$ display the same dislocation pattern reflecting the symmetry of the energy around this point. Note that there is no  spiral in the intereferograms since no hole has opened up. All correlators are evaluated with exact diagonalization for $L=15$ by setting $\textbf{r}' = (0,R)$ and radius $R=1$. The color bar is non-linear by setting $\mathrm{sgn}(G_{\mathrm{R,C}})\sqrt{ |G_{\mathrm{R,C}}|}$. Figure reprinted from~\cite{chetcuti_interfer}.
    }
    \label{fig:6intb}
\end{figure}

For a system with repulsive interactions, one is able to not only track but also observe the fractionalization through the self-heterodyne phase portrait. Remarkably, the fractional angular momentum, which corresponds to different spin excitations~\cite{yu1992persistent,chetcuti2021persistent}, is explicitly captured in the interferogram through the dislocations that are now dependent on the flux. Indeed, the dislocations are able to characterize the $N_{p}$ fractionalized parabolas due to the different types of spin excitations through the number and orientation of the dislocations --Figure~\ref{fig:6intb}. Specifically, there are $\lceil \frac{N_{p}}{2}\rceil +1$ different types of interference patterns (i.e., different number and orientation of the dislocations).

\begin{figure}[h!]
    \centering
    \includegraphics[width=0.9\linewidth]{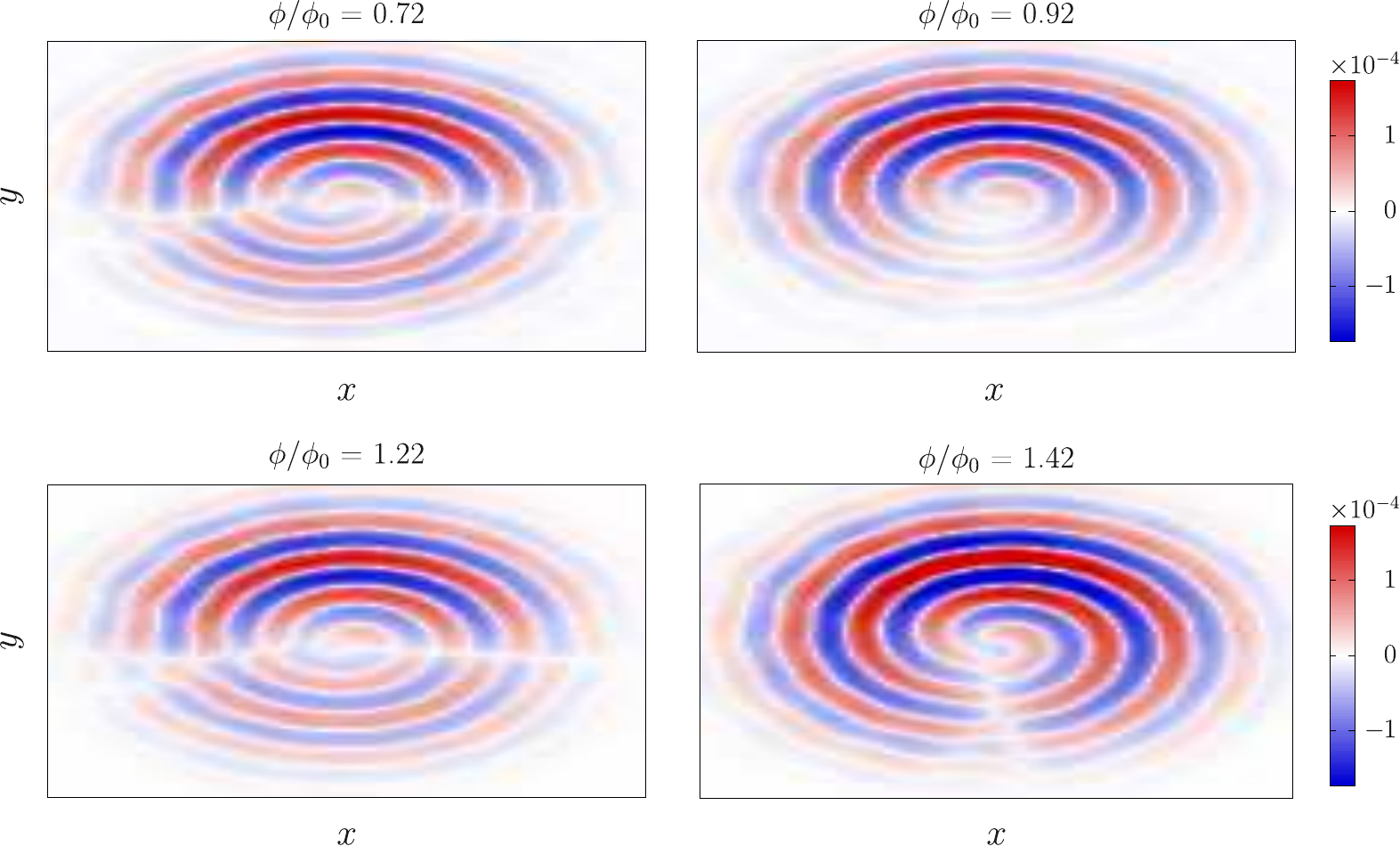}
    \caption{The interference $G_{\mathrm{R,C}}$ between ring and center for $N_{p}=4$ particles with SU(2) symmetry, is shown as a function of the effective magnetic flux $\phi$ at $U=5000$ and short time $t=0.026$. At $\phi = 0.92$, we observe spiral $B$ that disappears on going to the next parabola with $\phi = 1.22$. The spiral does emerge again at the next parabola with $\phi = 1.42$, which at this point exceeds the threshold imposed by fractionalization. Unlike the previous spiral, this spiral persists on going to the next parabolas. All correlators are evaluated with exact diagonalization for $L=15$ by setting $\textbf{r}' = (0,R)$ and radius $R=1$. The color bar is non-linear by setting $\mathrm{sgn}(G_{\mathrm{R,C}})\sqrt{ |G_{\mathrm{R,C}}|}$. Figure taken from~\cite{chetcuti_interfer}.}
    \label{fig:dsp}
\end{figure}

\noindent The characteristic spirals are also observed in interferograms at infinite repulsion. However, due to the presence of dislocations, it is very hard to deduce the order of the spirals. The emergence of spirals is not as clear-cut as in the zero interaction case since multiple spirals appear at different flux values --Figure~\ref{fig:dsp}. Firstly, we have a spiral that, for the sake of convenience, we will call spiral A. The appearance of this spiral coincides with the depression of the momentum distribution occurring at $\phi_{S_{\mathrm{A}}} = \phi_{H}+\frac{N_{p}-1}{2N_{p}}$. On increasing flux and going to larger angular momenta, spiral A gains more arms as in the zero interaction case. 

\noindent The second spiral, denoted as spiral B, emerges as we traverse to a parabola with angular momentum $\ell_{H}$. To be precise, it appears  
at $\phi_{S_{\mathrm{B}}} = \phi_{H} + \frac{1}{2N_{p}} + \Delta$ ($\phi_{S_{\mathrm{B}}} = \phi_{H} - \frac{1}{2N_{p}}$) for a diamagnetic (paramagnetic) system respectively, with $\Delta = -\frac{1}{2N_{p}}$ for an odd number of particles and zero otherwise. So basically, we are saying that spiral B corresponds to the one observed at zero interactions when the Fermi sphere has been displaced by $\big\lceil \frac{N_{p}}{2N}\big\rceil$. The extra term  $\frac{1}{2N_{p}}$ takes into account the change in the energy landscape brought on by the level crossings --Figure~\ref{fig:mess}. Furthermore, we note that spiral B only lasts for the period of that fractionalized parabola $\frac{1}{N_{p}}$ (see Figure~\ref{fig:dsp}). Accordingly, at larger values of the angular momenta, only one spiral pattern persists corresponding to spiral A. In the special case of $N_{p}=2$ SU(2) particles, we point out that the second term of $\phi_{S_{\mathrm{A}}}$ is $\frac{1}{2N_{p}}$ such that spiral A and B are one and the same. \\

\begin{figure}[h!]
    \centering
    \includegraphics[width=\linewidth]{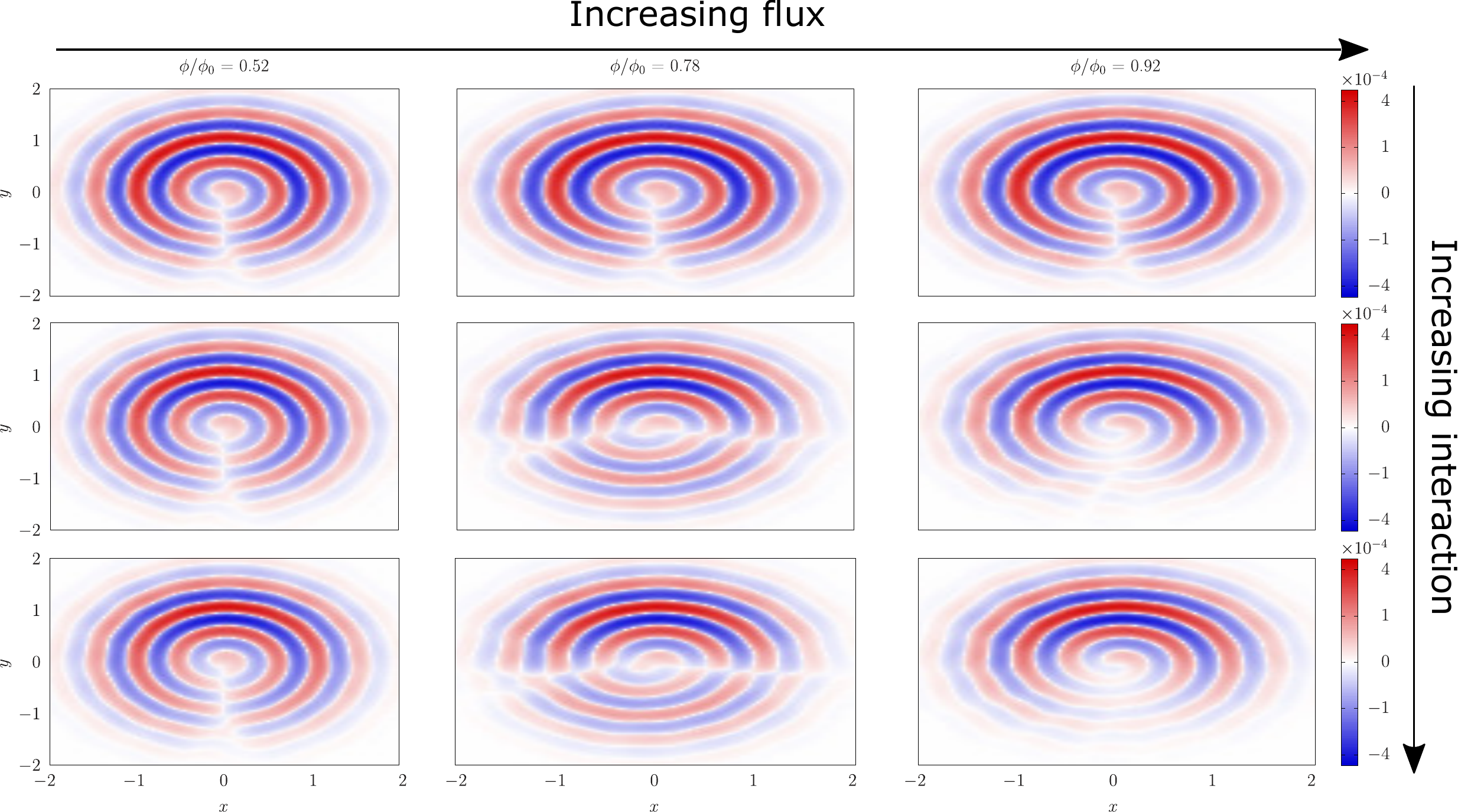}
    \caption{The interference $G_{R,C}$ between ring and center for $N_{p}=4$ particles with SU(2) symmetry against the effective magnetic flux $\phi$ at short time $t=0.026$ as a function of the interaction $U$. In this figure, one can clearly see how the number and orientation of the dislocations change as one increases the interaction. All correlators are evaluated with exact diagonalization for $L=15$ and repulsive $U=\{0,20,5000\}$ (panels are in descending order) by setting $\textbf{r}' = (0,R)$ and radius $R=1$. The color bar is non-linear by setting $\mathrm{sgn}(G_{\mathrm{R,C}})\sqrt{ |G_{\mathrm{R,C}}|}$. Figure adapted from~\cite{chetcuti_interfer}.
    }
    \label{fig:4int}
\end{figure}

\noindent For the intermediate interaction regime, the interference patterns capture the partial fractionalization in the system. On going from zero to strong interactions, as in Figure~\ref{fig:4int}, the dislocation patterns change both their number and orientation as the system undergoes fractionalization. These alterations in the interferograms are consistent with the second derivative of the momentum distribution,  further reinforcing the symbiosis between the two protocols.

\newpage
\subsubsection{Attractive interactions}

\begin{figure}[h!]
    \centering
    \includegraphics[width=\linewidth]{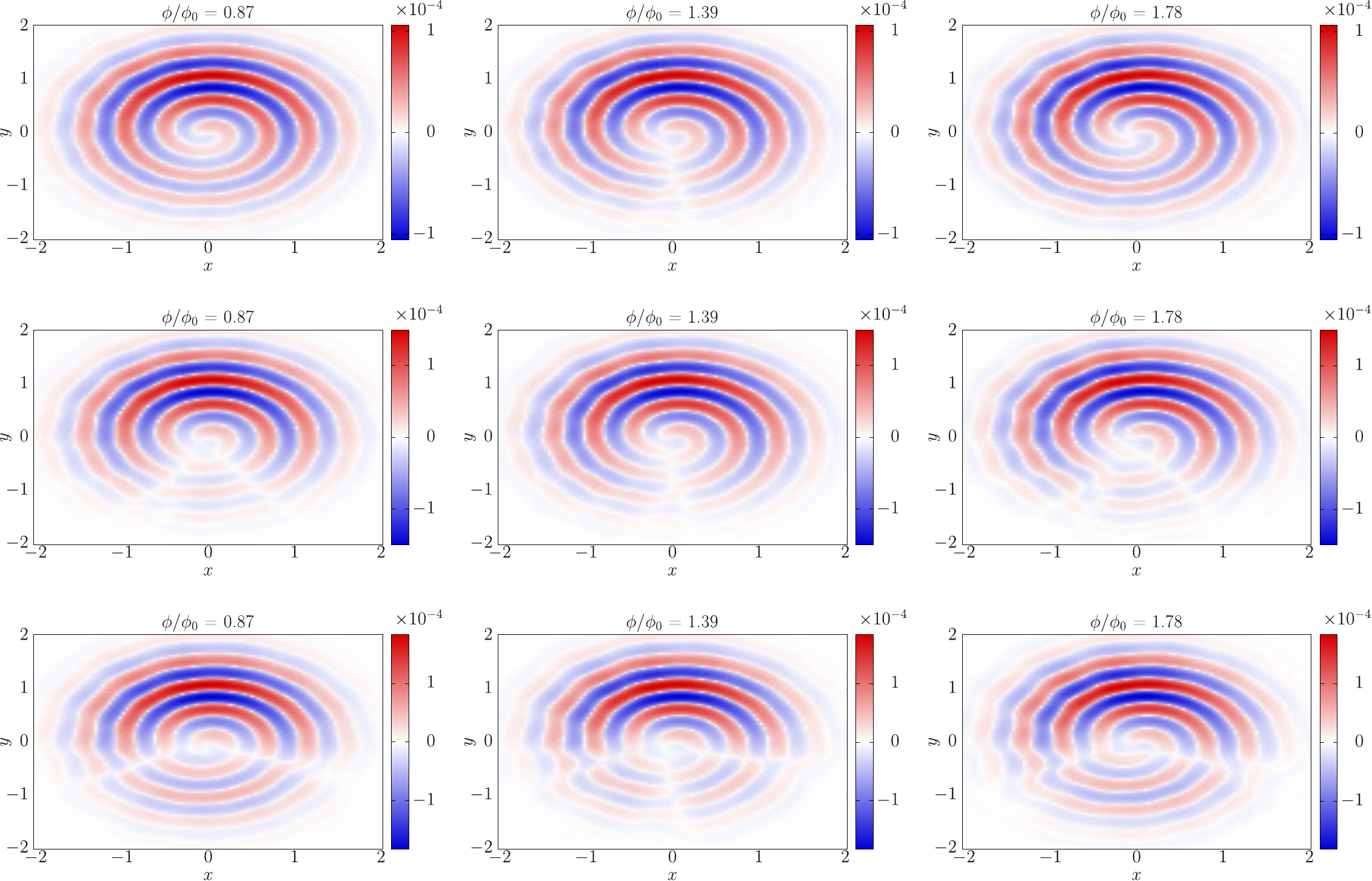}
    \caption{The interference $G_{\mathrm{R,C}}$ between ring and center for 
    $N_{p}=2$ (top), $N_{p}=4$ (middle) and $N_{p}=6$ (bottom) particles with SU(2) symmetry, is shown as a function of the flux $\phi$ at $U=-3$ and short time $t=0.025$. An extra dislocation appears for $\phi$ corresponding to a fractionalized parabola in comparison with Figure~\ref{fig:interU0}. For $N_{p}=4$, the fractionalized parabolas are the outer ones as opposed to the middle ones like $N_{p}=2$ and $N_{p}=6$ due to a difference in parity. All results are evaluated with exact diagonalization for $L=15$ by setting $\textbf{r}' = (0,R)$ and radius $R=1$.  The color bar is non-linear by setting $\mathrm{sgn}(G_{\mathrm{R,C}})\sqrt{ |G_{\mathrm{R,C}}|}$. Figure reprinted from~\cite{chetcuti_interfer}.}
    \label{fig:spiat}
\end{figure}

In contrast to the homodyne protocol, the self-heterodyne one provides direct information on the fractionalization of the persistent current: (i) the $N$ fractionalized parabolas are characterized by different dislocations, both in number and orientation; (ii) spirals emerge in the interferogram --Figure~\ref{fig:spiat}. Coincidentally, for $N$-body bound states the observation of the spiral occurs at $\phi_{S_{-}} = \phi_{H} +  \frac{N\! -\! 1}{2N}$, which would correspond to when one would expect momentum distribution depression. Just like in the repulsive case, there is a `delay' in visualizing the spirals. However, since the fractionalization is dependent only on $N$ and irrespective of $N_{p}$, the `delay' associated with it is uniform for different particle systems and is given by $\frac{N\! -\! 1}{2N}$. It is interesting to note that unlike the repulsive case, here we do not observe the emergence of the second spiral (previously called spiral B). \\

\noindent We remark that the mentioned results pertain only to SU(2) fermions. For SU($N>2$), an interferogram is not adequate to capture any information about the attractive system. The self-heterodyne interference patterns rely on the use of a two-body correlator, which is an accurate measure when one has bound  pairs as in SU(2) fermionic systems. However, bound states consisting of a larger number of particles, such as trions in the SU(3) case, probably require an $N$-body correlator to adequately describe the system.

\section{Discussions and Conclusions} \label{sec:conc}

In this chapter, persistent currents have been utilised to analyze and characterize different SU($N$) fermionic systems. To this end, we investigate the interference dynamics via both homodyne (momentum distribution) and heterodyne (co-expansion of two the ring and center) protocols by applying a combination of exact diagonalization, and whenever possible, DMRG~\cite{whitedmrg,itensor} and Bethe ansatz. Both of these read-out techniques can be  experimentally implemented within the cold atoms infrastructure and can be experimentally traced. These protocols, which have become a staple tool in bosonic experiments, are harder to implement for fermions on account of the lower condensate fractions and the high chemical potential that results in rapid expansion of the condensate~\cite{wright2022persistent}.

\paragraph{Free particle regime:}
For spinless fermions, the characteristic hole in the momentum distribution, reflecting coherent matter-wave flow, opens up when the effective magnetic flux displaces half of the Fermi sphere~\cite{pecci2021phase}. In the following, we will refer to such a feature as a `delay' in the value of the flux at which a hole occurs. For $N$-component fermions,  the Pauli exclusion principle relaxes and allows  more particles to inhabit a given level --Figure~\ref{fig:mom}. Accordingly, we find that a hole in the momentum distribution appears if the Fermi sphere is displaced less, precisely by the ceiling function $\lceil\frac{N_p}{2N}\rceil$. This is consistent with the fact that SU($N$) fermions (with $N_p\! <\!N$) resemble bosons for $N\rightarrow \infty$. \\

\noindent The features of the momentum distribution are observed to be consistent with the heterodyne interference images. In particular, holes in the momentum distribution correspond to spirals in the interferograms.  Note that, in contrast  with the bosonic case~\cite{andrews1997,eckel2014interferometric,corman2014quench,mathew2015self,haug2018readout}, the order of the spirals does not reflect the angular momentum quantization of the system. 
Furthermore, when the number of components $N$ divides the number of particles $N_{p}$, $\frac{N_{p}}{N}-1$ dislocations (radially segmenting lines) appear in the interferograms giving information about the number of particles present in the system --Figure~\ref{fig:interU0}. These observations still hold true in the case of small attractive or repulsive interactions~\cite{pecci2021phase}.
Moreover, we note that at zero interactions, the properties displayed by the homodyne and self-heterodyne phase portraits depend solely on $\frac{N_{p}}{N}$ --see Figures~\ref{fig:interU0} and~\ref{fig:sunint}. 

\paragraph{Repulsive regime:}

\noindent The persistent current fractionalizes with a reduced of $\phi_{0}/N_{p}$, as discussed in Chapter~\ref{chp:repcurr}.  Interestingly enough, the fractional values of the angular momentum are not displayed as holes in the momentum distribution as in the non-interacting case. Furthermore, as soon as the correlations depart from the non-interacting case,
the characteristic hole becomes a small depression (finite valued local minimum) in the momentum distribution at momentum $\mathbf{k}=0$ --Figure~\ref{fig:TOFrp}(\textbf{a}). 
Moreover, an additional `delay' characterising the specific fractionalization of the angular momentum is found. In other words, the depression occurs at a larger flux value than the one where the hole used to open up --Figure~\ref{fig:TOFrp}(\textbf{b}). Specifically, for a given system of $N_{p}$ particles, a momentum distribution depression appears at $\phi_{D} = \phi_{H} + \frac{N_{p}-1}{2N_{p}}$, where $\phi_{H}$ is the flux at which the hole opens up at zero interactions and {the second term accounts for the fractionalization `delay'.} Therefore, such a property makes the number of particles accessible by monitoring the actual value of the flux at which the depression occurs. At intermediate interactions, the system experiences only partial fractionalization. Nonetheless, this is captured by the homodyne protocol by tracking the momentum distribution depression at zero momenta --Figure~\ref{fig:TOFrp}(\textbf{b}). \\

\noindent Heterodyne interferograms embody the features of the fractionalization. Apart from observing the `delay' through the emergence of the spiral as in the zero interaction case, the angular momentum fractionalization can be tracked by monitoring the number and orientation of the dislocations: the presence of the spin excitations modifies the dislocations that are observed at zero interaction --see Figures.~\ref{fig:6intb} and~\ref{fig:4int}.  In contrast with the zero interaction case, the dislocations are now flux dependent, enabling us to monitor the spin excitations in the system through interference patterns.

\paragraph{Attractive regime:} As we explored in Chapter~\ref{chp:probe}, a system with attractive interactions also experiences fractionalization like its repulsive counterpart, albeit with a reduced period dependent on the number of components $N$. However, the fractionalization is not readily observed in the momentum distribution, at least not directly. Due to the reduced coherence that transpires from the formation of bound states, no depression is observed --Figure~\ref{fig:TOFat}(\textbf{a}). As such, one cannot monitor the  `delay' in the occurrence of the aforementioned depression. Nonetheless, by measuring the variance of the width of the momentum distribution, one can deduce the SU($N$) nature of the system through the number of steps depicted, but no information regarding the number of particles can be obtained --Figure~\ref{fig:TOFat}(\textbf{b}). In the intermediate regime of interaction, we observe that the characteristic depression gets `delayed' to larger values of the effective magnetic flux as the current gets  fractionalized --Figure~\ref{fig:maprp} (right panels). Eventually, as the interaction strength increases, the depression is smoothed out. \\

\noindent Self-heterodyne interference patterns for SU($N\! >\!2$) systems calculated via density-density correlators are found to be incapable of  providing observables to monitor the persistent current pattern. In fact, a   higher order correlator may be  required to capture the features of $N$-body bound states. In the case of SU(2) systems, we find that the fractionalization is characterized by a change in the number of dislocations in the interferograms --Figure~\ref{fig:spiat}. Remarkably, the flux values where a spiral emerges correspond to the ones that would result in a depression (that is suppressed in this regime) in the momentum distribution. As in the repulsive case, the spirals  experience a two-fold `delay' originating from the combined effect of  displacement of the Fermi sphere and the fractionalization. \\

\noindent Recently, self-heterodyne interferograms were experimentally observed for SU(2) attractive fermions in the context of the BCS-BEC crossover~\cite{roati2022imprinting}.  
Specifically, the interference fringes observed correspond to BEC side of the crossover. Our analysis, instead, pertains to the BCS regime, where our results predict a characteristic  `delay' that provides information on the structure of the Fermi surface. The interference patterns within the BCS regime still remain to be analysed experimentally, with the main challenge lying in the fast expansion stemming from the large momenta occupations of the fermions. One route to address present limitations involves using dilute density systems such that the particles do not occupy levels with high momenta. Another option would be to selectively address particles close to the Fermi surface and perform the expansion.\\

\noindent In summary, we have shown how one can characterize SU($N$) correlated matter-wave through homodyne and self-heterodyne interference patterns. In particular, by monitoring the number of dislocations at weak interactions (repulsive or attractive), we can gain knowledge on  $N_p/N$ --Figures~\ref{fig:interU0} and~\ref{fig:sunint} - this feature  provides the generalization of the work carried out in~\cite{pecci2021phase} to $N$-component fermions. Furthermore, going to the strongly interacting regimes, the number and configuration of the dislocations change, reflecting the persistent current fractionalization --Figures~\ref{fig:6intb} and~\ref{fig:4int}. To be specific, there are $\big\lceil \frac{N_{p}+1}{2}\big\rceil$ interference patterns with various dislocation numbers and orientations. This feature implies that, for repulsive interactions, we can access the number of particles $N_p$ (see~\cite{zhao2021heuristic} for characterization of SU($N$) systems through neural networks). For the repulsive case, the  experimental analysis could be carried out through quantum gas microscopy~\cite{bakr2009micro,mitra2017quantum,bloch2017quantum,selim2018spin}, which is an atom imaging technique capable of probing a quantum gas with a high degree of spatial resolution and single-atom sensitivity.


\chapter{Exact one-particle density matrix for SU(\textit{N}) fermionic matter-waves in the strong repulsive limit}\label{chp:ogata}

\noindent In low-dimensional many-body systems, quantum fluctuations are particularly pronounced, and therefore  even a weak interaction can lead to dramatic correlations. Such a simple fact makes the physics of one-dimensional many-body systems exotic and distinct from the physics of higher-dimensional systems~\cite{gogolin2004bosonization,buschreview}. The breakdown of the Fermi liquid paradigm and Luttinger liquid behaviour, the spin-charge separation in fermionic systems, elementary excitations with fractional statistics, and Haldane order are just some of the characteristic traits addressed in the last few decades of research on the subject~\cite{haldane1981,giamarchi2003quantum,takahashi2005thermodynamics,affleck1989quantum}. One-dimensional systems can be realized by confining the spatial degrees of freedom, as in quantum wires~\cite{datta1997electronic}, in chains of Josephson junctions~\cite{fazio2001quantum}, or in certain classes of polymers~\cite{baeriswyl1992overview}; in other instances, the dimensionality is constrained dynamically, as in carbon nanotubes~\cite{dresselhaus1995physics}, edge states in quantum Hall effect~\cite{cage2012quantum} or in metals with dilute magnetic impurities~\cite{hewson1997kondo}.  
With the advent of quantum technology seeking quantum correlations as a resource, the impact of one-dimensional physics has been considerably widened.  \\

\noindent Exact solutions of one-dimensional interacting quantum many-body systems play a particularly important role since their physics is often non-perturbative, with properties that are beyond the results obtained with approximations~\cite{gogolin2004bosonization}. As such, exact results, though rare and technically difficult to achieve, form a precious compass to get oriented in the physics of one dimension. In this chapter, we provide the exact expression for the two-point correlation matrix of SU($N$) fermions confined in one dimension and subjected to an external magnetic flux, determining the one-body density matrix in the limit of large repulsive interactions. Specifically, we analyze the momentum distribution of the particles, which despite being one of the simplest correlations, is able to reflect certain effects of the interaction~\cite{fradkin2013field}. On the technical side, we point out that, despite its simple expression, the momentum distribution can only be calculated numerically for a small number of particles and is even less accessible when considering the strongly correlated regimes. Even for integrable models, the momentum distribution is not managable, especially in the mesoscopic regime of finite but sufficiently large particle systems. The case of two-component fermions in the absence of a magnetic flux was discussed by Ogata and Shiba~\cite{OgataShiba90}. \\

\noindent The one-body density matrix  plays a crucial role in different schemes of time-of-flight expansions in cold atoms settings~\cite{roth2003superfluidity,amico2005quantum,amico2022,chetcuti2021probe,chetcuti_interfer}, as explored in Chapter~\ref{chp:repcurr}. The specific persistent current pattern that arises is produced as a result of specific transitions between suitable current states characterized by a different configuration in the spin degrees of freedom of the system~\cite{yu1992persistent,chetcuti2021persistent}. 
We will show that this process brings substantial complications in computing the ground-state to access the correlation matrix of the system for different magnetic fluxes.

\section{Model and Methods}\label{sec:modeland}

\noindent The system under consideration is that of $N_{p}$ fermions with SU($N$) symmetry residing on a mesoscopic ring-shaped lattice composed of $L$ sites and pierced by a synthetic magnetic field $\phi$. Such a system can be modeled by the one-dimensional Hubbard Hamiltonian for $N$-component fermions that reads 
\begin{equation}\label{eq:Ham2}
\mathcal{H} = -t\sum\limits_{j}^{L}\sum\limits_{\alpha=1}^{N}(e^{i\frac{2\pi\phi}{L}}c_{j,\alpha}^{\dagger}c_{j+1,\alpha} +\textrm{h.c.})+ U\sum\limits_{j}^{L}\sum\limits_{\alpha<\beta}n_{j,\alpha}n_{j,\beta},   
\end{equation}
fixing the energy scale with $t=1$ and setting $U>0$ as the interactions between the fermions are taken to be repulsive. Specifically, we are interested in the regime of dilute filling fractions such that model~\eqref{eq:Ham2} tends to the continuous limit, which is Bethe ansatz solvable (see Chapter~\ref{chp:repcurr}). Accordingly, within a given particle ordering $x_{Q_{1}}\leq\hdots \leq x_{Q_{N_{p}}}$, the eigenstates of the model~\eqref{eq:Ham2} can be expressed as
\begin{align}\label{eq:wavywavex}
 \Psi(x_{1},\hdots,x_{N_{p}};\alpha_{1},\hdots,\alpha_{N_{p}}) = \sum\limits_{P\in S_{N_{p}}}\mathrm{sign}(P)\mathrm{sign}(Q)\psi [Q;P]\exp\bigg(i\sum_{j=1}^{N_{p}} k_{Pj}x_{Qj}\bigg),
\end{align}
where $P$ and $Q$ are permutations introduced to account for the eigenstates' dependence on the relative ordering of the particle coordinates $x_{j}$ and quasimomenta $k_{j}$, with $\psi$ being the spin-dependent amplitude. The latter accounts for all different components of the system, which can be obtained by nesting the Bethe ansatz~\cite{sutherland1968}. As a result, the spin-like rapidities for each additional colour $\Lambda_{\beta_{l}}$ are all housed in $\psi$. \\

\begin{figure}[h!]
    \centering
    \includegraphics[width=\linewidth]{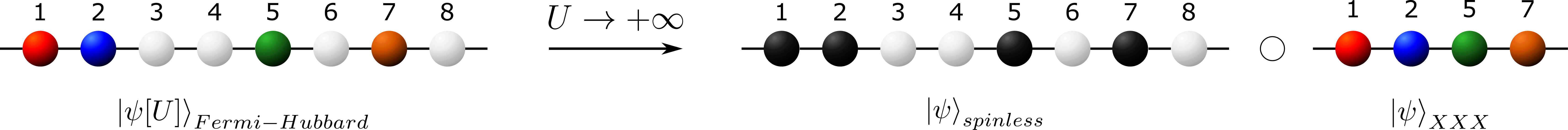}
    \caption{Schematic representation of the decoupling of the SU($N$) Fermi-Hubbard into the spinless and XXX Heisenberg Hamiltonians at infinite repulsive interactions $U$. On the left, the figure depicts the SU(4) Hamiltonian with one particle per colour and 4 empty sites (white). On the right, we have the spinless Hamiltonian with 4 fermions (black) and 4 empty sites. In addition, there is the SU(4) Heisenberg Hamiltonian with one spin in each orientation. Note that after the decoupling, the index corresponding to a given colour in the Heisenberg Hamiltonian changes in order to accommodate the new framework, but the arrangement of the colours in the original chain is maintained. The circle indicates the successive operation in a mathematical sense of a composition function: $f\circ g=f(g)$. Figure adapted from~\cite{osterloh2022exact}.}
    \label{fig:ogchain}
\end{figure}

\noindent Despite access to the energy spectrum being greatly simplified due to integrability, the calculation of the exact correlation functions remains a very challenging problem~\cite{korepin_bogoliubov_izergin_1993}, especially in the mesoscopic regime of large but finite number of particles and ring sizes~\cite{essler}. In what follows, we will be focusing on the strong repulsive limit where the correlation function becomes addressable. The simplification arises because the charge and spin degrees of freedom decouple, which occurs only for states with real quasimomenta~\cite{OgataShiba90,yu1992persistent}. This decoupling is manifested in the Bethe equations of the system. To start, let us recall that  $k_{j}/U$ goes to zero in the limit $U\rightarrow \infty$ and can thus be neglected~\cite{OgataShiba90,yu1992persistent}. Consequently, by taking the SU(3) case as an example we have that the product form Bethe ansatz equations~\eqref{eq:chargesu3rep}--\eqref{eq:spin2su3rep} in this limit read
\begin{equation}\label{eq:decchar}
e^{i (k_{j}L_{R}-2\pi\phi)} = \prod\limits_{\alpha = 1}^{M_{1}} \frac{2Q_{\alpha} -i }{2Q_{\alpha} +i } \hspace{3mm} j=1,\hdots,N_{p},
\end{equation}
\begin{equation}\label{eq:decspin1}
\prod\limits_{\beta \neq\alpha}^{M_{1}} \frac{(Q_{\alpha}-Q_{\beta}) +i }{(Q_{\alpha}-Q_{\beta}) -i } = \bigg[ \frac{2Q_{\alpha} +i }{2Q_{\alpha} -i } \bigg]^{N_{p}}\prod\limits_{a =1}^{M_{2}} \frac{2(Q_{\alpha}-P_{a}) +i }{2(Q_{\alpha}-P_{a}) -i }\hspace{3mm} \alpha=1,\hdots,M_{1},
\end{equation}
\begin{equation}\label{eq:decspin2}
\prod\limits_{b \neq a}^{M_{2}} \frac{P_{a}-P_{a} +i }{P_{a}-P_{b} -i } = \prod\limits_{\beta =1}^{M_{1}} \frac{2(P_{a}-Q_{\beta}) +i }{2(P_{a}-Q_{\beta}) -i}\hspace{3mm} a=1,\hdots,N_{2}, 
\end{equation}
defining $Q_{\alpha} = 2\frac{\Lambda_{\alpha}}{U}$ and $P_{a} = 2\frac{\lambda_{a}}{U}$ respectively. The Bethe ansatz equations for the charge and spin rapidities decouple, in that Equations~\eqref{eq:decspin1} and~\eqref{eq:decspin2} are independent of the quasimomenta $k_{j}$. Naturally, the same decoupling manifests itself in the wavefunction~\eqref{eq:wavywavex}. As an effect of the spin-charge decoupling, each wavefunction amplitude can be written as a product between a Slater determinant of spinless fermions $\mathrm{det}[\exp(i k_{j}x_{Q{j}})]$ and the spin wavefunction $\Phi (y_{1},\hdots,y_{M})$~\cite{OgataShiba90}
\begin{align}\label{eq:wave}
\Psi(x_{1},...,x_{N_{p}};\alpha_{1},...,\alpha_{N_{p}}) = \mathrm{sign}(Q)&\mathrm{det}[\exp(i k_{j}x_l)]_{jl}\Phi (y_{1},\hdots,y_{M}).
\end{align}
The logic in constructing the wavefunction through the nested Bethe ansatz is to reduce the problem into a simpler one by considering only the spin degrees of freedom (see Chapter~\ref{chp:tools}). The corresponding spin wavefunction amplitudes in the ansatz are those of the inhomogenous XXX Heisenberg model~\cite{deguchi2000thermodynamics,essler}. On account of the decoupling, the inhomogeneity $k_{j}$ in the spin wavefunction vanishes. Indeed, Equations~\eqref{eq:decspin1} and~\eqref{eq:decspin2} coincide with the Bethe ansatz equations of an isotropic SU(3) Heisenberg spin chain, up to a normalization constant~\cite{bethe,essler}. Consequently, in the limit of infinite repulsion the states of the SU($N$) Hubbard model can be written as 
\begin{equation}\label{eq:decoupling_wavefunction}
    \ket{\psi[U]}_{\rm Fermi-Hubbard}\stackrel{U\to +\infty}{\longrightarrow}
\ket{\psi\left[\psi_{XXX}\right]}_{\rm spinless},
\end{equation}
with a schematic of the decoupling occurring in the wavefunction depicted in Figure~\eqref{fig:ogchain}. It is important to emphasize that this is not a tensor product but corresponds to a composition of functions. Such a result holds for all $N\geq 2$ in the integrable regimes of model~\eqref{eq:Ham2} due to the nature of the nested Bethe ansatz procedure\footnote{For $N=1$ all the particles populate one colour such that $M_{1}=M_{2}=0$. Therefore, it is clear to see from the outset that there is no spin-charge decoupling, with such a case corresponding to that of a gas of one-component hard-core bosons integrable by the Lieb-Liniger Bethe ansatz~\cite{liebliniger}.}. Even though the fermionic and spin chain Hamiltonians share a similar wavefunction, we point out that their spectra are markedly distinct. The energy of model~\eqref{eq:Ham2} is obtained through the quasimomenta outlined in Equation~\eqref{eq:kincomm}, that of the Heisenberg spin chain is provided solely by the spin rapidities. Nevertheless, the energy correction to leading order in $1/U$ of the Hubbard model corresponds to the energy of the antiferromagnetic Heisenberg model with an effective coupling $J_{eff}$. This energy, that we elected to call the energy spin correction in Chapter~\ref{chp:repcurr}, satisfies the Heisenberg Bethe ansatz equations~\eqref{eq:decspin1} and~\eqref{eq:decspin2}. 

\noindent By splitting the problem into the spinless and Heisenberg models, the problem of obtaining the wavefunction is greatly simplified. For starters, the calculation of the Slater determinant for spinless fermions is independent of the number of components $N$. As such, one only needs to evaluate the SU($N$) Heisenberg for systems with fixed number of particles and sites but different $N$. Additionally, diagonalizing the Heisenberg model is less computationally intensive  compared to the Hubbard model due to the lower dimensionality of the Hilbert space. Consequently, systems with larger number of particles and ring sizes can be considered. \\

\noindent The XXX Heisenberg model can be constructed as a sum of permutation operators $\mathcal{H}_{XXX} = \sum_{i}P_{i,i+1}$~\cite{sutherland1975,capponi}, where $P_{i,i+1}$ can be expressed in terms of SU($N$) generators (see Appendix~\ref{app:Heisenberg}). These states permutes spins having the same orientation on sites $i$ and $i+1$. As mentioned previously, this spin chain model is also Bethe ansatz integrable for any SU($N\geq 2$). Nevertheless, as in all Bethe ansatz solvable models, accessing the explicit expression of the antiferromagnetic Heisenberg eigenstates remains quite challenging. As such, in our approach the quantum state is obtained by combining the Bethe ansatz analysis with the Lanczos numerical method. The procedure is described below:

\paragraph{(a) Finding the ground-state.}
Firstly, we note that for each non-degenerate ground-state of the Hubbard model, there exists a corresponding single eigenstate of the Heisenberg model. In principle, such a state-to-state correspondence could be obtained by identifying the spin quantum numbers labeling the states of the Hubbard model and inserting them into the Bethe ansatz equations of the Heisenberg model. However, as explained earlier such a procedure is quite involved when trying to access to the quantum states. Therefore, we employ a combination of Bethe ansatz and numerical methods: 
\begin{enumerate}
    \item The spin quantum numbers characterizing a given state in the Hubbard model are inserted into the Heisenberg Bethe ansatz, enabling us to calculated the correct energy. This energy is then matched with the numerically obtained spectrum of the antiferromagnet to pinpoint the corresponding eigenstates.
    \item Through the SU($N$) quadratic Casimir operators (see the Appendix~\ref{Casimir}), we characterize the total SU($N$)-spin of the states. These operators commute with the whole SU($N$) group and hence are constants of the motion of both the Heisenberg Hamiltonian and the SU$(N)$ Hubbard model. The square of the total spin operator is the Casimir operator for $N=2$. For zero flux, the state of the Hubbard model results to be non-degenerate. Therefore, this approach can uniquely characterize the states. However, for non-vanishing flux it turns out that the energy of the Heisenberg model is degenerate as is the Casimir value.
    \item In the SU(2) case, the degeneracy can be resolved by looking at the permutation operators $P_{j,j-1}$, which do not commute with the spin chain Hamiltonian by construction. For $N\!>\!2$ and non-vanishing flux, we do not have a general method. Moreover, the situation becomes more complicated due the $N\! -\! 1$ fold degeneracy stemming from the different sets of spin quantum numbers of the Hubbard model (see Chapter~\ref{chp:repcurr}), that manifests in the Heisenberg model as a consequence of the one-to-one correspondence. However, we note that degenerate states with the same Casimir value consist of different projections into the Heisenberg basis, which allow us to uniquely identify the correct ground-states to be taken at increasing flux.
\end{enumerate}

\noindent The states are chosen based on the parity of the species' occupation number. This comes around since we find that non-degenerate ground-states with odd and even number of particles per species correspond to different values of the Casimir operators and in turn to different representation of the SU($N$) algebra\footnote{It is worth noticing that this eigenvalue may be accidentally degenerate in the Heisenberg model.}. \\

\noindent At zero flux, the ground-state wavefunction of the Hubbard model for systems with $(2n)N$ fermions with integer $n$, is not a singlet as opposed to that of the antiferromagnetic Heisenberg model. In the case of two-component fermions, this issue was circumvented by considering anti-periodic boundary conditions for the Hubbard model, which results to be a singlet ground-state~\cite{OgataShiba90}. In contrast with the method presented in~\cite{OgataShiba90}, we do not modify the boundary conditions for model~\eqref{eq:Ham2}. Instead, we modify the spin quantum numbers as outlined in Chapter~\ref{chp:repcurr}, such that the non-degenerate triplet eigenstate of the Heisenberg model is selected. 

\paragraph{(b) The one-body density matrix.} In what follows, 
we apply the factorization~\eqref{eq:decoupling_wavefunction} to determine
the one-particle density operator through the calculation of the two-point correlation matrix of the  SU($N$) Hubbard model~\eqref{eq:Ham2}, together with its dependence on the flux $\phi$:
\begin{equation}
    \langle \Psi_\alpha(x)^\dagger \Psi_\alpha(x')\rangle=\sum_{l,j} w^*(x-x_l) w(x'-x_j) 
    \langle c_{l,\alpha}^{\dagger}c_{j,\alpha}\rangle ,
\end{equation}
where $\Psi_\alpha^\dagger(x)$ and $\Psi_\alpha(x)$ are the fermionic field operators satisfying $\{\Psi_\alpha^\dagger(x),\Psi_{\alpha'}(y)\} = \delta(x-y)\delta_{\alpha,\alpha'}$,  and  $w(x)$ are Wannier functions that we take to be independent of the specific $N$ component. 

\section{Momentum distribution}

\begin{figure}[h!]
    \centering
    \includegraphics[width=0.8\linewidth]{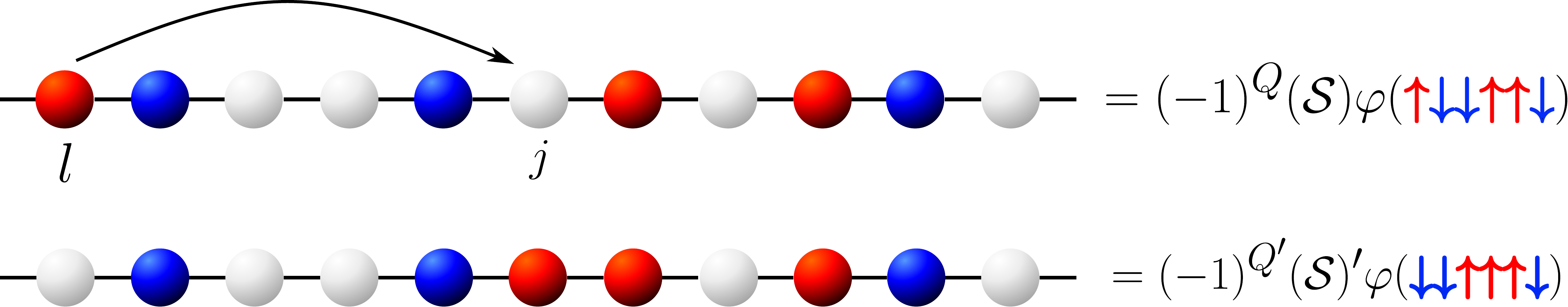}
    \caption{Schematic representation of the effect of $c_{l,\alpha}^{\dagger}c_{j,\alpha}$ on an SU(2) wavefunction. The upper part depicts the initial state in a given configuration with the corresponding decoupled wavefunction shown on the right. The bottom figure illustrates the final state and its corresponding wavefunction after performing the hopping action on the initial state. Figure reproduced from~\cite{OgataShiba90}.}
    \label{fig:spintr}
\end{figure}

\noindent The momentum distribution defining the occupation of fermions in the Fermi sphere is given by 
\begin{equation}\label{eq:mom}
    n_\alpha(k) = \frac{1}{L}\sum\limits_{j,l}^{L}e^{\imath k (r_{l}-r_{j})}  \langle c_{l,\alpha}^{\dagger}c_{j,\alpha}\rangle ,
\end{equation}
with $r_{j}$ denoting the position of the lattice sites in the ring’s plane. In the aforementioned limit of infinite repulsion, the correlation matrix can be recast as
\begin{equation}\label{eq:newmome}
\langle c_{l,\alpha}^{\dagger}c_{j,\alpha} \rangle = \sum\limits_{\{x\}}\mathrm{sign}(Q)\mathrm{sign}(Q')
(\mathcal{S})^{*}(\mathcal{S})'\omega(j\rightarrow l,\alpha),
\end{equation}
where $\{x\}$ is the spinless fermion configuration, $\mathcal{S}$ denotes the Slater determinant of the charge degrees of freedom $Q$ refers to the sign of the corresponding permutation. $\mathcal{S}$' and $Q$' are the same quantities but evaluated for the wavefunction of a fermion that moved from the $j$-th to the $l$-th site (see Figure~\ref{fig:spintr}). In the presence of flux, we note that apart from choosing the correct Heisenberg state, one has to account for the shift in quasimomenta $k_{j}$ induced by the spin quantum numbers through Equation~\eqref{eq:kincomm}. These quasimomenta are different from the momenta $k$ in the momentum distribution discussed here. Furthermore, we would like to emphasize that instead of calculating the Slater determinant for the continuous Gaudin-Yang-Sutherland model, we discretize it. Such an approach is necessary in order to keep track of the mapping between the spin wavefunctions of the Hubbard and Heisenberg models. The term $\omega(j\rightarrow l,\alpha)$ corresponds to the spin part of the wavefunction of the Hubbard model, taking into account the sum over all the spin configurations and any changes in $\Phi (y_{1},\cdots,y_{M})$. 

\noindent Before proceeding to evaluate Equation~\eqref{eq:newmome}, we note that $\omega(j\rightarrow l,\alpha)$ is independent of $\alpha$ such that $\omega(j\rightarrow l,\alpha)=\omega(j\rightarrow l)$. Moreover, in the limit of infinite repulsion, the spin wavefunction of the Hubbard model can be mapped to that of the Heisenberg, such that $\omega(j\rightarrow l) = 
\tilde{\omega}(j'\rightarrow l')$, where the tilde indicates the spin correlation function of the Heisenberg model.
In this mapping, we associate the $j'$th spin of the Heisenberg model to the fermion on the $j$th site of the Hubbard model, that after the hopping operation $c_{l}^{\dagger}c_{j}$, becomes the $l'$th spin corresponding to a fermion of the $l$th site --Figure~\eqref{fig:spintr}. 
We emphasize that the expression in Equation~\eqref{eq:newmome} is of the same form as for the SU(2) case~\cite{OgataShiba90}. The difference lies in the definition of $\tilde{\omega}(j'\rightarrow l')$, which encodes the SU($N$) character of the system: 
\begin{equation}\label{eq:heiscorr}
    \tilde{\omega} (j'\rightarrow l') \equiv \langle P_{l',l'-1}P_{l'-1,l'-2}\cdots P_{j'+1,j'}\rangle_{H_{XXX}}.
\end{equation}
This expression corresponds to the expectation value in the Heisenberg state of the SU($N$) permutation operator $P_{j',j'-1}$ that exchanges the $j'$th and $(j'-1)$th sites. With the states obtained as summarized in Section~\ref{sec:modeland},  we evaluate the momentum distribution $n(k)$ in Equation~\eqref{eq:mom}.

\begin{figure}[h!]
    \centering
    \includegraphics[width=0.7\linewidth]{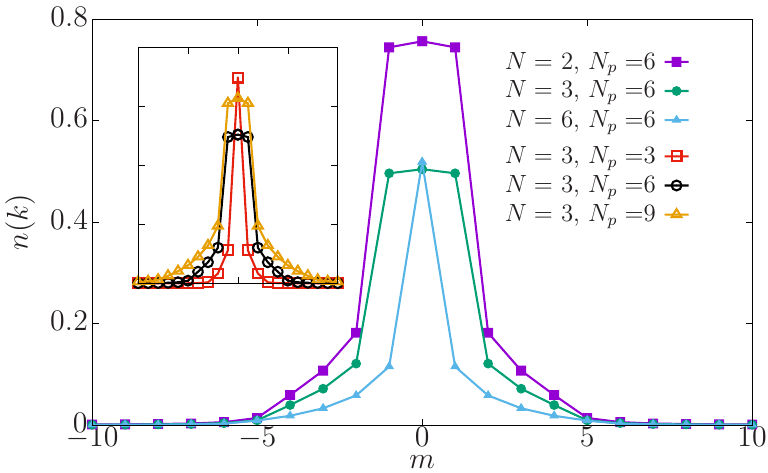}
    \caption{Main panel: Momentum distribution function $n(k)$ for a fixed number of particles $N_{p}=6$ but varying SU($N$) symmetry. Inset displays the momentum distribution for SU(3) systems with different number of particles. Both panels showcase the interplay between the occupation and the SU($N$) character of the system. The integers $m$ correspond to the momenta $2\pi m/L$ where the system size is fixed to $L=27$. Figure reprinted from~\cite{osterloh2022exact}.}
    \label{fig:nofkdiffsun}
\end{figure}
\begin{figure}[h!]
    \centering
    \includegraphics[width=\linewidth]{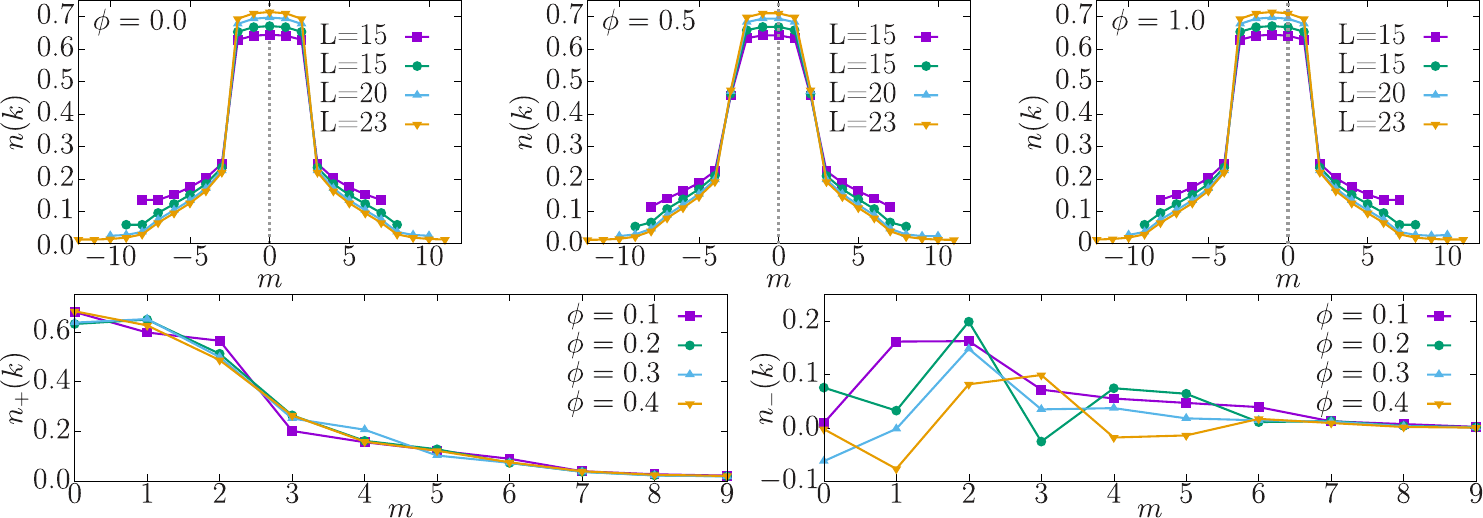}
    \caption{Top row: The momentum distribution function $n(k)$ for $N_{p}=10$ particles with SU(3) symmetry and various sites $L$. Essentially, the effective magnetic flux $\phi$ shifts the momentum distribution as expected. Bottom row: The symmetric (left) and anti-symmetric (right) components of the momentum distribution denoted as $n_{+}(k)[n(k-\Delta k/2)+n(-k-\Delta k/2)]/2$ and $n_{-}(k)=n(k-\Delta k/2)-n(-k-\Delta k)/2)$ respectively are plotted as a function of the flux $\phi$. These intermediate values of the flux, that correspond to the inner fractionalized parabolas in the energy as discussed in Chapter~\ref{chp:repcurr}, produces a momentum distribution that is non-symmetric. The integers $m$ correspond to the momenta $2\pi m/L$. Figure adapted from~\cite{osterloh2022exact}.}
    \label{fig:fluxmom}
\end{figure}
\noindent The momentum distribution in the absence of magnetic flux is presented for different SU($N$) in Figure~\ref{fig:nofkdiffsun}. For a fixed $N_{p}$ and increasing $N$, the momentum distribution is observed to be less broad and to be more peaked around $k=0$. This is to be expected since as $N\rightarrow\infty$, SU($N$) fermions emulate bosons in terms of level occupations~\cite{frahm1995,Pagano2014}. Conversely, for fixed SU($N$) and increasing number of particles, the momentum distribution reflects the fermionic statistics of the system, as it becomes broader due to the occupation of different momenta (see inset of Figure~\ref{fig:nofkdiffsun}). \\

\noindent Figure~\ref{fig:fluxmom} depicts the momentum distribution for an SU(2) symmetric system in the presence of an effective flux. In this case, the ground-state of the Hubbard model is characterized by level crossings to counteract the flux imparted to the system~\cite{yu1992persistent} (see Chapter~\ref{chp:repcurr}). Such level crossings correspond to different Heisenberg states, which can be obtained with the procedure introduced in Section~\ref{sec:modeland} by an appropriate change in spin quantum numbers. From the top row of Figure~\ref{fig:fluxmom}, it is clear that the effect of the magnetic flux manifests itself as a shift in the momentum distribution: the shift gets progressively larger with increasing flux. To capture how this happens precisely in the momentum distribution, we plot the symmetric and anti-symmetric components of the momentum distribution denoted as $n_{+}$ and $n_{-}$ respectively in the bottom panel of Figure~\ref{fig:fluxmom}.

\subsection{The Fermi gap for $U=\infty$}

In the thermodynamic limit at zero temperature $T=0$ and zero interactions $U=0$, the Fermi function drops from a finite value to zero at the Fermi momentum $k_{f}$. At finite interaction, states with $k>k_{f}$ can be occupied and, compared with the free fermion case, the gap at $k_{f}$ is reduced accordingly.  For SU($N$) symmetric particles, the maximum occupation of a single momentum level is $N$. 
For $N\rightarrow\infty$, the Fermi-distribution function should resemble a Bose-distribution. \\

\noindent Since the system under consideration deals with a finite number of particles and mesoscopic sized rings, it is far from the thermodynamic limit. In this regime, parity effects appear in $N_{p}/N$ for SU($N$) fermions. Therefore, we distinguish the two cases: odd occupations ($N_{p}/N = 2n+1$) and even occupations ($N_{p}/N=2n$). Defining the gap for odd occupations is straightforward: every single $k$-level up to the Fermi momentum is occupied at zero interactions. For example, in the case of two-component fermions with $N_{p}=6$,  all $k\in\{-1,0,1\}$ are fully occupied (see Chapter~\ref{chp:interfer}). Therefore, the Fermi gap $\Delta$ is defined as $f(k_f)-f(k_f+\Delta k)$, with $\Delta k=2\pi/L$, where $f(k)$ corresponds to the Fermi distribution function. The situation is different for even particles per species. In this case, the levels $|k_f|$ are only partially filled and this is visible even for $U=0$. For instance, two-component fermions with $N_{p}=4$, one can have either $k\in\{0,1\}$ or $k\in\{-1,0\}$. Therefore, the definition of the Fermi gap changes. In the case of an even number of particles per species, we  define the gap as
$f(k_f-\Delta k) - f(k_f+\Delta k)$. 
\begin{figure}[h!]
    \centering
    \includegraphics[width=0.9\linewidth]{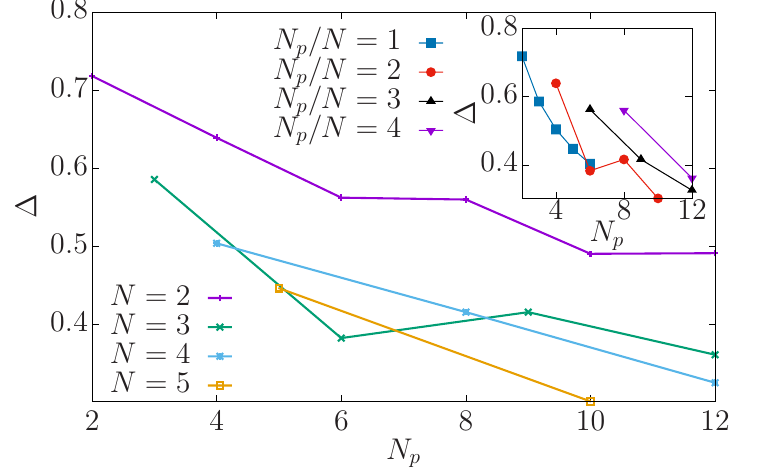}
\caption{Gap in the momentum distribution $\Delta$ as function of the number of particles $N_{p}$, in the limit of infinite repulsion. It is shown for different cases for dependencies of $N$ in SU($N$) (main panel) and for different values of $N_p/N$ from $1$ to $4$ (inset) with $L=27$. It can be seen that the gap decreases with growing $N$. The single exception is the case of SU(3) (see main panel). The particularities concerning the definition of the gap for finite number of particles are described in the main text.
}\label{fig:gap-at-27sites}
\end{figure}

\noindent In agreement with the above argument, we find that the gap is generically going down with $N$, but with a non-trivial dependence on $N_p$ and with parity effects for SU(2) and more pronounced for SU(3) --Figure~\ref{fig:gap-at-27sites}. The two particles per species in the SU($3$) symmetric case is clearly against the trend of decreasing gap with growing $N_p$, that is present for the other curves. Indeed, we point out on account of this behaviour of the SU(3) symmetric case, it would lead one to think that the case of $N_{p}/N=1$ for $N=6$, included only in the inset of the figure, seems to go against the general trend of the gap decreasing with increasing $N$. The non-monotonic behaviour displayed by the $N_{p}/N=2$ SU(3) symmetric case is quite peculiar, but as of yet we do not have a deeper understanding of this phenomenon. However, we note that expecting the gap to decrease for $N_p$ fixed with growing $N$,  would give a hint towards a parity effect of the number of components $N$, at least in the case $N_p=6$. In principle, Fermi gap need not follow a monotonic behavior. The expectation is that for each $N_p$ it has to eventually converge to zero as $N\to\infty$, since the Bose-distribution does not have one. Lastly, it is important to notice that in the systems considered in this paper, we never come below the ratio of $N_p/N=1$ because we fixed the occupation of each component to be equal.

\section{Interference dynamics in ultracold atoms}
In this section, we present a particular scenario in which  the exact one-body density matrix can be tested in the current state-of-the-art experimental observables in ultracold atom settings. Specifically, we consider the homodyne and self-heterodyne protocols explored in Chapter~\ref{chp:interfer}. For the sake of conveniency, we provide a brief recap of both protocols. 
\begin{figure}[h!]
    \centering
    \includegraphics[width=0.75\linewidth]{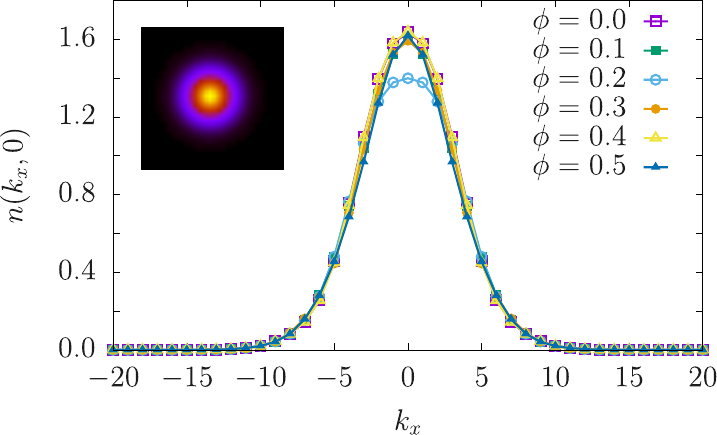}
    \caption{Cross-section of the momentum distribution $n^{(TOF)}(k_{x},0)$ in the limit of infinite repulsion for various values of the effective magnetic flux $\phi$. No depression is observed in the momentum distribution as the threshold imposed by the fermionic statistics and fractionalization has not been surpassed. Additionally, we note that the peak of the momentum distribution is non-monotonous as outlined in Chapter~\ref{chp:interfer}. The results are obtained using the exact one-particle density matrix for $L=15$. Figure adapted from~\cite{osterloh2022exact}.}
    \label{fig:hole10}
\end{figure}

\noindent The homodyne protocol consists in performing time-of-flight (TOF) imaging of the spatial density distribution of the atomic cloud: upon sudden release from its confinement potential, the atomic cloud expands freely, with the initially trapped atoms interfering with each other creating specific interference patterns. At long time expansions, the resulting inteference pattern corresponds to calculating the momentum distribution $n(\mathbf{k})^{(TOF)}$ defined as
\begin{equation}\label{eq:TOFmom}
    n^{\text(TOF)}_{\alpha}(\textbf{k}) = |w(\textbf{k})|^{2}\sum\limits_{j,l}^{L}e^{\imath \textbf{k}(\textbf{r}_{l}-\textbf{r}_{j})}\langle c_{l,\alpha}^{\dagger}c_{j,\alpha}\rangle ,
\end{equation}
where $w(\textbf{k})$ is the Fourier transform of the Wannier function, $\textbf{r}_{j}$ denotes the position of the lattice sites in the ring in the plane and $\textbf{k} = (k_x,k_y)$ are their corresponding Fourier momenta. \\

\noindent The self-heterodyne protocol follows the same procedure as the homodyne one, albeit with an additional condensate placed in the center of the system of interest, to act as a phase reference. Accordingly, as the center and the ring undergo free co-expansion in TOF, characteristic spirals emerge as the two systems interfere with each other. In order to observe the phase patterns in a second quantized setting, one needs to consider density-density correlators between the center and the ring
\begin{equation}\label{eq:TOFinter}
G_{R,C} = \sum\limits_{\alpha}\sum\limits_{j,l}I_{jl}(\textbf{r},\textbf{r}',t) \langle c_{l,\alpha}^{\dagger}c_{j,\alpha}\rangle ,
\end{equation}
where $I_{jl}(\textbf{r},\textbf{r}',t) = w_{c}(\textbf{r}',t) w_{c}^{*}(\textbf{r},t) w_{l}^{*}(\textbf{r}'-\textbf{r}_{l}',t) w_{j}(\textbf{r}-\textbf{r}_{j},t)$ the Wannier functions of the interfering terms. \\

\begin{figure}[h!]
    \centering
    \includegraphics[width=\linewidth]{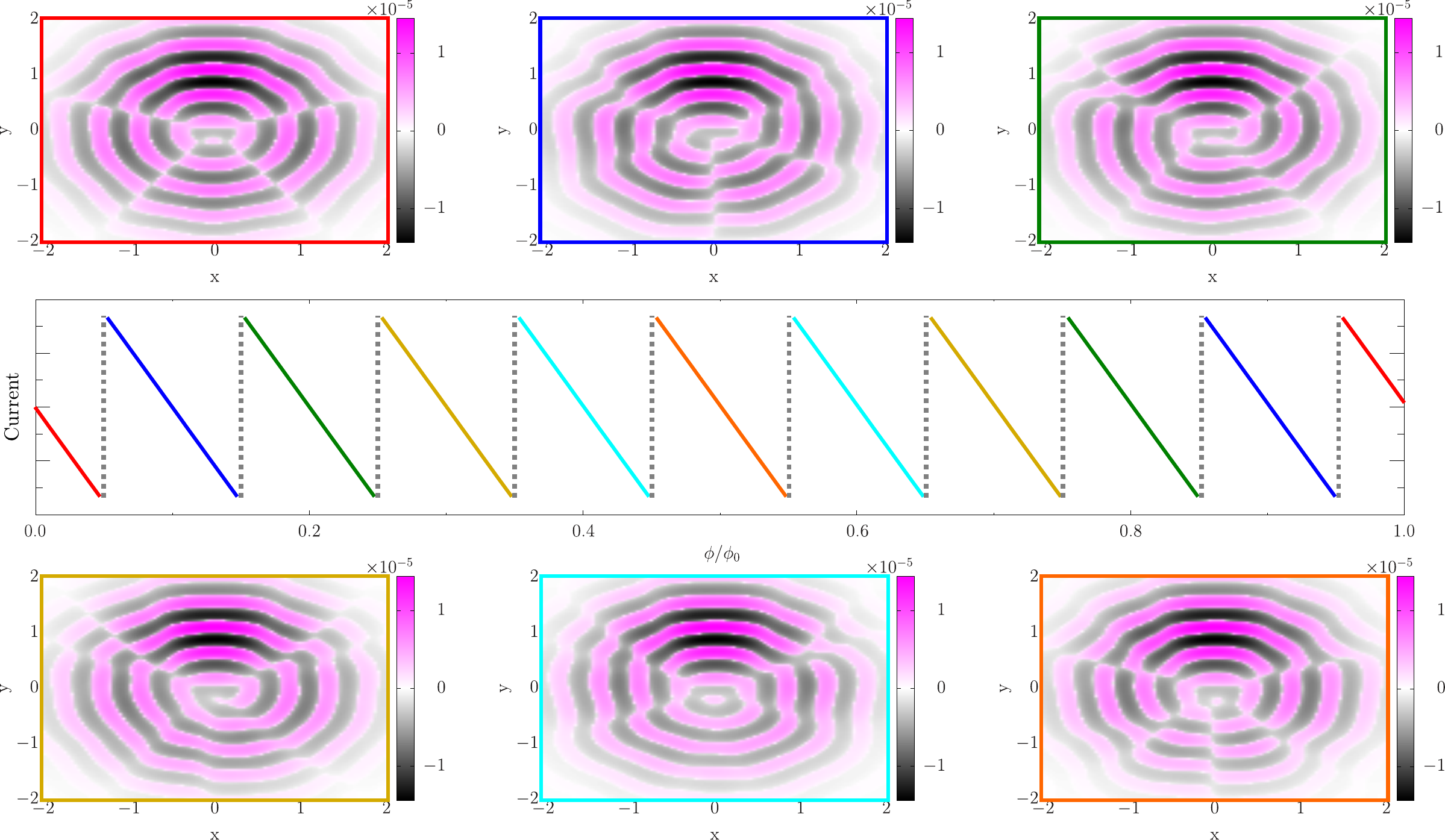}
    \caption{Interference $G_{\mathrm{R,C}}$ between ring and center for $N_{p}=10$ particles with SU(2) symmetry against the effective magnetic flux $\phi$ at short time $t=0.022$ in the limit of infinite repulsion. Middle panel is the schematic for the fractionalized persistent current for the corresponding system. The interference pattern corresponding to the first parabola (red), displays four dislocations as in the zero interaction case. On going to the next parabolas, that correspond to jumps in the persistent current, the number and orientation of the dislocations changes reflecting the angular momentum fractionalization. There is no spiral in the above interference patterns since no hole has opened up (see Figure~\ref{fig:hole10}). All correlators are evaluated using the exact one-particle density matrix for $L=15$ by setting $\textbf{r}' = (0,R)$ and radius $R=1$. The color bar is non-linear by setting $\mathrm{sgn}(G_{\mathrm{R,C}})\sqrt{ |G_{\mathrm{R,C}}|}$. Figure reprinted from~\cite{osterloh2022exact}.}
    \label{fig:spiral10}
\end{figure}

\noindent By exploiting the correlation matrix calculated in the previous sections, we can evaluate the exact interference images obtained through these two expansion protocols for two-component fermions --Figures~\ref{fig:hole10} and~\ref{fig:spiral10}. The characteristic traits of the interferograms as discussed in Chapter~\ref{chp:interfer} are readily observed. In particular, we highlight that the interference patterns in both figures correspond to $N_{p}=10$ SU(2) symmetric fermions, which is a system that is beyond the reach of both exact diagonalization and DMRG. 

\section{Conclusions and Outlook}

In this chapter, we develop a theoretical framework to calculate the exact one-particle density matrix of $N$-component fermions in the limit of strong repulsion using Bethe ansatz analysis of the SU($N$) Hubbard model.  By splitting the problem into the spinless fermionic and SU($N$) Heisenberg models, we manage to compute these observables for number of particles, ring sizes and number of components well beyond the current state-of-the-art tractable by numerical methods: On one hand, the number of particles and system size are well beyond exact diagonalization schemes; on the other hand, we remark that through Bethe ansatz the limit of infinite repulsion, a notoriously challenging limit for DMRG, could accessed. From the technical side, we note that our Bethe ansatz scheme agrees well with the numerics of the lattice model (for system parameters that could be accessed). Specifically, we are able to calculate the correlations of systems composed of 38 sites and 12 particles in the spinless configuration. Depending on $N$, this would correspond to a larger Hilbert space in the Hubbard model, such as $7.62\times 10^{12}$ for $N=2$. Exact diagonalization/Lanczos can only handle around 7 million. Therefore, there is no direct comparison between the two methods possible in this respect. Furthermore, we highlight that our system is far from being in the dilute regime of Equation~\eqref{eq:Ham2}. In spite of this, there is an excellent agreement between exact diagonalization and our scheme that is intrisically reliant on the system being Bethe ansatz integrable. Bethe ansatz integrability hinges on the fact that the scattering of more than two particles does not occur (Yang-Baxter factorization of the scattering matrix). In the infinite repulsive regime, the multiparticle scattering is suppressed since the probability of two particles interacting is vanishing. Therefore, despite the fact that we are not in the dilute limit condition, the system is indeed very close to be integrable and our method is able to accurately tackle the infinite repulsive limit of the SU($N$) Hubbard model. \\

\noindent The Fourier transform of the correlation matrix is the momentum distribution of the system. Despite being one of the simplest interesting correlation functions, the momentum distribution reflects the many-body character of the quantum state. In particular, we quantify exactly the dependence of the gap at the Fermi point on the different number of particles and components, confirming the general expectation that for a large number of fermionic components the Pauli exclusion principle relaxes. However, we find that the suppression of the gap for finite systems is non-monotonous. \\

\noindent Furthermore, we applied this scheme to the case in which SU($N$) matter can flow in ring-shaped potentials pierced by an effective magnetic flux. An additional complication in the calculation arises since the matter-wave states obey a complex dependence on the flux, ultimately leading to persistent currents with fractional quantization as discussed in Chapter~\ref{chp:repcurr}. In particular, we read-out such a phenomenon in terms of spin-states of the SU($N$) Heisenberg model. In this context, we provide an example where the developed theory allows us to calculate the readily available experimental observables homodyne~\cite{wright2022persistent} and self-heterodyne~\cite{roati2022imprinting} time-of-flight measurements introduced in Chapter~\ref{chp:interfer}. 

\newpage
\noindent Our exact results can be exploited to benchmark observables related  to the one-body density matrix of SU($N$) fermions in the strongly interacting limiting. Lastly, the developed theoretical framework opens the possibility to study more complicated correlation functions.


\chapter{Conclusions and Outlook}\label{chp:conc}

\noindent Atomtronics is the quantum technology of guided ultracold atoms~\cite{amico2021,amico2022}. After an intense activity devoted to atomtronic circuits operating with bosonic fluids, recent experiments have set the scene to explore the potential of atomtronic circuits comprised of matter-waves of fermionic nature~\cite{wright2022persistent,roati2022imprinting}. In this thesis, we considered fermionic atomtronic circuits 
made by a ring-shaped quantum gas of SU($N$) symmetric fermions threaded by an effective magnetic flux, starting a persistent current. 
Such matter-wave current is found to display specific quantization properties depending on the attributes of the $N$-component quantum fluid. \\

\noindent In Chapter~\ref{chp:repcurr}, we studied a matter-wave current of SU($N$) fermions subject to repulsive interactions for different filling fractions~\cite{chetcuti2021persistent}. At incommensurate fillings, we find that the angular momentum is quantized to fractional values fixed by the number of particles $N_{p}$. This is reflected in the persistent current, which has its period reduced by $1/N_{p}$. Although this phenomenon shares similarities with the one in attractive bosons, their sources are markedly different. While in \textit{attracting bosons}, the fractionalization arises from the formation of an $N_{p}$-body bound state, in \textit{repulsive fermions}, it originates due to the coupling between the spin and matter degrees of freedom mimicking an `attraction from repulsion'. On going to commensurate filling fractions, the angular momentum retains its integer quantization due to the opening of the Mott gap as the system transitions from the superfluid to the Mott phase. This is reflected in the persistent current by the smoothening out of its characteristic sawtooth shape. By performing finite-size scaling analysis, we demonstrate that the persistent current, despite its mesoscopic nature, can detect the onset to the Mott phase transition. \\

\noindent Next, our attention shifted to attractive SU($N$) fermionic matter in Chapter~\ref{chp:probe} to study the formation of complex bound states whose nature goes beyond the standard two-component fermion pairing~\cite{chetcuti2021probe}. In particular, we focus on SU(3) fermions that are able to form two- and three-body bound states, called colour superfluids (CSFs )and trions, respectively. These are the cold atoms analogues of mesons and hadrons in quantum chromodynamics (QCD). We find that for attractive interactions, the angular momentum displays a fractional quantization that is fixed by the number of SU($N$) components instead of $N_{p}$ as found in repulsive matter.
The persistent current exhibits a reduced periodicity by $1/r$, indicating the formation of an $r$-body bound state, irrespective of the number of particles in the system. Accordingly, we demonstrated that the persistent current
can distinguish between CSFs and trions. In addition, analysis of finite temperature effects reveals how thermal fluctuations can lead to the deconfinement of bound states: for weak attractions, we observe a crossover from a three-body bound state to free particles, marked by the reinstatement of the single-particle frequency of the persistent current. Such a crossover shares similarities with the Quark-Gluon plasma formation at large temperatures and small baryonic densities in QCD. We note that, to our knowledge, this study provides the first instance of a persistent current-based simulator of specific problems in QCD.\\

\noindent Inspired by the experimental know-how in the field, in Chapter~\ref{chp:interfer}, we analysed the persistent current of SU($N$) fermions through interference dynamics generated by homodyne and self-heterodyne protocols~\cite{chetcuti_interfer}. The presence of circulating current states with a given angular momentum is surveyed through the appearance of a characteristic hole/spiral in the interferograms.  In both cases, we find that the interference patterns display a distinctive dependence on the structure of the Fermi distribution and the particles' correlations. Accordingly, this enables us to monitor the fractional values of the angular momentum observed in the repulsive and attractive regimes discussed in Chapters~\ref{chp:repcurr} and~\ref{chp:probe}. Furthermore, our study demonstrates how the analysis of these interferograms provides information on both the number of particles and components, two quantities that are notoriously hard to extract experimentally. \\

\noindent Lastly, in Chapter~\ref{chp:ogata}, we worked out an exact Bethe ansatz scheme for the computation of the one-particle density matrix of SU($N$) matter in the limit of strong repulsive interactions, which is rather challenging to obtain numerically despite its simple expression~\cite{osterloh2022exact}. Through our developed framework, we are able to compute the momentum distribution, as well as the observables that are of direct interest to the expansion  protocols discussed in Chapter~\ref{chp:interfer}, for a larger number of particles and system sizes that is tractable with the current numerical infrastructure.  We expect that our exact results can be exploited to benchmark observables related  to the one-body density matrix of SU($N$) fermions in the strongly interacting limit. Finally, the theoretical framework we developed  opens the possibility of studying more complicated correlation functions.\\

\noindent In conclusion, this thesis has laid the foundation for the blueprints of matter-wave circuits based on SU($N$) fermionic platforms. Through our results, we provide support to the notion that persistent currents can be used as diagnostic tools to probe interacting quantum many-particle systems. Additionally, the specific quantization properties displayed by the persistent current in both the repulsive and attractive regimes are expected to lead to enhanced performances in rotation sensing, as was recently predicted for attracting bosons~\cite{naldesi2020enhancing,polo2021quantum}. In this direction, SU($N$) fermionic matter could realize SU($N$) atomic SQUIDs. An exciting direction opened up by the work carried out in this thesis is to explore atomtronics-enabled quantum simulators for high energy physics. Various scenarios can be investigated in this context, ranging from QCD to lattice gauge theories and quantum gravity. Another interesting avenue that can be considered moving forward is that of quantum transport in fermionic systems. Although extensively studied in solid state platforms, fermionic transport could be revisited with a new twist in the field of atomtronics owing to the high degree of control and versatility of cold atoms.

\begin{appendices}

\chapter{Mapping the SU(\textit{N}) Hubbard model to the Gaudin-Yang Sutherland model}\label{sec:mappinggys}

\noindent In this Appendix, we present the derivation of the Gaudin-Yang-Sutherland model as the continuous limit of vanishing lattice spacing of the SU($N$) Hubbard model. \\

\noindent The density of fermions in the lattice denoted by $D$ can be expressed as $D = N_{p}/(L\Delta )$ with $N_{p}$ being the number of particles, $L$ denoting the number of sites and $\Delta$ corresponding to the lattice spacing. The filling factor $\nu$ is related to the lattice spacing in the following manner
\begin{equation}\label{eq:hh1}
\nu = \frac{N_{p}}{L} = \frac{N_{p}}{\Delta L} \Delta .
\end{equation}
Therefore, in the continuous limit of vanishing lattice spacing $\Delta\rightarrow 0$ and finite particle density $\frac{N_{p}}{\Delta L}$, the filling fraction must be  accordingly small. 
For the anti-commutation relations to hold in the continuous limit, the fermionic operators have to be rescaled such that
\begin{equation}\label{eq:hh2}
c_{j,\alpha}^{\dagger} = \sqrt{\Delta}\Psi^{\dagger}_{\alpha}(x_{j}) \hspace{20mm} n_{j,\alpha} = \Delta\Psi^{\dagger}_{\alpha}(x_{j})\Psi_{\alpha}(x_{j})\hspace{10mm}\mathrm{with}\hspace{2mm}x_{j} = j\Delta
\end{equation}
where $\Psi^{\dagger}$ is the creation field operator for a fermion with colour $\alpha$, and obeys the standard anti-commutation relations $\{\Psi_{\alpha}(x),\Psi^{\dagger}_{\beta}(y)\} = \delta_{\alpha,\beta}\delta (x-y)$ and $\{\Psi^{\dagger}_{\alpha}(x),\Psi^{\dagger}_{\beta}(y)\} = 0$. \\

\noindent Utilizing Equation~\eqref{eq:hh2}, 
the SU($N$) Hubbard model in Equation~\eqref{eq:fhmsun}
is mapped onto the Fermi gas quantum field theory in the following way
\begin{equation}\label{eq:hh3}
\mathcal{H}_{\mathrm{SU}(N)} = t\Delta^{2}\mathcal{H}_{FG}-2N_{p},   
\end{equation}
where the Fermi gas quantum field Hamiltonian reading
\begin{equation}\mathcal{H}_{FG} = \int \bigg[(\partial_{x}\Psi^{\dagger}_{\alpha})(\partial_{x}\Psi_{\alpha}) + c\sum\limits_{\alpha <\beta}^{N}\Psi^{\dagger}_{\alpha}\Psi^{\dagger}_{\beta}\Psi_{\beta}\Psi_{\alpha}\bigg],\end{equation}
with $\alpha$ and $\beta$ denoting different SU(\textit{N}) colours, $N_{p}$ corresponds to the number of particles  and having an interaction strength $c = \frac{U}{t\Delta}$. The Fermi gas field theory is the quantum field theory of the Gaudin-Yang-Sutherland model. This can be demonstrated through eigenstates of $\mathcal{H}_{FG}$, which can be written as 
\begin{equation}
    \ket{\psi (\lambda )} = \sum\limits_{\alpha_{1} \hdots\alpha_{N_{p}}}^{N} \int \chi (\textbf{x}|\lambda )\Psi^{\dagger}_{\alpha_{1}} (x_{1})\hdots\Psi^{\dagger}_{\alpha_{N_{p}} }(x_{N_{p}})\ket{0}\textrm{d}\textbf{x}.
\end{equation}
One can prove that $\chi (\textbf{x}|\lambda )$ are eigenfunctions of the Gaudin-Yang-Sutherland Hamiltonian~\cite{sutherland1968,decamp2016high}
\begin{equation}\label{eq:hh4}
  \mathcal{H}_{GYS} = -\sum\limits_{\alpha=1}^{N}\sum\limits_{i=1}^{N_{\alpha}} \frac{\partial^{2}}{\partial x_{i,\alpha}^{2}} + c\sum\limits_{i<j}\sum\limits_{\alpha,\beta}\delta (x_{i,\alpha} - x_{j,\beta}),
\end{equation}
where $N_{\alpha}$ is the number of particles with colour $\alpha$. The Gaudin-Yang-Sutherland model is Bethe ansatz integrable for all $N$ unlike its lattice counterpart, whose integrability is preserved only for $N=2$~\cite{frahm1995}.  In the case of $N\!>\!2$, the lattice regularization of model~\eqref{eq:hh4} spoils the integrability of the model for the same reasons as the Bose-Hubbard case~\cite{amico_korepin}.  Nonetheless, the mapping that was sketched out enables us to compare our numerical results obtained for the SU($N$) Hubbard model having large system sizes and small filling fractions, with the exact solution provided by the Bethe ansatz of the Gaudin-Yang-Sutherland model. 

\chapter{SU(\textit{N}) Heisenberg model }\label{app:Heisenberg}

\noindent At integer filling fractions of one particle per site and large repulsive interactions, the SU($N$) Hubbard model reduces to an effective spin model, which through second order perturbation theory is found to correspond to the SU($N$) Heisenberg model~\cite{manmana2011n,cazalilla_2014,capponi}. Indeed, the low-lying excited states of the original fermionic model~\eqref{eq:fhmsun} result to be captured by the spin model. The SU(2) Heisenberg model is a sum of permutation operators
\begin{equation}\label{eq:Heisa}
    \mathcal{H}_{XXX}=\sum_{i} P_{i,i+1}=\sum_i (\mathds{1} + \vec{\sigma}_{i+1}\cdot \vec{\sigma}_i)/2,
\end{equation}
with $\vec{\sigma}_i$ corresponding to the Pauli matrices, the three generators of the SU$(2)$ Lie algebra. In the case of the SU($N$) Heisenberg model, the Hamiltonian can be constructed in a similar fashion~\cite{sutherland1975,capponi}. In general we obtain for the generators $\lambda_i$ of the SU($N$) 
\begin{equation}\label{eq:sunHeisa}
    P_{i,i+1}=\frac{1}{N}\mathds{1} + \frac{1}{2}\vec{\lambda}_{i} \cdot \vec{\lambda}_{i+1},
\end{equation}
which acts on sites $i$ and $i+1$ permuting the SU($N$) states. 

\section{Details about the SU(\textit{N}) Generators}\label{Generators}

\noindent The generators of the Lie algebra SU($N$) group are analogues of the Pauli matrices in SU(2). Taking SU(3) as an example, we have six non-diagonal generators
\begin{align}\label{SU2-subalgebra}
&\lambda_1 = \begin{pmatrix}0 & 1 & 0 \\
                         1 & 0 & 0\\
                         0 & 0 & 0\end{pmatrix}\;  
\lambda_2 =  \begin{pmatrix}0 & -i & 0 \\
                         i & 0 & 0\\
                         0 & 0 & 0\end{pmatrix}\; 
\lambda_3 = \begin{pmatrix}0 & 0 & 1 \\
                         0 & 0 & 0\\
                         1 & 0 & 0\end{pmatrix}\; \nonumber \\  
&\lambda_4 =  \begin{pmatrix}0  & 0 & -i \\
                         0 & 0 & 0\\
                         i & 0 & 0\end{pmatrix}\;
\lambda_5 = \begin{pmatrix}0 & 0 & 0 \\
                         0 & 0 & 1\\
                         0 & 1 & 0\end{pmatrix}\;
\lambda_6 = \begin{pmatrix}0 & 0 & 0 \\
                         0 & 0 & -i\\
                         0 & i & 0\end{pmatrix},
\end{align}
that together with two diagonal generators
\begin{equation}\label{additional-generators}
\lambda_7 = \begin{pmatrix}1 & 0 & 0 \\
                         0 & -1 & 0\\
                         0 & 0 & 0\end{pmatrix}\;  
\lambda_8 =  \frac{1}{\sqrt{3}}\begin{pmatrix}1 & 0 & 0 \\
                         0 & 1 & 0\\
                         0 & 0 & -2\end{pmatrix},
\end{equation}
comprise the Gell-Mann matrices that are the matrix representation of the SU(3) Lie algebra group. For generalization purposes, the generators were grouped by defining $\lambda_{2p-1/2p}$, $p=1,\dots ,\frac{N(N - 1)}{2}$ which are analogues to the $\sigma_{x/y}$ that operate between the different subspaces of SU(3) which are $(i,j)$, $i<j$. Here, both run from 1 to $3$. We decided to group the elements of the diagonal Cartan basis at the end as $\lambda_7$ and $\lambda_8$, which differs from the standard Gell-Mann matrices, but is eases the generalisation. For the extension to SU($N$), one has to consider the $N(N\! -\! 1)/2$ elements $\lambda_i$, which would correspond to $\sigma_{x/y}$ in some space $(i,j)$, where $i<j\in\{1,\dots ,N\}$. Additionally, the corresponding diagonal Cartan elements need to be taken into account. There are $N\! -\! 1$ Cartan elements that can be constructed via the following formula $\lambda_{N^{2}-(N+1)+m} = \mathrm{diag}\{1,\dots,1, -(m-1),0,\dots ,0\}/\sqrt{m(m-1)/2}$ where $m = 2,\cdots, N$; the $1/0$ occurs $(m-1)/(N-m)$ times, respectively.

\section{Casimirs of SU(\textit{N}) fermions}
\label{Casimir}

\noindent Whereas in SU(2) we have a single Casimir operator, for SU($N$) groups, we are faced with $N\! -\! 1$ Casimirs. Out of these Casimirs, we are only interested in the quadratic Casimir 
\begin{equation}\label{eq:casy}
    C_{1} = \frac{1}{4}\sum\limits_{i=1}^{N^{2}-1}\lambda_{i}^{2},
\end{equation}
as it relates to the total spin quantum number $S^{2}$, which is necessary for us to classify the Heisenberg eigenstates. An issue that presents itself is to calculate the Casimir in different SU($N$) representations. In the following, we sketch out the procedure to write the quadratic Casimir operator for SU(3) and SU(4) in the SU(2) representations. \\

\noindent We start by looking at the SU(3) case, where its representations $\Lambda(n_1,n_2)$ are labeled by integer numbers which correspond to the simple Cartan elements $(h_1,h_2)$: $\Lambda(n_1,n_2)=\vec{n}\cdot\vec{h}$. The elements are given by
\begin{equation}\label{Cartan-SU3a}
h_1=(\lambda_{3},\lambda_{8})\cdot (1,0)^T \implies h_1=:(\sigma_z)_{1,2}, 
\end{equation}
\begin{equation}\label{Cartan-SU3b}
h_2 = (\lambda_{3},\lambda_{8})\cdot \bigg(-\frac{1}{2},\frac{\sqrt{3}}{2}\bigg)^T \implies h_2=:(\sigma_z)_{2,3}\;  .
\end{equation}

\noindent To calculate the quadratic Casimir values for these representations $\Lambda$, we need the Cartan matrix
\begin{equation}\label{eq:carty}
C_h=2\left(\frac{\vec{h}_i\cdots \vec{h}_j}{|\!|h_i|\!|^2}\right)_{ij}=\begin{pmatrix}2 & -1 \\
-1 & 2\end{pmatrix},
\end{equation}
defined by the Killing form $(\lambda_j,\lambda_k):=K(\lambda_j,\lambda_k)=\frac{1}{8}\tr \lambda_j\lambda_k =\frac{1}{4} \delta_{jk}$ (see~\cite{CornwellGroups}, chapter 12 for the evaluation of the Casimir). Evaluating the quadratic Casimir, we get that 
\begin{equation}\label{eq:labbs}
3 C_1 = (\Lambda,\Lambda+\delta)=\left[\vec{n}C^{-1}_h+\vec{\delta}\right]\vec{n}^T\\
=\sum_{i=1}^2 n_i(n_i+3) + n_1 n_2 ,
\end{equation}
giving the value of $4/3$ for the fundamental representations $(1,0)$ and $(0,1)$. Here, $\vec{\delta}=\frac{1}{2}\sum_{h\in\Delta_+} h =(2,2)^T$ ( see~\cite{CornwellGroups,WPfeifer-LieAlgs,GeorgiHEPGroups} for the positive roots $\Delta_+$. These are the two simple roots together with their sum, $h_1+h_2$. If one introduces half-integer values as in the SU(2) representation for each $n_i$ such that ($n_i = 2J_{i}$), we obtain
\begin{equation}\label{eq:casysu3}
C_1(\Lambda)=\frac{4}{3} \bigg[\sum_i J_i\bigg(J_i+\frac{3}{2}\bigg) + J_1J_2\bigg] . 
\end{equation} 
\noindent Likewise for SU(4), the representations $\Lambda(n_1,n_2,n_3)$ of SU(4) are labeled by the Cartan elements $(h_1,h_2,h_3)$: $\Lambda(n_1,n_2,n_3)=\vec{n}\cdot\vec{h}$, which are given by
\begin{equation}\label{Cartan-SU4a}
h_1 =(\lambda_{13},\lambda_{14},\lambda_{15})\cdot (1,0,0)^T \implies \vec{h}_1=:(\sigma_z)_{1,2} ,
\end{equation}
\begin{equation}\label{Cartan-SU4b}
h_2=(\lambda_{13},\lambda_{14},\lambda_{15})\cdot \bigg(-\frac{1}{2},\frac{\sqrt{3}}{2},0\bigg)^T \implies\vec{h}_2=:(\sigma_z)_{2,3},
\end{equation}
\begin{equation}\label{Cartan-SU4c}
h_3 =(\lambda_{13},\lambda_{14},\lambda_{15})\cdot \bigg(0,-\frac{1}{\sqrt{3}},\sqrt{\frac{2}{3}}\bigg)^T \implies\vec{h}_3=:(\sigma_z)_{3,4} .
\end{equation}
The corresponding Cartan matrix reads
\begin{equation}
C_h=\begin{pmatrix}2 & -1 & 0\\
-1 & 2 & -1\\
0 & -1 & 2\end{pmatrix}.
\end{equation}
Upon evaluating the quadratic Casimir as in Equation~\eqref{eq:casysu3}, we have that
\begin{equation}\label{eq:casysu4}
   2 C_1(\Lambda)=(n_1+2n_2+n_3)^2+n_1\left(2n_1+\frac{3}{4}\right) + n_3\left(2n_3+\frac{3}{4}\right) + n_2 . 
\end{equation}
with $\vec{\delta}=\frac{1}{2}\sum_{h\in\Delta_+} h =(3,4,3)^T$. 
These are the three simple roots together with $h_1+h_2$, $h_2+h_3$, and $h_1+h_2+h_3$. Introducing half-integer values as for SU(2), we obtain 
\begin{equation}
C_1(\Lambda)=2J_2\bigg(J_2+\frac{1}{4}\bigg) + \sum_{i=1,3} 3J_i\bigg(2 J_i+\frac{1}{4}\bigg)  + 4\sum_{i<j}J_iJ_j ,
\end{equation}
leading to the value of $15/8$ for the fundamental representations $(1,0,0)$ and $(0,0,1)$, and $5/4$ for the representation  $(0,1,0)$.


\chapter{Bethe ansatz of the Gaudin-Yang Hamiltonian}\label{sec:betheb}

\noindent In this Appendix, we focus on the Bethe ansatz of the Gaudin-Yang model, which is the continuum limit of Equation~\eqref{eq:fhmsun} with $N=2$. Then, we present the Lieb-Wu equations for a system consisting of $N_{p}$ particles with $M$ flipped spins. \\

\noindent Consider $N_{p}$ two-component fermions with mass $m$ residing on a one-dimensional ring of size $L_{R}$ that interact via a delta interaction of strength $c$\footnote{Here, we take $c>0$. The model is still integrable for $c<0$, but one needs to do additional steps by introducing the string solutions, which will be covered in detail in Chapter~\ref{chp:probe}}. The model describing such a system, called the Gaudin-Yang Hamiltonian, is 
\begin{equation}\label{eq:gaudinsu2}
    \mathcal{H}_{GY} = -\frac{\hbar^{2}}{2m}\sum\limits_{j=1}^{N_{p}}\frac{\partial^{2}}{\partial x_{j}^{2}} + 2c\sum\limits_{i<j}\delta (x_{i}-x_{j}),
\end{equation}
with $x_{j}$ denoting the coordinate of the $j$-th particle. In what follows, the energy scale is given by $\hbar=m=1$. The corresponding stationary Schrödinger equation reads
\begin{equation}\label{eq:schrodinger}
     -\frac{\hbar^{2}}{2m}\sum\limits_{j=1}^{N_{p}}\frac{\partial^{2}\Psi}{\partial x_{j}^{2}} + 2c\sum\limits_{i<j}\delta (x_{i}-x_{j})\Psi = E\Psi,
\end{equation}
for an arbitrary many-body wavefunction $\psi (x_{1},\hdots,x_{N_{p}})$. \textit{Bethe's hypothesis} supposes that in a given sector $Q$ defined by $0<x_{Q1}<x_{Q2}<\hdots< x_{QN_{p}}<L_{R}$, 
the solution to Equation~\eqref{eq:schrodinger} is a combination of plane waves such that
\begin{equation}\label{eq:wavefuun}
    \Psi (x_{1},\hdots,x_{N_{p}}) = \sum\limits_{P\in S_{N_{p}}} \mathrm{sign}(P)\mathrm{sign}(Q)\psi [Q;P] \exp \bigg( \imath \sum\limits_{j=1}k_{Pj}x_{Qj}\bigg),
\end{equation}
where $\psi [Q;P]$ being an $N_{p}!\times N_{p}!$ matrix housing all the amplitude coefficients, with $P$ and $Q$ being permutations of the distinct momenta $k_{j}$ and the position of the particles $x_{j}$ respectively. The summation over $P$ covers all the permutations of the numbers $1,\hdots,N_{p}$, forming the symmetric group $S_{N_{p}}$. The permutation $Q$ is introduced to take into account the identity of the particles, seeing as they are distinguishable\footnote{The order of a particle $x_{j}$ in a sequence is dictated by the value that $Qj$ adopts. Here, we choose the sector of ascending order.}. The quantities $\mathrm{sign}(P)$ and $\mathrm{sign}(Q)$ are the +1 ($-1$) for permutations with an even (odd) number of exchanges respectively.  \\

\noindent In order for Equation~\eqref{eq:schrodinger} to be satisfied, we require that at the boundary $X$ between two different sectors of $Q$, the wavefunction in both regions has to ``match'' such that 
\begin{equation}\label{eq:boundary}
    \psi|_{X=0^{+}} = \psi|_{X=0^{-}},
\end{equation}
where $X = x_{Q(j+1)}-x_{Qj}$ for $j\in\{1,\hdots , N_{p}-1\}$. Additionally, because of the delta interaction, its first derivative 
is discontinuous
\begin{equation}\label{eq:disconfer}
    \frac{\partial\psi}{\partial X}\bigg|_{X = 0^{+}} - \frac{\partial\psi}{\partial X}\bigg|_{X = 0^{-}} = 2c \psi|_{X=0}.
\end{equation}
\begin{figure}[h!]
    \centering
    \includegraphics[width=0.8\linewidth]{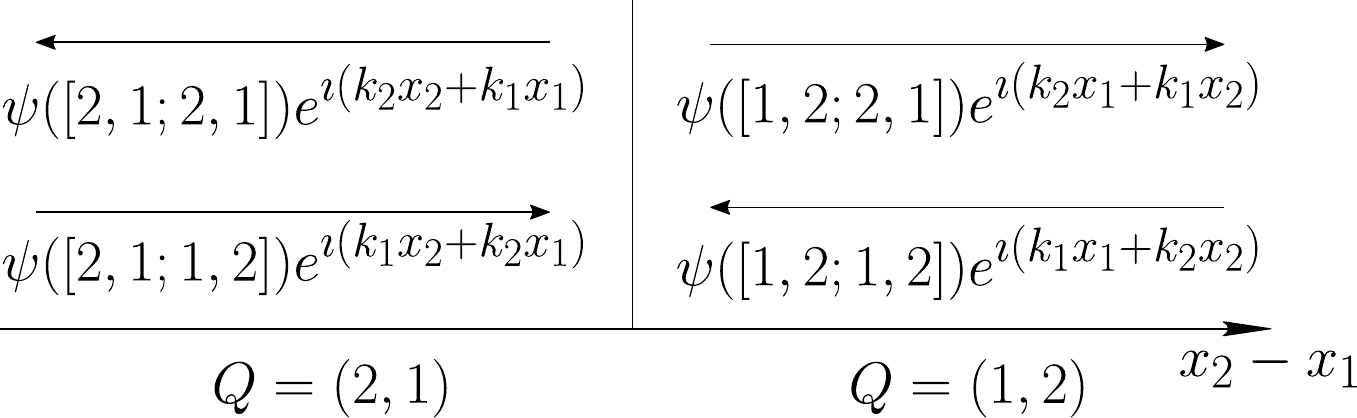}
    \caption{Pictorial representation of the scattering process of two particles. The figure is split into two sectors $Q=(12)$ for $x_{2}>x_{1}$ and $Q = (21)$ for $x_{1}<x_{2}$. Both the incoming and outgoing waves are shown and are in line with Equation~\eqref{eq:scattering}. }
    \label{fig:scatter}
\end{figure}

\noindent Applying the continuity boundary condition (Equation~\eqref{eq:boundary})on our wavefunction yields 
\begin{equation}\label{eq:conticont}
    \psi [Q;P] - \psi [Q;P'] = \psi [Q';P'] - \psi [Q';P],
\end{equation}
where $P' = P(j,j+1)$ and $Q' = Q(j,j+1)$ are permutations that interchange the elements $j$ and $j+1$\footnote{To put this into perspective, for the two-body problem the continuity equation would read as $[1,2;1,2] - [1,2;2,1] = [2,1;2,1] - [2,1;1,2]$. In this manner, the boundary condition becomes very intuitive as upon exchanging two fermions, be it their quasi-momenta or their position, a change in sign is acquired due to their anti-symmetry.}.
For the discontinuity condition defined in Equation~\eqref{eq:disconfer}, we are left with
\begin{equation}\label{eq:ferfer}
    \imath (k_{j+1} - k_{j})(\psi [Q;P] -\psi [Q';P] + \psi [Q;P'] -\psi [Q';P']) = 2c(\psi [Q;P] -\psi [Q;P']).
\end{equation}
Substituting the relation from the continuity equation into the latter, we obtain  
\begin{equation}\label{eq:scattering}
    \psi [Q;P'] = -\frac{\imath c}{k_{j+1}-k_{j}-\imath c}\psi [Q;P] + \frac{k_{j+1}-k_{j}}{k_{j+1}-k_{j}-\imath c}\psi [Q';P],
\end{equation}
which can be interpreted in terms of the scattering between two bodies. When two distinguishable particles scatter such as the spin-$\frac{1}{2}$ fermions with opposite spins that are under consideration, the wavefunction will either be transmitted or reflected. Consequently, this leads us to define a scattering matrix $S$,
\begin{equation}\label{eq:scatmat}
    S_{ji}^{ab} = R_{ji}\mathds{1} + T_{ji}\hat{P}^{ab},
\end{equation}
where $\hat{P}^{ab}$ is the permutation operator that exchanges two particles $a$ and $b$. Here, we have opted to define the scattering matrix in the reflection representation as the reflection coefficients lie on the diagonal. The reflection and transmission coefficients, denoted by $R_{j+1,j}$ and $T_{j+1,j}$ respectively, are expressed as
\begin{equation}\label{eq:reflectran}
    R_{j+1,j} = -\frac{\imath c}{k_{j+1}-k_{j}-\imath c} \hspace{5mm} T_{j+1,j} = \frac{k_{j+1}-k_{j}}{k_{j+1}-k_{j}-\imath c}.
\end{equation}
Note that in the case of reflection, the particles only exchange momenta such that $Q$ remains unchanged, whilst in transmission both $P$ and $Q$ change (see Figure~\ref{fig:scatter}). Naturally, $R_{j+1,j}+T_{j+1,j}=1$ since the scattering process is purely elastic because of integrability. \\

\noindent For a given $N_{p}$ in a fixed configuration $P$ with permuting $Q$, the scattering process can be written in the following form.
\begin{equation}\label{eq:scatmata}
    \begin{pmatrix}
        \psi [Q^{1};P]\\
        \psi [Q^{2};P]\\
        \vdots \\
        \psi [Q^{N_{p}!};P]
    \end{pmatrix} =  (R_{ji} + T_{ji}\hat{P}^{ab})   \begin{pmatrix}
        \psi [Q^{1};P']\\
        \psi [Q^{2};P']\\
        \vdots \\
        \psi [Q^{N_{p}!};P']
    \end{pmatrix},
\end{equation}
where the scattering operator having a dimensionality of $N_{p}!\times N_{p}!$ acts on the amplitude vector of size $N_{p}!$. The permutation operator $\hat{P}^{ab}$ is an $N_{p}!\times N_{p}!$ matrix relating the two fixed configurations of $P = (1,\hdots,a,b,\cdots,N_{p})$ and $P'=(1,\hdots,b,a,\cdots,N_{p})$; whilst $Q^{i}$ corresponds to the permutations of the particles' identities, with $i$ running over all the possible configurations. In total, there are $(N_{p}-1)(N_{p}!)$ equations having the same form as Equation~\eqref{eq:scatmata}, but there are $(N_{p}!)^{2}$ amplitude coefficients\footnote{The factor $N_{p}-1$ comes out from having $N_{p}-1$ equations of the same form as Equation~\eqref{eq:scattering} on account of the boundary condition defined in Equation~\eqref{eq:ferfer}.}. As such, for these equations to be mutually consistent for any set of distinct quasi-momenta $k_{j}$, we require that this set of equations leads to a unique set of coefficients. This amounts to saying that the scattering process should be irrespective of the two-body scattering sequence that we adopt. To verify that this is the case, it is enough to consider the three particle problem. \\

\begin{figure}[h!]
    \centering
    \includegraphics[width=0.5\linewidth]{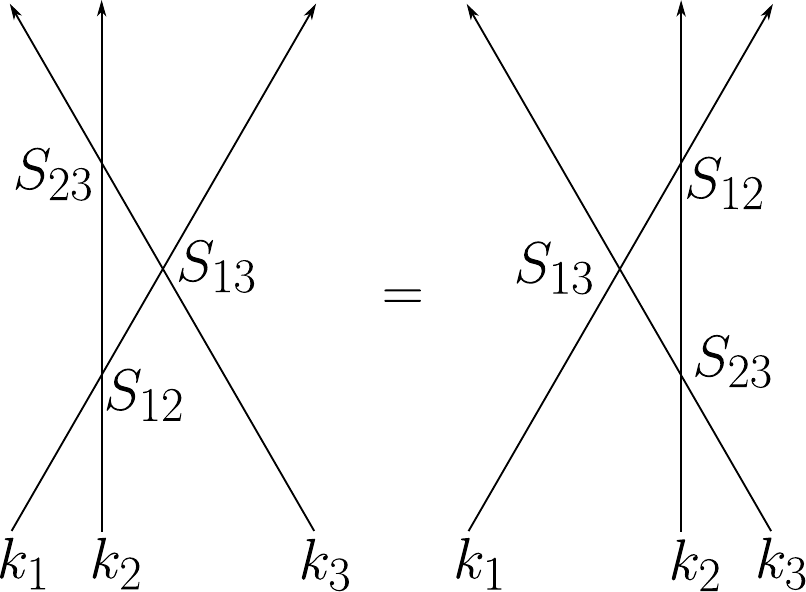}
    \caption{Graphical representation of the Yang-Baxter equation for the three-body problem. Starting from an arbitrary initial state, the final state is reached in both cases irrespective of the sequence of two-body scatterings, thereby ensuring the consistency of our approach.}
    \label{fig:yang}
\end{figure}

\noindent Starting from an initial configuration $ijk$, there are two possible ways to reach the final configuration $kji$ as depicted in Figure~\ref{fig:yang}. In such a case, we see that the scattering process does not depend on the sequence that we consider. The corresponding consistency equation in terms of the scattering matrices
\begin{equation}\label{eq:const}
     S_{jk}^{ab}S_{ik}^{bc}S_{ij}^{ab}= S_{ij}^{bc}S_{ik}^{ab}S_{jk}^{bc},
\end{equation}
leads to the Yang-Baxter equation, a sufficient condition for Bethe ansatz integrable systems. By the Yang-Baxter equation, the scattering happening in many-body systems is factorized in a suitable sequence of two-particles scattering. Through Equation~\eqref{eq:const} as well as noting that 
\begin{equation}\label{eq:stufff}
    S_{ji}^{ab}S_{ji}^{ab}=1; \hspace{7mm} S_{ij}S_{kl} = S_{kl}S_{ij} \hspace{2mm} i\neq j \neq k \neq l,
\end{equation}
one can straightforwardly check for arbitrary $N_{p}$ that the scattering operator characterizing our system is consistent.

\noindent The goal is to obtain the amplitude coefficients $\psi [Q;P]$ by solving the scattering matrix equation~\eqref{eq:scatmata},  Gaudin~\cite{gaudin} and Yang~\cite{yangcontinuum} figured out that the problem can be reduced to one with a smaller dimensionality by splitting it into two: the spatial part and the spin part. For the spatial part, the $N_{p}$ particles are taken to be indistinguishable. In this manner,  the problem can be treated in a similar way as bosons described by the Lieb-Liniger model, allowing us to calculate the quasi-momenta $k_{j}$ associated to the spatial wavefunction by the coordinate Bethe ansatz\footnote{For bosons, the problem is simplified since one needs to take into account only the exchange of momenta of the particles, which is symmetric under exchange, due to their indistinguishability. So then, after applying the continuity and discontinuity conditions, one needs to apply periodic boundary conditions to find that a particle acquires a phase factor after undergoing $N_{p}-1$ body scatterings as we will see later.}. Naturally, this is justified since in the case of same component fermions, the amplitudes of the scattering matrix are equal with a change in sign by antisymmetry~\cite{mcguire_rep}. With this out of the way, we can focus on the spin part of the wavefunction that handles the scattering between fermions in different spin projections. By exploiting the fermions' antisymmetric nature, we have that for a given configuration of $P$, only $C_{M}^{N_{p}}$ coefficients need to be calculated~\cite{takahashi2005thermodynamics}. So, it is sufficient for us to keep track of the wavefunction amplitudes for a given set of positions of the $M$ down spins. Consequently, we may express the wavefunction amplitudes as
\begin{equation}\label{eq:spinjust}
     \Phi(y_{1},y_{2},\hdots, y_{M};P) = \mathrm{sign}(Q)\psi [Q;P]  ,
\end{equation}
where $\Phi(y_{1},y_{2},\hdots, y_{M};P)$ is the spin wavefunction; $y_{1}<y_{2}<\hdots < y_{M}$ are the corresponding coordinates of the $M$ down spins being a subset of the particle positions $x_{j}$ such that $y_{i}\subset \{1,\hdots,N_{p}\}$\footnote{To put this into perspective, if we have five fermions with two down spins arranged in the following manner: $\uparrow\uparrow\downarrow\uparrow\downarrow$ with their corresponding coordinates being $(x_{1},x_{2},x_{3},x_{4},x_{5}) = (1,3,4,6,8)$. The position of both down spins is $y_{1}=3$ and $y_{2}=5$ and it is related to the position of the particle $x_{j}$.}. Through this observation, Yang constructed a generalized Bethe hypothesis, known as the \textit{Bethe-Yang hypothesis}, which assumes the following structure for the spin wavefunction~\cite{yangcontinuum}
\begin{equation}\label{eq:betheyangx}
    \Phi(y_{1},y_{2},\hdots,y_{M}) = \sum\limits_{\pi\in S_{M}} A(\pi) \prod\limits_{n = 1}^{M}F_{P}(\Lambda_{\pi n};y_{n}),
\end{equation} 
where $\pi$ corresponds to the permutations of the $M$ down spins forming the symmetric group $S_{M}$, with $A(\pi)$ and $F_{P}$ being coefficients that need to be determined. Here, we see the introduction of the quantity $\Lambda_{\alpha}$, which corresponds to the spin rapidities for each down spin $y_{\alpha}$. These rapidities adopt a similar role to the quasi-momenta in the spatial part: they characterize the evolution of the spins throughout the scattering processes. \\

\noindent The form of the function $F_{P}$ was obtained by McGuire for both repulsive and attractive interactions~\cite{mcguire_rep,mcguire_att}, by looking at the scattering matrix of $N_{p}$ particles with one flipped spin: 
\begin{equation}\label{eq:FF}
    F_{P}(y,\Lambda) = \prod\limits_{j=1}^{y-1}(k_{Pj}-\Lambda +\imath c')\prod\limits_{l = y+1}^{N_{p}}(k_{Pl}-\Lambda -\imath c'),
\end{equation}
which can be straightforwardly checked by looking at the two- and three-body problems with $M=1$~\cite{takahashi2005thermodynamics,mcguire_rep}. In the case of an arbitrary number of spin downs, we start by noting the following relations between the spin wavefunction amplitudes $\tilde{\Phi} = \Phi(y_{1},\hdots,y_{l},\hdots,y_{M};P)$~\cite{takahashi2005thermodynamics}:
\begin{align}\label{eq:genbounds}
    &\tilde{\Phi} = \Phi(y_{1},..,y_{l},..,y_{M};P') \hspace{2mm}\mathrm{if}\hspace{2mm} \hspace{2mm} \forall l, y_{l} \neq j,j+1 \hspace{4mm}\mathrm{or} \hspace{1mm} \mathrm{if} \hspace{2mm} y_{l} = j, y_{l+1}=j+1, \\
    &\tilde{\Phi} = -R_{j+1,j}\Phi(y_{1},..,y_{l},..,y_{M};P') + T_{j+1,j}\Phi(y_{1},..,y_{l}+1,..,y_{M};P) \hspace{2mm}\mathrm{if}\hspace{2mm} y_{l}=j, y_{l+1}\neq j+1,\nonumber \\
    &\tilde{\Phi} = -R_{j+1,j}\Phi(y_{1},..,y_{l},..,y_{M};P') + T_{j+1,j}\Phi(y_{1},..,y_{l}-1,..,y_{M};P) \hspace{2mm}\mathrm{if}\hspace{2mm} y_{l}=j+1, y_{l-1}\neq j, \nonumber
\end{align}
where $P,P'$ are elements of the symmetric group $S_{N_{p}}$ with $P' = P(j,j+1)$, and $j\in\{1,\hdots, N_{p}\}$ corresponds to the identifier of the particle. The first condition implies that the spin wavefunction remains unchanged if none of the down spins are involved in the interchange of momenta. This also holds if any neighbouring down spins are involved, as this exchange is already accounted for in the scattering process of the spatial part. The second (third) relation embodies the Pauli exclusion principle, in that for a particle to hop right (left), its intended destination cannot be occupied by another spin-down fermion\footnote{Note that the permutation symmetry of the particles is taken into account by the change in sign of $R_{j+1,j}$.}. By imposing these conditions on the generalized Bethe hypothesis in Equation~\eqref{eq:betheyangx}, as well as putting $P' = P(y_{l},y_{l}+1)$ and $\pi ' = \pi (l,l+1)$ one obtains an expression for the amplitudes $A(\pi)$~\cite{takahashi2005thermodynamics}
\begin{equation}\label{eq:AAf}
    A(\pi) = \epsilon_{\pi}\prod\limits_{j<l}(\Lambda_{\pi j}-\Lambda_{\pi l}-\imath c),
\end{equation}
where $\epsilon_{\pi} = \pm 1$ depending on the parity of the permutations.
With this, the complete expression for the spin wavefunction in terms of the charge $k_{j}$ and spin $\Lambda_{\alpha}$ rapidities for an arbitrary number of spin downs can be constructed. \\

\noindent The remaining step in determining the Bethe ansatz equations step is to apply periodic boundary conditions
$\psi(x_{1},x_{2},\hdots, x_{N_{p}}) = \psi(x_{N_{p}}-L_{R},x_{1},x_{2},\hdots, x_{N_{p}-1})$. In the more general case of a given particle, this amounts to saying that a particle with a given $k_{j}$ makes a full turn around the ring after undergoing scattering with $(N_{p}-1)$ particles, acquiring a phase factor:
\begin{equation}\label{eq:merry}
    \psi [Q;P]e^{\imath k_{j}L_{R}} = S_{j+1,j}\hdots S_{N_{p},j}S_{1,j}\hdots S_{j-1,j}\psi [Q;P],
\end{equation}
which can be re-cast in the following form for the $N_{p}$-th,
\begin{align}\label{eq:periodb}
\centering
    \Phi (y_{1},y_{2},\hdots,y_{M};P)e^{\imath k_{PN_{p}}L_{R}} &= \Phi (y_{1}+1,y_{2}+1,\hdots, y_{M}+1;PN_{p},P1,P(N_{p}-1)),\nonumber\\
    \Phi(y_{1},\hdots, y_{M-1},N_{p}+1;P) &= \Phi (1,y_{1},y_{2},\hdots, y_{M-1};P). 
\end{align}
There are two different situations that need to be considered: particle $N_{p}$ can either be a spin up $y_{M}\neq N_{p}$ or a spin down $y_{M}=N_{p}$~\cite{mcguire_rep,takahashi2005thermodynamics}. Keeping this in mind and substituting the complete expression for the Bethe-Yang hypothesis, gives us two sets of coupled equations
\begin{equation}\label{eq:BAkg}
e^{\imath k_{j}L_{R}} = \prod\limits_{\alpha =1}^{M}\frac{ k_{j}-\Lambda_{\alpha}+\imath c'}{ k_{j}-\Lambda_{\alpha}-\imath c'}  \hspace{4mm} j=1,\hdots ,N_{p},
\end{equation}
\begin{equation}\label{eq:BAlg}
\prod_{\substack{\beta = 1 \\\beta\neq\alpha}}^{M} \frac{\Lambda_{\alpha} - \Lambda_{\beta} +\imath c}{\Lambda_{\alpha} - \Lambda_{\beta} -\imath c}  = \prod\limits_{\beta = 1}^{N_{p}} \frac{\Lambda_{\alpha} -  k_{j} +\imath c'}{\Lambda_{\alpha} -  k_{j} -\imath c'} \hspace{4mm}\alpha = 1,\hdots , M.
\end{equation}
These are the Bethe ansatz equations for the Gaudin-Yang model~\eqref{eq:gaudinsu2}. Upon comparing these equations with those of the Lieb-Liniger model for bosons, one can fully appreciate the power of the Bethe-Yang hypothesis: only $M$ extra equations were added in contrast to the one-component case.  \\

\noindent As we mentioned previously, the Gaudin-Yang model is the continuum limit of the Hubbard model and so it is natural to expect that the Bethe equations of the latter have a similar structure~\cite{essler}. Indeed, if we repeat the whole procedure again for the Hubbard model~\eqref{eq:fhmsun} with $N=2$, we observe that the scattering equation~\eqref{eq:scattering} and the spin wavefunction (Equations~\eqref{eq:betheyangx},~\eqref{eq:FF} and~\eqref{eq:AAf}) are of the same form, with the added difference that $k_{j}$ is replaced by $\sin k_{j}$~\cite{deguchi2000thermodynamics}. The introduction of $\sin k_{j}$ accounts for the different dispersion relation and it comes out from the regularisation in constructing the lattice theory from a continuous one~\cite{essler,amico_korepin}.  In their logarithmic form, the Lieb-Wu equations for the Hubbard model read as~\cite{liebwu},
\begin{equation}\label{eq:BAk}
k_{j}L - \sum\limits_{\alpha = 1}^{M}\theta (\sin k_{j}-\Lambda_{\alpha}) = 2\pi I_{j} \hspace{4mm} j=1,\hdots ,N_{p}, 
\end{equation}
\begin{equation}\label{eq:BAl}
 \sum\limits_{\beta}^{M}\theta (2\Lambda_{\alpha} - 2\Lambda_{\beta}) - \sum\limits_{j=1}^{N_{p}}\theta (\Lambda_{\alpha}-\sin k_{j})   = 2\pi J_{\alpha} \hspace{4mm}\alpha = 1,\hdots , M,
\end{equation}
where $\theta (x-y)= -2\arctan\big(\frac{x-y}{u}\big)$ with $u = U/4$. The equations are parameterised by two sets of quantum numbers denoted by $I_{j}$ and $J_{\alpha}$, which are called the charge and spin quantum numbers respectively~\cite{takahashi2005thermodynamics}. By construction, each set of quantum numbers needs to be mutually distinct, since the wavefunction would vanish if this was not the case~\cite{deguchi2000thermodynamics}. Through manipulation of the quantum numbers, one gains access to the whole spectrum of the system and as such, they are instrumental in understanding the underlying many-body physics of the model.


\chapter{Spinon configurations in the Bethe ansatz of strongly repulsive SU($N$) fermions}

\noindent In this Appendix, we present in detail how the charge and spin quantum numbers need to be changed in order to capture the fractionalization of the energy and in turn the persistent current presented in Chapter~\ref{chp:repcurr}. \\

\noindent To minimize the energy at a given flux value $\phi$, one requires that the summation over the spin rapidities $X$ satisfies the degeneracy point equation having the form~\cite{kusmar, yu1992persistent}
\begin{equation}\label{eq:quanrep1}
\frac{2w - 1}{2N_{p}}\leq \phi +D \leq \frac{2w+1}{2N_{p}},\hspace{5mm} \textrm{where} \hspace{5mm} X = -w,
\end{equation}
with $w$ only being allowed to have integer or half-integer values due to the nature of the spin rapidities, and $D = \frac{I_{\mathrm{max}}+I_{\mathrm{min}}}{2}$. \\

\noindent Consider the case of $N_{p}=3$ SU(3) symmetric fermions. There are three sets of quantum numbers: one pertaining to the charge momenta $I_{j}$ and the other two belonging to the spin momenta denoted as $J_{\alpha_{1}}$ and $J_{\alpha_{2}}$ introduced in Chapter~\ref{chp:repcurr}. The ground-state configuration for such a system is given as $I_{j} = \{-1,0,1\}$, $J_{\alpha_{1}} = \{-0.5,0.5\}$ and $J_{\alpha_{2}} = \{0\}$\footnote{The form for these quantum numbers comes out from taking the logarithm of the product Bethe ansatz equations. The expression for these quantum numbers can be found in~\cite{sutherland1968}.}. The spin quantum numbers for all the values of the flux per Equation~\eqref{eq:quanrep1} is as follows 
\begin{table}[h!]
\centering
\begin{tabular}{|c| c| c| c|}
\hline
Magnetic flux & $J_{\alpha_{1}}$  &  $J_{\alpha_{2}}$  &$X$\\
\hline
$0.0-0.1$&$\{-0.5,0.5\}$ &$\{0\}$  &$0$ \\
$0.2-0.5$&$\{-1.5,0.5\}$&$\{0\}$ &$-1$ \\
$0.6-0.8$&$\{-0.5,1.5\}$&$\{0\}$&$+1$  \\
$0.9-1.0$&$\{-0.5,0.5\}$ &$\{0\}$ & $0$ \\
\hline
\end{tabular}
\caption{\label{T:quantrep1}Spin quantum number configurations with the flux for $N_{p} = 3$ SU(3) symmetric fermions.}
\end{table}

\noindent As can be seen from Table~\eqref{T:quantrep1}, in cases where $X=0$, the spin quantum number configuration is different from the ground-state one and `holes' are introduced such that the spin quantum number configurations are no longer consecutive, with the set of $I_{j}$ remaining unchanged. There are two notable points worthy of mention. The first is that one could have chosen a different way to arrange the set of quantum numbers. An alternative arrangement is given by Table~\eqref{T:quantrep2}. The target value $X$ is reached via a different configuration, which in turn leads to a degenerate state. Such a phenomenon is a characteristic property of SU($N$) systems that is not present for SU(2). As $N$ increases, the number of degenerate states that are present in the system increases due to the various Bethe quantum number configurations that one can adopt.

\begin{table}[h!]
\centering
\begin{tabular}{|c| c| c| c|}
\hline
Magnetic flux & $J_{\alpha_{1}}$  &  $J_{\alpha_{2}}$  &$X$\\
\hline
$0.0-0.1$&$\{-0.5,0.5\}$ &$\{0\}$  &$0$ \\
$0.2-0.5$&$\{-0.5,0.5\}$&$\{-1\}$ &$-1$ \\
$0.6-0.8$&$\{-0.5,0.5\}$&$\{+1\}$&$+1$  \\
$0.9-1.0$&$\{-0.5,0.5\}$ &$\{0\}$ & $0$ \\
\hline
\end{tabular}
\caption{\label{T:quantrep2}Alternative spin quantum number configurations with the flux for $N_{p} = 3$ with SU(3) spin with $M_{1} =2$ and $M_{2} = 1$.}
\end{table}

\noindent The other point concerns the value of $X$ for $\phi = 0.6-0.8$ and $\phi = 0.9-1.0$. According to Equation~\eqref{eq:quanrep1}, $X$ should be equal to $-2$ and $-3$ respectively. However, this is not the case. The reason behind this is due to the fact that the degeneracy equation has to be applied within a specific flux range that depends on the parity of the system: for a flux in the interval of $-0.5$ to $0.5$ for $N_{p}=N(2n+1)$ and in the range of  $\phi = 0.0$ to $1.0$ in the case of $N_{p}=N(2n)$. The ground-state energy of the system is given by a series of parabolas in the absence of interactions as discussed in Chapter~\ref{chp:persis}. These parabolas each have a well defined angular momentum $l$. They intersect at the degeneracy points, which is parity dependent, and are shifted with respect to each other by a Galilean translation~\cite{blochsuper}. Consequently, when the magnetic flux piercing the system falls outside the range outlined previously, one needs to change the $I_{j}$ quantum numbers in order to offset the increase in angular momentum $l$ that one obtains on going to the next energy parabola. Note that in the limit of strong repulsion, the angular momentum of the system increases at $\phi = \big(s \pm \frac{1}{2N_{p}} + \delta\big)$ with $s$ being (half-odd) integer in the case of (diamagnetic) paramagnetic systems, with $\delta =\mp \frac{1}{2N_{p}}$ for an odd number of particles. \\

\noindent For positive $\phi$ one requires that the $I_{j}$ quantum numbers need to all be shifted by one to the left. For example in the case considered above for $\phi >0.5$, the $I_{j}$ go from $\{-1,0,1\}$ to $\{-2,-1,0\}$  for $0.5 < \phi < 1.5$. On going to the next parabola, they would need to be shifted again by one to the left. In the case of negative $\phi$, the shift occurs to the right. \\

\noindent Note that there are other combinations of the quantum numbers, not outlined in Tables~\ref{T:quantrep1} and~\ref{T:quantrep2}, whose total sum reaches the target value of $X$. However, these configurations do not give the lowest value for the energy as the ones mentioned, even though the value of $X$ is the same. At infinite $U$, the system is solely dependent on the value of $X$ and not on the arrangement of the spin quantum number configuration. Consequently, the system is highly degenerate. This is also observed in the SU(2) case. However, as mentioned in Chapter~\ref{chp:repcurr}, the degeneracy is lifted on going to large but finite $U$ and one is left with only one combination that gives the lowest energy in the case of SU(2) systems. On the other hand, for SU($N$) systems whilst this degeneracy is also lifted, they also benefit from an extra `source' of degeneracy due to the different configurations of the Bethe quantum numbers as shown in Tables~\ref{T:quantrep1} and~\ref{T:quantrep2}.


\chapter{Jordan-Wigner Transformation for SU(\textit{N}) fermions}\label{sec:vqejw}

\begin{figure}[h!]
\centering
    \includegraphics[width=\linewidth]{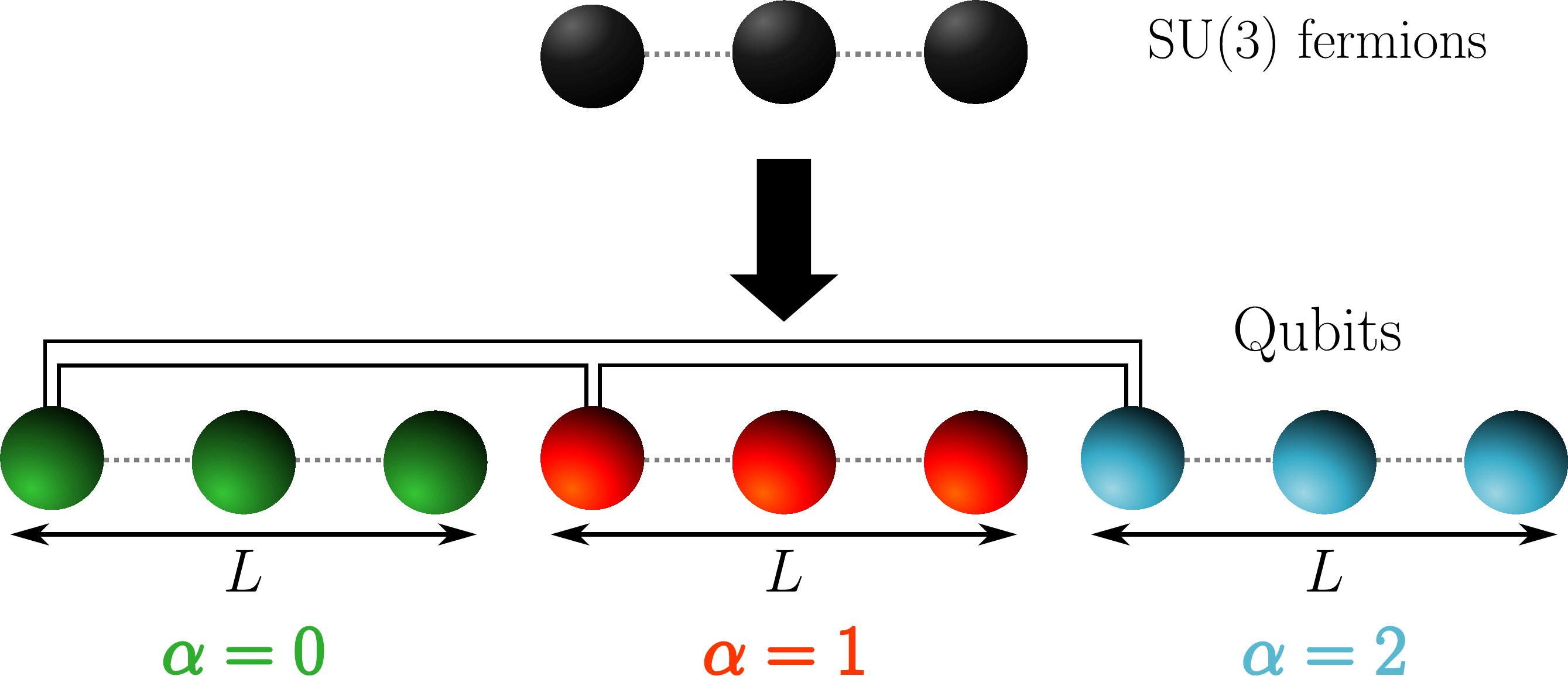}
    \caption{Mapping of a three site linear chain of the SU(3) Hubbard model (black) to $N$ chains of $L$ qubits, where $N=L=3$. The dashed grey lines depict the hopping term between qubits pertaining to the same colours, whilst the black solid lines correspond to the interactions between qubits in different colours. Figure adapted from~\cite{vqe}.}
    \label{fig:vqe mapping}
\end{figure}

\noindent The SU($N$) Hubbard Hamiltonian describing a ring of $L$-sites threaded by an effective magnetic flux $\phi$ reads
\begin{align}
    \mathcal{H} = -t\sum_{j}^{L}\sum\limits_{\alpha}^{N}\left(e^{\imath \frac{2 \pi \phi}{L}} c_{j, \alpha}^\dagger c_{j+1,\alpha} +\text{h.c.} \right) + U\sum_{j}^{L} \sum\limits_{\alpha < \beta}n_{j, \alpha}n_{j, \beta} ,
    \label{eq:Ham}
\end{align}
To map the fermionic model onto spins (qubits), we need to make use of the Jordan-Wigner transformation, which is characterised as a non-local transformation. Originally devised to map spinless fermions onto spins~\cite{1928JW}, the Jordan-Wigner transformation was extended for two-component fermions~\cite{Shastry1986, Reiner_2016} and now we apply it to the general SU($N$) case. By introducing $N$ sets of Pauli operators, one for every colour $\alpha$ of the fermionic atom, the mapping assumes the following form:
\begin{equation}\label{eq:E.map}
    c_{i,\alpha}^\dagger=\prod_{j<n}\sigma_j^z\sigma_n^+ \,,\, c_{i,\alpha}=\prod_{j<n}\sigma_j^z\sigma_n^- \,,\, (i, s) \xrightarrow{} n = i + \alpha L \,,
\end{equation}
where $\sigma^\pm = (\sigma^x \pm \imath\sigma^y) / 2$ are the ladder operators acting on $\ket{0}_n$ and $\ket{1}_n$ that represent the absence and presence respectively, of a fermion with colour $\alpha$ on site $j$. It is straightforward to show that the mapping in Equation~\eqref{eq:E.map} preserves fermionic commutation rules irrespective of the colour $\alpha$. Equation~\eqref{eq:E.map} implies that a fermionic operator of colour $\alpha$ acting on site $j$ is mapped onto a spin-$\frac{1}{2}$ operator $\sigma^+_n$ acting on qubit $n=j+ \alpha L$. Consequently, a system of SU($N$) fermions comprised of $L$ sites is mapped onto $NL$ qubits --Figure~\ref{fig:vqe mapping}. In the following, without loss of generality, we adopt an increasing order for $\alpha$, such that $n<n'$ for fermionic operators with $\alpha<\beta$. \\

\noindent Let us start by noting that $\sigma^z_n \sigma^\pm_n = \mp \sigma^\pm_n$ and $\sigma^\pm_n \sigma^z_n = \pm \sigma^\pm_n$ through the commutation rules of the Pauli matrices. For the nearest-neighbor hopping terms, we find that
\begin{align}~\label{eq:hopsy}
    c^\dagger_{j, \alpha}c_{j+1, \alpha} 
    &= \prod_{m<n}\sigma_{m}^z\sigma_n^+\prod_{k<n+1}\sigma_k^z\sigma_{n+1}^- \implies 
    c^\dagger_{j, \alpha}c_{j+1, \alpha}= \sigma^+_n\sigma^-_{n+1}.
\end{align}
Likewise, the Hermitian conjugate term can be expressed as
\begin{align}
    c^\dagger_{j+1, \alpha}c_{j, \alpha} 
    &= \prod_{m<n+1}\sigma_{m}^z\sigma_{n+1}^+\prod_{k<n}\sigma_k^z\sigma_{n}^- \implies 
    c^\dagger_{j+1, \alpha}c_{j, \alpha}= -\sigma^+_{n+1}\sigma^-_{n}.
\end{align} 
Special care needs to be taken at the boundary condition when $j=L$ and $j+1 = 1$. In this case, we have that
\begin{align}
    c^\dagger_{L, \alpha}c_{1, \alpha} 
    &= \prod_{m<L+sL}\sigma_m^z\sigma_{L+\alpha L}^+ \prod_{k<1 + \alpha L}\sigma_k^z\sigma_{1 + \alpha L}^- = \sigma^-_{1+\alpha L}\left(\sum_{m=1+\alpha L}^{L+\alpha L}\sigma^z_m\right)\sigma^+_{L+1+\alpha L} \,,
\end{align}
and similarly for the Hermitian conjugate terms. The term in brackets corresponds to $\exp(\imath\pi \hat{N}_{\alpha} +1)$ where $\hat{N}_{p}$ is the total particle number operator per colour~\cite{lieb}. For an odd (even) number of particles of a given species, the exponential becomes $+1$ ($-1$) respectively. As such, we need to impose the following parity boundary condition on the Hamiltonian
\begin{equation}
    P_{i,\alpha} = \begin{cases}
        -1 \,,\, \text{if } i = L - 1 \text{ and } N_{\alpha} \text{ is even,} \\
        +1 \,,\, \text{otherwise.}
        \end{cases}
\end{equation}
Now, since $n_{j, \alpha} = c^\dagger_{j, \alpha}c_{j, \alpha}$, then
\begin{equation}
    n_{j, \alpha} = \prod_{m<n}\sigma_j^z\sigma_n^+ \prod_{k<n}\sigma_k^z\sigma_n^- = \sigma^+_n \sigma^-_n \,,
\end{equation}
which can be rewritten as
\begin{equation}~\label{eq:intit}
    n_{j, \alpha} = \dfrac{1 - \sigma^z_n}{2} \,.
\end{equation}
Therefore, putting Equations~\eqref{eq:hopsy} and~\eqref{eq:intit} we find the original SU($N$) Hubbard model, after application of the Jordan-Wigner transformation, gets mapped to
\begin{align}
    H= -t\sum\limits_{j, \alpha} P_{j,\alpha}\left(e^{\imath \frac{2 \pi \phi}{L}} \sigma_{j+\alpha L}^+\sigma_{j+1+\alpha L}^-+ \text{h.c.}\right) + \frac{U}{4}\sum\limits_{j, \alpha < \beta}\left( 1-\sigma_{j+\alpha L}^z \right)\left( 1-\sigma_{j+\beta L}^z \right). 
\end{align}


\chapter{The formation of symmetric and asymmetric trions}\label{Appendix:trionformation}

\noindent In this Appendix, we investigate all the different ways that a trion can be formed by tuning the different interactions between the colours. 
\begin{figure}[h!]
    \centering
    \includegraphics[width=\linewidth]{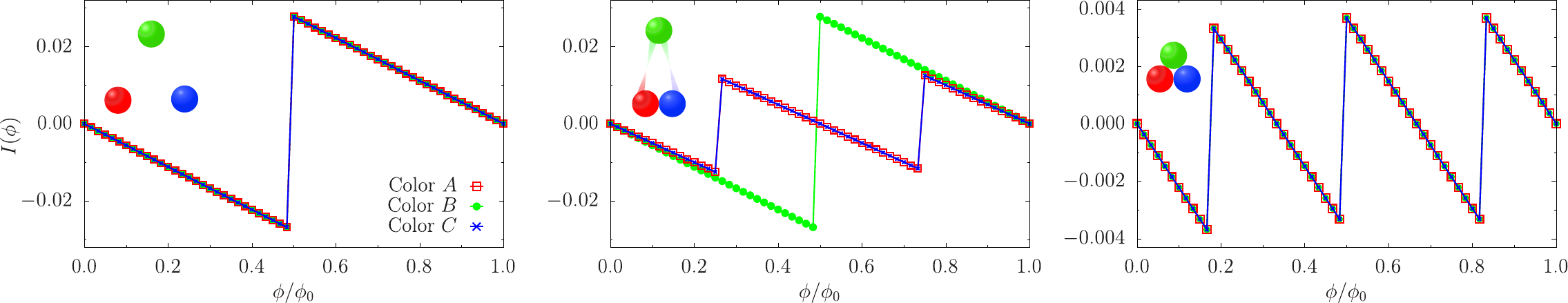}
    \put(-305,72){(\textbf{\textbf{a}})}
    \put(-163,72){(\textbf{\textbf{b}})}
    \put(-17,72){(\textbf{\textbf{c}})}
    \caption{Persistent current $I(\phi)$ against flux $\phi/\phi_{0}$ for the three main phases of SU(3) fermions: (\textbf{a}) unpaired, (\textbf{b}) csf and (\textbf{c}) trion. Results were obtained with exact diagonalization for $N_{p}=3$ and $L=15$. The lines are meant as a guide to the eye for the reader.}
    \label{fig:m1}
\end{figure}

\begin{figure}[h!]
    \centering
    \includegraphics[width=\linewidth]{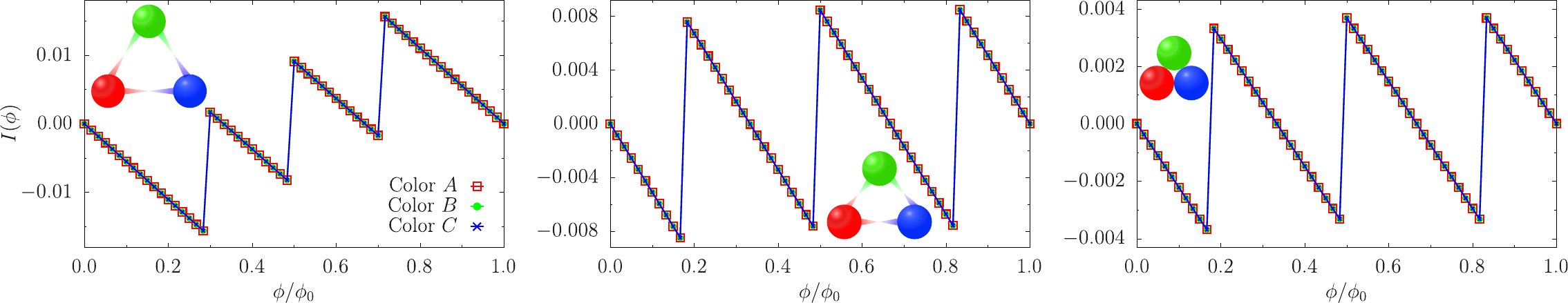}
    \put(-305,72){(\textbf{\textbf{a}})}
    \put(-161,72){(\textbf{\textbf{b}})}
    \put(-17,72){(\textbf{\textbf{c}})}
    \caption{Persistent current $I(\phi)$ against flux $\phi/\phi_{0}$ for the formation of a trion with symmetric interactions $|U| = |U_{AB}|=|U_{BC}|=|U_{AC}|$ for: (\textbf{a}) $|U|=0.5$, (\textbf{b}) $|U|=1$ and (\textbf{c}) $|U|=3$. Results were obtained with exact diagonalization for $N_{p}=3$ and $L=15$. The lines are meant as a guide to the eye for the reader.}
    \label{fig:m2}
\end{figure}
\noindent The first path to forming a trion is displayed in Figure~\ref{fig:m2} and it is achieved by using symmetric interactions, i.e., equal interactions between the colours. Starting from a system with zero interactions like the one depicted in Figure~\ref{fig:m1}(\textbf{a}), the persistent current fractionalizes with increasing interaction and experiences a periodicity change from $\phi_{0}$ to $\phi_{0}/N$ upon formation of the trion --Figure~\ref{fig:m2} (\textbf{c}). 

\begin{figure}[h!]
    \centering
    \includegraphics[width=\linewidth]{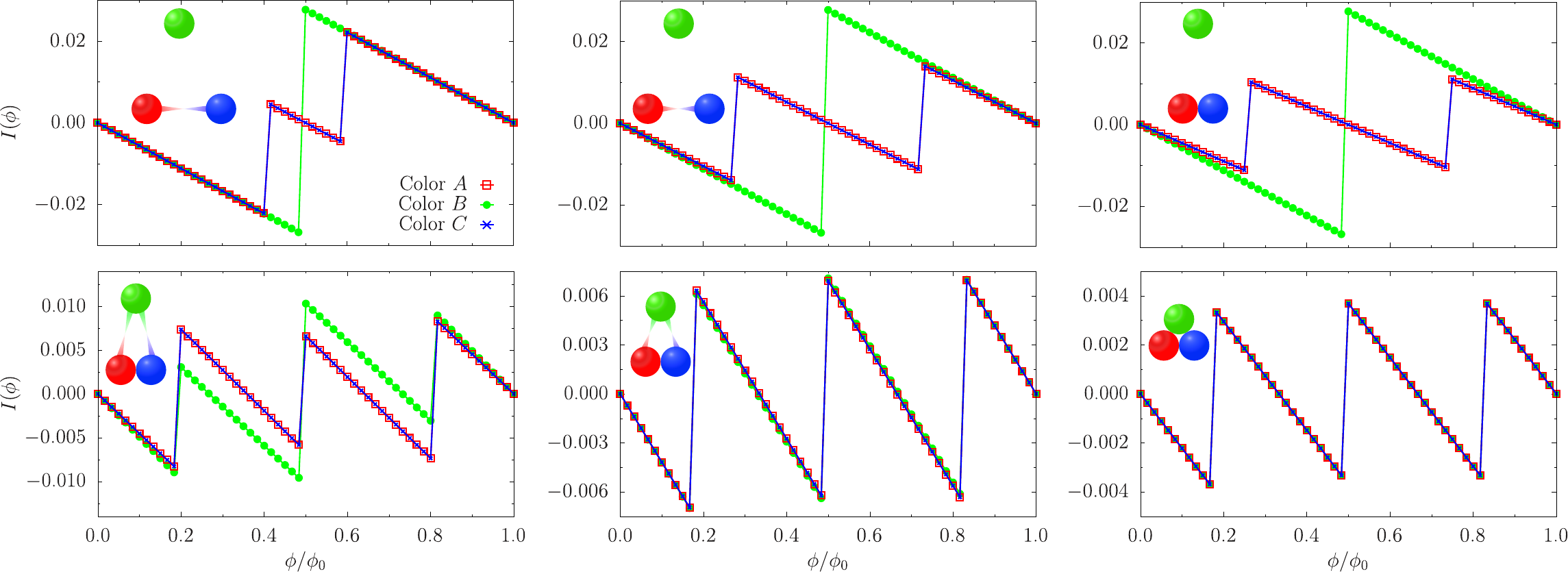}
    \put(-305,145){(\textbf{\textbf{a}})}
    \put(-161,145){(\textbf{\textbf{b}})}
    \put(-17,145){(\textbf{\textbf{c}})}
    \put(-305,70){(\textbf{\textbf{d}})}
    \put(-159,70){(\textbf{\textbf{e}})}
    \put(-17,70){(\textbf{\textbf{f}})}
    \caption{Persistent current $I(\phi)$ against flux $\phi/\phi_{0}$ for the formation of a trion by forming a CSF as an intermediate step. Starting from $|U_{AC}|=0.5$ in (\textbf{a}), one starts increasing the interactions between colours $A$ and $C$ until one forms a CSF in (\textbf{c}) with $|U_{AC}|=3$. The interactions $|U_{AB}|$ and $|U_{BC}|$ are kept at a value of 0.01. In the bottom panel, the interactions between colour $B$ to both colours $A$ and $C$ are increased from $|U_{AB}|=|U_{BC}|=0.5$ in (\textbf{d}), to $|U_{AB}|=|U_{BC}|=3$ in (\textbf{f}) at which a trion is  formed. Results were obtained with exact diagonalization for $N_{p}=3$ and $L=15$. The lines are meant as a guide to the eye for the reader.}
    \label{fig:m3}
\end{figure}

\begin{figure}[h!]
    \centering
    \includegraphics[width=1\linewidth]{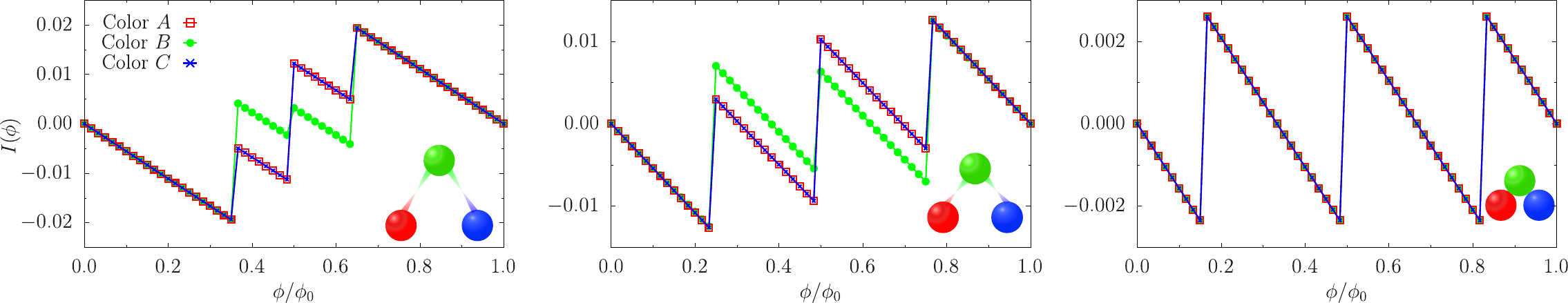}
    \put(-305,72){(\textbf{\textbf{a}})}
    \put(-163,72){(\textbf{\textbf{b}})}
    \put(-17,72){(\textbf{\textbf{c}})}
    \caption{Persistent current $I(\phi)$ against flux $\phi/\phi_{0}$ for the formation of a trion with asymmetric interactions. The interaction is increased between colours $A$ and $B$, and between $B$ and $C$ denoted by $|U_{AB}|$ and $|U_{BC}|$ respectively. At (\textbf{a}) the interactions $|U_{AB}|=|U_{BC}|=0.5$ are increased, until one forms a trion in (\textbf{c}) with $|U_{AB}|=|U_{BC}|=6$. The interaction $|U_{AC}|$ is kept at a constant value of 0.01 throughout. Results were obtained with exact diagonalization for $N_{p}=3$ and $L=15$. The lines are meant as a guide to the eye for the reader.}
    \label{fig:m4}
\end{figure}

\noindent Another way is to break SU(3) symmetry by forming the CSF bound state during an intermediate step. In this case, we take one interaction between two colours, say $|U_{AC}|$, to be significantly larger than the other ones denoted by $|U_{AB}|$ and $|U_{BC}|$. On increasing the interaction $|U_{AC}|$, the persistent current of colours $A$ and $C$  starts to fractionalize reflecting the formation of the two-body bound state --Figures~\ref{fig:m3} (\textbf{a}) to (\textbf{c}). Subsequently increasing interactions $|U_{AB}|$ and $|U_{BC}|$ to match $|U_{AC}|$, causes further the persistent current of colour $B$ to fractionalize 
such that the current obtains a reduced period of $\phi_{0}/N$ for all three species.  \\

\noindent Lastly, a trion can be formed by choosing asymmetric interactions between the colours. At variance with the formation of a CSF, we take two interactions $|U_{AB}|$ and $|U_{BC}|$ to be significantly larger than $|U_{AC}|$ --Figure~\ref{fig:m4}. In this case, the persistent current fractionalizes and eventually achieves the tri-partite periodicity. Interestingly enough, one has formed a trion by breaking SU(3) symmetry. However, we remark that this `asymmetric' trion can still be distinguished from an actualy trion seeing as the energy of the former is higher (less stable) than that on an actual trion, which in turn is reflected in the current. Furthermore, the species-wise persistent current of the `asymmetric' trion is not equivalent in all three colours. This is to be expected since one colour is interacting twice as much as the other two.


\chapter{Time-of-flight momentum distribution in the free particle regime}\label{sec:momder}

\noindent In cold atoms systems, the persistent current is experimentally observed through time-of-flight (TOF) imaging. This entails looking at the momentum distribution of the gas, which is one of the few observables that can be experimentally probed~\cite{iskin2009momentum}. Here, we will give the derivation for the momentum distribution. \\

\noindent Starting from the expression for the one-body correlator $n(\textbf{r},\textbf{r}',t)$ defined as 
\begin{equation}\label{eq:amom1}
  \langle n(\textbf{r},\textbf{r}')\rangle  =  \langle \Psi^{\dagger}(\textbf{r})\Psi (\textbf{r}') \rangle,
\end{equation}
where $\textbf{r}$ is position, $\Psi^{\dagger}(\textbf{r})$ and $\Psi (\textbf{r})$ are the fermionic creation and annihilation field operators. Expanding the field operators in the basis set of the single-band Wannier functions  $w_{j}(\textbf{r})$ such that $\Psi (\textbf{r}) = \sum_{j}^{L}w(\textbf{r}-\textbf{r}_{j})c_{j}$, the two-point correlator has the following expression
\begin{equation}\label{eq:amom2}
    \langle n(\textbf{r},\textbf{r}')\rangle = \sum\limits^{L}_{j,l}w^{*}(\textbf{r}-\textbf{r}_{l})w(\textbf{r}'-\textbf{r}'_{j})\langle c_{l}^{\dagger}c_{j}\rangle,
\end{equation}
with $w_{j}(\textbf{r}-\textbf{r}_{j})$ being the Wannier function localised at site $j$ with position $r_{j}$ and $L$ being the number of lattice sites. If we consider the free expansion in time $t$ of the particle density $n(\textbf{r},\textbf{r}',t)$, it is still defined as in Equation~\eqref{eq:amom2}, but the time dependence is encoded in the expansion of the Wannier function $w_{j}(\textbf{r},t)$ that reads
\begin{equation}\label{eq:amom3}
    w(\textbf{r}-\textbf{r}_{j},t) = \frac{1}{\sqrt{\pi}}\frac{\eta_{j}}{\eta_{j}^2 + \imath\omega_{0}t}\exp\bigg\{-\frac{(\textbf{r}-\textbf{r}_{j})^{2}}{2(\eta_{j}^{2} + \imath\omega_{0}t)} \bigg\}.
\end{equation}
where $\eta_{j}$ is the width of the center at the $j$-th site and $\omega_{0} = \frac{\hbar}{m}$ with $\hbar$ and $m$ denoting Planck's constant and the particles' mass respectively, both of which are set to 1 in this calculation. Note that we have taken the zeroth order approximation of the Wannier function and the harmonic approximation. By letting the density distribution to expand for large values of time, one obtains the momentum distribution $n(\textbf{k})$. The momentum distribution  is defined as the Fourier transform of the one-body correlator $n (\textbf{r},\textbf{r}')$,
\begin{equation}\label{eq:amom4}
    \langle n(\textbf{k})\rangle = \int e^{\imath \textbf{k}(\textbf{r}-\textbf{r}')} \langle \Psi^{\dagger}(\textbf{r})\Psi (\textbf{r}') \rangle \textrm{d}\textbf{r}\textrm{d}\textbf{r}',
\end{equation}
where $\textbf{k}$ is the momentum. One can verify that $\lim_{t\rightarrow\infty}\langle n(\textbf{r},\textbf{r}',t) \rangle \approx \langle n(\textbf{k})\rangle$, by taking the limit $t\rightarrow\infty$ of Equation~\eqref{eq:amom3} and performing a Taylor expansion. \\

\noindent Substituting the expression for the field operators into Equation~\eqref{eq:amom4}, the expression for $n(\textbf{k})$ reads
\begin{equation}\label{eq:amom5}
    \langle n(\textbf{k})\rangle = \int e^{\imath\textbf{k}(\textbf{r}-\textbf{r}')}\sum\limits_{j,l}^{L}[w^{*}(\textbf{r}-\textbf{r}_{l})w(\textbf{r}'-\textbf{r}'_{j})\langle c_{l}^{\dagger}c_{j}\rangle ] \textrm{d}\textbf{r}\textrm{d}\textbf{r}'.
\end{equation}
Utilising the change of variables $\textbf{R} = \textbf{r}-\textbf{r}_{j}$ and $\textbf{R}' = \textbf{r}'-\textbf{r}_{l}'$, we arrive to 
\begin{equation}\label{eq:amom6}
    \langle n(\textbf{k})\rangle  = \int e^{\imath\textbf{k}(\textbf{R}-\textbf{R}')}\sum\limits_{j,l}^{L}e^{\imath\textbf{k}(\textbf{r}_l-\textbf{r}_j')}[w^{*}_{l}(\textbf{R})w_{j}(\textbf{R}')\langle c_{l}^{\dagger}c_{j}\rangle ] \textrm{d}\textbf{R}\textrm{d}\textbf{R}',
\end{equation}
which by making use of the fact that $w(\textbf{k}) = \int w(\textbf{R})e^{\imath \textbf{k}\cdot\textbf{R}}\textrm{d}\textbf{R}$, can be further simplified to give
\begin{equation}\label{eq:amom7}
    \langle n(\textbf{k})\rangle = |w(\textbf{k})|^{2}\sum\limits_{j,l}^{L}e^{\imath\textbf{k}(\textbf{r}_{l}-\textbf{r}'_{j})}\langle c_{l}^{\dagger}c_{j}\rangle ,
\end{equation}
with $w(\textbf{k})$ being the Fourier transform of $w(\textbf{R})$.
Finally, we write our ring configuration explicitly as 
\begin{equation}\label{eq:amom8}
    \langle n(\textbf{k})\rangle = |w(\textbf{k})|^{2}\sum\limits_{j,l}^{L}e^{\imath [k_{x}r (\cos (\frac{2\pi l}{L})-\cos (\frac{2\pi j}{L})) + k_{y}r(\sin (\frac{2\pi l}{L})-\sin (\frac{2\pi j}{L}))]}\langle c_{l}^{\dagger}c_{j}\rangle ,
\end{equation}
where the momentum vector can be written as $\textbf{k} = (k_{x},k_{y})$ and polar coordinates were utilized $\textbf{r} = (r\cos\theta ,r\sin\theta)$, with $\theta = \frac{2\pi}{L}l$. \\

\noindent Now, we introduce the creation operator and its Fourier transform~\cite{essler}
\begin{equation}\label{eq:amom9}
 c_{l}^{\dagger} = \frac{1}{\sqrt{L}}\sum\limits_{k}^{L}e^{-\imath kl}c_{k}^{\dagger},
\end{equation}
as well as the Fourier transform of its counterpart the annihilation operator
\begin{equation}\label{eq:amom10}
 c_{j} = \frac{1}{\sqrt{L}}\sum\limits_{k'}^{L}e^{\imath k'j}c_{k'}.
\end{equation}
Therefore, the one-body correlator $\langle c_{l}^{\dagger}c_{j}\rangle$ in Fourier space reads
\begin{equation}\label{eq:amom11}
    \langle c_{l}^{\dagger}c_{j}\rangle = \frac{1}{L}\sum\limits_{k,k'}^{L} e^{-\imath kl + \imath k'j} \langle c_{k}^{\dagger}c_{k'} \rangle .
\end{equation}
Furthermore, at zero interactions we have that $\langle c_{k}^{\dagger}c_{k'}\rangle =\delta_{k,k'}$. Consequently, the expression for the momentum distribution reads
\begin{equation}\label{eq:amom12}
    \langle n(\textbf{k})\rangle =\frac{1}{L} |w(\textbf{k})|^{2}\sum\limits_{j,l}^{L}e^{\imath [k_{x}r (\cos (\frac{2\pi l}{L})-\cos (\frac{2\pi j}{L})) + k_{y}r(\sin (\frac{2\pi l}{L})-\sin (\frac{2\pi j}{L}))]}\sum\limits_{\{q\}}e^{-\frac{2\pi\imath}{L}(l-j)q} ,
\end{equation}
where we made use of the fact that for free fermions $k=\frac{2\pi}{L} q$ with $q$ being the quantum number labeling the Fermi sphere levels. At this point, by setting $k_{x} = |\textbf{k}|\sin\phi$ and $k_{y} = |\textbf{k}|\cos\phi$ we have that
\begin{equation}\label{eq:amom13}
    \langle n(\textbf{k})\rangle \propto \frac{1}{L}\sum\limits_{\{q\}}\bigg|\sum\limits^{L}_{l}e^{\imath r(B\sin\phi\cos\theta_{l} +B\cos\phi\sin\theta_{l})}e^{-\frac{2\imath\pi l}{L}q}\bigg|^{2} ,
\end{equation}
where $B = \sqrt{2|\textbf{k}|^{2}}$. Using the trigonometric identity $\sin(A+B)  = \sin A\cos B + \cos A\sin B$, the expression is simplified even further and reads
\begin{equation}\label{eq:amom14}
    \langle n(\textbf{k})\rangle \propto \frac{1}{L}\sum\limits_{\{q\}}\bigg|\sum\limits^{L}_{l}e^{\imath rB\sin (\phi+\theta_{l})}e^{-\frac{2\imath\pi l}{L}q}\bigg|^{2}  .
\end{equation}

\noindent The expression in Equation~\eqref{eq:amom14} is a \textit{q}-th order Bessel function of the first kind~\cite{pecci2021phase}
\begin{equation}\label{eq:amom15}
J_{q}(x) = \frac{1}{2\pi}\int\limits_{-\pi}^{\pi}e^{\imath (x\sin\tau - q\tau) },
\end{equation}
where $x = rB$ and $\tau = \frac{2\pi l}{L}$ . It is important to stress here that replacing the sum by an integral is only an approximation, which becomes valid in the thermodynamic limit. As a result, by setting
\begin{equation}\label{eq:amom16}
J_{q}(B) \approx \sum\limits^{L}_{l} e^{\imath rB\sin (\phi+\theta_{l})}e^{-\frac{2\imath\pi l}{L}q}  , 
\end{equation}
the momentum distribution reads as
\begin{equation}\label{eq:amom17}
    \langle n(\textbf{k})\rangle \propto \frac{1}{L}\sum\limits_{\{q\}} |J_{q}(\textbf{k})|^{2}.
\end{equation}
This enables us to study the momentum distribution analytically, by considering it as a summation of different Bessel functions as was carried out in~\cite{pecci2021phase} and generalized to SU($N$) in the main text. It is important to note that the Equation~\eqref{eq:amom17} only holds at zero interactions. In the case of interacting particles $\langle c_{k}^{\dagger}c_{k'}\rangle \neq \delta_{k,k'}$. As such, we no longer have an analytical expression for the momentum distribution and different behaviours are observed in the interacting regimes as reported in this paper. The scheme developed in Chapter~\ref{chp:ogata} allows us to calculate the momentum distribution exactly in the limit of strong repulsive interactions. Note that the derivation for the momentum distribution carried in this section is done for spinless fermions. In the case of SU($N$) fermions, which in the free particle regime one can treat as $N$ chains of spinless fermions, Equation~\eqref{eq:amom17} would need to be multiplied by $N$.


\chapter{Self-heterodyne interference between a ring and a quantum degenerate gas in the free particle regime}\label{sec:interder}

\noindent In this Appendix, we consider the density-density correlator $G (\textbf{r},\textbf{r}',t)$ for a ring and an additional site in the center that was employed in Chapter~\ref{chp:interfer} to model the self-heterodyne protocol. The two-body correlator is defined in the following way
$G(\textbf{r},\textbf{r}',t) = \sum_{\alpha,\beta}^{N}\langle n_{\alpha}(\textbf{r},t)n_{\beta}(\textbf{r}',t)\rangle$ .
The density operator is defined as $n(\textbf{r},t) = \psi^{\dagger} (\textbf{r},t)\psi (\textbf{r},t)$ where $\psi^{\dagger} = (\psi^{\dagger}_{R} + \psi^{\dagger}_{C})$ being the field operator of the whole system of the ring and the center, denoted by $R$ and $C$ respectively. Initially, the ring and the center are decoupled until they are released from their confinement potential. Thus, at time $t=0$ the ground-state can be seen as a product state $|\phi\rangle = |\phi\rangle_{R}\otimes|\phi\rangle_{C}$. \\

\noindent Assuming free expansion for $t\geq 0$, i.e., negligible particle-particle interactions, the density-density correlator gets markedly simplified as the number of terms can be reduced. To start, terms comprised of an odd number of creation or annihilation operators have an expectation value of zero since the number of particles in the system has to be conserved. The only remaining terms are ones consisting of an equal number of creation-annihilation pairs. Consequently, the expression for the density-density correlator reads
\begin{align}\label{eq:ash2}
    \sum\limits_{\alpha,\beta}^{N}\langle n_{\alpha}(\textbf{r},t)n_{\beta}(\textbf{r}',t)\rangle &= \sum\limits_{\alpha,\beta}\langle n_{\alpha}(\textbf{r},t)n_{\beta}(\textbf{r}',t)\rangle_{R} + \langle n_{\alpha}(\textbf{r},t)n_{\beta}(\textbf{r}',t)\rangle_{C} \nonumber\\
    &+ \sum\limits_{\alpha,\beta}[\langle n_{\alpha}(\textbf{r},t)\rangle_{R} \langle n_{\beta}(\textbf{r}',t)\rangle_{C} + \langle n_{\beta}(\textbf{r},t)\rangle_{C} \langle n_{\alpha}(\textbf{r}',t)\rangle_{R} ]\nonumber \\
    &+ \sum\limits_{\alpha,\beta}\langle\phi_{C}|\psi_{C,\alpha}^{\dagger}(\textbf{r}) \psi_{C,\beta} (\textbf{r}') |\phi_{C}\rangle [\delta (\textbf{r}-\textbf{r}')\delta_{\alpha\beta} - \langle\phi_{R}|\psi_{R,\beta}^{\dagger}(\textbf{r}') \psi_{R,\alpha} (\textbf{r}) |\phi_{R}\rangle ] \nonumber \\
    &+ \sum\limits_{\alpha,\beta}[\delta (\textbf{r}-\textbf{r}')\delta_{\alpha,\beta} - \langle\phi_{C}|\psi_{C,\alpha}^{\dagger}(\textbf{r}) \psi_{C,\beta} (\textbf{r}') |\phi_{C}\rangle]\langle\phi_{R}|\psi_{R,\beta}^{\dagger}(\textbf{r}') \psi_{R,\alpha} (\textbf{r}) |\phi_{R}\rangle .
\end{align}
The first four terms in Equation~\eqref{eq:ash2} do not give rise to any interference patterns. Indeed, it is the cross-terms between the ring and the center (last two terms) that give rise to interference. Therefore, taking these two terms and decomposing the wavefunction into the Wannier states yields 
\begin{equation}\label{eq:ash3}
    G_{R,C}(\mathbf{r},\mathbf{r}',t) = \sum\limits_{\alpha,\beta}^{N}\sum\limits_{j,l=1}^{L}I_{jl}(\textbf{r},\textbf{r}',t)\big[ N_{0,\alpha} \delta_{\alpha,\beta}(\delta_{jl} - \langle\phi_{R}|c_{l,\alpha}^{\dagger}c_{j,\alpha} |\phi_{R}\rangle ) + \delta_{\alpha,\beta}(1-N_{0,\alpha})\langle\phi_{R}|c_{l,\alpha}^{\dagger}c_{j,\alpha} |\phi_{R}\rangle  \big ], 
\end{equation}
where we defined the Wannier functions of the interference terms as 
\begin{equation}
  I_{jl}(\textbf{r},\textbf{r}',t) = w_{c}(\textbf{r}',t) w_{c}^{*}(\textbf{r},t) w^{*}(\textbf{r}'-\textbf{r}_{l}',t) w(\textbf{r}-\textbf{r}_{j},t),
\end{equation}
and $N_{0} = \langle\phi_{C}|c_{0,\beta}^{\dagger}c_{0,\beta} |\phi_{C}\rangle$ defines the expectation value of the number operator in the center, which in the current protocol is always equal to one. Consequently, the second term in Equation~\eqref{eq:interx} does not contribute to the interference pattern. Note that one of the summations over the number of components is removed due to the Kronecker delta $\delta_{\alpha\beta}$ that arises due to the colour conservation nature of the Hamiltonian describing the system. To enhance the visibility of the spirals, we neglect the Kronecker delta in the first term of Equation~\eqref{eq:ash3} such that
\begin{equation}\label{eq:ash4}
    G_{R,C} = \sum\limits_{\alpha}\sum\limits_{j,l}I_{jl}(\textbf{r},\textbf{r}',t) \langle c_{l,\alpha}^{\dagger}c_{j,\alpha}\rangle .
\end{equation}
In the non-interacting regime, the interference can be computed explicitly. First, we recast the expression of the Wannier functions in Equation~\eqref{eq:amom3} in the following way:
\begin{equation}\label{eq:ash5}
    w(\textbf{r}-\textbf{r}_{j},t) = \frac{1}{\sqrt{\pi}}\frac{\eta_{j}}{\eta_{j}^2 + \imath\omega_{0}t}\frac{\eta_{j}^2 - \imath\omega_{0}t}{\eta_{j}^2 - \imath\omega_{0}t}\exp\bigg\{-\frac{(\textbf{r}-\textbf{r}_{j})^{2}}{2(\eta_{j}^{2} + \imath\omega_{0}t)}\frac{\eta_{j}^2 - \imath\omega_{0}t}{\eta_{j}^2 - \imath\omega_{0}t} \bigg\},
\end{equation}
which can be re-written to give
\begin{equation}\label{eq:ash6}
    w (\textbf{r}-\textbf{r}_{j},t) = \frac{1}{\sqrt{\pi}}
    \frac{\eta_{j}(\eta_{j}^2 - \imath\omega_{0}t)}{\eta_{j}^4 + \omega_{0}^{2}t^{2}}\exp\bigg\{-\frac{(\textbf{r}-\textbf{r}_{j})^{2}n_{j}^{2}}{2(\eta_{j}^4 + \omega_{0}^{2}t^{2})} \bigg\}\exp\bigg\{\frac{(\textbf{r}-\textbf{r}_{j})^{2}\imath\omega_{0}t}{2(\eta_{j}^4 + \omega_{0}^{2}t^{2})} \bigg\}.
\end{equation}
By setting $\mathcal{A}(\tau) = \frac{1}{\sqrt{\pi}}\frac{\eta_{j}(\eta_{j}^{2}-\imath\tau)}{b(\tau)}$, $b(\tau) = \eta_{j}^{4} + \tau^{2} $ and $\tau = \omega_{0}t$, we have that the Wannier functions take the following form:
\begin{equation}\label{eq:ash7}
    w (\textbf{r}-\textbf{r}_{j},t) = \mathcal{A}(\tau)\exp\bigg\{-\frac{(\textbf{r}-\textbf{r}_{j})^{2}\eta_{j}^{2}}{2b(\tau)} \bigg\}\exp\bigg\{\frac{(\textbf{r}-\textbf{r}_{j})^{2}\imath \tau}{2b(\tau)} \bigg\}.
\end{equation}
Therefore, the interference between the ring and center in Equation~\eqref{eq:ash4} can be written explicitly as
\begin{align}\label{eq:ash8}
    G_{R,C}(\mathbf{r},\mathbf{r}',t)  = &\frac{1}{L}|\mathcal{A}(\tau)|^{4}\exp\bigg\{-\frac{(r^{2}+r'^{2})\eta_{0}^{2}}{2b(\tau)} \bigg\}\exp\bigg\{ \frac{\imath\tau (r^{2}-r'^{2})}{2b(\tau)}\bigg\}\times \nonumber \\
    &\sum\limits_{l,j}^{L} \exp\bigg\{ -\frac{(\mathbf{r}-\mathbf{r}_{l})^{2}\eta_{l}^{2}}{2b(\tau)}\bigg\}\exp\bigg\{ \frac{\imath\tau (\mathbf{r}-\mathbf{r}_{l}^{2})}{2b(\tau)}\bigg\} \exp\bigg\{ -\frac{(\mathbf{r}-\mathbf{r}_{j})^{2}\eta_{l}^{2}}{2b(\tau)}\bigg\}\exp\bigg\{ \frac{-\imath\tau (\mathbf{r}-\mathbf{r}_{j}^{2})}{2b(\tau)}\bigg\}\times\nonumber \\
    &\sum\limits_{\{n\}}\exp\bigg\{ -\frac{2\imath\pi}{L}n(l-j)\bigg\},
\end{align}
by making use of a result obtained in Appendix~\ref{sec:momder}, $\langle c_{l}^{\dagger}c_{j}\rangle = \frac{1}{L}\sum_{\{n\}}e^{-\frac{2\imath\pi}{L}n(l-j)}$ that holds only in the non-interacting regime. After subsequent arrangement, the interference term can be expressed as
\begin{equation}\label{eq:ash9}
    G_{R,C}(\mathbf{r},\mathbf{r}',t) = \frac{1}{L}\sum\limits_{\{n\}}I_{n}(\mathbf{r})I_{n}^{*}(\mathbf{r'}), 
\end{equation}
where we have defined 
\begin{equation}\label{eq:ash10}
    I_{n}(\mathbf{r}) = |\mathcal{A}(\tau)|^{2}\exp\bigg\{-\frac{(r^{2})\eta_{j}^{2}}{2b(\tau)} \bigg\}\exp\bigg\{-\frac{\imath\tau(r^{2})}{2b(\tau)} \bigg\}\sum\limits_{j=1}^{L}\exp\bigg\{-\frac{(\mathbf{r}-\mathbf{r}_{j})^{2}\eta_{j}^{2}}{2b(\tau)}\bigg\}\exp\bigg\{\frac{\imath\tau(\mathbf{r}-\mathbf{r}_{j})^{2}}{2b(\tau)} -\frac{2\imath\pi}{L}nj \bigg\}.
\end{equation}
The term $I_{n}(\mathbf{r})$ corresponds to a spiral in the $x$-$y$ plane, with the number of spirals it exhibits depending on the quantum number of the level it occupies $n$.  Note that in the derivation sketched above, we choose the number of components $\alpha=1$. For $N$-component fermions, Equation~\eqref{eq:ash9} acquires a factor of $N$.


\chapter{Derivation for the exact one-body correlation function in the limit of infinite repulsion}\label{eq:andreasdermom}

\noindent In the Chapter~\ref{chp:ogata}, we outlined how the spin and charge degrees of freedom decouple yielding a simplified form of the Bethe ansatz wavefunction, that at infinite repulsion reads
\begin{equation}\label{eq:adeva1}
   \Psi(x_{1},\hdots,x_{N_{p}};\alpha_{1},\hdots,\alpha_{N_{p}}) = \mathrm{sign}(Q)\mathrm{det}[\exp(i k_{j}x_l)]_{jl} \Phi (y_{1},\hdots,y_{M}).
\end{equation}
Here, we are going to show how to evaluate the Slater determinant of the charge degrees of freedom and the corresponding spin wavefunction in the presence of an effective magnetic flux.

\section{Slater determinant}\label{sec:coorr}

\noindent To calculate the Slater determinant of spinless fermions, we need to start by noting that quasimomenta can be expressed as
\begin{equation}\label{eq:slatts}
    k_{j} = -(N_{p}-1 + \ell)\frac{\pi}{L} + (j-1)\Delta k + k_{0} + \frac{X}{N_{p}}, \hspace{5mm} j = 1,\hdots,N_{p}
\end{equation}
where $\Delta k = \frac{2\pi}{L}$, $X$ denotes the sum over the spin quantum numbers and $\ell$ is the angular momentum. $k_{0}$ is a constant shift can be $0$ or $-\frac{\pi}{L}$ for systems with $(2n)N$ and $(2n+1)N$ fermions respectively, that will henceforth be termed as paramagnetic and diamagnetic\footnote{The expression for the quasimomenta is obtained from the decoupling of the Bethe ansatz equations outlined in Chapter~\ref{chp:repcurr}.}. Through Equation~\eqref{eq:slatts}, we can re-write the Slater determinant in the following form
\begin{equation}\label{eq:slatty}
    \mathrm{det}[\exp(i k_{j}x_l)]_{jl} = \exp(i k_{1}r_{cm}N_{p})\mathrm{det}\begin{pmatrix} 1&y_{1}&y_{1}^{2}&\cdots &y_{1}^{N_p-1} \\
    1&y_{2}&y_{2}^{2}&\cdots&y_{2}^{N_p-1} \\ 1&y_{3}&y_{3}^{2}&\cdots&y_{3}^{N_p-1} \\ \vdots&\vdots&\vdots&\ddots&\vdots\\ 1&y_{N_p}&y_{N_p}^{2}&\cdots&y_{N_p}^{N_p-1}\end{pmatrix},
\end{equation}
with $r_{cm}$ denoting $\sum_i x_i/N_p$ which we refer to as the 
{\em center of mass}. The matrix elements of the determinant are of the form $y_{m}^{j-l} = \exp(i (k_{j}-k_{l})r_{m})$, whereby we made use of the fact that all the quasimomenta are equidistant. By noting that the matrix in Equation~\eqref{eq:slatty} has the same structure of the Vandermonde matrix~\cite{OgataShiba90}, we can express the Slater determinant as
\begin{equation}\label{eq:slat}
    \mathrm{det}[\exp(i k_{j}x_{Q{j}})] = \exp(i k_{1}r_{cm}N_{p})\prod\limits_{1\leq i<j\leq n}(\exp(i \Delta kr_{j})-\exp(i \Delta kr_{i})),
\end{equation}
which upon simplification reads
\begin{equation}\label{eq:sla}
    \mathrm{det}[\exp(i k_{j}x_{Q{j}})] = \exp(i k_{1}r_{cm}N_{p})\prod\limits_{1\leq i<j\leq n}\exp\bigg(i \Delta k\frac{r_{j} +r_{i}}{2}\bigg)\prod\limits_{1\leq i<j\leq n}\bigg(2i\sin\frac{\Delta k (r_{j}-r_{i})}{2} \bigg).
\end{equation}
This expression can be further simplified by noticing that 
\begin{equation}\label{eq:las}
    \prod\limits_{1\leq i<j\leq n}\exp\bigg(i \Delta k\frac{r_{j} +r_{i}}{2}\bigg) =\exp\bigg(\frac{i \Delta k}{2}[r_{cm}N_{p}^{2}   - r_{cm}N_{p}]\bigg),
\end{equation}
that in conjunction with Equation~\eqref{eq:slatts} reduces Equation~\eqref{eq:slat} into
\begin{equation}\label{eq:fsl}
    \mathrm{det}[\exp(i k_{j}x_{Q{j}})] = \exp\bigg(i \bigg[k_{0} +\frac{X}{N_{p}}-\ell\Delta k\bigg] r_{cm}N_{p}\bigg)\prod\limits_{1\leq i<j\leq n}\bigg(2i\sin\frac{\Delta k (r_{j}-r_{i})}{2} \bigg).
\end{equation}
\noindent In the presence of an effective magnetic flux, the variables $X$ and $\ell$ need to be changed in order to counteract the increase in flux. For the spin quantum numbers, the shift needs to satisfy the degeneracy point equation introduced in Chapter~\ref{chp:repcurr}
\begin{equation}\label{eq:numbers}
    \frac{2w-1}{2N_{p}} \leq \phi + D\leq\frac{2w+1}{2N_{p}} \hspace{5mm}\mathrm{where}\hspace{2mm} X = -w,
\end{equation}
with $\phi$ ranging from 0.0 to 1.0 and $D$ being 0 \big[$-\frac{1}{2}$\big] for diamagnetic [paramagnetic] systems. Upon increasing $\phi$, the angular momentum of the system increases at $\phi = \big(s \pm \frac{1}{2N_{p}} + \delta\big)$ with $s$ being (half-odd) integer in the case of (diamagnetic) paramagnetic systems, with $\delta =\mp \frac{1}{2N_{p}}$ for an odd number of particles.

\section{Resolving degeneracies of the spin wavefunction}\label{app:spinwave}

\noindent  As $U\rightarrow+\infty$, all the spin configurations of the model are degenerate. The reason is that the energy contribution from the spin part of the wavefunction $E_{\mathrm{spin}}$ is of the order of $\frac{t}{U}$. However, there is no spin degeneracy observed in the Hubbard model: the ground-state is non-degenerate for SU($2$), except for special points in flux with an eigenstate crossing. Hence, a single  state has to be chosen properly to match with the Hubbard model. Due to the symmetry of both models, we choose the square of the total spin, $\vec{S}_{\rm tot}^{2}$, or quadratic Casimir operator $C_1$ to label the eigenstates. 
The selected eigenstates of both models need to have the same value for this operator. This benchmarking with the Hubbard model was only utilized for small system sizes in order to understand what 
representations of the Heisenberg model we have to choose. \\

\noindent The resulting composition from spinless fermionic and Heisenberg Hamiltonians results in a translationally invariant model only in cases where these states match. To this end, we use this as a control mechanism. As already explained in Chapter~\ref{chp:ogata}, the spin wavefunction $\Phi (y_{1},\hdots,y_{M})$ is obtained by performing exact diagonalization Lanczos methods of the one-dimensional anti-ferromagnetic Heisenberg model. \\

\noindent {\em a) Zero flux--} The ground-state with an odd/even number of particles per species for the Hubbard model corresponds to different values of the Casimir operator, and, therefore to other representations of the SU($N$) algebra. For an odd number of species, the Casimir operator corresponds to a singlet state for all SU($N$).
The ground-state of the anti-ferromagnetic Heisenberg model instead is always a singlet and non-degenerate for all SU($N$). Hence, this state will be chosen for an odd number of particles per species $N_{p}/N$. However, a different state needs to be chosen for an even number of particles per species. In the case of $N=2$, the required state is the next non-degenerate excited triplet-state (of total spin $J=1$, $\vec{J}^2=J(J+1)$). This corresponds to an $n=2J$-representation (see Appendix~\ref{Casimir}). \\

\noindent For SU($N\! >\! 2$), i.e., $N=3$ and $N=4$, the first non-degenerate state corresponds to a Casimir eigenvalue $C_1=6$. 
Examples are the $10$-dimensional representations $(n_1,n_2)=(3,0)$ for SU(3) and correspondingly
$(n_1,n_2,n_3)=(4,0,0)$ for SU(4). The numbers $n_i$ in the SU(3) representations correspond to the numbers $p$ and $q$ frequently used in SU(3) representations in the mathematical literature and high energy physics; there, they represent the number of (anti-)quarks. The dimension of a representation $(n_1,n_2)$ of SU(3) is $d(n_1,n_2)=(n_1+1)(n_2+1)(n_1+n_2+2)/2$. 
Both representations for SU(3) and SU(4) 
have a Casimir value of $C_1=6$.
We assume that this representation will be $(N,0,\dots,0)$ for SU($N$). This state takes the role of the non-degenerate triplet state of SU(2)
in the zero field ground-state for an even species number occupation.
\\

\noindent {\em b) Non-zero flux--}
For strong repulsive interactions, 
a fractionalization of the persistent current in the model is observed, as discussed in Chapter~\ref{chp:repcurr}. The fractionalization appears since formerly higher excited states are bent by the field to be the ground-state. A unique method to identify these states would be to utilize the SU($N$) Heisenberg Bethe equations, which have the same spin quantum number configurations as their Hubbard counterparts. In this manner, we are guaranteed that the corresponding eigenstates obtained from the Heisenberg model correspond to the ground-state of the Hubbard model. However, this method is rather tedious to achieve the whole state.  This is particularly true because the Bethe ansatz gives direct solutions only for the highest weight states, and we work at an equal occupation of each species: the resulting state is then obtained by applying sufficiently often the proper lowering operators of SU($N$). \\

\noindent In the case of paramagnetic systems, i.e.,  an even number of particles per species, the central fractionalized parabola (centered around $\phi = 0.5$), corresponds to a singlet state. This parabola results to be non-degenerate for the Heisenberg and is therefore easily distinguished. As such, 
one obtains the corresponding states for the outer and central fractionalized parabolas in a straightforward manner for arbitrary SU($N$). For finite field and degenerate ground-states of the Hubbard model, we do not have a general procedure to choose the states for SU($N\! >\! 2$). Therefore to describe this method, we start by considering SU(2) and then apply it to a system with $N_{p}=3$ SU(3) symmetric fermions. \\

\noindent For SU(2), the remaining fractionalized parabolas (i.e., excluding the two outer parabolas and the central one)
have a common spin value of $J=1$. This, in turn, results in a two-fold degeneracy in the spin-$\frac{1}{2}$ Heisenberg model for a given sum of the spin quantum numbers $|X|$. Hence, the relevant states for two of the parabolas of a given $|X|$ are superpositions of these degenerate eigenstates of the Heisenberg Hamiltonian. These states can be separated by different eigenvalues for $P_{j,j-1}$, which is
part of the Heisenberg model but not commute with it. We choose to call both eigenstates of this permutation operator  $\ket{\psi_{1/2}}$.
The states $\ket{\psi_{l/r}}$ corresponding to the inner branches of the fractionalization are obtained from the two spin- and energy-degenerate states $\ket{\psi_{1/2}}$ as  
\begin{equation}\label{eq:leftright}
    \ket{\psi_{l/r}}:=\frac{1}{\sqrt{2}}\left(\ket{\psi_1} \pm i \ket{\psi_2} \right)\, .
\end{equation}

\noindent It is worth mentioning that these states correspond to fractionalized parabolas that are marked with a change from a singlet state in the absence of flux to a triplet state (on account of the fractionalization) with each of the basis elements being non-zero. This happens here gradually via intermediate triplet states excluding certain basis states. As an example, we take an SU(2) state with $N_{p}=6$ particles to explain this better, that goes from one parabola at zero interactions to six parabolic-wise segments upon fractionalization. Since this state has $3$ particles per species ($\uparrow$ or $\downarrow$), the parabolas at zero flux start from a singlet and arrive to a triplet state in the center parabola.
The singlet state is made of three distinct configurations: a) $\ket{111000}\pm \mbox{ cyclic permutations}$, b) $\ket{101010}-\ket{010101}$, and c) the possible remaining configurations with alternating sign (singlet state). This is mediated via fractionalized states where component a) is missing in the first inner parabola and additionally component b) vanishes for the second parabola. The triplet has the same components as the singlet state but without alternating signs.\\

\noindent For SU(2) the corresponding states that belong to the fractionalized parabolas are the triplet representation, as well. However, this changes for $N\! >\! 2$. The representations are modified for the intermediate parabolas. In the case of SU(3), we obtain $(n_1,n_2)=(1,1)$ as the $8$-dimensional representation (instead of $(n_1,n_2)=(3,0)$), which governs the intermediate parabola. For SU(4), it is $(1,2,0)$ instead of $(4,0,0)$ (see Appendix~\ref{Casimir}).
The Casimir $C_1$ has values $3$ and $4$, respectively.
These representations take the role of the (degenerate) triplet state of SU(2). \\

\noindent In the case of non-vanishing flux threading a ring of SU$(N\! >\!2)$ symmetric fermions, the ground-states of the Hubbard Hamiltonian~\eqref{eq:fluxHub} belonging to a given $|X|$, is $N \! -\! 1$-fold degenerate coming from the $N\! -\! 1$ sets of spin quantum numbers. This degeneracy holds for the inner fractionalized parabolas. As a consequence of its one-to-one correspondence with the Hubbard model, these degeneracies are manifested in the Heisenberg model, in addition to the two-fold degeneracy mentioned previously for both parabolas with equal values for $|X|$. In order for this extra degeneracy to be resolved, we make certain coefficients of the wavefunction in the Heisenberg basis vanish by corresponding superpositions of the degenerate states. This has been motivated by former observations in SU(2) (see discussion above). \\

\noindent To get a better idea of the method, here we exemplify the case of 3 particles in SU(3).
There are only two possible values for $|X|$ in this case, and each parabola is two-fold degenerate in the Hubbard model.
The degeneracy of the Heisenberg model is hence 4-fold.
So, the distinct states have to be selected from a remaining 
two-fold degeneracy. 
The zeroth parabola is in the singlet state of SU(3) that belongs to $C_1=0$, for which every component of the wavefunction is non-zero.
Both two-fold degenerate inner parabolas have $C_1=3$
and correspond in one case to the positive or negative permutation of the species number only;
in the second degenerate case, they correspond to configurations $\{\ket{021},\ket{102}\}$ and $\{\ket{120},\ket{201}\}$ as the only non-zero component.
These are the states $\{\ket{\psi_1},\ket{\psi_2}\}$ that are to be superposed by formula \eqref{eq:leftright}. The direct way to obtain the corresponding state of the Heisenberg model is via the Bethe ansatz wavefunction for the same spin quantum numbers of the Hubbard model.
The degeneracies amount to $2(N\! -\! 1)$-fold for the SU($N$) Heisenberg model. These are distinguished by the eigenstates of the permutation operator $P_{j,j+1}$ up to a remaining $(N\! -\! 1)$-fold degeneracy.

\end{appendices}

\backmatter

\end{document}